\documentclass[11pt,reprint,showpacs,showkeys,amsfonts,amssymb,amsmath,nofootinbib]{article}
\usepackage{bm,amsmath,epsf,epsfig,latexsym,amssymb,cite,lmodern}
\usepackage{graphics,psfrag}
\usepackage{hyperref}
\newcommand{\bg}{\begin{align}}
\newcommand{\eeg}{\end{align}}
\newcommand{\be}{\begin{equation}}
\newcommand{\ee}{\end{equation}}
\newcommand{\ba}{\begin{eqnarray}}
\newcommand{\ea}{\end{eqnarray}}

\newcommand{\nn}{\nonumber}

\begin{document}

\thispagestyle{empty}

\vspace{2cm}
\begin{center}
{\Large{\bf Nucleon-Nucleon scattering from the dispersive $N/D$ method: next-to-leading order study}}
\end{center}
\vspace{.5cm}

\begin{center}
{\Large  Zhi-Hui~Guo$^{1,2}$, J.~A.~Oller$^2$ and G.~R\'{\i}os$^{2}$}
\end{center}

\begin{center}
$^1${\it Department of Physics, Hebei Normal University, 050024 Shijiazhuang, P.R. China.}\\
$^2${\it  Departamento de F\'{\i}sica. Universidad de Murcia, E-30071 Murcia. Spain.}
\end{center}
\vspace{1cm}

\begin{abstract}
We consider nucleon-nucleon ($NN$) interactions from Chiral Effective Field Theory applying the $N/D$ method. The dynamical input 
is given by the discontinuity of the $NN$  partial-wave amplitudes across the left-hand cut (LHC) calculated in Chiral Perturbation Theory (ChPT)
 by  including one-pion exchange (OPE), once-iterated OPE and leading irreducible two-pion exchange (TPE).
 We discuss both uncoupled and coupled partial-waves.  
We show algebraically that the resulting integral equation has a unique solution when the input is taken only from OPE because 
it is of the Fredholm  type with a squared integrable kernel and an inhomogeneous term.
  Phase shifts and mixing angles are typically 
rather well reproduced, and a clear improvement  of the 
results obtained previously with only OPE is manifest. We also show that the
 contributions to the  discontinuity across the LHC
are amenable to a chiral expansion.  Our method also establishes correlations between the $S$-wave effective ranges and scattering lengths 
based on unitarity, analyticity and chiral symmetry.
\end{abstract}


\newpage
\section{Introduction}\label{sec:intro}

The application of ChPT, the low-energy effective field theory of QCD, to the problem of nuclear forces was elaborated in Ref.~\cite{weinn} and first put in practice 
in Ref.~\cite{ordo94}. Its application to $NN$ scattering has reached nowadays  a sophisticated and phenomenologically successful status \cite{ordo94,entem,epen3lo,phillipssw,pavon06,frederico:1999,ollernpa}. See Refs.~\cite{revvankolck,revbeane,revbedaque,revepe1,revepe2,revmachleidt} for related reviews. In particular,  Refs.~\cite{entem,epen3lo}  take the next-to-next-to-next-to-leading order (N$^3$LO)  potential and reproduce $NN$ phase shift data up 
to $E_{lab}\sim 200$~MeV accurately, with $E_{lab}$ the laboratory-frame kinetic energy.

However, the use of the $NN$ potential calculated in ChPT up to some order in a Lippmann-Schwinger equation, as originally 
proposed in Ref.~\cite{weinn}, is known to yield regulator dependent results. That is, the chiral counterterms present in the potential are not able to reabsorb all the ultraviolet divergences that result in the solution of the Lippmann-Schwinger equation  \cite{kswa,nogga,pavon06,entem08,birse,eiras,phillipssw,pavon11,longyang,gegelia}.
  Stable results with the $NN$ potential determined from OPE are obtained in Refs.~\cite{nogga,phillipssw} for 
$\Lambda\! <4$~GeV, where $\Lambda$ is a three-momentum cut-off.
 This is achieved by promoting  counterterms from higher to lower orders in the partial waves with attractive $1/r^3$ tensor force generated by  OPE \cite{nogga}. 
The extension of these ideas to higher orders in the chiral potential is undertaken 
in Refs.~\cite{pavon11,longyang} by treating perturbatively subleading contributions to the $NN$ potential beyond OPE. 
 When the limit $\Lambda\to\infty$ is taken in the Lippmann-Schwinger equation 
it results that  only one counterterm is operative for 
attractive singular potentials and none for the repulsive singular ones \cite{entem08,entem12,pavon,phillipssw}. 
This  scheme is too rigid from the point of view of effective field theory which implies deficiencies in the 
description of some $NN$ partial waves compared with data 
as well as the loss of  order-by-order improvement in the predictions in those cases. This has been 
recently analyzed in detail in Ref.~\cite{entem12} up to N$^3$LO. 
On the other hand, 
it has been shown in Ref.~\cite{gegelia2012} that OPE is renormalizable in manifestly Lorentz 
covariant baryon ChPT, while this is not the case when the heavy-baryon expansion is used as shown in Ref.~\cite{eiras}.

Regulator dependence can also be avoided by employing dispersion relations (DRs), that involve only convergent integrals once enough  subtractions are taken. 
This technique was recently applied in Refs.~\cite{paper1,paper2} employing the $N/D$ method \cite{chew} and OPE.
 Refs.~\cite{paper1,paper2} argued that this method could be applied to higher orders in the chiral expansion by calculating perturbatively in ChPT  the discontinuity of a partial wave amplitude, 
which is $2 i$ times its imaginary part,  along the LHC.
 Within ChPT, this discontinuity stems from  multi-pion exchanges and it constitutes, together with the subtraction constants, the input required to solve the $N/D$ method. 
We want to investigate  explicitly 
the chiral expansion of the discontinuity along the LHC  and extend the calculations in Refs.~\cite{paper1,paper2} by including the leading irreducible and reducible TPE,  as calculated by Ref.~\cite{peripheral} in ChPT. 
 One of the main aims of the work is to show quantitatively that the referred chiral expansion of the discontinuity of 
a $NN$ partial wave amplitude   is meaningful. 
 The leading contribution to this imaginary part is OPE, ${\cal O}(p^0)$, and both of the subleading ones, once-iterated OPE and irreducible TPE, 
have typically similar sizes, as we show below, and could be booked in the chiral counting as the latter,
 which is explicitly ${\cal O}(p^2)$. 
It is also shown that further contributions to the discontinuity along the LHC by increasing 
the numbers of pion ladders in $NN$ reducible diagrams are more suppressed because they contribute only deeper  in the complex plane 
and move further away from the low-energy physical region. 

Another novelty in the present work compared with Refs.~\cite{paper1,paper2} is the way that $NN$ partial 
waves with orbital angular momentum ($\ell$) $\ell\geq 2$  and the mixing partial waves with
 total angular momentum $(J)$ $J\geq 2$ are treated in order to fulfil the right threshold behavior, 
which requires that they vanish as $A^\ell$ and $A^J$, respectively, for $A\to 0$. 
Here and in the following we denote the center-of-mass-frame  (c.m.) three-momentum squared by $A$. 
This is done by taking at least $\ell$ or $J$ subtractions 
in the appropriate DRs. We see below that at most only one of the resulting 
subtraction constants is necessary to be fitted to data, while the others are fixed to their perturbative values. 
This comprises the so called principle of maximal smoothness, which simplifies considerably 
the description of higher $NN$ partial waves.

The $N/D$ method was used in Refs.~\cite{wong1,scotti63,scotti,oteo:1989} to study $NN$ scattering.  Ref.~\cite{wong1} was 
restricted to  $NN$ $S$-waves and took only OPE as input along the LHC. Refs.~\cite{scotti63,scotti} also included other heavier mesons as source for the discontinuity, in line with the meson theory of nuclear forces. 
 Ref.~\cite{oteo:1989} modeled the LHC discontinuity by OPE and one or two ad-hoc poles. No attempt was made in these works to offer a systematic procedure to improve the calculation of the discontinuity along the LHC. The main novelty that 
modern chiral effective field theory of nuclear forces can offer to us in connection with the $N/D$ method consists precisely 
in calculating systematically such input discontinuity. This is the point that we want to elaborate further in the present research.
  Importantly, we also show that we achieve a reproduction of $NN$ phase shifts and mixing angles  in good agreement with the Nijmegen partial-wave 
analysis (PWA) \cite{Stoks:1994wp}, that offers a clear improvement compared with that obtained in Refs.~\cite{paper1,paper2} with only OPE.
 We also mention Ref.~\cite{lutz} where the $N/D$ method is  used in connection with ChPT and $NN$ scattering. It is important 
 to stress that  we 
do not  perform any truncation of the LHC and we keep its full extent in all the dispersive integrals considered in the  c.m. three-momentum squared complex plane, while this is not the case in Ref.~\cite{lutz}. In the latter reference the dispersive integrals 
along the LHC are cut at the c.m. three-momentum squared value $-9 M_\pi^2/4$. Because of this truncation  Ref.~\cite{lutz} 
does not resolve soft pion-exchange contributions involving a center-of-mass (c.m.) three-momentum squared smaller 
than $-9 M_\pi^2/4$ (whose square root in modulus is just 1.5~$M_\pi$) from short-range physics. This is avoided 
by construction in our framework where we keep the full extent of the integrals along the pertinent cuts, as 
required by analyticity.

The contents of the paper are organized as follows. After this introduction we explain the formalism and deduce the proper integral equations (IEs) for the uncoupled waves in Sec.~\ref{unformalism}. The expansion of the discontinuity along the LHC in powers of three-momentum and pion masses (the so-called chiral expansion) and in the number of pions 
exchanged is discussed in Sec.~\ref{delta}. 
We also show in this section that the IEs when this discontinuity is given in terms of OPE have a unique solution,
 and discuss some necessary conditions for having a solution when considering higher order correction to $\Delta(A)$.
 The method is applied to the $^1S_0$ and the uncoupled $P$-waves in Secs.~\ref{1s0} and \ref{pw}, respectively. 
The constraints that result from requiring the proper threshold behavior for partial waves 
with $\ell \geq 2$  are the contents of Sec.~\ref{leq2}. 
Then, the numerical results for the uncoupled $D$, $F$, $G$ and $H$ waves are considered in Secs.~\ref{dw}--\ref{hw}. 
 We quantify the different contributions to the discontinuity of  a partial wave across the LHC in Sec.~\ref{cont_da}, where it is shown quantitatively the dominance of OPE and the subleading role of TPE. Section~\ref{sec:formalism} provides the extension of the formalism to the coupled-partial-wave case. This is then applied to the systems $^3S_1-{^3D_1}$, Sec.~\ref{sec:deuteron}, $^3P_2-{^3F_2}$, Sec.~\ref{3pf2},  $^3D_3-{^3G_3}$,  Sec.~\ref{3dg3}, $^3F_4-{^3H_4}$, Sec.~\ref{3fh4} and to $^3G_5-{^3I_5}$ in Sec.~\ref{3gi5}. Conclusions and outlook are then provided in Sec.~\ref{sec:conclusions}.

\section{The $N/D$ method: uncoupled waves}
\label{unformalism}

A $NN$ partial wave amplitude in the three-momentum squared plane (that we call the $A$ plane) has two disjoint cuts. 
The right-hand cut (RHC) is due to the intermediate states in $NN$ scattering and then it extends from threshold ($A=0$) up to $A=\infty$. 
It comprises  the elastic cut with two-nucleon intermediate states, as well as the inelastic cuts, whose lighter thresholds 
are due to $n$-pion production giving contribution for $A^2\geq n^2 M_\pi^2/4+ n m M_\pi$, with $M_\pi$ the pion mass and $m$ the 
nucleon mass. There is also the LHC which lower energy contributions are due to the exchange of $n$ pions for $A\leq -M_\pi^2 n^2/4$, so that OPE  extends for $A\leq -M_\pi^2/4$, TPE for $A\leq -M_\pi^2$, and so on. 

We first start by considering the uncoupled $NN$ partial waves. Below, in Sec.~\ref{sec:formalism},  we present the generalization to coupled waves.

The two cuts present in a given $NN$ partial wave, $T_{J\ell S}(A)$, with $S$ the total spin, $\ell$ the orbital angular momentum and $J$ the total angular momentum, can be separated by writing it as the quotient of a numerator, $N_{J\ell S}(A)$,
 and a denominator, $D_{J\ell S}(A)$, function
\begin{align}
T_{J\ell S}(A)&=\frac{N_{J\ell S}(A)}{D_{J\ell S}(A)}~,
\label{tdef}
\end{align}
such that $N_{J\ell S}(A)$ has only LHC while $D_{J\ell S}(A)$ has only RHC.
 This is the essential point of the $N/D$ method, first introduced in Ref.~\cite{chew} to study $\pi\pi$ scattering.

In the rest of this section we skip the subscripts $J\ell S$ since we always refer to a definite $NN$ partial wave. In addition, since all the functions involved in Eq.~\eqref{tdef} are real at least in a finite interval along the real axis, they fulfill the 
Schwartz reflection principle
\begin{align}
f(z^*)=f(z)^*~.
\end{align}
As a result their discontinuity across a cut along the real axis is given entirely by the knowledge of the imaginary 
part of the function, because $f(z+i 0^+)-f(z-i 0^-)=2i \mathrm{Im} f(z+i 0^+)$, with $z\in \mathbb{R}$. 
 
Elastic unitarity in our normalization requires
\begin{align}
\mathrm{Im} T(A)=\frac{m\sqrt{A}}{4\pi}|T|^2~~,~~A \geq 0~.
\label{unit}
\end{align}
In the following we designate by
\begin{align}
\rho(A)=\frac{m \sqrt{A}}{4\pi}~~,~~A\geq 0~,
\label{rhodef}
\end{align}
 the phase-space factor in Eq.~\eqref{unit}. This equation has a simpler expression when 
given as the imaginary part of the inverse of the partial wave along the RHC,\footnote{Inelastic channels due to (multi-)pion production are not included in our low-energy analysis.}
\begin{align}
\mathrm{Im} \frac{1}{T}&=-\rho(A)~~,~~A\geq 0~.
\label{unitinv}
\end{align}
Equation~\eqref{unitinv}, together with Eq.~\eqref{tdef}, translates to the following 
equation for $\mathrm{Im} D(A)$ along the RHC,
\begin{align}
\mathrm{Im} D(A)=-\rho(A) N(A)~,~A>0~.
\label{disd}
\end{align} 

On the other hand, the discontinuity of a $NN$ partial wave along the LHC is denoted  by
\begin{align}
T(A+i 0^+)-T(A-i 0^+)=2 i \Delta(A)~~,~~\Delta(A)\equiv \mathrm{Im} T(A+i0^+)~~,~~
A\leq -\frac{M_\pi^2}{4}~~.
\label{dislhc}
\end{align}
From  Eqs.~\eqref{tdef} and \eqref{dislhc} this in turn implies the following result for $\mathrm{Im} N(A)$ along the LHC
\begin{align}
\mathrm{Im} N(A)= \Delta(A) D(A)~~,~~A\leq -\frac{M_\pi^2}{4}~.
\label{disn}
\end{align}

Next, we want to make use of Eqs.~\eqref{disd} and \eqref{disn} to write down the dispersive integrals for $D(A)$ and $N(A)$, respectively. For that we need to take into account the high-energy behavior of these functions. 
The relation  in our normalization between the $S$- and $T$-matrix in partial waves, as follows from Eq.~\eqref{unit}, is
\begin{align}
S(A)=1+2i\rho(A)T(A)~,
\end{align}
with $S(A)$ the $S$-matrix element. Inverting the previous equation it follows that $T(A)={\cal O}(A^{-\frac{1}{2}})$ 
at high-energies, $A\in {\mathbb R}$ and $A\to \infty$, because $S(A)={\cal O}(1)$ along the RHC. Let us assume that $D(A)={\cal O}(A^{n_0})$ for $A\to \infty$ then, because
\begin{align}
N(A)=T(A) D(A)~,
\end{align}
it follows that for real $A$ and $A\to +\infty$, $N(A)={\cal O}(A^{n_0-\frac{1}{2}})$. Since $N(A)$ has only LHC this limit is also 
valid for any other direction in the $A$ plane for $A\to \infty$, according to the the Sugawara and Kanazawa theorem \cite{barton,suga}. As a result of the high-energy behavior of $N(A)$ and $D(A)$, if we divide simultaneously both functions by $(A-C)^n$,~$n>n_0$, we can write down unsubtracted DRs for the new functions $\widehat{D}(A)$ and $\widehat{N}(A)$ defined as
\begin{align}
\widehat{D}(A)&=\frac{D(A)}{(A-C)^n}~,\nn\\
\widehat{N}(A)&=\frac{N(A)}{(A-C)^n}~.
\label{primef}
\end{align}
 To avoid unnecessary complications in the technical derivations we take $-M_\pi^2/4<C<0$, and 
 the following DRs, on account of Eqs.~\eqref{disd} and \eqref{disn}, result
\begin{align}
\widehat{D}(A)&=\sum_{i=1}^n \frac{\widetilde{\delta}_i}{(A-C)^i}-\frac{1}{\pi}\int_0^\infty dq^2\frac{\rho(q^2)\widehat{N}(q^2)}{q^2-A}~,\nn\\
\widehat{N}(A)&=\sum_{i=1}^n \frac{\widetilde{\nu}_i}{(A-C)^i}+\frac{1}{\pi}\int_{-\infty}^L dk^2 \frac{\Delta(k^2)\widehat{D}(k^2)}{k^2-A}~,
\label{undr}
\end{align} 
with
\begin{align}
L&=-\frac{M_\pi^2}{4}~.
\label{defL}
\end{align}

Coming back to our original functions $D(A)$ and $N(A)$ by multiplying both sides of Eq.~\eqref{undr} by $(A-C)^n$, it results 
\begin{align}
D(A)&=\sum_{i=1}^n \widetilde{\delta}_i (A-C)^{n-i}-\frac{(A-C)^n}{\pi}\int_0^\infty dq^2\frac{\rho(q^2)N(q^2)}{(q^2-A)(q^2-C)^n}~,\nn\\
N(A)&=\sum_{i=1}^n \widetilde{\nu}_i (A-C)^{n-i}+\frac{(A-C)^n}{\pi}\int_{-\infty}^L dk^2\frac{\Delta(k^2)D(k^2)}{(k^2-A)(k^2-C)^n}~.
\label{standardr0}
\end{align}
It is convenient to relabel the coefficients in the polynomial term of the previous equation and define 
$\delta_i\equiv \widetilde{\delta}_{n-i+1}$ and $\nu_i\equiv \widetilde{\nu}_{n-i+1}$ so that  Eq.~\eqref{standardr0} is rewritten as
\begin{align}
D(A)&=\sum_{i=1}^n \delta_i (A-C)^{i-1}-\frac{(A-C)^n}{\pi}\int_0^\infty dq^2\frac{\rho(q^2)N(q^2)}{(q^2-A)(q^2-C)^n}~,\nn\\
N(A)&=\sum_{i=1}^n \nu_i (A-C)^{i-1}+\frac{(A-C)^n}{\pi}\int_{-\infty}^L dk^2\frac{\Delta(k^2)D(k^2)}{(k^2-A)(k^2-C)^n}~.
\label{standardr}
\end{align}
and we recover standard $n$-time subtracted DRs. In the previous equation one has to take the limit $A+i0^+$ for real values of $A$ along the integration intervals. Since it is possible to divide simultaneously $N(A)$ and $D(A)$ by a constant, because only its ratio is relevant for obtaining $T(A)$, we normalize the function  $D(A)$ in the following as 
\begin{align}
D(0)=1~.
\label{normd}
\end{align}
In this way, one of the subtraction constants $\delta_i$ in Eq.~\eqref{standardr} is superfluous.

In summary, as a result of the discussion in this section, we can state the following conclusion: If there exists an  $N/D$ representation of the on-shell $NN$ partial wave, Eq.~\eqref{tdef}, then the  functions $D(A)$ and $N(A)$ must satisfy  $n$-time subtracted DRs, Eq.~\eqref{standardr}, for $n$ large enough.

To solve Eq.~\eqref{standardr} it is useful to insert the expression for $N(A)$ into that of $D(A)$ and then we end with the following IE for $D(A)$,
\begin{align}
D(A)&=\sum_{i=1}^n \delta_i (A-C)^{n-i}-\sum_{i=1}^n \nu_i\frac{(A-C)^n}{\pi}\int_0^\infty dq^2\frac{\rho(q^2)}{(q^2-A)(q^2-C)^{n-i+1}}\nn\\
&+\frac{(A-C)^n}{\pi^2}\int_{-\infty}^L dk^2\frac{\Delta(k^2)D(k^2)}{(k^2-C)^n}\int_0^\infty dq^2\frac{\rho(q^2)}{(q^2-A)(q^2-k^2)}~.
\label{inteq1}
\end{align}
Notice that on the right-hand side (r.h.s.) of the previous equation $D(A)$ is only needed along the LHC. We solve numerically this IE by discretization and determine $D(A)$ for $A\leq -M_\pi^2/4$. Once this is known, we can then calculate $D(A)$ and $N(A)$ for any other values of $A$ making use of the DRs in Eq.~\eqref{inteq1} and the one in the last line of Eq.~\eqref{standardr}, respectively.\footnote{For a large enough number of subtractions, typically three or more, it is more advantageous numerically to solve the IEs in the form corresponding to Eq.~\eqref{undr}. In this way, one avoids having too large numbers for large values of $A$ that could cause problems to the numerical subroutines for inverting matrices.}

  The integrals along the RHC in Eq.~\eqref{inteq1} can be done algebraically. We define the function $g(A,k^2)$ as 
\begin{align}
g(A,k^2)&\equiv \frac{1}{\pi}\int_0^\infty dq^2\frac{\rho(q^2)}{(q^2-A)(q^2-k^2)}=\frac{i m / 4\pi}{\sqrt{A+i0^+}+\sqrt{k^2+i0^+}}~.
\label{gdef}
\end{align}
Here, the $+i0^+$ is relevant for calculating this function when needed in the dispersive integrals above.
In terms of $g(A,k^2)$ one also has
\begin{align}
\frac{\partial^{i-1}g(A,C)}{\partial C^{i-1}}=\frac{(i-1)!}{\pi}\int_0^\infty dq^2\frac{\rho(q^2)}{(q^2-A)(q^2-C)^i}~.
\label{derg}
\end{align}

It is also clear that once the $D(A)$ and $N(A)$ are expressed in the form of standard DRs, Eq.~\eqref{standardr}, 
 it is not really necessary to take the subtraction point $C$ with the same value for both functions. In practice we take $C=0$ for the function  $N(A)$. For the function  $D(A)$ we always take  one subtraction at $C=0$, because then it is straightforward to impose the normalization condition Eq.~\eqref{normd}. Let us stress that DRs are independent of the value taken for the subtraction point since a change in $C$ would be reabsorbed in a change of the values of the subtraction constants  \cite{spearman}, $ \delta_i$ for $D(A)$ and $\nu_i$ for $N(A)$.

\section{The input $\Delta(A)$ function}
\label{delta}

\begin{figure}
\begin{center}
\includegraphics[angle=0, width=.6\textwidth]{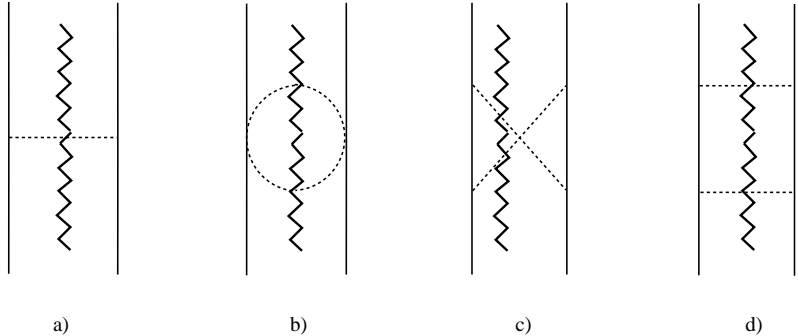}
\end{center}
\caption[pilf]{\protect \small From left to right OPE and TPE diagrams. The solid lines are nucleons, the dashed ones are pions and the angular lines indicate the way the diagram should be cut to contribute to $\Delta(A)$.
\label{nnbar}
}
\end{figure}

In Refs.~\cite{paper1} and \cite{paper2} the input function $\Delta(A)$ was calculated 
from OPE. We now extend this calculation and determine $\Delta(A)$ including as well leading TPE, both 
irreducible TPE and once-iterated OPE from the results of Ref.~\cite{peripheral}.\footnote{Leading  TPE means that the  vertices employed in the calculation are the lowest-order ones in the chiral expansion that stem from the ${\cal O}(p)$ $\pi N$ Lagrangian.} The relevant Feynman diagrams are depicted schematically in Fig.~\ref{nnbar}, where the solid lines are nucleons, the dashed lines are pions and the angular lines indicate how each diagram should be cut to give contribution to $ \Delta(A)$. 
From left to right in Fig.~\ref{nnbar}, the first diagram is OPE, the second and third ones correspond to irreducible TPE, while the last one is once-iterated OPE. This latter diagram contains both irreducible and reducible contributions, explicitly separated in Ref.~\cite{peripheral}. 

Notice that the calculation of the imaginary part of the diagrams in Fig.~\ref{nnbar} along the LHC is finite. When cutting the loop diagrams for TPE, as indicated in Fig.~\ref{nnbar}, an extra Dirac-delta function originates (beyond those required by energy-momentum conservation) that reduces the momentum integration to a finite domain.

It is known since long \cite{weinn} that $NN$ irreducible diagrams are amenable to a chiral expansion. This source of $\Delta(A)$ could then be calculated perturbatively and improved order by order in the chiral expansion in a systematic way. In the standard chiral counting \cite{weinn}, 
OPE is ${\cal O}(p^0)$ and leading irreducible TPE is ${\cal O}(p^2)$.

\begin{figure}
\begin{center}
\includegraphics[angle=0, width=.1\textwidth]{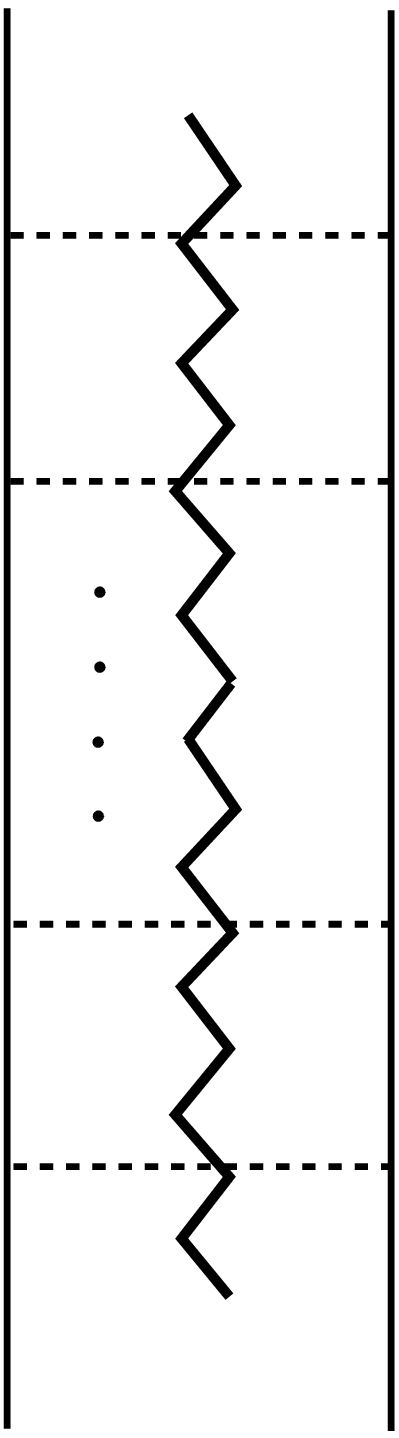}
\end{center}
\caption[pilf]{\protect {\small $NN$ reducible diagram with $n$-time iterated OPE. The meaning of the lines 
is the same as in Fig.~\ref{nnbar}. 
 The vertical dots indicate extra pion ladders. This diagram only contributes to $\Delta(A)$ for $A\leq -n^2 M_\pi^2/4$.}
\label{ntimesope}
}
\end{figure}

Regarding  the $NN$ reducible diagrams, they give contribution to $\Delta(A)$ by cutting the OPE ladders. Indeed, an 
$n$-time iterated OPE diagram, see Fig.~\ref{ntimesope}, contributes only to $\Delta(A)$ by putting on-shell
all  the $n$ pion lines, that is, for $A\leq -n^2 M_\pi^2/4$. This is obvious if we keep in mind the fact that the Schr\"odinger propagator for each of the $NN$ intermediate states cannot be cut because it is proportional to $1/(A-\mathbf{q}^2)$, with $\mathbf{q}$ a three-momentum that stems from the linear combination of loop and external three-momenta, and $A<0$. Then, from Cutkosky rules, the cutting of just one pion line requires to cut the rest of lines because no nucleon line can be cut for $A<0$ and the angular line in Fig.~\ref{ntimesope} must go through all the pion ladders, as shown in the figure.\footnote{We have explicitly checked this conclusion for twice and three-time iterated OPE.} This establishes a natural hierarchy of pion ladders at low energies, because by adding one extra  ladder we move deeper in the LHC and then further away from the low-energy physical region that has $A>0$.  For a given $NN$ reducible diagram with $n$ pion ladders we can also consider its chiral corrections, which will be relatively suppressed  by higher orders in the chiral expansion with respect to the simplest diagram with $n$ pion ladders, depicted in Fig.~\ref{ntimesope}, which is calculated from the lowest-order ${\cal L}_{\pi N}$ Lagrangian. 

Then, increasing both the number of pions exchanged and the chiral order of the calculation reduce the weight of a diagram to $\Delta(A)$ at low energies. 
As we discuss in more detail below in Sec.~\ref{cont_da}, irreducible and reducible TPE diagrams typically contribute with a similar size to $\Delta(A)$, so that we book the relative suppression of increasing the number of pion ladders by one in a reducible $NN$ diagram as ${\cal O}(p^2)$, the same amount  as
  an irreducible loop calculated with lowest order $\pi N$ vertices counts in the chiral expansion.
 In this way, in order to proceed with the calculation of $\Delta(A)$, for a given 
 $NN$ irreducible Feynman diagram we count its chiral order
 in the standard manner \cite{weinn} and book its contribution to $\Delta(A)$ according to the latter. 
For a $NN$ reducible diagram, a leading two-pion ladder (calculated with the lowest-order $\pi N$ vertex) counts as ${\cal O}(p^2)$ and 
 every extra leading pion ladder introduces additionally two extra powers of momentum in the chiral counting. 
On top of that, we add the chiral order corresponding to other parts of the diagram that are $NN$ 
irreducible, as well as the increase in the chiral order due to perturbative corrections to the leading calculation of pion 
ladders. The result of this addition is the final chiral order corresponding to the considered $NN$ reducible diagram to $\Delta(A)$.

\subsection{Subtractions in the  IE for $D(A)$ and the chiral power of $\Delta(A)$}
\label{IEtheory}

As discussed above, the input function $\Delta(A)$ is calculated up to some chiral order in ChPT. 
 The higher the order of the calculation the higher is the maximum divergence of $\Delta(A)$ for $A\to\infty$. 
Indeed, the latter typically diverges except for the OPE case in which it vanishes at least as $1/A$ for $A\to \infty$.
 This can be explicitly checked with the expressions given in Appendix \ref{app.delta} 
of $\Delta(A)$ obtained from OPE in all the partial waves studied in this work. 
At NLO the function $\Delta(A)$ diverges at most linearly in $A$ in the limit $A\to\infty$, while at 
 N$^2$LO it does at most as $|A|^{3/2}$ in the same limit.
 If we generally set that $\Delta(A)\to A^\alpha$ for $A\to \infty$, with $\alpha$ a real number, 
we have typically an increase in the value of $\alpha$ with the chiral order. 
 Thus, it is  an interesting question to settle  whether there is a relation 
between the chiral order up to which $\Delta(A)$ is calculated and 
the minimum number of subtractions needed in the IE for $D(A)$ along the LHC ($A<L$), Eq.~\eqref{inteq1}, 
in order to have a well-defined solution. 
 One should not expect any restriction on the 
maximum number of subtractions by the requirement that the IE is mathematically meaningful. 
Indeed, we show below that when $\Delta(A)$ is calculated only from OPE one 
 has always a unique solution, no matter how large is the number of subtraction taken, 
because it can be reduced to a Fredholm IE 
of the second kind where the associated kernel and the inhomogeneous term are quadratically integrable.

Let us first take  the once-subtracted DRs for $D(A)$ and $N(A)$. We assume that 
$\Delta(A)= \lambda (-A)^\gamma$ and we demonstrate  the important result that the
 once-subtracted DRs has a unique solution for $\gamma<-1/2$.
 Taking in Eq.~\eqref{inteq1} $C=0$, $n=1$ and $\delta_1=1$, 
so as to fulfill the normalization condition Eq.~\eqref{normd}, we have
\begin{align}
\label{delta.da.0}
D(A)&=1-\nu_1\frac{A}{\pi}\int_0^\infty dq^2\frac{\rho(q^2)}{q^2(q^2-A)}
+\frac{A}{\pi^2}\int_{-\infty}^L dk^2 \frac{D(k^2)\Delta(k^2)}{k^2}\int_0^\infty dq^2\frac{\rho(q^2)}{(q^2-k^2)(q^2-A)}~.
\end{align}
 The integrals along the RHC 
can be done explicitly taking into account Eq.~\eqref{gdef}, and then Eq.~\eqref{delta.da.0} simplifies to
\begin{align}
D(A)=1+\nu_1\frac{m\sqrt{-A}}{4\pi}+\frac{m A}{4\pi^2}\int_{-\infty}^L dk^2\frac{D(k^2)\Delta(k^2)}{k^2(\sqrt{-A}+\sqrt{-k^2})}~.
\label{delta.da.1}
\end{align}

Now, we substitute $\Delta(k^2)$ by its
 explicit expression given above and introduce the dimensionless variables
\begin{align}
x&=L/k^2~,\nn\\
y&=L/A~.
\label{ad.var}
\end{align}
  Equation~\eqref{delta.da.1} becomes now
\begin{align}
\hat{D}(y)=1+\nu_1\frac{m (-L)^\frac{1}{2}}{4\pi y^\frac{1}{2}}
+\frac{\lambda m}{4\pi^2}(-L)^{\gamma+\frac{1}{2}}\int_0^1 \frac{dx}{x^{\gamma+\frac{1}{2}} y^{\frac{1}{2}}} 
\frac{\hat{D}(x)}{\sqrt{x}+\sqrt{y}}~,
\label{delta.da.2}
\end{align}
where we have denoted by $\hat{D}(y)$ the function $D(L/y)\equiv \hat{D}(y)$. 
In order to symmetrize the previous IE with respect to $x$ and $y$,  we multiply $\hat{D}(y)$ by $y^{-\frac{\gamma}{2}}$, and define the new function
\begin{align}
\widetilde{D}(y)=y^{-\frac{\gamma}{2}} \hat{D}(y)~.
\end{align}
The IE satisfied by this function follows straightforwardly from Eq.~\eqref{delta.da.2} and it 
reads
\begin{align}
\widetilde{D}(y)&=y^{-\gamma/2}+ y^{-\frac{\gamma+1}{2}} \nu_1\frac{m (-L)^\frac{1}{2}}{4\pi}
+\frac{\lambda m}{4 \pi^2}(-L)^{\gamma+\frac{1}{2}}\int_0^1 dx
\frac{\widetilde{D}(x)}{(x y)^{\frac{\gamma +1 }{2}}(\sqrt{x}+\sqrt{y})}~.
\label{delta.sym.1}
\end{align}
The symmetric kernel in this IE  is
\begin{align}
K(y,x)=\frac{1}{(x y)^{\frac{\gamma +1 }{2}}(\sqrt{x}+\sqrt{y})}~,
\label{kern.sym}
\end{align}
which is quadratically integrable for $\gamma<-1/2$, as well as the inhomogeneous term $y^{-\gamma/2}+ y^{-\frac{\gamma+1}{2}} \nu_1\frac{m (-L)^\frac{1}{2}}{4\pi}$. 
  As a result of the Fredholm theorem 
\cite{Tricomi} one can guarantee that  Eq.~\eqref{delta.sym.1} has always a unique solution 
as far as the constant 
\begin{align}
\beta=(-L)^{\gamma+\frac{1}{2}} \lambda m/(4 \pi^2)
\label{ie.beta.def}
\end{align}
 is not an eigenvalue of the kernel $K(x,y)$ given above. 
 Note  that the coefficient $\lambda$ is a function of 
 low-energy physical constants determined within an error interval, so that an infinitesimal change in those 
constants will make that the resulting $\beta$ is no longer an eigenvalue of the kernel, because there are no 
accumulation points of the eigenvalues in the finite domain \cite{Tricomi}. Then we can state:

\vskip 5pt
{\it Proposition 1: The solution for the once-subtracted IE for $D(A)$ along the LHC always exists and it is unique for 
$\gamma<-1/2$.}
\vskip 5pt

Indeed, for any number of subtractions, the resulting IE in 
terms of the dimensionless variables $x$ and $y$, once it is symmetrized as  
in the discussion above, has the same kernel  as in Eq.~\eqref{kern.sym}. This can be understood 
easily because if one subtraction is added then we have an extra factor $A$ in front of the LHC dispersive integral and 
another $1/k^2$ inside it. In terms of the  variables $x$ and $y$ this implies the factor $L/y \cdot x/L$. Thus, 
 the $L$'s cancel each other and we have one extra $y$ and one less $x$ in the denominator. As a result, 
when symmetrizing by considering $\widetilde{D}(y)$, one has to multiply $\hat{D}(y)$ by 
 one extra $y$, and the same kernel is obtained because the extra $x$ is eaten by $\hat{D}(x)$
 in order to become $\widetilde{D}(x)$ 
inside the integration. In addition, 
the inhomogeneous term in the process of adding more subtraction does not become more singular because 
the extra factor of $1/y$ that appears in the last subtraction terms added is canceled by the extra factor 
$y$ when ending with the symmetric IE for $ \widetilde{D}(y)$.  As a result Proposition 1 can be generalized to 
any number of subtractions:

\vskip 5pt
{\it Proposition 2: The solution for the any-time-subtracted IE to calculate $D(A)$ along the LHC always exists and it is unique for 
$\gamma<-1/2$.}
\vskip 5pt

It is important to realize that when calculating $D(A)$ with $\Delta(A)$ from OPE, one has that $\Delta(A)=\sum_{i=0}^n \alpha_i (-A)^{\gamma-i}$
 with $\gamma\leq -1$ and $n$ is 
a finite natural number. As a result, the kernel that is obtained proceeding as done previously with $\lambda=\alpha_0$ is given  
 by Eq.~\eqref{kern.sym} times the polynomial $1+\sum_{i=1}^n  (y/L)^{i} \alpha_i/\alpha_0$, which does not affect the fact that it is a 
quadratically integrable kernel for $\gamma\leq -1/2$. On the other hand, the inhomogeneous term is the same as before
 and  it is quadratically integrable as well, by the same arguments as given above. As a result, we obtain the following 
important  result:

\vskip 5pt
{\it Proposition 3: When $\Delta(A)$ is given at LO from OPE, the resulting IE for calculating $D(A)$ along the LHC 
 always has a unique solution for any number of subtractions.}
\vskip 5pt

This theorem on the existence of a unique solution of the $N/D$ method when $\Delta(A)$ is restricted to its leading contribution 
from OPE,  contrasts with the situation found by solving the Lippmann-Schwinger equation with the OPE potential. The latter  
 has a singular behavior diverging as $1/r^3$ for $r\to 0$ in the triplet waves  and its solution 
 does not follow the standard procedure for non-singular potential in Quantum Mechanics. 
In order to obtain cut-off independent results when using this potential in a Lippmann-Schwinger equation one 
needs two S-wave counterterms, one for each $NN$  S-wave  \cite{frederico:1999,savage:2002,nogga}, as well as in any other partial wave 
for which the  tensor force is attractive \cite{nogga}. The addition of these last counterterms for $P$- and higher partial waves violates
 the naive chiral power counting. The situation that emerges after  resumming relativistic corrections in the nucleon propagator is discussed 
in Ref.~\cite{gegelia2012}.

 However, $\Delta(A)$ for $A\to \infty$ diverges typically as $A$ at next-to-leading order (NLO), and $\gamma> 1$ for higher orders 
in the chiral expansion of $\Delta(A)$.  Once  $\gamma\geq -1/2$ the symmetric kernel
 $K(x,y)$, Eq.~\eqref{kern.sym}, is not quadratically integrable so that 
we cannot apply the Fredholm theorem.
 In order to proceed we study the limit $A\to-\infty$ and take the leading diverging behavior for $\Delta(A)$ as $\lambda (-A)^\gamma$ when $A\to -\infty$. 
 Next, let us integrate in Eq.~\eqref{delta.sym.1} with $x\geq \varepsilon$,  $\varepsilon>0$,  
 taking at the end the limit $\varepsilon\to 0^+$. To keep the integration limits between 0 and 1 for this case as well,
 we introduce the new variables $t=(x-\varepsilon)/(1-\varepsilon)$ and $u=(y-\varepsilon)/(1-\varepsilon)$ and denote the solution 
to the resulting IE as $\widetilde{D}_\varepsilon(u)$. The new IE reads 
\begin{align}
\widetilde{D}_\varepsilon(u)&=(1-\varepsilon)^{-\frac{\gamma}{2}}(u+\frac{\varepsilon}{1-\varepsilon})^{-\frac{\gamma}{2}}
\left(1+(1-\varepsilon)^{-\frac{1}{2}}(u+\frac{\varepsilon}{1-\varepsilon})^{-\frac{1}{2}}\nu_1\frac{m(-L)^\frac{1}{2}}{4\pi}\right)\nn\\
&+\frac{\lambda m}{4\pi^2}(-L)^{\gamma+\frac{1}{2}}\int_0^1 dt 
\frac{ \widetilde{D}_\varepsilon(t)\,(1-\varepsilon)^{-\gamma-\frac{1}{2}}}{\left[(t+\frac{\varepsilon}{1-\varepsilon})
(u+\frac{\varepsilon}{1-\varepsilon})\right]^\frac{\gamma+1}{2}\left(\sqrt{t+\frac{\varepsilon}{1-\varepsilon}}
+\sqrt{u+\frac{\varepsilon}{1-\varepsilon}}\right)}~.
\label{delta.ep.1}
\end{align}
With the modified kernel $K_\varepsilon(u,t)$ given by
\begin{align}
K_\varepsilon(u,t)&=\frac{(1-\varepsilon)^{-\gamma-\frac{1}{2}}}{\left[(t+\frac{\varepsilon}{1-\varepsilon})
(u+\frac{\varepsilon}{1-\varepsilon})\right]^\frac{\gamma+1}{2}\left(\sqrt{t+\frac{\varepsilon}{1-\varepsilon}}
+\sqrt{u+\frac{\varepsilon}{1-\varepsilon}}\right)}~.
\label{kern.ep.1}
\end{align}
This kernel is now quadratically integrable. Let us denote the inhomogeneous term by $f(u)$, namely,
\begin{align}
f(u)&=(1-\varepsilon)^{-\frac{\gamma}{2}}(u+\frac{\varepsilon}{1-\varepsilon})^{-\frac{\gamma}{2}}
\left(1+(1-\varepsilon)^{-\frac{1}{2}}(u+\frac{\varepsilon}{1-\varepsilon})^{-\frac{1}{2}}\nu_1\frac{m(-L)^\frac{1}{2}}{4\pi}\right)~.
\label{def.feu}
\end{align}
 The solution of Eq.~\eqref{delta.ep.1}, according to the Fredholm theorem,
 can be given in terms of  the resolvent kernel $H_\varepsilon(u,t)$ as:\footnote{We assume again that we are not in the 
 unlikely situation in which $\beta$ is an eigenvalue of 
the kernel $K_\varepsilon(u,t)$. If this is not the case,  
we change infinitesimally the physical constants, e.g. $g_A$ or $M_\pi$, so that the resulting $\beta$ is not longer  
an eigenvalue because the eigenvalues have no accumulation point in the finite domain.} 
\begin{align}
\widetilde{D}_\varepsilon(u)=f(u)
+\frac{\lambda m}{4\pi^2}(-L)^{\gamma+\frac{1}{2}}\int_0^1 dt \, H_\varepsilon(u,t)f(t)
~.
\label{dep.sol}
\end{align}
It is important to remark that the kernel $K_\varepsilon(u,t)$ given in Eq.~\eqref{kern.ep.1} is positive definite. 
The calculation of $H_\varepsilon(u,t)$ applying the  Neumann series gives
\begin{align}
H_\varepsilon(u,t)&=\sum_{n=1}^\infty \beta^{n-1} K_{\varepsilon;n}(u,t)~,\nn\\
K_{\varepsilon;n+1}(u,t)&=\int_0^1 dv K_\varepsilon(u,v)K_{\varepsilon;n}(v,t)~,~(n\geq 1)~,\nn\\
K_{\varepsilon;1}(u,t)&\equiv K_\varepsilon(u,t)~.
\label{ns.hep}
\end{align}
Since $K_\varepsilon(u,t)>0$ for $u,~t\in[0,1]$ it follows from the previous equation that this is also the case 
for $K_{\varepsilon;n}(u,t)$, $n \geq 1$.  
 For the $P$- and higher partial waves $ \nu_1=0$, so that $f(u)$ is a positive-definite function. The same can be 
said for the S-waves with scattering length ($a_S$) less than zero, because $\nu_1=-4\pi a_S/m$ is then a positive quantity. 
From Eq.~\eqref{ns.hep} it follows that the resolvent kernel $H_\varepsilon(u,t)>0$  if $\beta>0$ (that is equivalent to $\lambda>0$, see Eq.~\eqref{ie.beta.def})\footnote{The Neumann series converges in the 
$\beta$-complex plane inside the circle which radius is the smallest of the moduli of the
eigenvalues of $K_\varepsilon(u, t)$. The analytical extrapolation in $\beta$ of Eq. (32) is needed beyond this circle in the 
$\beta$-complex plane.
However, we trust this statement for any $\beta$ because it is valid at any order in perturbation theory.}. In this case 
one also  has that  
$\widetilde{D}_\varepsilon(u)\geq (1-\varepsilon)^{-\frac{\gamma}{2}}(u+\frac{\varepsilon}{1-\varepsilon})^{-\frac{\gamma}{2}}$ which, 
in terms of the original function  $D(A)$,   implies that $D(A)\geq 1$. Taking into account this bound it is clear that 
the original once-subtracted IE for $D(A)$, Eq.~\eqref{delta.da.1}, has no solution in the limit $\epsilon\to 0^+$ because 
the last integral in Eq.~\eqref{delta.da.1} does not converge as soon as $\gamma\geq 1/2$. We have then arrived to the 
following result:

\vskip 5pt
 {\it Proposition 4: For $\gamma\geq 1/2$ the $P$- and higher partial waves, as well as for $S$-waves with $a_S<0$, to have $\lambda<0$ is a necessary condition 
for the existence of solution for the once-subtracted IE satisfied by $D(A)$.} 
\vskip 5pt

Note that once $\lambda<0$ we can have a cancellation between the two 
terms in the r.h.s. of Eq.~\eqref{dep.sol},
 which are needed in order to achieve a vanishing $D(A)$ in the limit $A\to -\infty$ for 
$\gamma\geq 1/2$. In our numerical procedure for the once-subtracted IE with $\lambda<0$ and $\gamma=1$ (as corresponds to 
the NLO case) we have always found the solution having a perfectly stable $\epsilon\to 0^+$ limit.

 In the case with $\lambda>0$ by including more subtractions the inhomogeneous term in Eq.~\eqref{dep.sol} changes so that 
it is no longer positive definite and finally a solution can be obtained. This is apparent by looking at the inhomogeneous 
term in  Eq.~\eqref{inteq1}  for $A<L$. All the subtraction constants $\delta_i$ 
with $i\geq 2$ and $\nu_j$ ($j=1,~2,\ldots$) have a priori not definite sign. In addition, the monomials $(A-C)^n$ 
 for $n$ odd and $A\to -\infty$ are negative, while the right-hand-cut integrals multiplying the constants $\nu_i$ 
 are positive definite for negative 
$A$ and $C$.

Adding more subtraction could be also motivated not only by having an IE with a well-defined solution but also 
by the interest of enhancing the information in the low-energy region,  
so that the results are less sensitive to higher energies. 
 The inhomogeneous term $f(u)$, Eq.~\eqref{def.feu}, is a power expansion with integer or half-integer powers of $u$ and by 
including more subtractions we increase the number of terms in the expansion. As a result 
$f(u)$ contains more and more information that controls the low-energy limit $u\to 1$, both in $f(u)$ itself as well 
as  in the integral in Eq.~\eqref{dep.sol}. A question arises about whether it is possible to ascribe some kind 
of (chiral) power counting that indicates the minimum number of subtraction constants that we should include 
in the IE, in harmony  with the chiral order in which $\Delta(A)$ is evaluated.  In the so-called Weinberg scheme \cite{weinn,ordo94,entem,epen3lo} the number 
of counterterms included in the calculation of the potential is fixed by naive chiral power counting. The final consistency 
of this scheme, once the potential is iterated in a Lippmann-Schwinger equation,
 has been discussed for long in the literature, as discussed in the Introduction, and other schemes are proposed that 
differ mostly in the treatment of the local counterterms 
\cite{kswa,nogga,pavon06,entem08,birse,eiras,phillipssw,pavon11,longyang,gegelia}.  It is beyond the scope of the 
present research at this stage to ascribe 
a chiral power to the subtraction constants present in our equations. 
 Nevertheless, we decide the number of subtractions to be included in each IE so that:
\begin{itemize}
\item[i)] We have an IE giving rise to stable solutions at low energies, $\sqrt{A}\lesssim 500$~MeV. Stable here means 
that the results are independent of the lower limit of integration along the LHC.
\item[ii)] We require that our description of the Nijmegen phase shifts and mixing angles are not worse than the one obtained 
 by solving the Lippmann-Schwinger equation within the Weinberg scheme at NLO. That is, when the Lippmann-Schwinger equation 
is solved with a three-momentum cut-off that is fine tuned to data employing the NLO chiral potential given by 
 OPE and leading TPE \cite{epen3lo}.
\end{itemize}

\section{Uncoupled waves: $^1 S_0$}
\label{1s0} 
In this section we study the $^1 S_0$ partial wave. We first take once-subtracted DRs, $n=1$, and the IE for $D(A)$, Eq.~\eqref{inteq1}, with $C=0$ reads
\begin{align}
D(A)&=1-\nu_1 A  g(A,0)+\frac{A}{\pi}\int_{-\infty}^L dk^2\frac{\Delta(k^2)D(k^2)}{k^2}g(A,k^2)~,
\label{onceD}
\end{align}
with $N(A)$, Eq.~\eqref{standardr}, given by
\begin{align}
N(A)&=\nu_1+\frac{A}{\pi}\int_{-\infty}^Ldk^2\frac{\Delta(k^2)D(k^2)}{k^2(k^2-A)}~.
\label{onceN}
\end{align}
We have one free parameter $ \nu_1$ that can be fixed in terms of the $^1S_0$ scattering length $a_s$ by taking into account the effective range expansion for an S-wave, that reads in our normalization:
\begin{align}
\frac{4\pi}{m}\frac{D}{N}=-\frac{1}{a_s}+\frac{1}{2}r_s A-i\sqrt{A}+{\cal O}(A^2)~,
\label{efr1}
\end{align}
with $r_s$ the $^1S_0$ effective range. Since for $A=0$ we have  $N(0)=\nu_1$ and $D(0)=1$, it follows that
\begin{align}
\nu_1=-\frac{4\pi a_s}{m}~.
\label{1s0nu1}
\end{align}
The experimental value for the $^1S_0$ scattering length is $a_s=-23.76\pm 0.01$~fm \cite{epen3lo}.

\begin{figure}
\begin{center}
\includegraphics[angle=0, width=.5\textwidth]{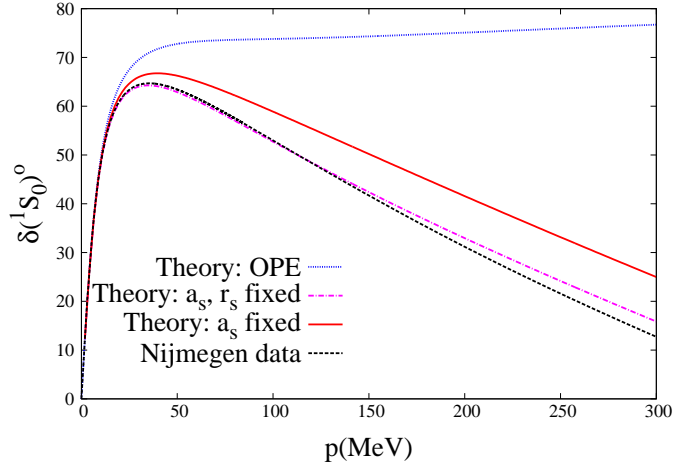}
\end{center}
\caption[pilf]{\protect {\small (Color online.) Phase shifts of the $^1S_0$ $NN$ partial wave as a function of the 
c.m. three-momentum $p$ expressed in MeV.  The (red) solid line is our results with $n=1$, Eq.~\eqref{onceD}, with only $a_s$ fixed to its experimental value. The (magenta) dash-dotted line corresponds to $n=2$, Eq.~\eqref{twiceD}, with $a_s$ and $r_s$ fixed to their experimental value and $\nu_2$ fitted to data, Eq.~\eqref{fit1s0}.  In addition, the (blue) dotted line is the OPE result of Ref.~\cite{paper1}. Finally, the Nijmegen PWA phase shifts are shown by the (black) dashed line.}
\label{fig:1fp1s0}
}
\end{figure}

The phase shifts obtained by solving the IE of Eq.~\eqref{onceD} are shown in Fig.~\ref{fig:1fp1s0} as a function of the c.m.
 three-momentum $p=\sqrt{A}$. The 
(red) solid line  corresponds to our results from Eqs.~\eqref{onceD} and \eqref{1s0nu1} with $\Delta(A)$ calculated up-to-and-including ${\cal O}(p^2)$ contributions, and they are compared with the neutron-proton ($np$) $^1S_0$ phase shifts of the Nijmegen PWA \cite{Stoks:1994wp}  (black dashed line) and with the OPE results  of Ref.~\cite{paper1}  (blue dotted line). As we see, there is a clear improvement when including TPE.

We can also  predict $r_s$ by expanding the left-hand side  of Eq.~\eqref{efr1} up-to-and-including ${\cal O}(A)$, it results
\begin{align}
r_s&=\frac{m}{2\pi^2 a_s}\int_{-\infty}^L dk^2\frac{\Delta(k^2)D(k^2)}{(k^2)^2}\left\{\sqrt{-k^2}-\frac{1}{a_s}\right\}~.
\label{rs1s0}
\end{align}
Our calculation of $D(A)$ at ${\cal O}(p^2)$ gives the numerical result
\begin{align}
r_s=2.32~\hbox{fm}~,
\label{rs.1s0}
\end{align}
 close already to  its experimental value $r_s=2.75\pm 0.05$~fm or the value $r_s=2.670$~fm determined in Ref.~\cite{pavon:2008} for the NijmII potential.  

It is important to stress that Eq.~\eqref{rs1s0}  exhibits a clear correlation between the effective range and the scattering length for the $^1S_0$ partial wave. This correlation, first noticed in Ref.~\cite{pavon06}, can be written as 
\begin{align}
r_s&=\alpha_0+\frac{\alpha_{-1}}{a_s}+\frac{\alpha_{-2}}{a_s^2}~,
\label{rs.le}
\end{align}
where the coefficients $\alpha_{0,-1,-2}$ are independent of the scattering length $a_s$. This follows because $D(A)$ satisfies 
the linear IE Eq.~\eqref{onceD}, that we now rewrite as $L[D(A)]=1+a_s\,\frac{4\pi}{m}A g(A,0)$,
with the linear operator $L[D(A)]$ defined as 
\begin{align}
L[D(A)]&=D(A)-\frac{A}{\pi}\int_{-\infty}^L dk^2\frac{\Delta(k^2)D(k^2)}{(k^2)^2}g(A,k^2)~.
\end{align}
The solution $D(A)$ can be split as the sum of two terms $D_0(A)+a_s D_1(A)$, with $D_{0,1}(A)$ independent of  $a_s$ and satisfying
\begin{align}
L[D_0(A)]&=1~,\nn\\\
L[D_1(A)]&=\frac{4\pi}{m}A g(A,0)~.
\end{align}
Substituting $D(A)=D_0(A)+a_s D_1(A)$ into Eq.~\eqref{rs1s0} we then have the following expressions for the coefficients
\begin{align}
\alpha_0&=\frac{m}{2\pi^2}\int_{-\infty}^L dk^2\frac{\Delta(k^2)D_1(k^2)}{(k^2)^2}\sqrt{-k^2}~,\nn\\
\alpha_{-1}&=\frac{m}{2\pi^2}\int_{-\infty}^L dk^2\frac{\Delta(k^2)}{(k^2)^2}\left[D_0(k^2)\sqrt{-k^2}-D_1(k^2)\right]~,\nn\\
\alpha_{-2}&=-\frac{m}{2\pi^2}\int_{-\infty}^L dk^2\frac{\Delta(k^2)D_0(k^2)}{(k^2)^2}~.
\end{align}
Notice that in our formalism this correlation between $r_s$ and $a_s$ stems from unitarity and analyticity and it makes 
sense as long as the once-subtracted DR, Eq.~\eqref{onceD}, exists. 
 Our  NLO solution gives the numerical values:
\begin{align}
\alpha_0&=2.10~\rm{fm}~,\nn\\
\alpha_{-1}&=-4.89~\rm{fm}^2~,\nn\\
\alpha_{-2}&=5.46~\rm{fm}^3~.
\label{alfas.1s0}
\end{align}
The $^1S_0$ effective range was predicted by the knowledge of the 
scattering length and the chiral TPE potential in the first entry of Ref.~\cite{pavon06} by renormalizing the Lippmann-Schwinger equation 
with boundary conditions and imposing the hypothesis of orthogonality of the wave functions determined with
 different energy.\footnote{Since the potentials involved are singular this orthogonality condition does not follow like in the case of a regular potential but must be imposed, which is a working assumption of the formalism of Ref.~\cite{pavon06}.}  The 
numerical values for the coefficients obtained in Ref.~\cite{pavon06} when the TPE potential is calculated at NLO are: $\alpha_0=2.122$~fm, 
$\alpha_{-1}=-4.889$~fm$^2$ and $\alpha_{-2}=5.499$~fm$^3$, resulting in $r_s=2.29$~fm. These numbers, obtained by a completely 
independent method from ours,  are 
indeed in remarkably good agreement  with our results for $r_s$, Eq.~\eqref{rs.1s0}, and with the coefficients 
in Eq.~\eqref{alfas.1s0}. 
 The same reference also calculated 
these coefficients including subleading TPE  up to N$^2$LO with the result:
 $\alpha_0=2.59\sim 2.67$~fm, $\alpha_{-1}=-5.85\sim (-5.64)$~fm$^2$ and $\alpha_{-2}=5.95 \sim 6.09$~fm$^3$. The intervals of numerical values arise from the values taken for the $c_i$ counterterms of the ${\cal O}(p^2)$ $\pi N$ Lagrangian \cite{pavon06}. 
 We should also stress that our derivation of 
 the correlation of Eq.~\eqref{rs.le} is based on basic properties of the $NN$ partial wave amplitudes, namely, analyticity, unitarity
 and chiral symmetry. This is an important result, that also reinforces the assumption of orthogonality of the wave functions
 employed in Ref.~\cite{pavon06}.
 Regarding the phase shifts, our results by fixing only $a_s$ to experiment, solid line in 
Fig.~\ref{fig:1fp1s0},  are also quite similar to those obtained with the NLO TPE potential in Ref.~\cite{pavon06}. 
This is also the case when comparing with the phase shifts calculated  in the third entry of Ref.~\cite{phillipssw} by
 making use of a chiral potential 
with NLO TPE plus a contact term that is fixed in terms of the  experimental scattering length $a_s$. 
This reference obtains the value $r_s\simeq 2.26~$fm  (which  is
 extracted approximately from the Fig.~2 of Ref.~\cite{phillipssw}, 
because $r_s$ is not given explicitly there), that is also quite similar to our result in Eq.~\eqref{rs.1s0}.

Next,  we consider the twice-subtracted DRs, $n=2$, 
\begin{align}
\label{twiceD}
D(A)&=1+\delta_2 A-\nu_1\frac{A(A+M_\pi^2)}{\pi}\int_0^\infty dq^2\frac{\rho(q^2)}{(q^2-A)(q^2+M_\pi^2)q^2}
-\nu_2 A(A+M_\pi^2) g(A,-M_\pi^2)\nn\\
&+\frac{A(A+M_\pi^2)}{\pi^2}\int_{-\infty}^L dk^2\frac{\Delta(k^2)D(k^2)}{(k^2)^2}\int_0^\infty dq^2\frac{\rho(q^2)q^2}{(q^2-A)(q^2+M_\pi^2)(q^2-k^2)}~, \\
\label{twiceN}
N(A)&=\nu_1+\nu_2 A+\frac{A^2}{\pi}\int_{-\infty}^L dk^2\frac{\Delta(k^2)D(k^2)}{(k^2-A)(k^2)^2}~,
\end{align}
where, the extra subtraction in the  function $D(A)$ is taken at $C=-M_\pi^2$, while for the function $N(A)$ the two subtractions are taken at $ C=0$, see Eq.~\eqref{standardr}. The subtraction constant $\nu_1$ is also given by Eq.~\eqref{1s0nu1}. Next we fix $\delta_2$ in terms of $r_s$. For that, according to Eq.~\eqref{efr1}, one needs to expand 
\begin{align}
\frac{4\pi}{m}\frac{D}{N}+i\sqrt{A}
\end{align}
up-to-and-including ${\cal O}(A)$ terms. In this expansion, one should consider carefully the combination of the  first integral on the r.h.s. of Eq.~\eqref{twiceD} with  $im \sqrt{A}/4\pi$ 
\begin{align}
&-\frac{A(A+M_\pi^2)}{\pi}\int_0^\infty dq^2\frac{\rho(q^2)}{(q^2-A)(q^2+M_\pi^2)q^2}+\frac{i m \sqrt{A}}{4\pi}
\nn\\
&=\frac{A(A+M_\pi^2)}{M_\pi^2}\left[g(A,-M_\pi^2)-g(A,0)\right]+\frac{i m \sqrt{A}}{4\pi}
=\frac{m A}{4\pi M_\pi}~.
\end{align}
For the rest of terms the expansion is straightforward because the limit $A=0$  can be taken directly inside the integrals.
 One ends with the following expression for $\delta_2$,
\begin{align}
\delta_2&=\frac{a_s}{M_\pi}(1-\frac{1}{2}r_s M_\pi)
+\frac{\nu_2}{\nu_1}\bigg[ 1 + \nu_1 M_\pi^2 g(0,-M_\pi^2) \bigg]
-\frac{M_\pi^2}{\pi}\int_{-\infty}^L dk^2\frac{\Delta(k^2)D(k^2)}{(k^2)^2}g(k^2,-M_\pi^2)~,
\label{delta2.1s0}
\end{align}                    
which is then substituted in Eq.~\eqref{twiceD} and
 our final expression for the twice-subtracted DR of $D(A)$ results:
\begin{align}
\label{1s0.3sb}
 D(A)&=1+A\left\{\frac{a_s}{M_\pi}(1-\frac{1}{2}r_s M_\pi) 
+ \frac{\nu_2}{\nu_1}\left[ 1+\nu_1 M_\pi^2 g(0,-M_\pi^2) \right]\right\}\nn\\
&-A(A+M_\pi^2)\left[\nu_2 g(A,-M_\pi^2) - \nu_1 \frac{g(A,-M_\pi^2)-g(A,0)}{M_\pi^2}\right]\\
&+\frac{A}{\pi}\int_{-\infty}^L \!\! dk^2 \frac{\Delta(k^2)D(k^2)}{(k^2)^2}
\left\{\frac{A+M_\pi^2}{k^2+M_\pi^2}\left[ k^2 g(A,k^2)+M_\pi^2g(A,-M_\pi^2)\right]
 - M_\pi^2 g(k^2,-M_\pi^2)\right\}~.\nn
\end{align}
            The subtraction constant $\nu_2$ is fitted to the $np$ Nijmegen PWA phase shifts
 for $\sqrt{A}\leq 150$~MeV.\footnote{Since Ref.~\cite{Stoks:1994wp} does not provide errors we always perform a
 least square fit,  without weighting.} The best value is 
\begin{align}
\label{fit1s0}
\nu_2=0.24~M_\pi^{-4}~,
\end{align}
 and the resulting curve is shown by the (magenta) dash-dotted line in Fig.~\ref{fig:1fp1s0}. We see that this curve follows closely the experimental phase shifts. It is also interesting to remark that the results with $n=1$, that were able to predict the experimental value for $r_s$ rather closely, can be exactly reproduced, as expected, in terms of the twice-subtracted DRs with $\nu_2=-1.346~M_\pi^{-4}$.                                 

As we did before for the once-subtracted DR results, we compare  our phase shifts from the twice-subtracted DRs, 
Eq.~\eqref{1s0.3sb}, with the ones obtained in 
Ref.~\cite{phillipssw} but now when the NLO TPE potential is supplied with two counterterms, one of them 
 associated with an energy- or momentum-dependent local term. The (magenta) dash-dotted line 
in Fig.~\ref{fig:1fp1s0} runs much closer to data than the just mentioned results of Ref.~\cite{phillipssw} which, for 
the case with a momentum-dependent local term  in the potential, quickly become cutoff independent
 once $\Lambda>900$~GeV.
 Nevertheless, 
in this case we have to say that the twice-subtracted DRs contain three free parameters, while only two free parameters 
 are involved in Ref.~\cite{phillipssw}. An interesting point to discuss is that 
in the case of the twice-subtracted DRs  with $\Delta(A)$ calculated 
at NLO we are able to implement the exact experimental value for $r_s$, while this is not possible  
in Ref.~\cite{phillipssw} when solving the Lippmann-Schwinger equation with the NLO TPE potential 
 supplied with a  momentum-dependent local term. This limitation is also discussed in Ref.~\cite{pavon06} 
and it is connected to the Wigner bound that limits the impact of short-range physics included in energy-independent potentials 
on physical observables \cite{cohen}.\footnote{Ref.~\cite{phillipssw} also considers
 the case of adding to the NLO TPE an energy-dependent local term. The authors of Ref.~\cite{phillipssw} can then 
reproduce the experimental value for $r_s$, but they 
obtain phase shifts that show a strong oscillatory dependence with the actual value taken for the cutoff 
$\Lambda$.} In the twice-subtracted DR case of the $N/D$ method fixing $r_s$ to experiment is 
straightforward since it only implies a linear equation, Eq.~\eqref{delta2.1s0}, that allows us to determine $\delta_2$ 
in terms of the experimental values of $r_s$ and $\nu_2$. To better see how the experimental value of $r_s$ is implemented in 
Eq.~\eqref{1s0.3sb} let us particularize it for $A<L$,   because once $D(A)$ is solved along the 
LHC  everything is then calculated in terms of DRs involving $D(A)$ along this domain. 
 Equation~\eqref{1s0.3sb} simplifies to 
\begin{align}
D(A)&=1-a_s \sqrt{-A}\left(1-\frac{1}{2}r_s \sqrt{-A}\right)
+\left( \sqrt{-A} - \frac{1}{a_s}\right)\frac{m}{4\pi}\nu_2 A  \\ 
&+\frac{A}{\pi}\int_{-\infty}^L \!\! dk^2 \frac{\Delta(k^2)D(k^2)}{(k^2)^2}
\left\{\frac{A+M_\pi^2}{k^2+M_\pi^2}\left[ k^2 g(A,k^2)+M_\pi^2g(A,-M_\pi^2)\right]
 - M_\pi^2 g(k^2,-M_\pi^2)\right\}~.\nn
\end{align}
In the once-subtracted DR case, Eq.~\eqref{onceD}, the inhomogeneous term is just $1-a_s\sqrt{-A}$. 
 Note that the factor $1-r_s \sqrt{-A}/2$  multiplying $-a_s \sqrt{-A}$  cannot be considered as 
a correction  because  $r_s M_\pi\sim 1$. This is to be expected because 
including one extra subtraction in Eq.~\eqref{onceD} implies a reshuffling of the dispersive integral,
which has typically the same size as the counterterms.\footnote{The latter would change by contributions 
from the dispersive integral by just changing the subtraction point, so that the final result would be independent of the subtraction 
point chosen.}

\section{Uncoupled $P$-waves}
\label{pw} 
In this section we discuss the application of the method to the  uncoupled $P$-waves. 

\subsection{$^3P_0$ wave}
\label{3p0}
\begin{figure}
\begin{center}
\includegraphics[angle=0, width=.5\textwidth]{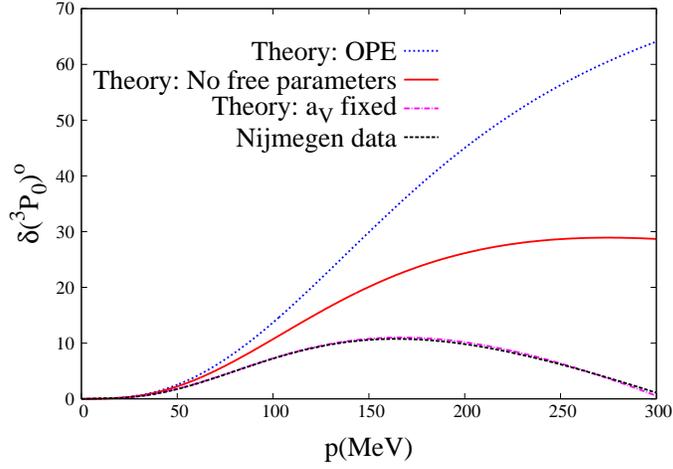}
\end{center}
\caption[pilf]{\protect {\small (Color online.) Phase shifts of the $^3P_0$ $NN$ partial wave. The (red) solid line corresponds to our results with $n=1$, Eq.~\eqref{onceDNl}, the (magenta) dash-dotted line is our results with $n=2$, Eq.~\eqref{twiceDNl}. The (blue)  dotted line is the OPE result from Ref.~\cite{paper1} and the (black) dashed line is the Nijmegen PWA phase shifts, which almost coincides with the $n=2$ result.}
\label{fig:3p0}
}
\end{figure}

For the $^3 P_0$ uncoupled wave we also consider first the once-subtracted DR already used 
for the $^1S_0$ case, i.e. Eqs.~\eqref{onceD} and \eqref{onceN}. The only important difference is that for $P$- and higher orbital-angular-momentum partial waves we have the threshold behavior  $T_{J \ell S}(0)=0$ at $A=0$, so that $\nu_1=0$. Hence, for $\ell\geq 1$ Eqs.~\eqref{onceD} and \eqref{onceN} reduce to
\begin{align}
D_{J\ell S}(A)&=1+\frac{A}{\pi}\int_{-\infty}^L dk^2\frac{\Delta(k^2)D_{J\ell S}(k^2)}{k^2}g(A,k^2)~,\nn\\
N_{J\ell S}(A)&=\frac{A}{\pi}\int_{-\infty}^L dk^2\frac{\Delta(k^2)D_{J\ell S}(k^2)}{k^2(k^2-A)}~.
\label{onceDNl}
\end{align}
Notice that there are no free subtraction constants in Eq.~\eqref{onceDNl} and the emerging results are then predictions of our approach. In Fig.~\ref{fig:3p0} we show our results by the (red) solid line. We see that this curve is much closer to data than the OPE result of Ref.~\cite{paper1} given by the (blue) dotted line, 
so that the correction is in the right direction.

Next, we consider the twice-subtracted DRs of Eqs.~\eqref{standardr} and \eqref{inteq1} but now with $ \nu_1=0$ and $C=0$. They 
can be written as 
\begin{align}
\label{twiceDNl}
D(A)&=1+\delta_2 A-\nu_2 A^2 g(A,0)
+\frac{A^2}{\pi}\int_{-\infty}^L dk^2\frac{\Delta(k^2)D(k^2)}{(k^2)^2}g(A,k^2)~,\nn \\
N(A)&=\nu_2 A+\frac{A^2}{\pi}\int_{-\infty}^L dk^2\frac{\Delta(k^2)D(k^2)}{(k^2-A)(k^2)^2}~.
\end{align}
In this equation  we take  $C=0$ for all the subtractions, because   no infrared divergences are  
 generated in the integrals along the RHC for $\nu_1=0$. Notice that this was not the case for the $^1S_0$ partial wave 
because of the first integral on the r.h.s. of Eq.~\eqref{twiceD}.  
 The subtraction constant $\nu_2$ can be fixed straightforwardly to the experimental scattering volume\footnote{That we define as 
$a_V= \lim_{A\to 0^+} \delta(A)/A^{3/2}$, with $\delta(A)$ the phase shifts.} 
\begin{align}
\nu_2&=\frac{4\pi a_V}{m}~.
\end{align}
For the $^3P_0$ partial wave we have $a_V=0.890~M_\pi^{-3}$, a value that is derived from the Nijmegen PWA phase shifts \cite{Stoks:1994wp}. Finally, the subtraction constant $\delta_2$ is fitted to data with the value
\begin{align}
\delta_2&=-0.30~M_\pi^{-2}~.
\label{fit3p0}
\end{align}
 The resulting curve is shown by the (magenta) dash-dotted line in Fig.~\ref{fig:3p0}, that perfectly agrees with the phase shifts of \cite{Stoks:1994wp} (given by the black dashed line). The reproduction of the data is so good that the fit is completely insensitive to the upper limit of $\sqrt{A}$ fitted, in the range shown in the figure.

\subsection{$^3 P_1$ wave}
\label{3p1}

This partial wave illustrates our conclusion in Sec.~\ref{unformalism} with respect to the fact that $n$ should be large enough in order to write down meaningful DRs for $D(A)$ and $N(A)$, Eqs.~\eqref{standardr}. Here, the once-subtracted DR, Eq.~\eqref{onceDNl},  does not have solution.\footnote{The numerical outcome depends on the
 lower limit of integration  when discretizing the IE for $D(A)$.} The reason is because for $^3P_1$ the asymptotic behavior of $\Delta(A)$ for $A\to -\infty$ corresponds to $\lambda (-A)$ with $\lambda>0$, so that the once-subtracted DR should not converge in this case as shown by the proposition 4 in 
Sec.~\ref{IEtheory}

\begin{figure}
\begin{center}
\includegraphics[angle=0, width=.5\textwidth]{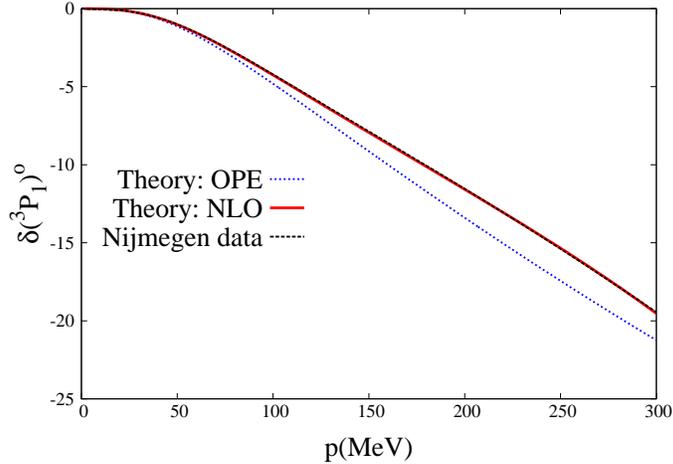}
\end{center}
\caption[pilf]{\protect {\small (Color online.) Phase shifts of the $^3P_1$ $NN$ partial wave. The (red) solid line corresponds to our results with three-time-subtracted DRs, with $a_V$ fixed and  $\delta_2$ and $\delta_3$ fitted, Eq.~\eqref{fit3p1}. The (blue)  dotted line is the OPE result from Ref.~\cite{paper1} and the (black) dashed line is the Nijmegen PWA phase shifts.}
\label{fig:3p1}
}
\end{figure}

We finally need to take three subtractions in order to have a meaningful IE without dependence in the lower limit  of integration along the LHC. The twice-subtracted DRs do not provide stable results either.  For the function $D(A)$ one subtraction is taken at $C=0$ and the other two at $C=-M_\pi^2$, while all of them are 
taken at $C=0$ for $N(A)$. The three-time subtracted DRs are then given by
\begin{align}
\label{three.times}
D(A)&=1+\delta_2 A+\delta_3 A^2-\nu_2\frac{A(A+M_\pi^2)^2}{\pi}\int_{0}^\infty dq^2\frac{\rho(q^2)}{(q^2+M_\pi^2)^2(q^2-A)}\nn\\
&-\nu_3 \frac{A(A+M_\pi^2)^2}{\pi}\int_0^\infty dq^2 \frac{q^2 \rho(q^2)}{(q^2+M_\pi^2)^2(q^2-A)}\nn\\
&+\frac{A(A+M_\pi^2)^2}{\pi^2}\int_{-\infty}^L dk^2\frac{\Delta(k^2)D(k^2)}{(k^2)^3}\int_0^\infty dq^2\frac{(q^2)^2 \rho(q^2)}{(q^2+M_\pi^2)^2(q^2-k^2)(q^2-A)}
\end{align}
The subtraction constant $\nu_2$ is fixed in terms of the $^3P_1$ scattering volume, $a_V=-0.543~M_\pi^{-3}$ from the the Nijmegen PWA \cite{Stoks:1994wp}, 
according to the expression $\nu_2=4\pi a_V/m$,  already used for the $^3P_0$ wave. We then 
fit the subtraction constants $\delta_2$ and $\delta_3$ while $\nu_3$ is finally fixed to zero. We have checked that the resulting fit is stable 
if we release $\nu_3$ from zero and, since we are able to reproduce perfectly the data, as shown by the (red) solid line in Fig.~\ref{fig:3p1}, we do 
not need to release $\nu_3$.  The resulting fit is
\begin{align}
\nu_3&=0^*~,\nn\\
\delta_2&\simeq (2.5\sim 3.0)~M_\pi^{-2},\nn\\
\delta_3&\simeq (0.2\sim 0.3)~M_\pi^{-4}~.
\label{fit3p1}
\end{align}
Here, the asterisk indicates that $\nu_3$ is fixed to zero. 
The final result, that overlaps data (black dashed line), is indicated by the (red) solid line in Fig.~\ref{fig:3p1}. 
The blue dotted line is the OPE result of Ref.~\cite{paper1} that was obtained in terms of a once-subtracted DR.

\subsection{$^1 P_1$ wave}
\label{1p1}

Now we consider the singlet uncoupled wave $^1 P_1$, where $\lambda<0$ so that we expect to have a solution for the once-subtracted IE.
 Indeed, this is the case.
 As usual, we discuss  first the once-subtracted DR and then the twice-subtracted case.
 The former has no free parameters. For the latter case the scattering volume, $a_V=-0.939~M_\pi^{-3}$, is used to fix $\nu_2$ and $\delta_2$ is fitted to data.
 However, now the fit is not very sensitive to this subtraction constant,
 which is determined only within a large interval of positive values from 0.8 up to 27~$M_\pi^{-2}$, depending on the upper limit for $\sqrt{A}$ taken in the fit.
 
\begin{figure}
\begin{center}
\includegraphics[angle=0, width=.5\textwidth]{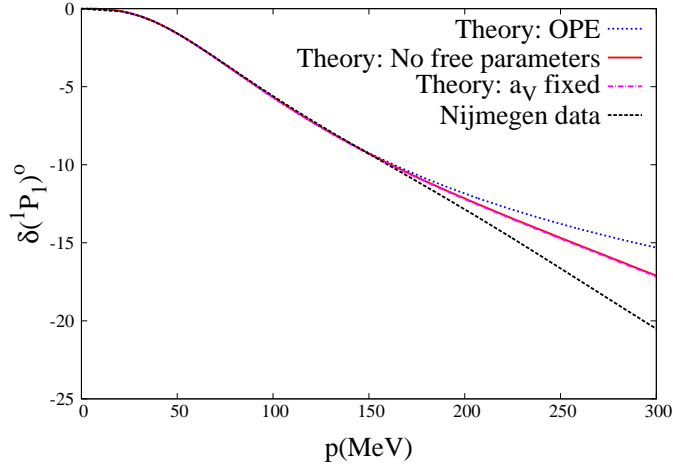}
\end{center}
\caption[pilf]{\protect {\small (Color online.) Phase shifts of the $^1P_1$ $NN$ partial wave. The (red) solid line corresponds to our results with a once-subtracted DR, $n=1$. The (magenta) dash-dotted line represents the case of a  twice-subtracted DR, $n=2$, with $a_V$ fixed and $\delta_2$ fitted. The (blue)  dotted line is the OPE result from Ref.~\cite{paper1} and the (black) dashed line is the Nijmegen PWA phase shifts.}
\label{fig:1p1}
}
\end{figure}

Our results for once- and twice-subtracted DRs are almost identical, as can be seen by comparing the (red) solid and (magenta) dash-dotted lines in Fig.~\ref{fig:1p1}, respectively. Both curves are overlapping and reproduce  the data fairly well for $\sqrt{A}<200$~MeV.

\section{Uncoupled waves:  $\ell\geq 2$ }
\label{leq2}

A partial wave amplitude with  $\ell\geq 2$ should vanish as $A^\ell$ in the limit $A\to 0$. This behavior is not directly implemented by the DR  
 Eq.~\eqref{standardr}, unless some constraints are imposed. The right threshold behavior can be achieved by taking the subtraction point $C=0$ in 
$N(A)$ and then imposing $\nu_i=0$ for $i=1,\ldots,\ell$ in Eq.~\eqref{standardr}.
 In this way, since $T=N/D$ and $D(0)=1$, one has that $T\to A^\ell$ for $A\to 0$, being necessary to consider at least $\ell$-time subtracted DRs. In practice we take the minimum number of subtractions,  $n=\ell$, and $C=0$ in Eq.~\eqref{standardr}, so that we end with the equations
\begin{align}
\label{highd}
D(A)&=1+\sum_{i=2}^{\ell}\delta_i A^{i-1}+\frac{A^\ell}{\pi}\int_{-\infty}^L dk^2\frac{\Delta(k^2)D(k^2)}{(k^2)^\ell}g(A,k^2)~,\\
\label{highn}
N(A)&=\frac{A^\ell}{\pi}\int_{-\infty}^\ell dk^2\frac{\Delta(k^2)D(k^2)}{(k^2)^\ell (k^2-A)}~.
\end{align}
The $\delta_i$ are free parameters that  are proportional to derivatives of the function $D(A)$ at $A=0$,\footnote{Strictly speaking, they correspond to the derivatives from the left of $D(A)$ at $A=0$, that is, the limit $A\to 0^-$ is the proper one in order to avoid 
the branch cut singularity in $D(A)$ due to the onset of the unitarity cut for $A>0$.}  namely:
\begin{align}
\delta_n=(n-1)!\, D^{(n-1)}(0)~,~n\geq 2~,
\label{deltai.dv}
\end{align}
with 
\begin{align}
D^{(n)}(0)=\frac{\partial^n D(A)}{\partial A^n}\bigg|_{A=0}~.
\end{align}
 Nevertheless, as $\ell$ increases, rescattering effects giving rise to the unitarity cut are less important  because of the centrifugal barrier and then $D(A)\simeq 1$ for $A$ in the range of interest here. 
This manifests in the fact that the $\delta_i$ can be taken equal to zero except the one 
 with the largest subscript, $i=\ell$, that is fitted to data providing a good reproduction of the latter in most of the cases. 
This is the situation that corresponds to the smoothest $D(A)$ in the low-energy region, and we refer to it as the ``principle of maximal smoothness''. 
This rule  stems from our study of $NN$ partial wave amplitudes with $\ell\geq 2$, and it holds not only in the uncoupled waves but it is also applicable to the coupled ones.
 Even if we released all the $\delta_i$ there is no any significant improvement in the reproduction of data with respect to that obtained when only $\delta_\ell\neq 0$.
 It is also shown below that if we insist on using the once-subtracted DRs, Eq.~\eqref{onceDNl},
 the resulting phase shifts are very similar 
to those obtained with $\ell$ subtractions  for the partial waves with $\ell\geq 3$.
 The reason is that for  partial waves  with $\ell$ high enough 
 the strict violation of the threshold behavior for such higher  partial waves is a rescattering effect that restricts indeed to 
very low energies and it is more an artifact of academic interest. 
In turn, this is a reflection of the general trend of $NN$ partial waves of becoming quite 
perturbative typically for $\ell\gtrsim 3$ as obtained in Ref.~\cite{peripheral}
 by studying perturbatively the $\ell\geq 2$ waves within the one-loop approximation of  baryon ChPT.

Another method to guarantee the right behavior at threshold was developed in Ref.~\cite{paper1}
 without the need of  increasing the number of subtraction constants. We refer to this reference for 
further details. The neat result is that the $D(A)$ function should satisfy the set of constraints
\begin{align}
\int_{-\infty}^L dk^2\frac{\Delta(k^2)D(k^2)}{(k^2)^\lambda}=0~~,~~(\lambda=2,\ldots,\ell)
\label{cddc}
\end{align}
with $\ell\geq 2$. In order to fulfill them a set of $\ell-1$ CDD poles \cite{cdd} are included in the $D(A)$ function, whose residues are adjusted by imposing Eq.~\eqref{cddc}.
 The final expressions are Eq.~\eqref{highn}, that is the same as here, and a different equation for $D(A)$
\begin{align}
D(A)&=1
+\frac{A}{\pi}\int_{-\infty}^L dk^2 \frac{\Delta(k^2)D(^2)}{k^2}g(A,k^2)+\frac{A\sum_{n=0}^{\ell-2} c_n A^n}{(A-B)^{\ell-1}}~,
\label{highcddD}\end{align}
with $B$ corresponding to the position of the CDD poles that is finally sent to infinity. Notice that at low energies ($A\ll B$) the addition of the CDD poles reduce to change the function $D(A)$ by a polynomial of degree 
$\ell-1$. In this sense, this method based on the constraints Eq.~\eqref{cddc} and the addition of the CDD poles,
 Eq.~\eqref{highcddD}, is a particular case at low energies of the most general solution with $\ell$ subtractions, Eq.~\eqref{highd}. We do not use further the method of Ref.~\cite{paper1} because 
unless $\Delta(A)$ vanishes fast enough in the infinite, e.g. like $1/A$ in OPE, it implies to use integrals along the LHC that grow with powers of $B\to \infty$,
 which makes very hard its numerical manipulation. In particular, standard numerical subroutines used to invert a matrix and find the numerical solution of the IE, do not provide 
the right answer for large $B$. This is the situation at NLO because $\Delta(A)\to A$ for $A\to \infty$, and it would be even worse if higher orders in the chiral expansion of $\Delta(A)$ were implemented. In the present formalism, based on the Eqs.~\eqref{highd} and \eqref{highn} above, 
we can skip  the aforementioned sum rules of Eq.~\eqref{cddc} because, by construction, we satisfy the threshold behavior by having 
included $\ell$ subtractions. The price to pay is that now we have $\ell-1$ free $\delta_i$, precisely the number of sum rules to be fulfilled in the 
formalism of Ref.~\cite{paper1}. For a more general presentation on the proliferation of these sum rules with increasing $\ell$ for a once-subtracted DR 
of a $NN$ partial wave, the interested reader is referred to 
the book by Barton \cite{barton}. 

Note that in the case of solving a Lippmann-Schwinger equation with a $NN$ potential $V$ the right threshold behavior is always
 implemented because of the series $T=V+VGV+VGVGV+\ldots$, where the left and right most $V$'s take 
care of proving the right power of $A$ when $A\to 0$. This is explicitly used in the subtractive method of Ref.~\cite{phillipssw}.

\begin{figure}
\begin{center}
\begin{tabular}{cc}
\includegraphics[width=.4\textwidth]{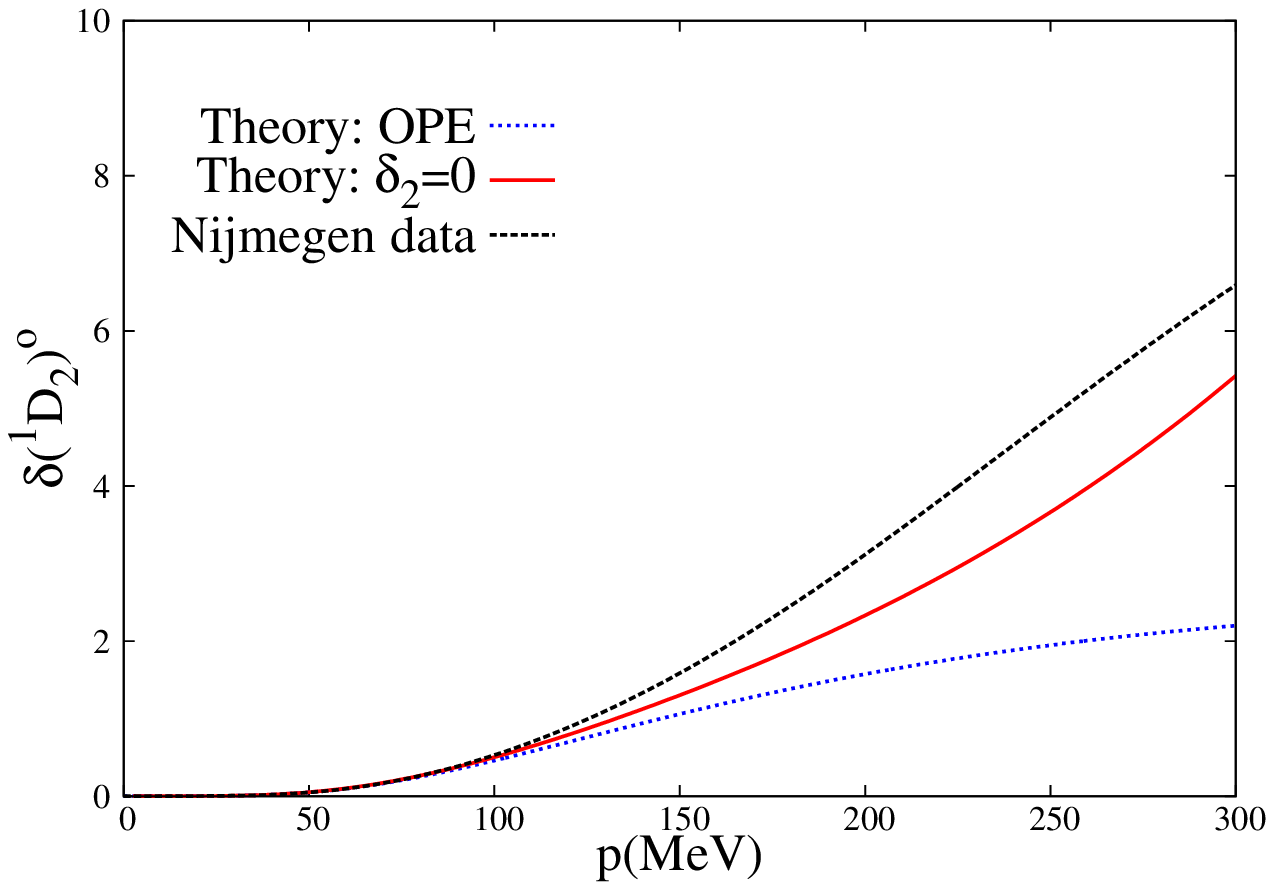} & 
\includegraphics[width=.4\textwidth]{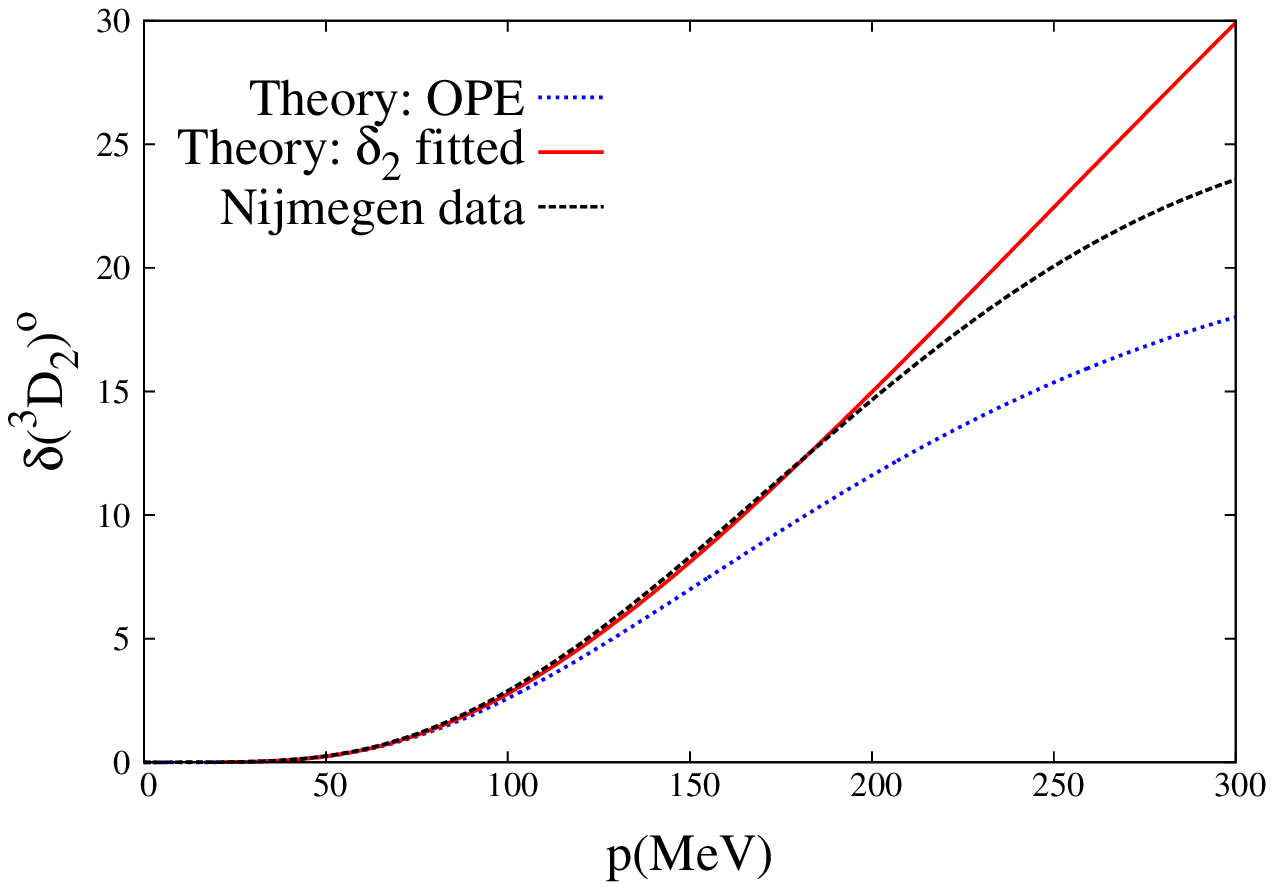}  
\end{tabular}
\caption[pilf]{\protect {\small (Color online.) Phase shifts for $^1D_2$ (left panel) and $^3D_2$ (right panel). 
$^1D_2:$ The (red) solid  line represents the NLO results with
  $\delta_2=0$. 
 $^3D_2$: The (red) solid line corresponds to $\delta_2$ fitted, Eq.~\eqref{fit3d2}.
 The OPE result from Ref.~\cite{paper1} is the (blue) dotted lines. 
The Nijmegen PWA  is the (black) dashed lines.}
\label{fig:dw} }
\end{center}
\end{figure}

\section{Uncoupled waves: $D$-waves}
\label{dw}

For the $D$-waves one has to solve  Eq.~\eqref{highd} with $\ell=2$.
 For the case of the singlet $^1D_2$ partial wave a fit to data is not appropriate here because 
it produces negative values of $\delta_2$, that in turn give rise to a resonance in the low-energy region, just a bit above the energy range fitted. 
 To avoid the resonance behavior we then impose that $\delta_2\gtrsim 0$. The (red) solid curve in Fig.~\ref{fig:dw} corresponds
 to $\delta_2=0~M_\pi^{-2}$.  Though there is a clear improvement compared with the OPE results of Ref.~\cite{paper1} (blue dotted line),  higher order corrections are still needed to provide an accurate reproduction of data.

We follow the same steps  for the $^3D_2$ partial wave. In this case we observe  a  numerical behavior not seen before when solving Eq.~\eqref{highd}. There is a dependence on the lower limit of integration along the LHC that can be reabsorbed, however, in the value of the free parameter $\delta_2$. In this way, the resulting phase shifts below $\sqrt{A}=300~$MeV are stable under changes in the lower limit of integration. The phase shifts with $\sqrt{A}<200$~MeV are fitted with 
\begin{align}
\delta_2=-0.18^{+0.02}_{-0.01}~M_\pi^{-2}~,
\label{fit3d2}
\end{align}
 where the errors show the variation in this parameter when the lower limit of integration varies from $-4^2$ to $-187^2$~GeV$^2$. The phase shifts obtained are the (red) solid line in the right panel of Fig.~\ref{fig:dw} (the other lines obtained with the different values mentioned for the lower limit of integration overlap each other and cannot be distinguished in the scale of the figure.)  We also see  a clear improvement in the reproduction of  data when moving from OPE to TPE, specially for $\sqrt{A}<200$~MeV.

\section{Uncoupled waves: $F$-waves}
\label{fw}

\begin{figure}
\begin{center}
\begin{tabular}{cc}
\includegraphics[width=.4\textwidth]{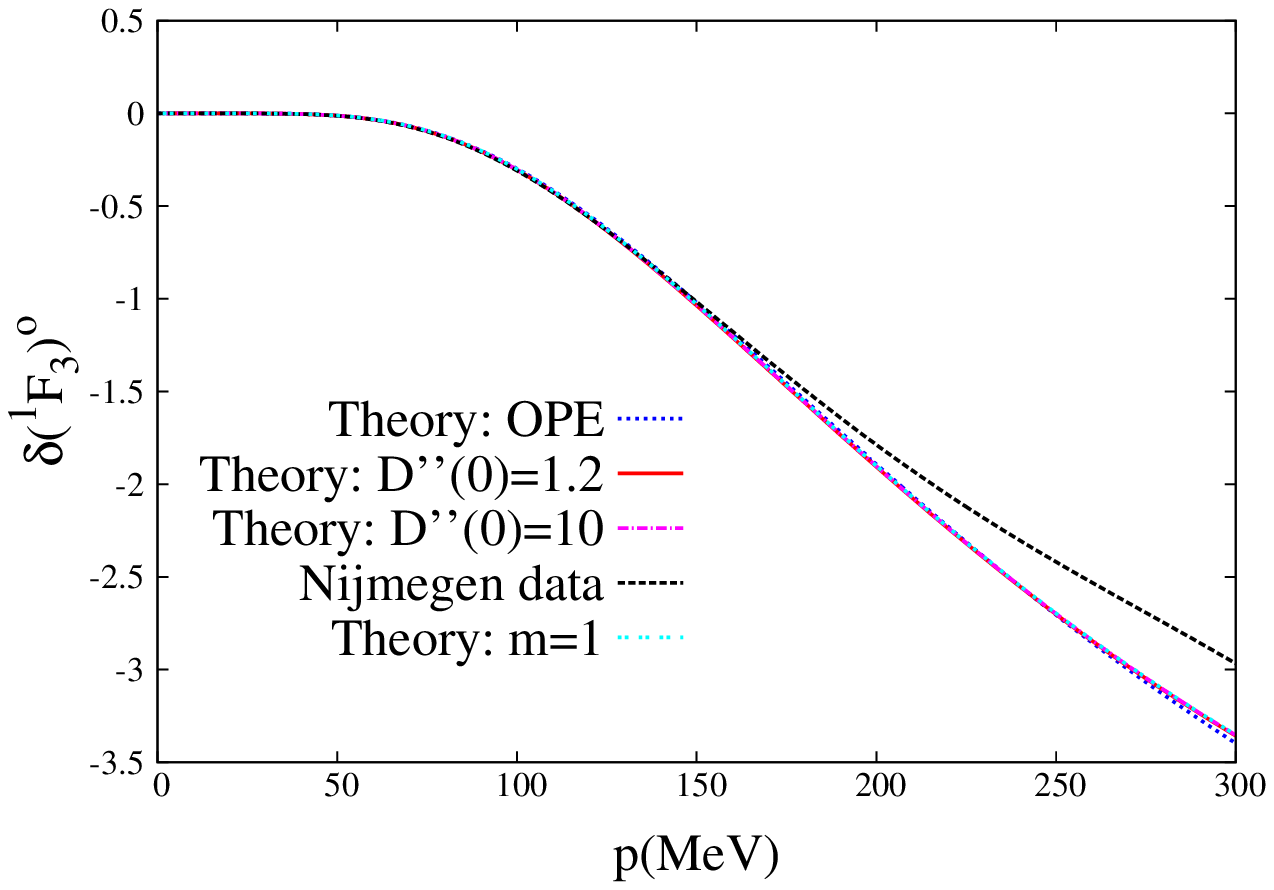} & 
\includegraphics[width=.4\textwidth]{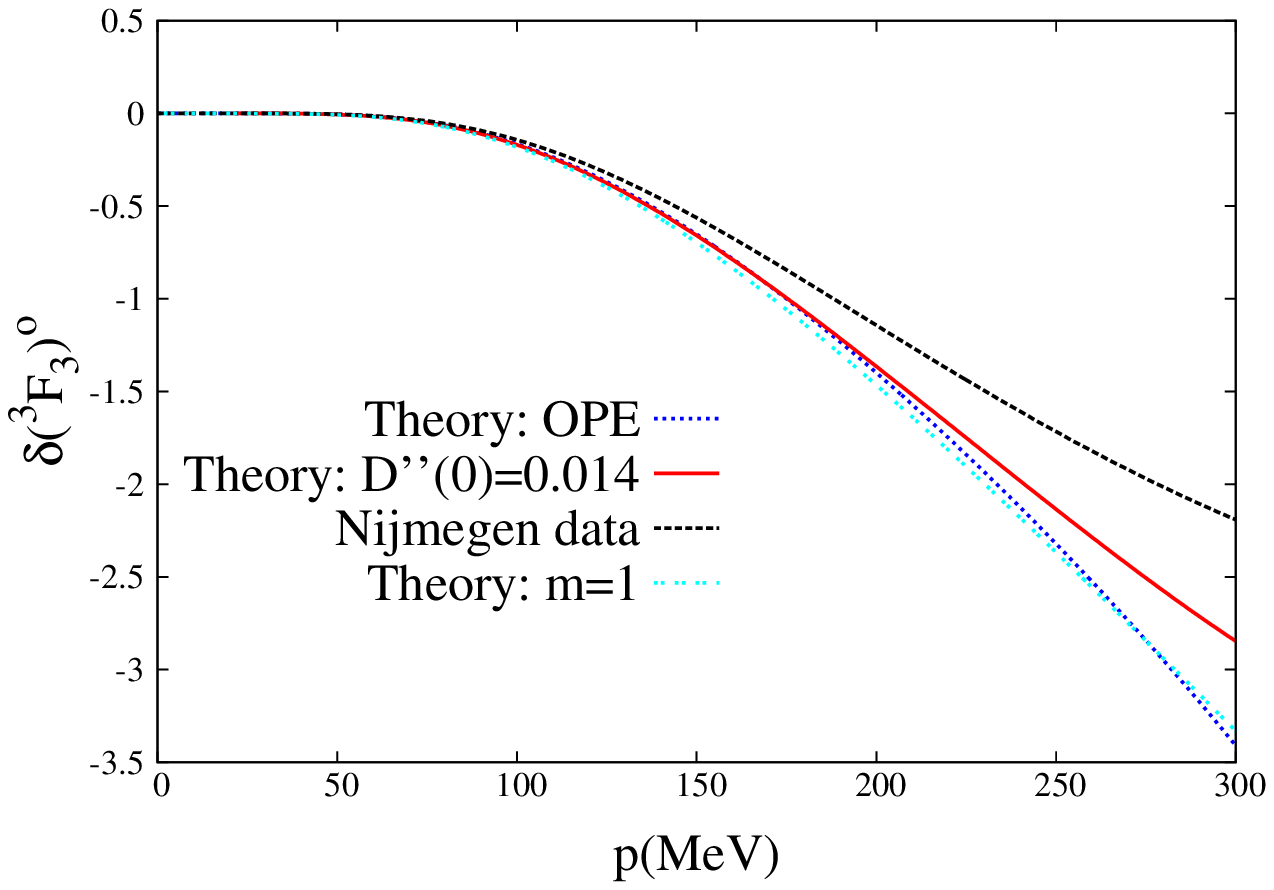}  
\end{tabular}
\caption[pilf]{\protect {\small (Color online.) Phase shifts for $^1F_3$ (left panel) and $^3F_3$ (right panel). 
$^1F_3:$ The (red) solid and (magenta) dash-dotted lines correspond to the NLO results with $D''(0)=1.7$ and $10~M_\pi^{-4}$. 
$^3F_3$: The (red) solid line is for $D''(0)=0.014~M_\pi^{-4}$. In both cases, the once-subtracted DR phase shifts, from  Eq.~\eqref{onceDNl}, are 
given by the (cyan) double-dotted lines. The OPE result from Ref.~\cite{paper1} is the (blue) dotted line. 
The Nijmegen PWA  is the (black) dashed line.}
\label{fig:fw} }
\end{center}
\end{figure}

Here we study the uncoupled $F$ waves, namely, $^1F_3$ and $^3F_3$.
 For these waves Eq.~\eqref{highd} is applied with $\ell=3$ and it requires three subtractions, with two free parameters $\delta_2$ and $\delta_3$, proportional to $D'(0)$ and $D''(0)$, in that order, according to Eq.~\eqref{deltai.dv}.  In the following we use the derivatives $D^{(n)}(0)$ as free parameters, which we consider more natural parameters for the polynomial in front of the integral in Eq.~\eqref{highd}. 

The partial wave $^1F_3$ is quite insensitive to $D'(0)$ and slightly dependent on $D''(0)$, which is required to be positive for a better reproduction of data. 
 We fix $D'(0)=0$ in the following, and show  
by the (red) solid line  in the left panel of Fig.~\ref{fig:fw} 
the outcome with $D''(0)=1.2~M_\pi^{-4}$, the resulting value of a fit to data up to $\sqrt{A}=150$~MeV. In turn, the (magenta) dash-dotted line corresponds to take $D''(0)=10~M_\pi^{-4}$. Despite the large variation in the value of $D''(0)$ the two lines overlap  each other, which clearly shows how little  the results depend on the actual values of $D''(0)$.  The OPE results are quite similar as the NLO ones. 

The $^3F_3$  partial wave is also insensitive to $D'(0)$, but the fit clearly prefers a value for $D''(0)$ around $0.014~M_\pi^{-4}$. 
The outcome at NLO is shown by the (red) solid line. One observes a clear improvement in the reproduction of data from OPE to NLO.

The fact that for both waves we only need to fit $D''(0)$, with $D'(0)$ fixed to zero,  illustrates the principle of maximal smoothness for $D(A)$ 
for high $\ell$. 
One can check  whether the $F$-waves could be already treated in perturbation theory. For that we propose to use the once-subtracted DR, Eq.~\eqref{onceDNl}, that has no the right threshold behavior which requires a partial wave to vanish as $A^3$ when $A\to 0$.  The origin for the failure to reproduce the proper threshold behavior stems from the resummation of the right-hand-cut undertaken by the $D(A)$ function. For a perturbative wave (Born approximation) unitarity requirements should be of little importance.  The outcome from Eq.~\eqref{onceDNl}  is shown by the (cyan) 
 double-dotted lines in Fig.~\ref{fw}, which run very close to the (red) solid lines, our 
NLO results that implement by construction the correct threshold behavior.  The $^3F_3$ wave seems less perturbative than the $^1F_3$, because the once-subtracted DR provides results that are more different compared with the full results. Had we applied the once-subtracted DR for the $D$-waves the outcome would have been very different from the results discussed in Sec.~\ref{dw} and shown in 
Fig.~\ref{fig:dw} (particularly for the $^3D_2$ that would not even match the correct sign). This indicates that the $F$-waves can be treated in good approximation in perturbation theory, while this is not the case for the $D$ waves yet. 
 A similar conclusion was also reached in Ref.~\cite{peripheral} by its perturbative study of $NN$ scattering 
 in one-loop baryon ChPT. 

\section{Uncoupled waves: $G$-waves}
\label{gw}
\begin{figure}
\begin{center}
\begin{tabular}{cc}
\includegraphics[width=.4\textwidth]{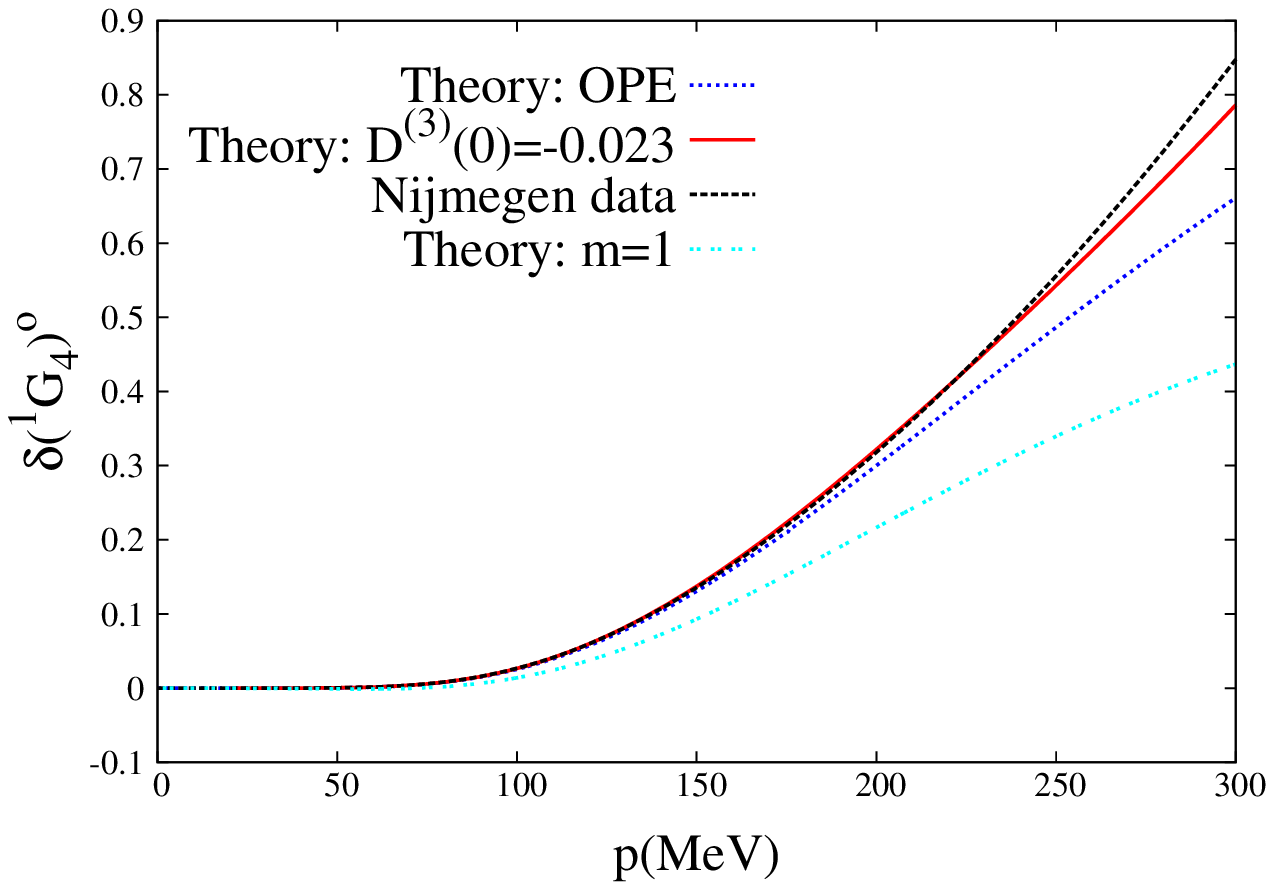} & 
\includegraphics[width=.4\textwidth]{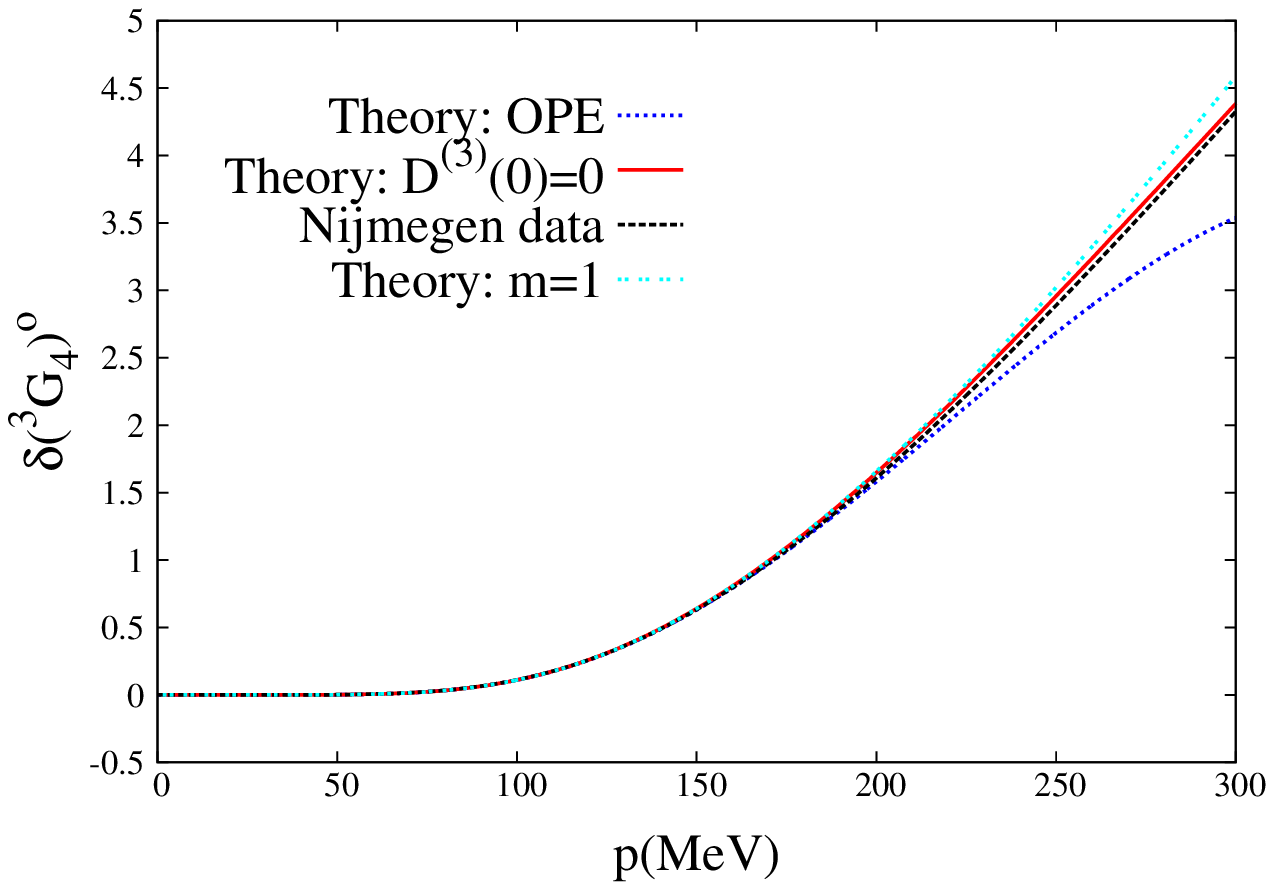}  
\end{tabular}
\caption[pilf]{\protect {\small (Color online.) Phase shifts for $^1G_4$ (left panel) and $^3G_4$ (right panel).
Full results are the (red) solid lines. The once-subtracted DR phase shifts, from  Eq.~\eqref{onceDNl}, are 
given by the (cyan) double-dotted lines. The OPE result from Ref.~\cite{paper1} is the (blue) dotted line. 
The Nijmegen PWA is the (black) dashed line.}
\label{fig:gw} }
\end{center}
\end{figure}

We now proceed to discuss the $G$-waves and solve Eq.~\eqref{highd} with $\ell=4$. The situation here follows the general rule discussed in Sec.~\ref{leq2},
 so that it is enough to release only $D^{(3)}(0)$, which acts then as  
 the active degree of freedom, with the other $D^{(i)}(0)$, with $i=1,~2$,
 fixed to zero. 
From the best fits obtained with one free parameter $D^{(3)}(0)$, if we release further the other two parameters, $D'(0)$ and $D''(0)$, no improvement is obtained. 
For the partial  wave $^1G_4$ we obtain from the fit to data the value $D^{(3)}(0)\simeq -0.031~M_\pi^{-6}$. However, for the $^3G_4$ wave the
 fit cannot pin down a precise value for  $D^{(3)}(0)$, which is finally fixed to zero.
In both waves the reproduction of data is very good as shown in Fig.~\ref{fig:gw} by the (red) solid line. 
 The $^3G_4$ wave is the most perturbative one,
  as one can see by the fact that the once-subtracted DR results (shown by the cyan double-dotted lines in Fig.~\ref{fig:gw}) are clearly
 closer to the full results than for the $^1G_4$ case.

\section{Uncoupled waves: $H$-waves}
\label{hw}
\begin{figure}
\begin{center}
\begin{tabular}{cc}
\includegraphics[width=.4\textwidth]{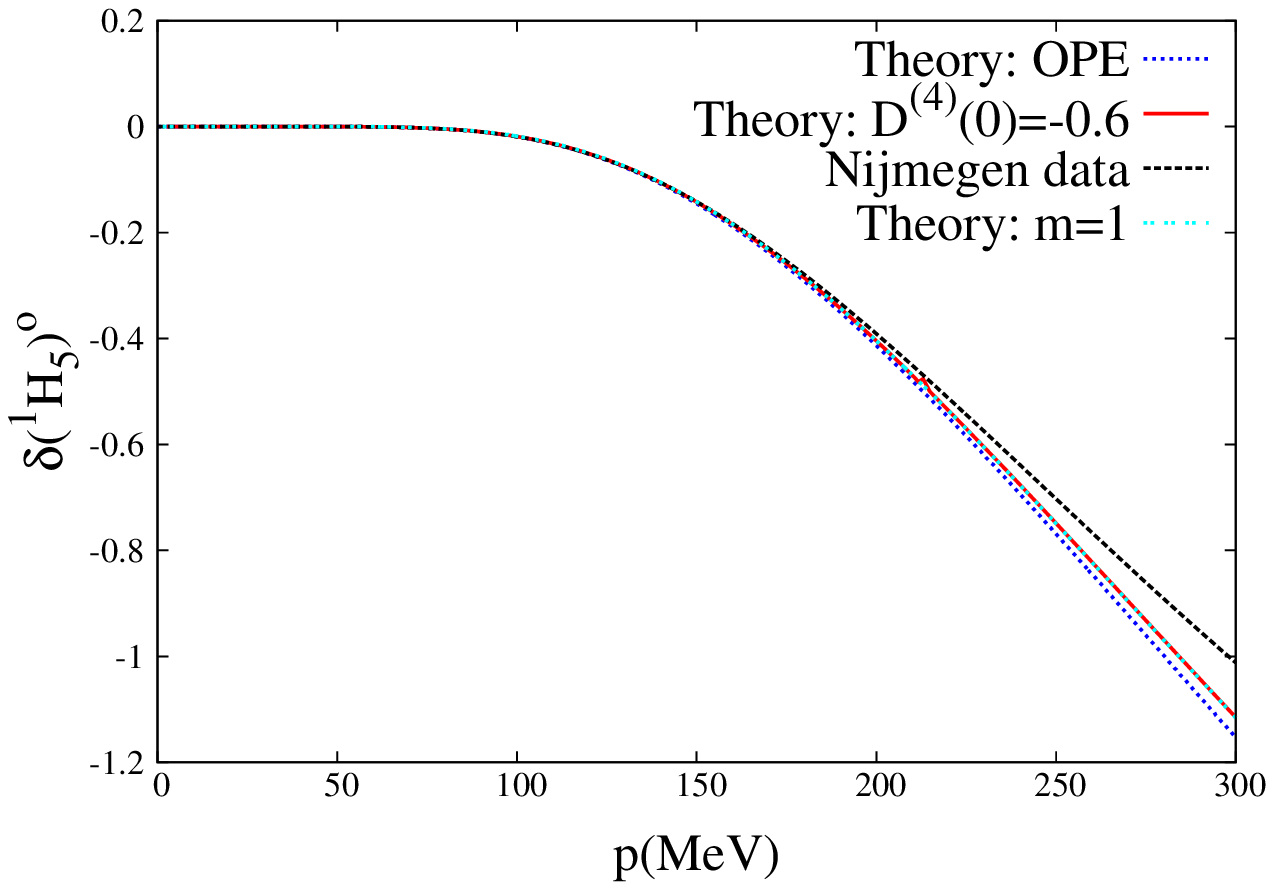} & 
\includegraphics[width=.4\textwidth]{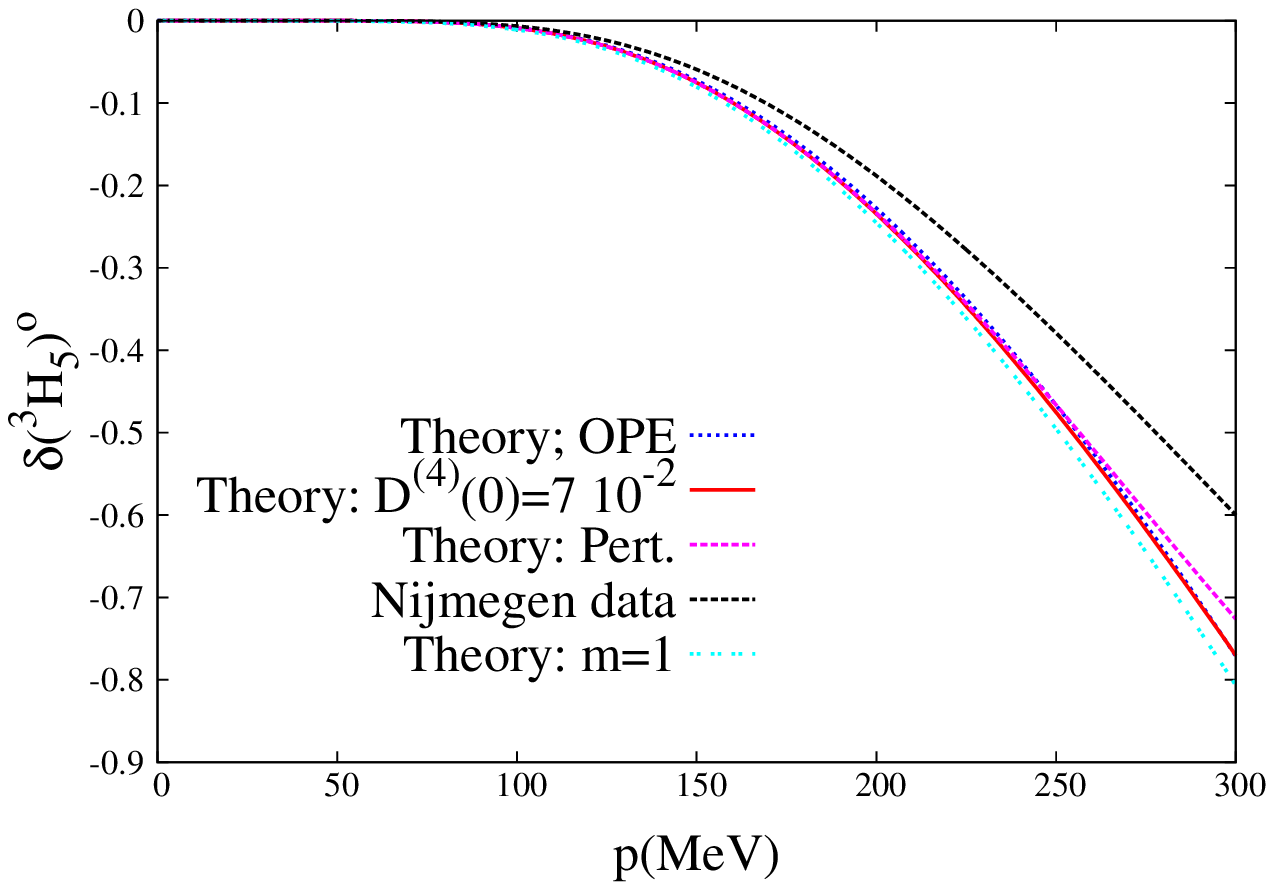}  
\end{tabular}
\caption[pilf]{\protect {\small (Color online.) Phase shifts for $^1H_5$ (left panel) and $^3H_5$ (right panel).
Full results are the (red) solid lines. The once-subtracted DR phase shifts, from  Eq.~\eqref{onceDNl}, are 
given by the (cyan) double-dotted lines. The OPE result from Ref.~\cite{paper1} is the (blue) dotted line. 
The Nijmegen PWA  is the (black) dashed line.}
\label{fig:hw} }
\end{center}
\end{figure}

The same rule of Sec.~\ref{leq2} regarding the number of active free parameters is observed here for $\ell=5$ as in the case of the 
 $F$- and $G$-waves, so that we only release $D^{(4)}(0)$. 
For the $^1H_5$ the best results are obtained with $D^{(4)}(0)= -0.6~M_\pi^{-8}$, corresponding to the (red) solid line in the left panel of Fig.~\ref{fig:hw}.
 The results reproduce the Nijmegen PWA phase shifts fairly well. The lines obtained from OPE \cite{paper1} and 
by employing a once-subtracted DR run close to our full ones at NLO. 
For the $^3H_5$ wave we obtain the best value $D^{(4)}(0)\simeq 0.7\cdot 10^{-2}~M_\pi^{-8}$, that gives rise to results  slightly better than 
by fixing it directly to zero. The phase shifts obtained  are shown by the (red) solid line 
 in the right panel of Fig.~\ref{fig:hw}. They are quite close 
to the Nijmegen PWA phase shifts in the range shown, note also the small absolute value of the phase shifts.\footnote{For $5\leq J\leq 8$ the Nijmegen PWA phase shifts \cite{Stoks:1994wp} are those obtained from the $NN$ potential model of Ref.~\cite{obe}.}   
The once-subtracted DR and OPE results are very similar between them and run rather close to the full results, 
indicating the perturbative nature of the $H$-waves.

\section{Quantifying contributions to $ \Delta(A)$}
\label{cont_da}

For any  partial  wave there is always a term, corresponding to the last line in Eq.~\eqref{inteq1}, that gives the nested contribution of the LHC to the function $D(A)$. This type of integration along the LHC is the proper one to ascertain the relative size of the different contributions to $\Delta(A)$, because any scattering quantity can be calculated once the  function $D(A)$ is known along 
the LHC.  It is then not  illuminating to look directly at the relative sizes of the different contributions to $\Delta(A)$, but better one should look at the amount that they  contribute to the integral along the LHC. Since this integration involves the very same function that we want to calculate, we evaluate it by substituting  $D(k^2)\to 1$, although any other constant value would be equally valid to ascertain relative differences. In this way, we can then perform an a  priori  quantitative study about the importance of the different contributions in $\Delta(A)$ when solving Eq.~\eqref{inteq1}. 

At the practical level we have used Eq.~\eqref{inteq1} with changes in its form because of different selections of the subtraction point $C$, as explained above. We display in Eq.~\eqref{quanty} the integrals used for each wave to quantify the weight in our results of the different contributions to $\Delta(A)$. All the integrals require two or more subtractions so that they are convergent, due to the fact that at NLO $\Delta(A)$ diverges at most as $ A$ for $A\to \infty$.
 Indeed twice- or more subtracted DRs have been used in all the partial waves in Secs.~\ref{1s0}--\ref{hw}.
\begin{align}
\ell \leq 1~:~ & \frac{A(A+M_\pi^2)}{\pi^2}\int_{-\infty}^L dk^2\frac{\Delta(k^2)}{(k^2)^2}\int_0^\infty dq^2
\frac{q^2\rho(q^2)}{(q^2-A)(q^2-k^2)(q^2+M_\pi^2)}~,\nn\\
\ell\geq 2~:~ & \frac{A^\ell}{\pi^2}\int_{-\infty}^L dk^2\frac{\Delta(k^2)}{(k^2)^\ell}\int_0^\infty dq^2
\frac{\rho(q^2)}{(q^2-A)(q^2-k^2)}~.
\label{quanty}
\end{align}

\begin{figure}
\begin{center}
\begin{tabular}{cc}
\includegraphics[width=.4\textwidth]{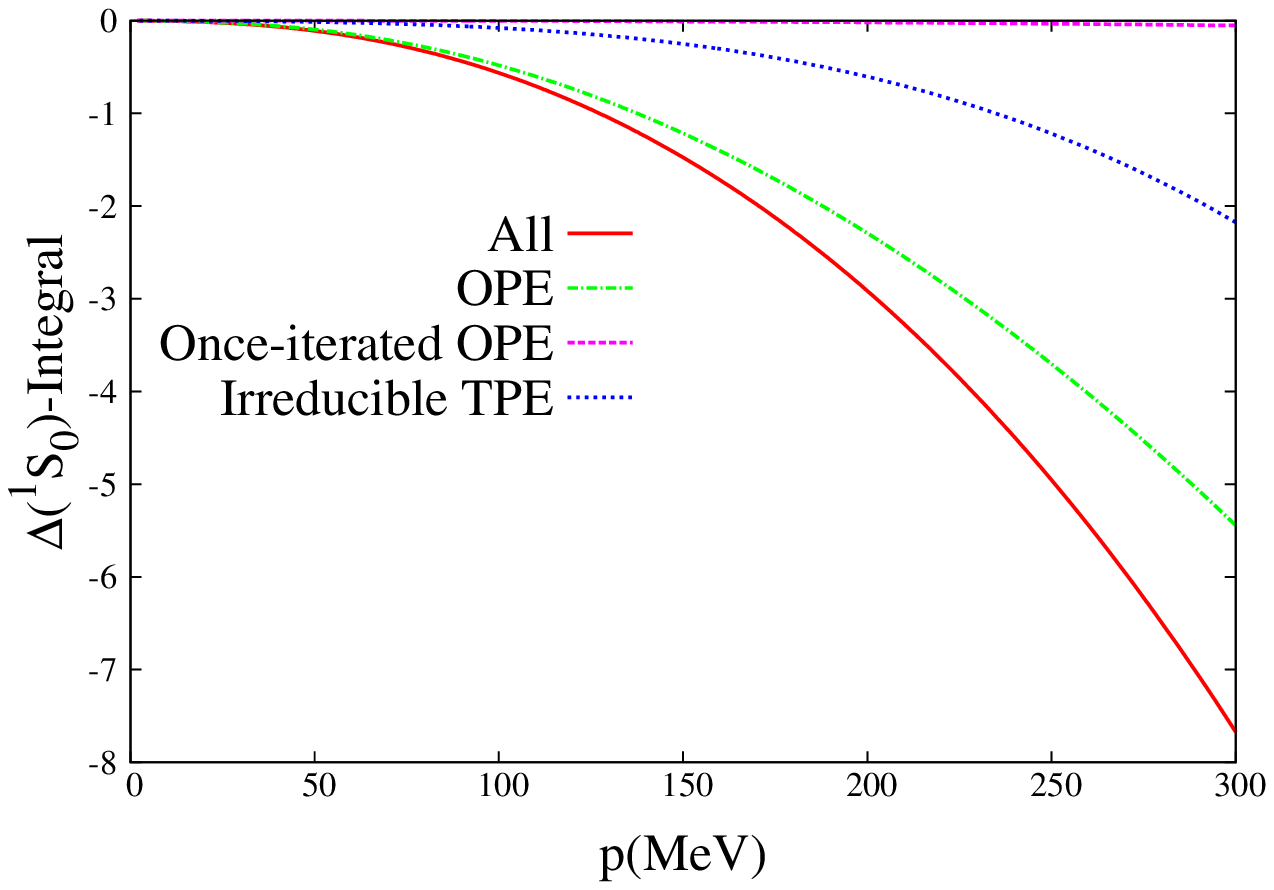} & 
\includegraphics[width=.4\textwidth]{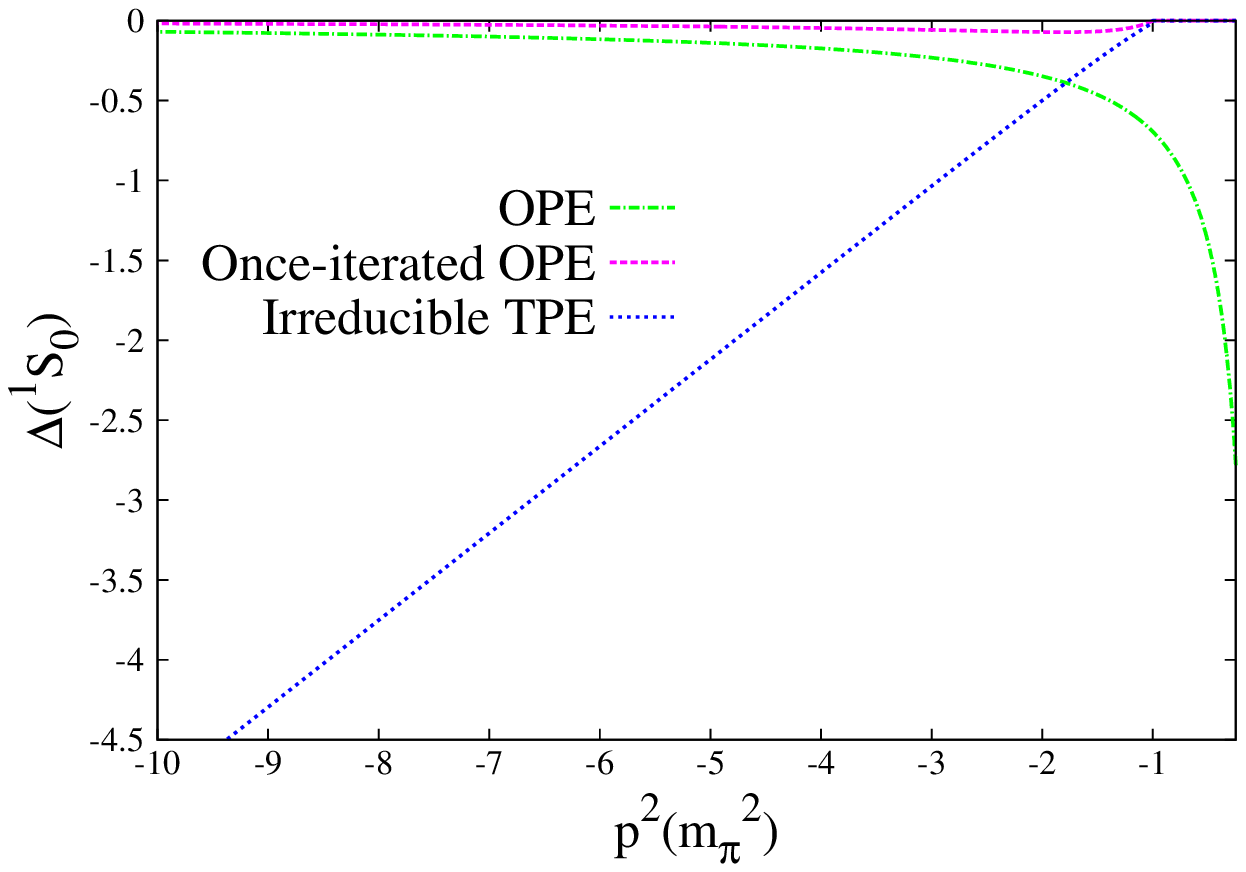}
\end{tabular}
\caption{ {\small (Color online.) Left panel: different contributions to the integral in Eq.~\eqref{quanty} with $\ell=0$. Right panel: contributions to $\Delta(A)$. These contributions comprise irreducible TPE (blue dotted line), reducible TPE (magenta dashed line) and OPE (green dash-dotted line). The total result, only shown for the left panel, is the (red) solid line.}
\label{fig:1s0quanty}}
\end{center}
\end{figure}

Let us analyze first the case of the $^1S_0$. For that we show in the left panel of Fig.~\ref{fig:1s0quanty} the corresponding integral in Eq.~\eqref{quanty}, while in the right panel we plot directly $\Delta(A)$. In both cases we distinguish between  OPE (green dash-dotted line), irreducible TPE (blue dotted line) and reducible TPE (magenta dashed line). The total result is given for the integral (left panel) and it corresponds to the (red) solid line. We see that the integral is 
clearly dominated by the OPE contribution, despite  the irreducible TPE contribution overpasses OPE in $\Delta(A)$ at around $-2 M_\pi^{-2}$. The next contribution in importance is irreducible TPE and the least important by far is reducible TPE. The latter contribution is so much suppressed because for the $^1S_0$ it is proportional to $m_\pi^4$.  

  The dominance of OPE in the integral at low energies along the RHC is because: i) It starts to contribute the soonest in all of them; ii) the integrand in Eq.~\eqref{quanty} is enhanced at low three-momenta by the factor $1/(k^2)^2$ for $\ell\leq 1$. Because of these reasons every contribution to $\Delta(A)$ that involves the exchange of a larger number of pions should be increasingly suppressed.  Let us recall that precisely the threshold for each contribution to $\Delta(A)$  controls its exponential suppression  for large radial distances in the $NN$ potential, as $\exp(- n M_\pi r)$ for an $n$-pion exchange contribution. Notice also that one can see clearly in the right panel of 
Fig.~\ref{fig:1s0quanty} that  OPE increases  very fast in absolute value towards its threshold, at $-M_\pi^2/4$. This is because OPE at low energies has a typical value for its derivative proportional to $1/A^2$, which implies a large relative change between the onset of OPE and that of TPE. We can say from the left panel of Fig.~\ref{fig:1s0quanty} that it is justified  to calculate perturbatively the different contributions to $\Delta(A)$ for the $^1S_0$.

\begin{figure}
\begin{center}
\begin{tabular}{cc}
\includegraphics[width=.4\textwidth]{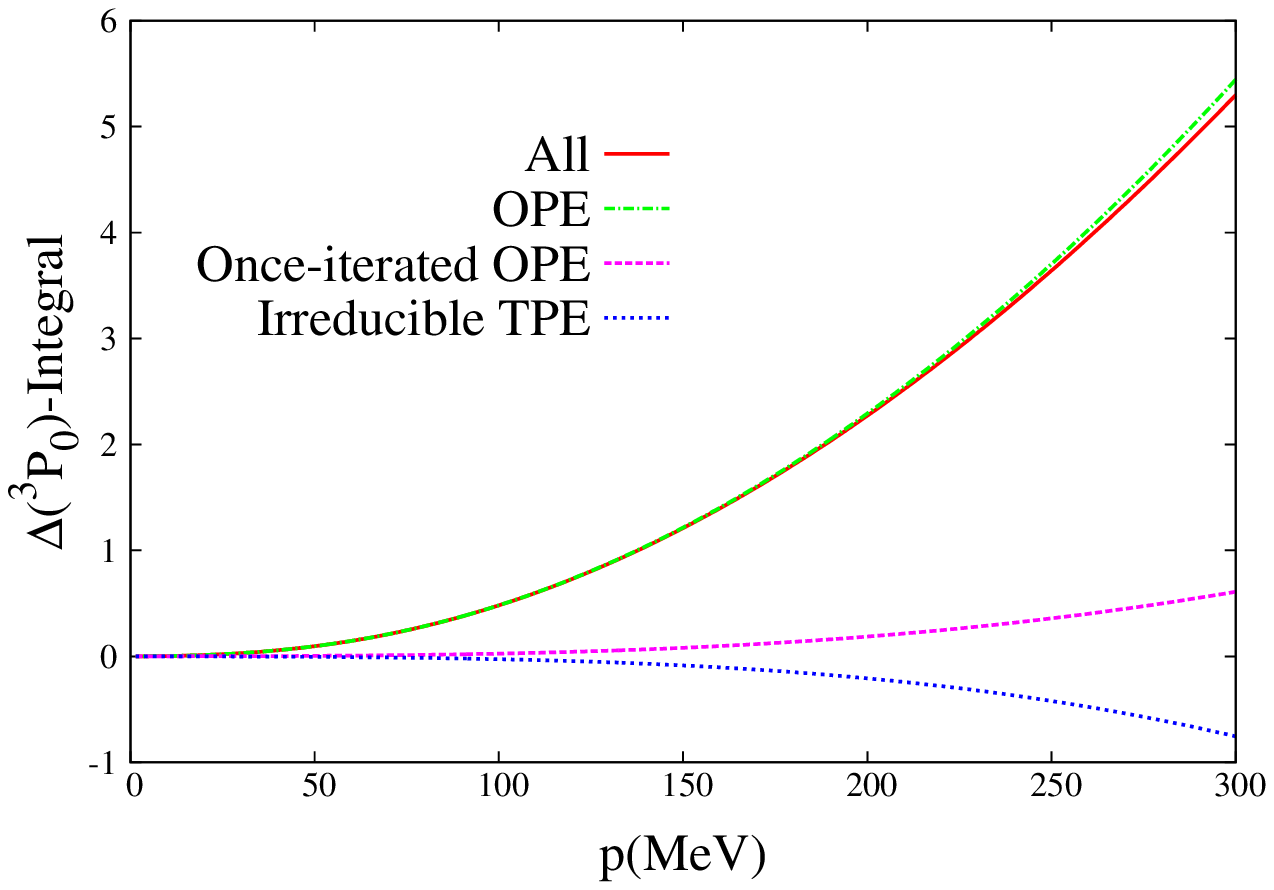} & 
\includegraphics[width=.4\textwidth]{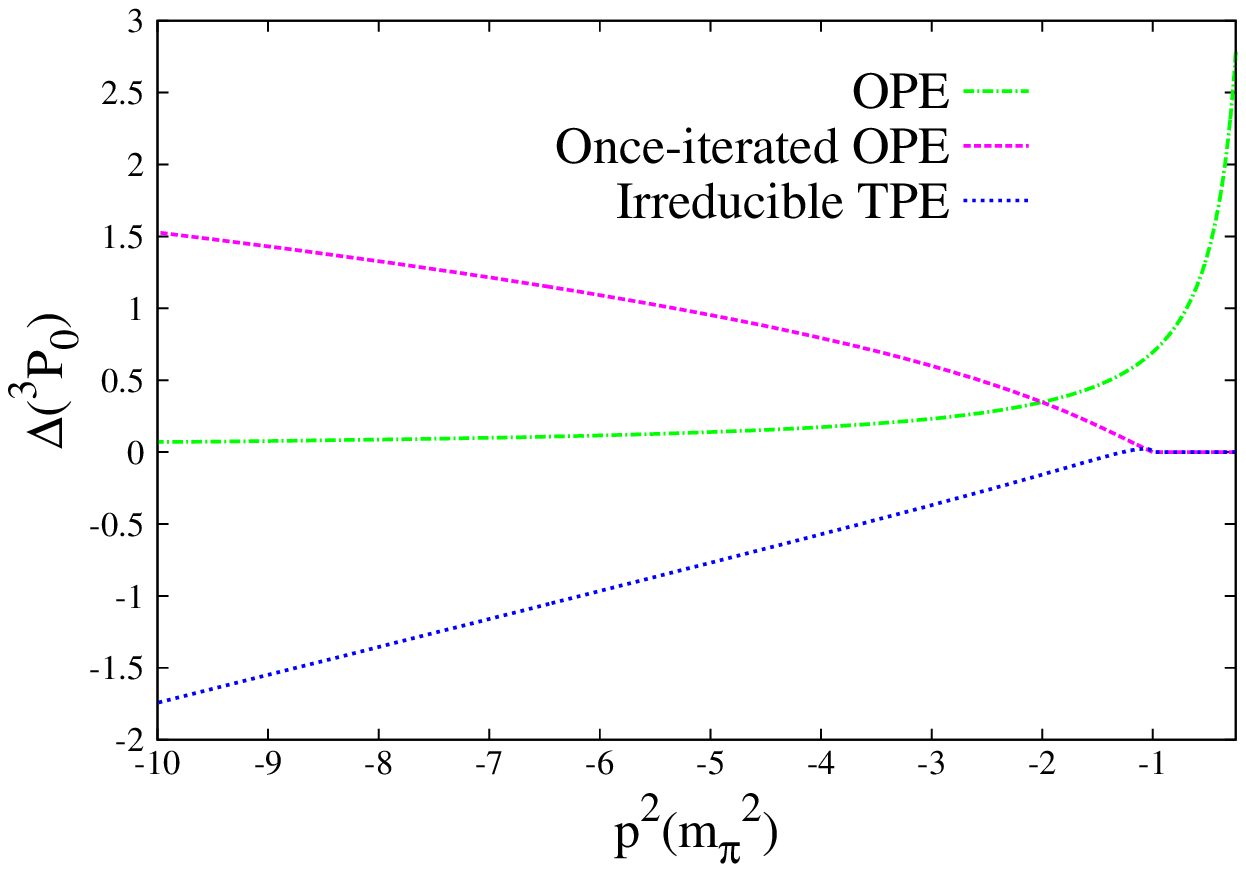}\\
\includegraphics[width=.4\textwidth]{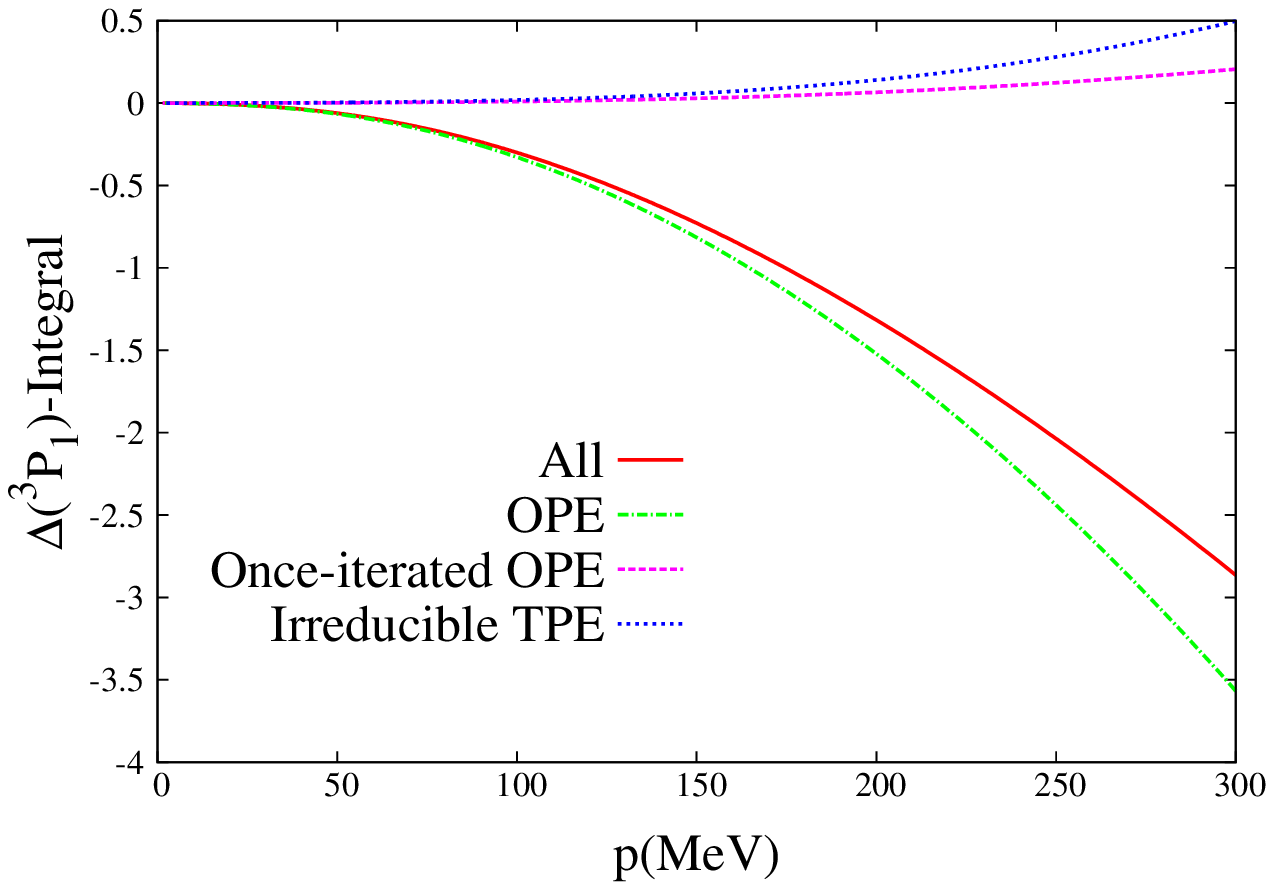} & 
\includegraphics[width=.4\textwidth]{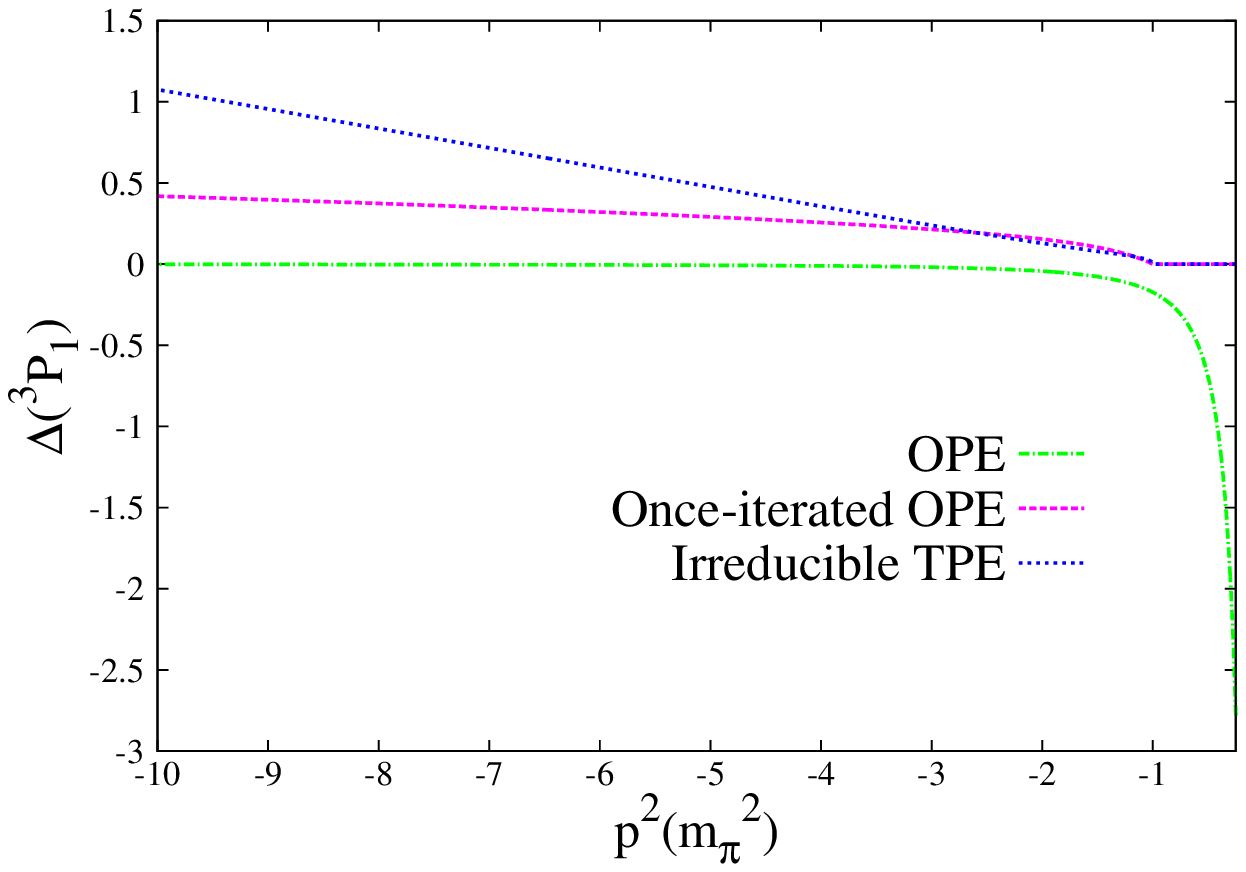}\\
\includegraphics[width=.4\textwidth]{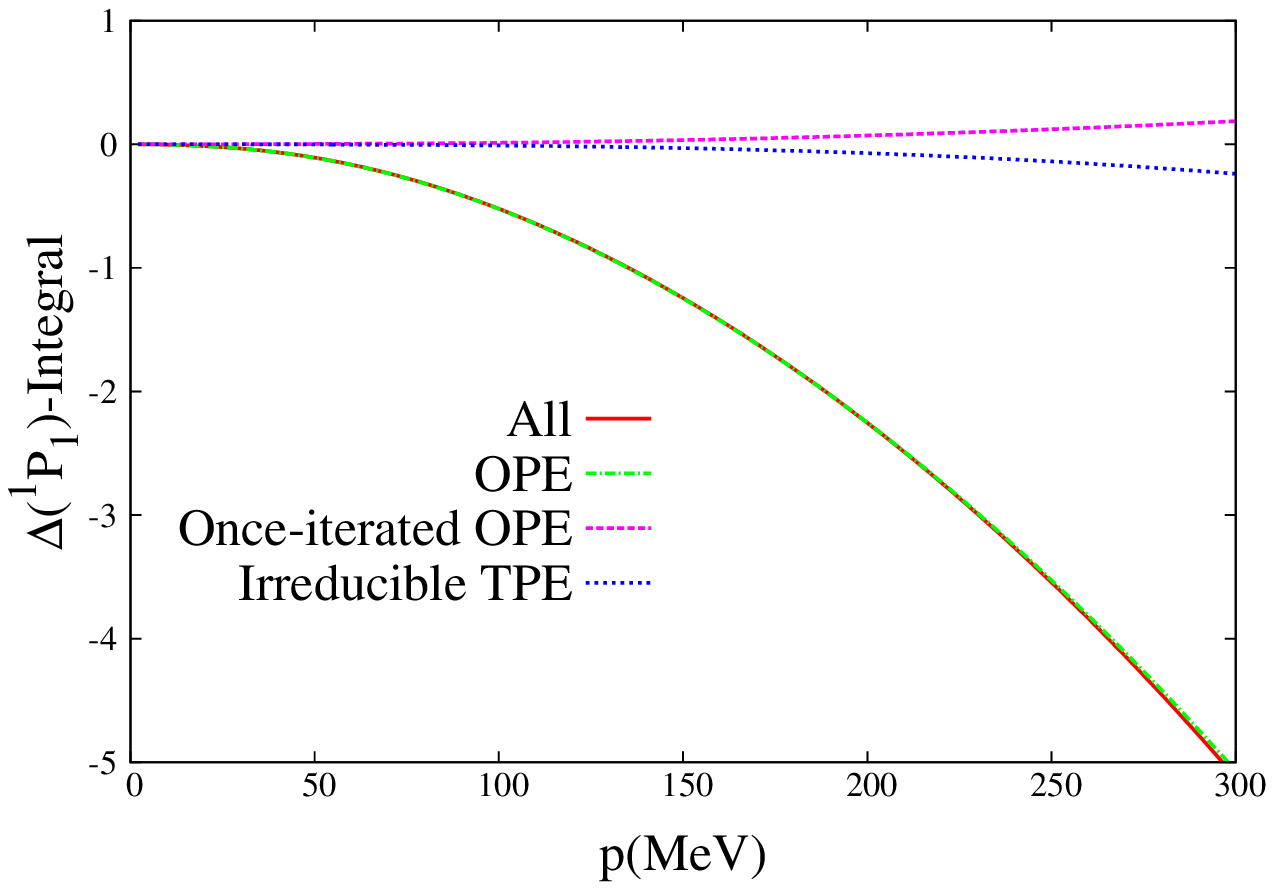} & 
\includegraphics[width=.4\textwidth]{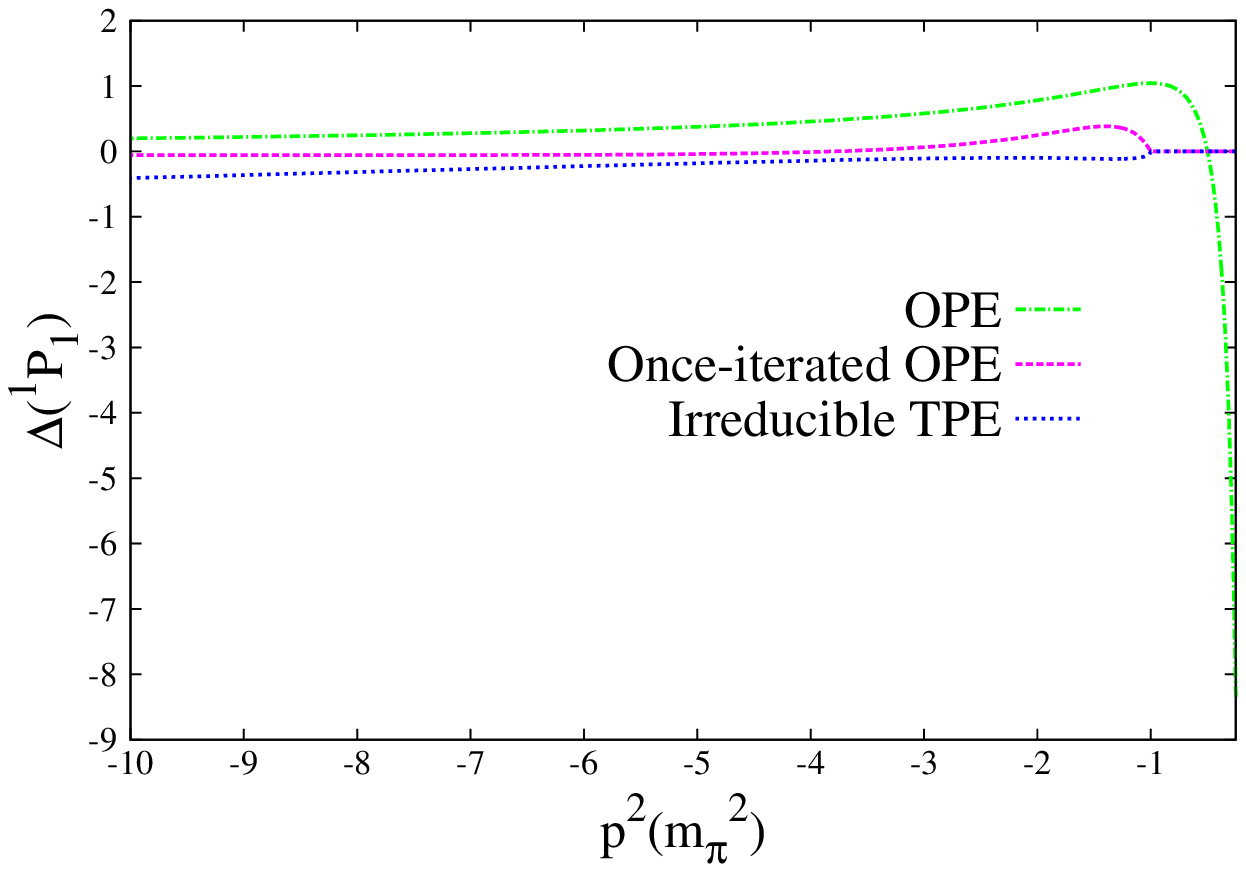}\\
\end{tabular}
\caption{ {\small (Color online.) Left panels: different contributions to the integral in Eq.~\eqref{quanty} with $\ell=1$. Right panels: Contributions to $\Delta(A)$. From top to bottom we show the $^3P_0$, $^3P_1$ and $^1P_1$ partial waves, respectively. The meaning of the lines is the same as in  Fig.~\ref{fig:1s0quanty}.}
\label{fig:pwquanty}}
\end{center}
\end{figure}

The case of the $P$-waves is shown in Fig.~\ref{fig:pwquanty}. From top to bottom we show the partial waves $^3P_0$, $^3 P_1$ and $^1P_1$, in that order. The left panels show the integral in Eq.~\eqref{quanty} and the right ones the different contributions to $\Delta(A)$. The notation is the same as used in Fig.~\ref{fig:1s0quanty}. 
 By comparing the (green) dash-dotted and (red) solid lines in the left panels of Fig.~\ref{fig:pwquanty} one clearly observes 
the dominance of the OPE contribution. 
For the $^3P_0$ wave both irreducible and reducible TPE are sizable but tend to cancel mutually. The actual extent of this cancellation could be sensitive to the exact values of the function $D(k^2)$ (substituted by 1 in the integral along the LHC in Eq.~\eqref{quanty}). We also observe that the irreducible and reducible TPE contributions are typically of similar size as a global picture for the $P$-waves. The pattern of results shown for the integral again suggests that a perturbative treatment for the different contributions to $\Delta(A)$, in the form discussed in Sec.~\ref{delta}, is meaningful.

\begin{figure}
\begin{center}
\begin{tabular}{cc}
\includegraphics[width=.4\textwidth]{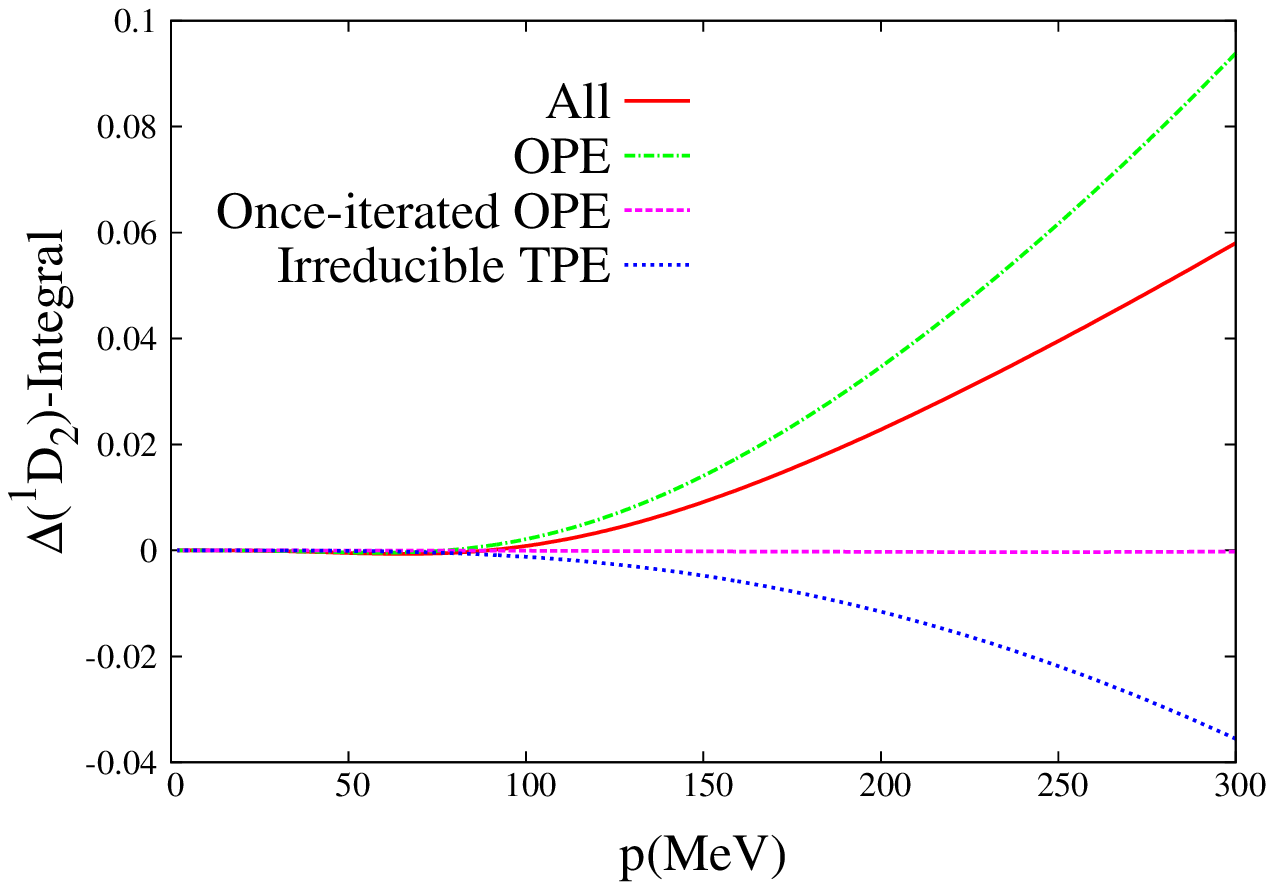} & 
\includegraphics[width=.4\textwidth]{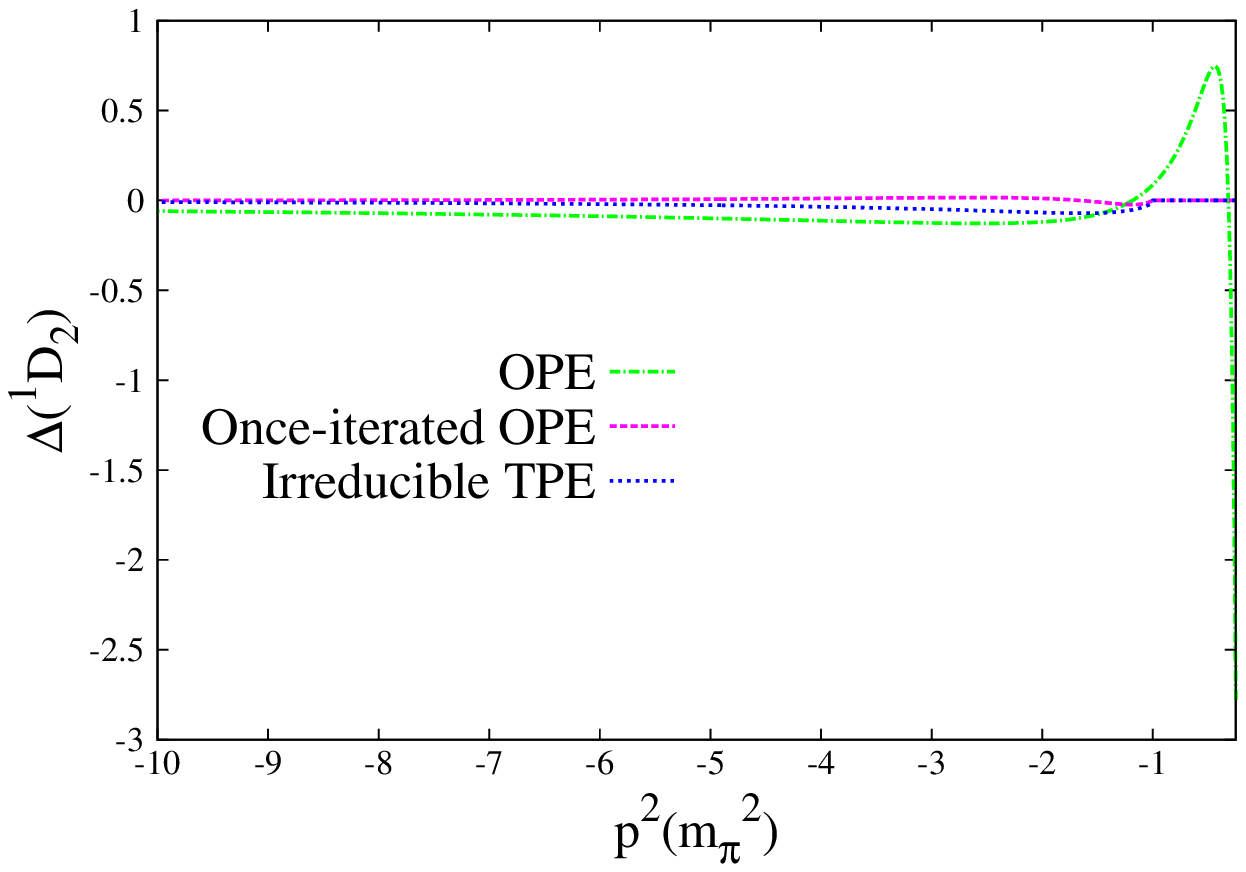}\\
\includegraphics[width=.4\textwidth]{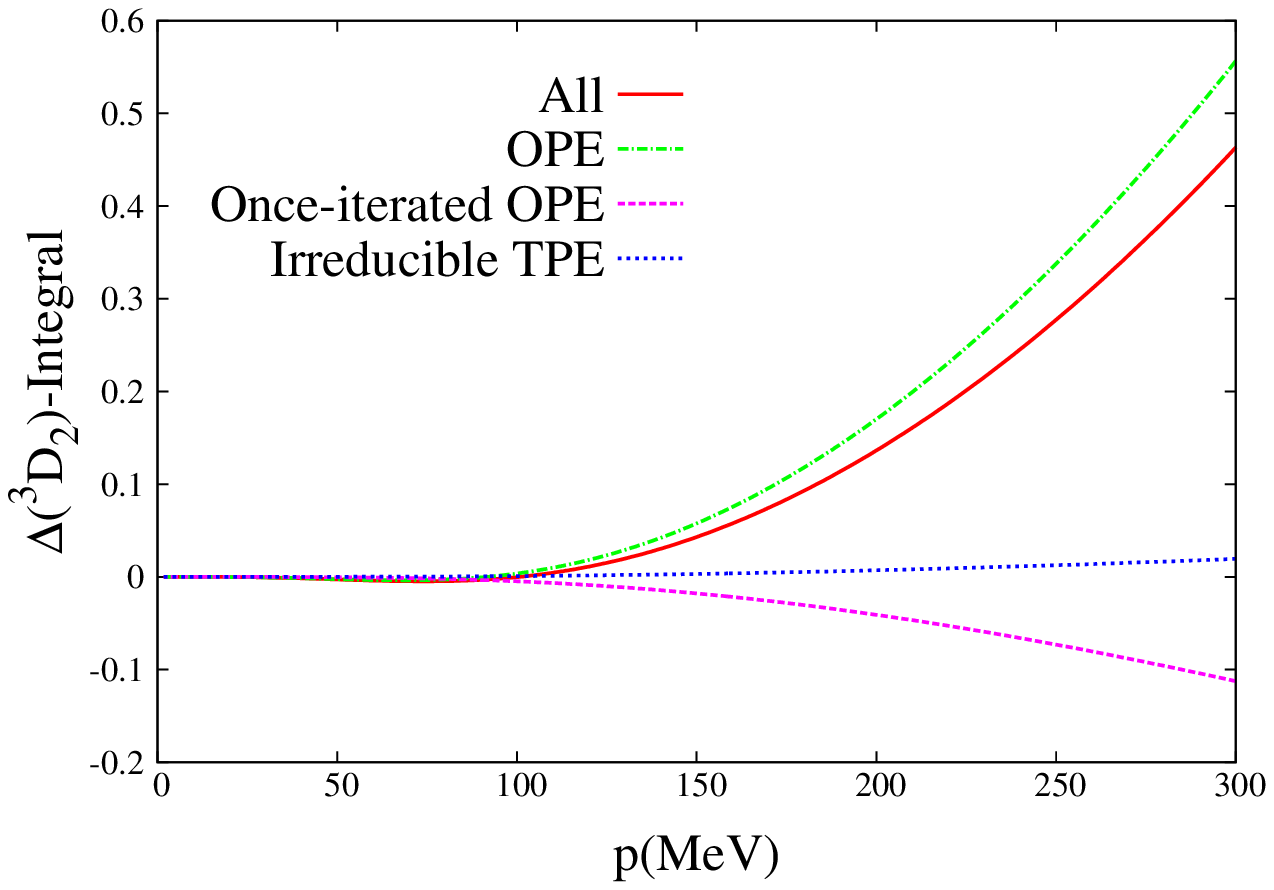} & 
\includegraphics[width=.4\textwidth]{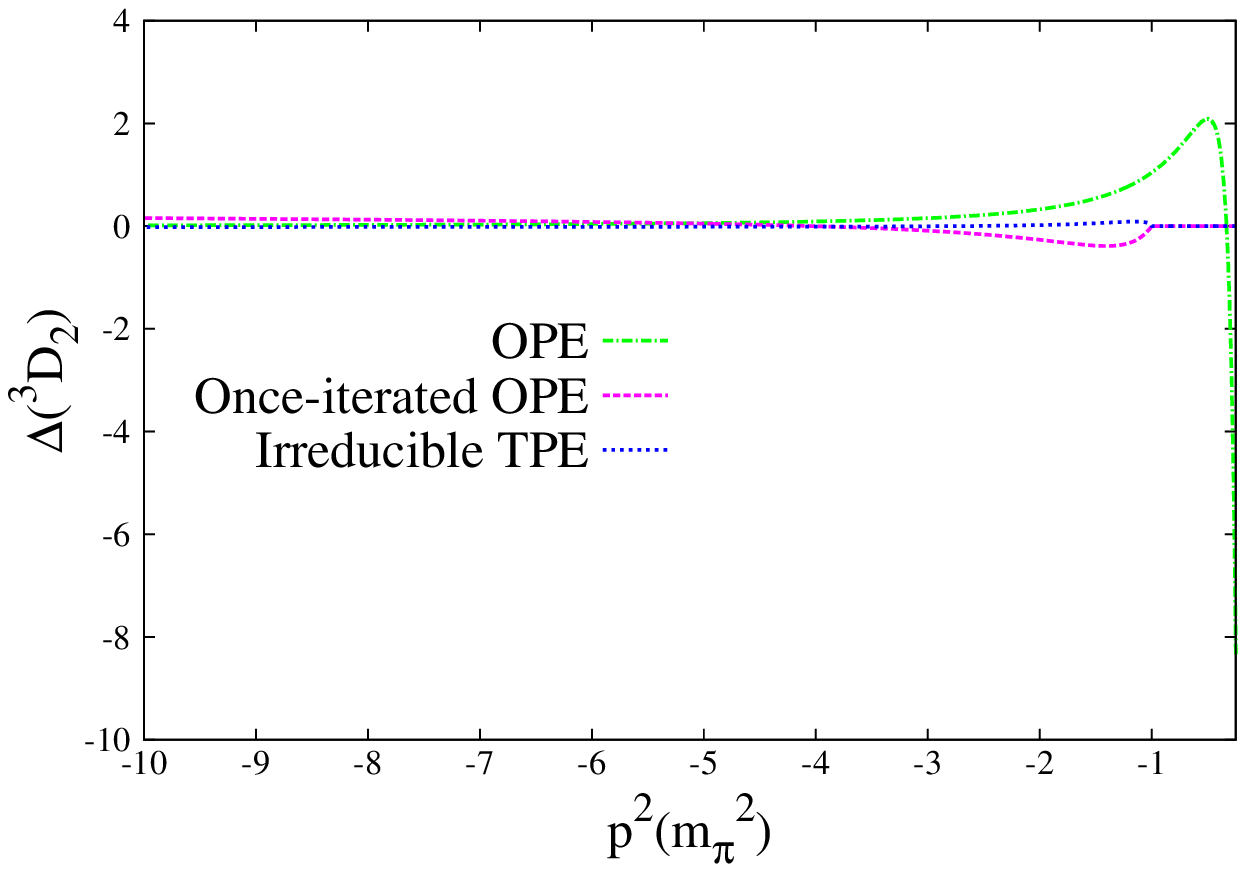}
\end{tabular}
\caption{ {\small (Color online.) Left panels: different contributions to the integral in Eq.~\eqref{quanty} with $\ell=2$. Right panels: Contributions to $\Delta(A)$. From top to bottom we show the $^1D_2$ and $^3D_2$ partial waves, respectively. 
The meaning of the lines is the same as in Fig.~\ref{fig:1s0quanty}.}
\label{fig:dwquanty}}
\end{center}
\end{figure}

The corresponding curves for the $D$-waves, $\ell=2$ in Eq.~\eqref{quanty}, are shown in Fig.~\ref{fig:dwquanty}. Again we observe a clear dominance of OPE in the integral of Eq.~\eqref{quanty}. For the $^1D_2$ wave the irreducible TPE is lager than the reducible contribution, but for the $^3D_2$ the situation is reversed. So we conclude that typically they should be considered 
 of similar size, as argued in Sec.~\ref{delta}.

\begin{figure}
\begin{center}
\begin{tabular}{cc}
\includegraphics[width=.4\textwidth]{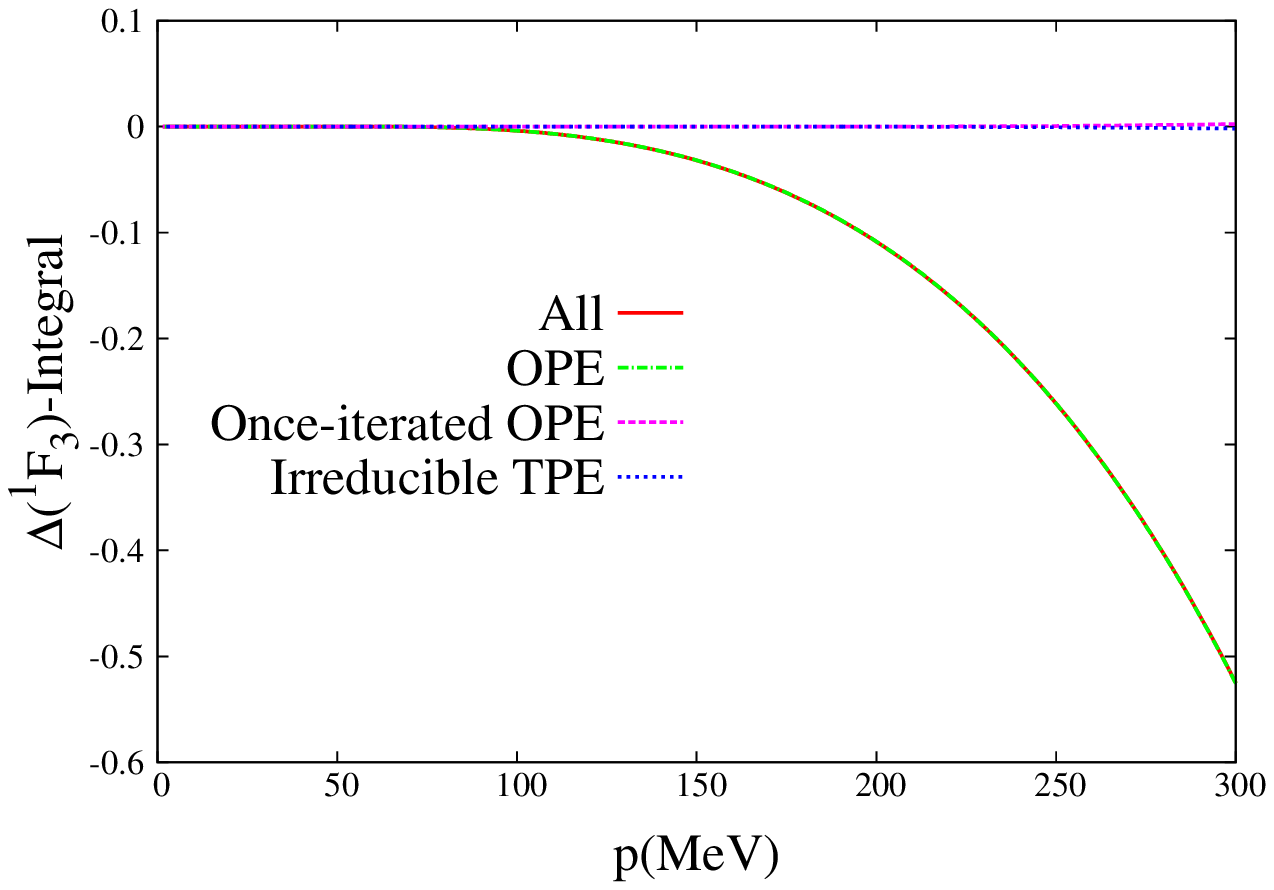} & 
\includegraphics[width=.4\textwidth]{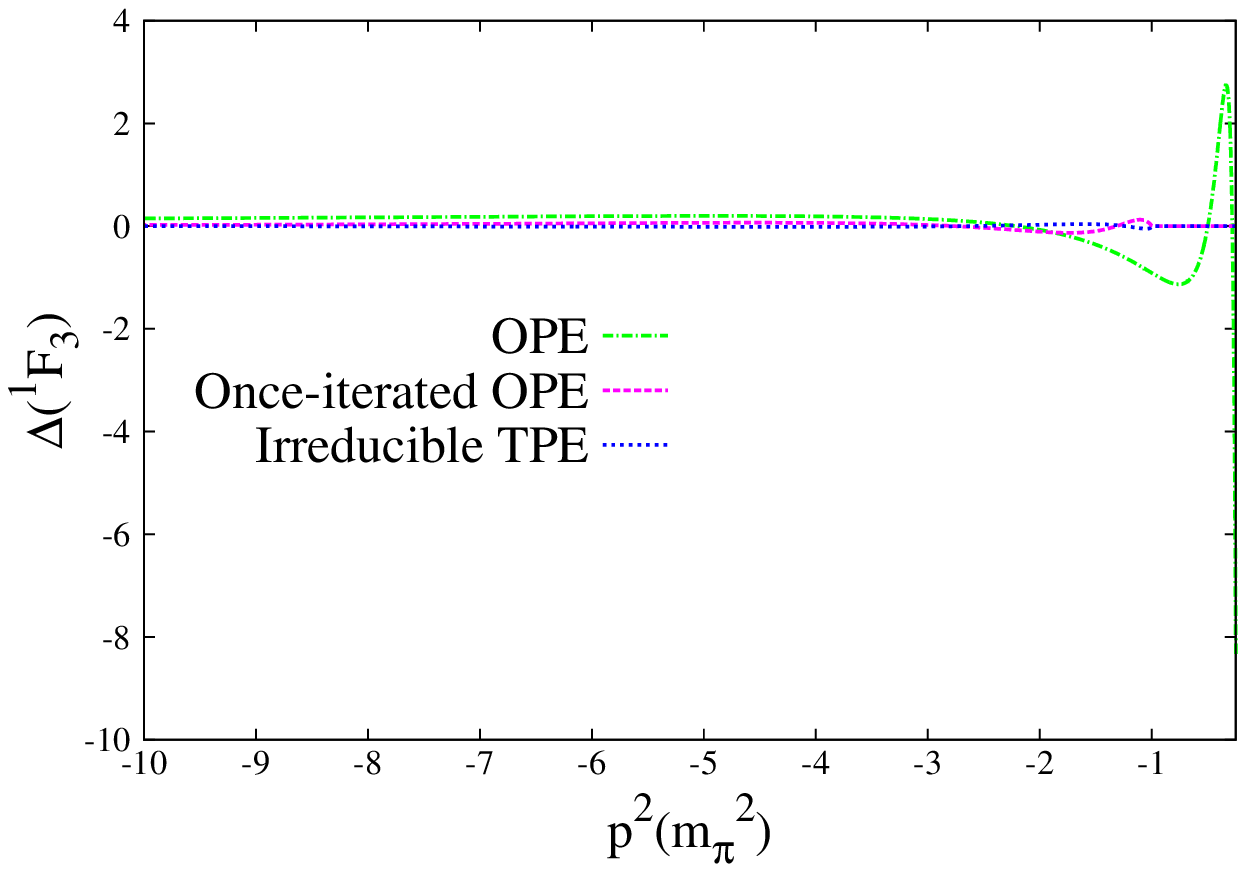}\\
\includegraphics[width=.4\textwidth]{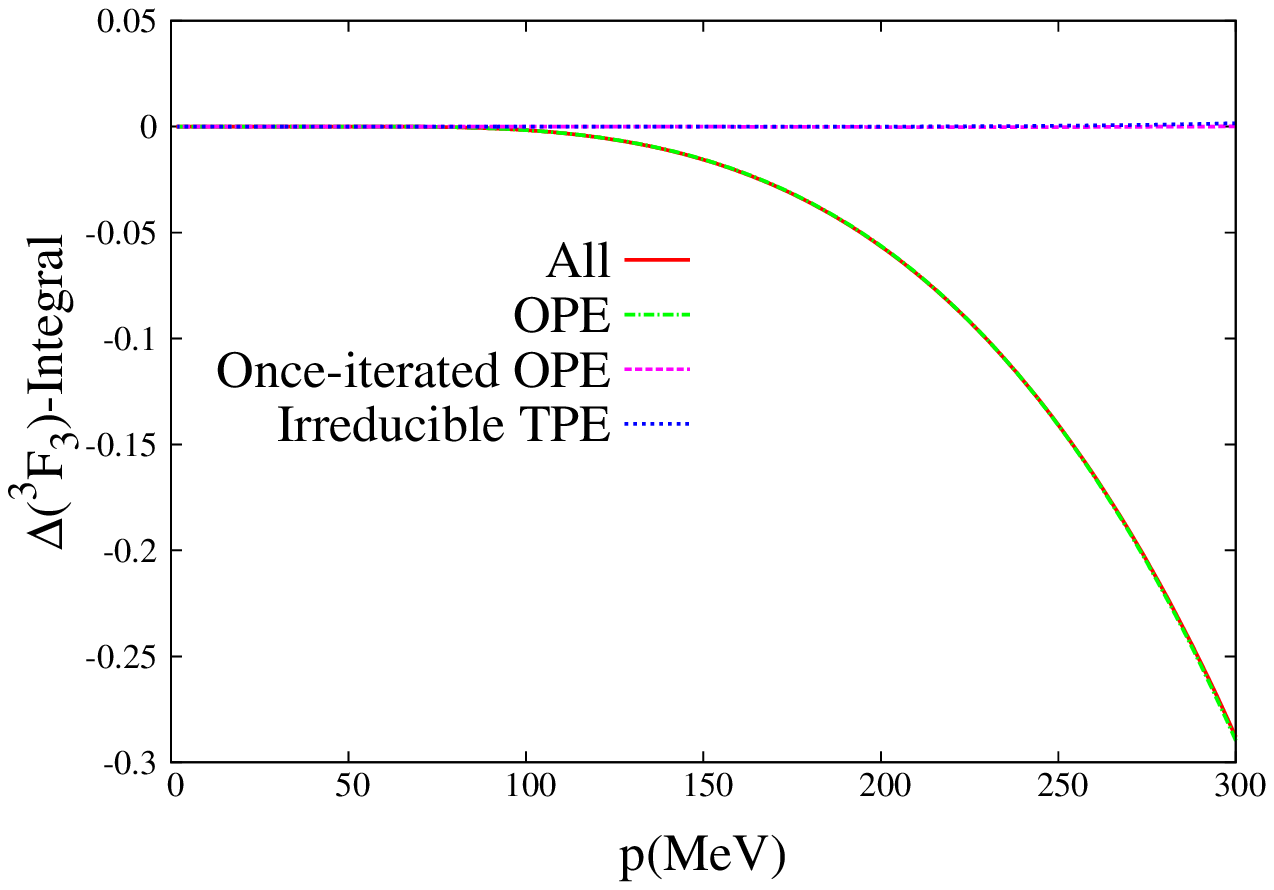} & 
\includegraphics[width=.4\textwidth]{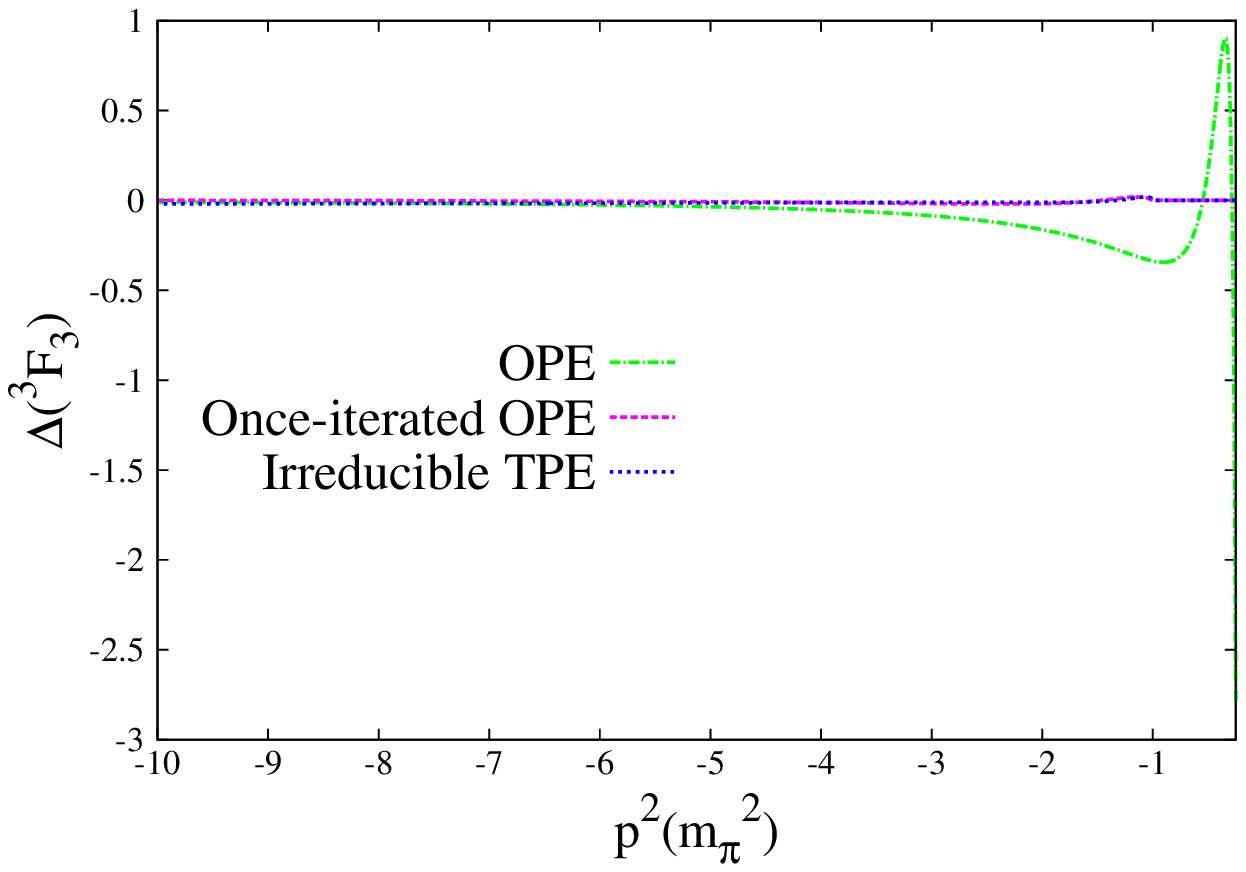}
\end{tabular}
\caption{ {\small (Color online.) Left panels: different contributions to the integral in Eq.~\eqref{quanty} with $\ell=3$. Right panels: Contributions to $\Delta(A)$. From top to bottom we show the $^1F_3$ and $^3F_3$ partial waves, respectively. The meaning of the lines is the same as in Fig.~\ref{fig:1s0quanty}.}
\label{fig:fwquanty}}
\end{center}
\end{figure}

\begin{figure}
\begin{center}
\begin{tabular}{cc}
\includegraphics[width=.4\textwidth]{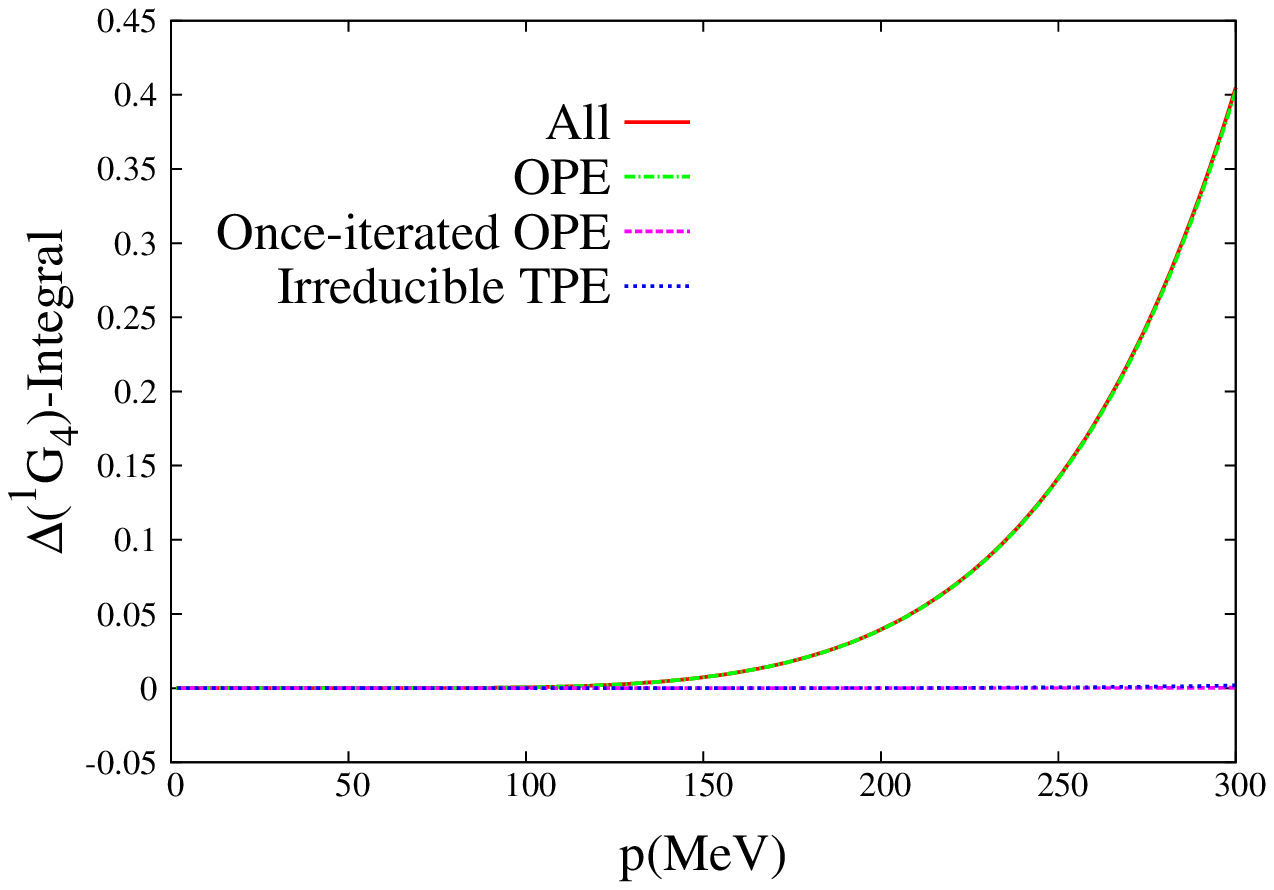} & 
\includegraphics[width=.4\textwidth]{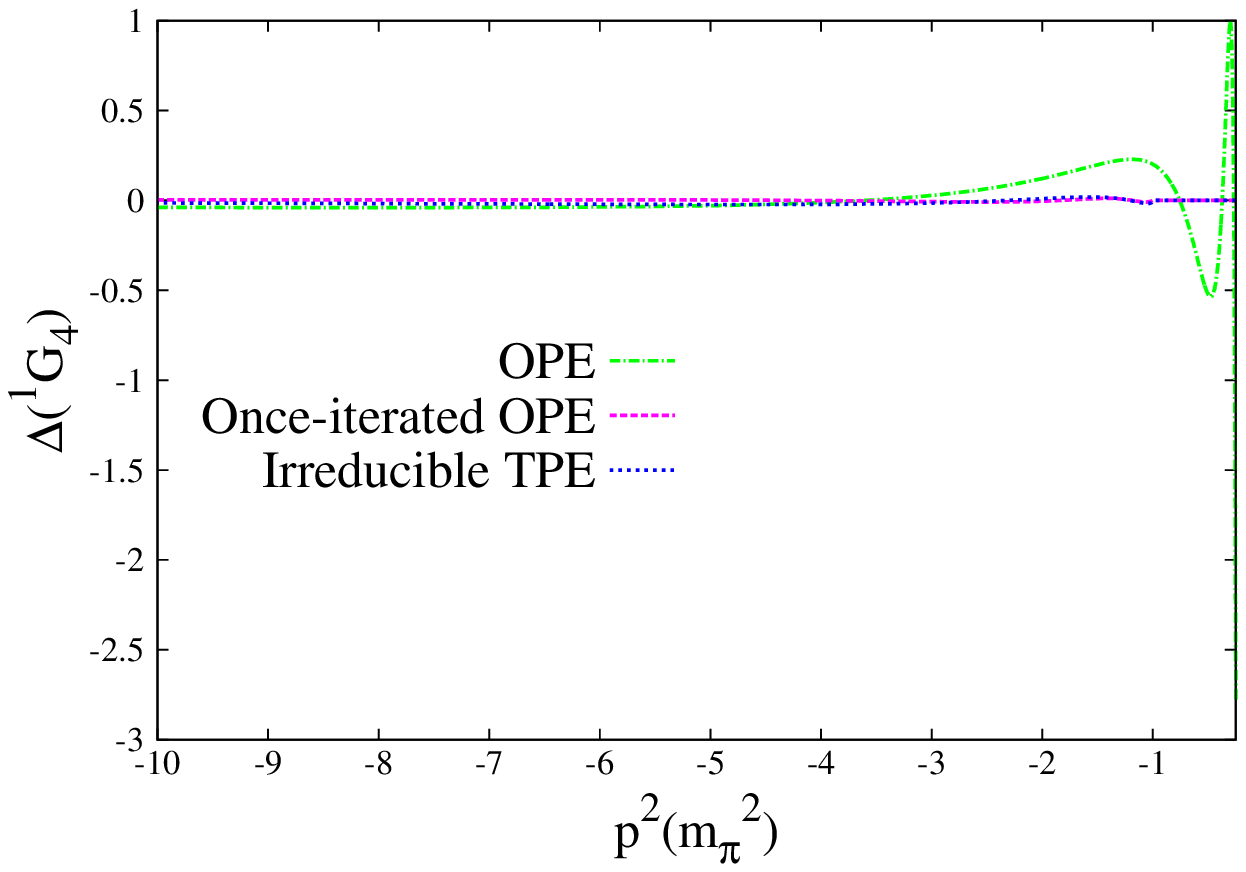}\\
\includegraphics[width=.4\textwidth]{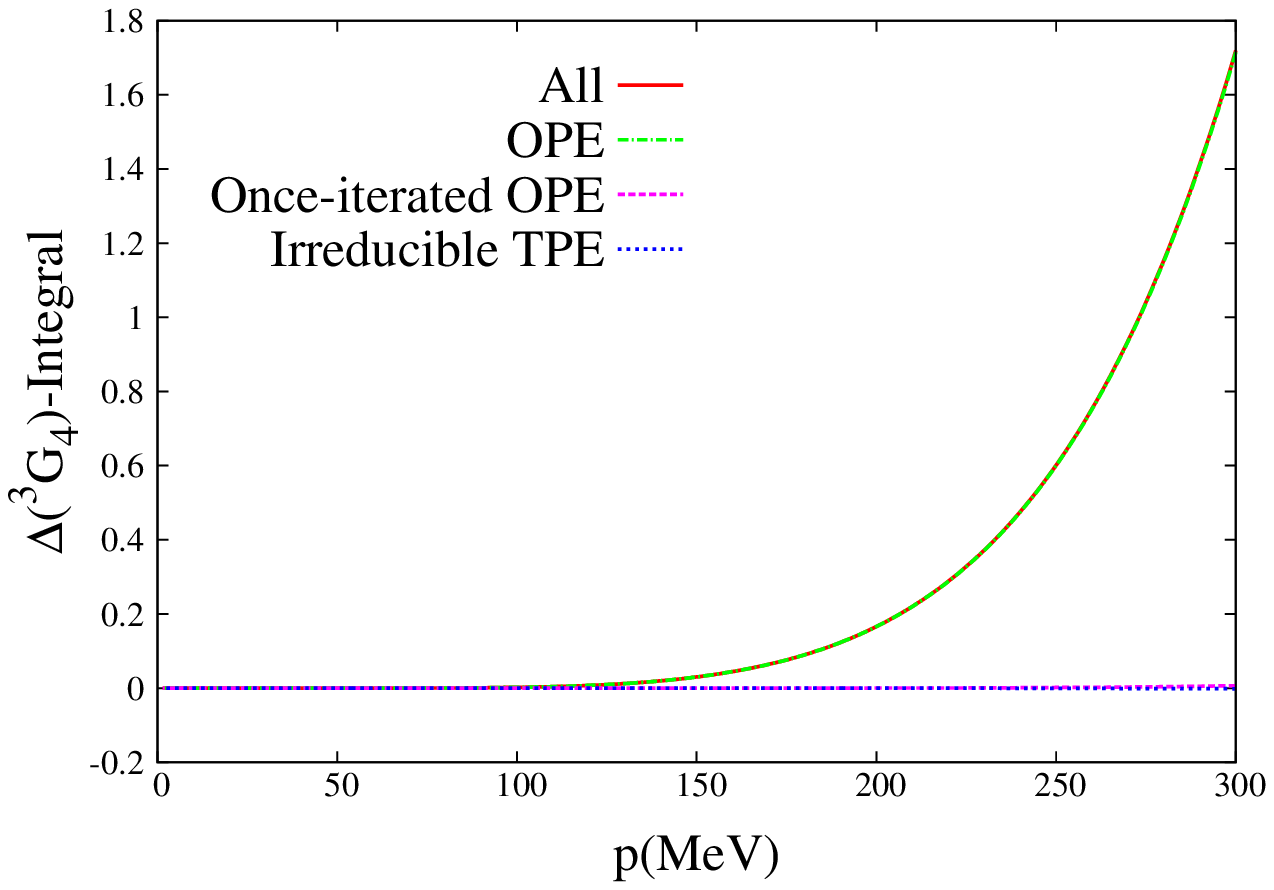} & 
\includegraphics[width=.4\textwidth]{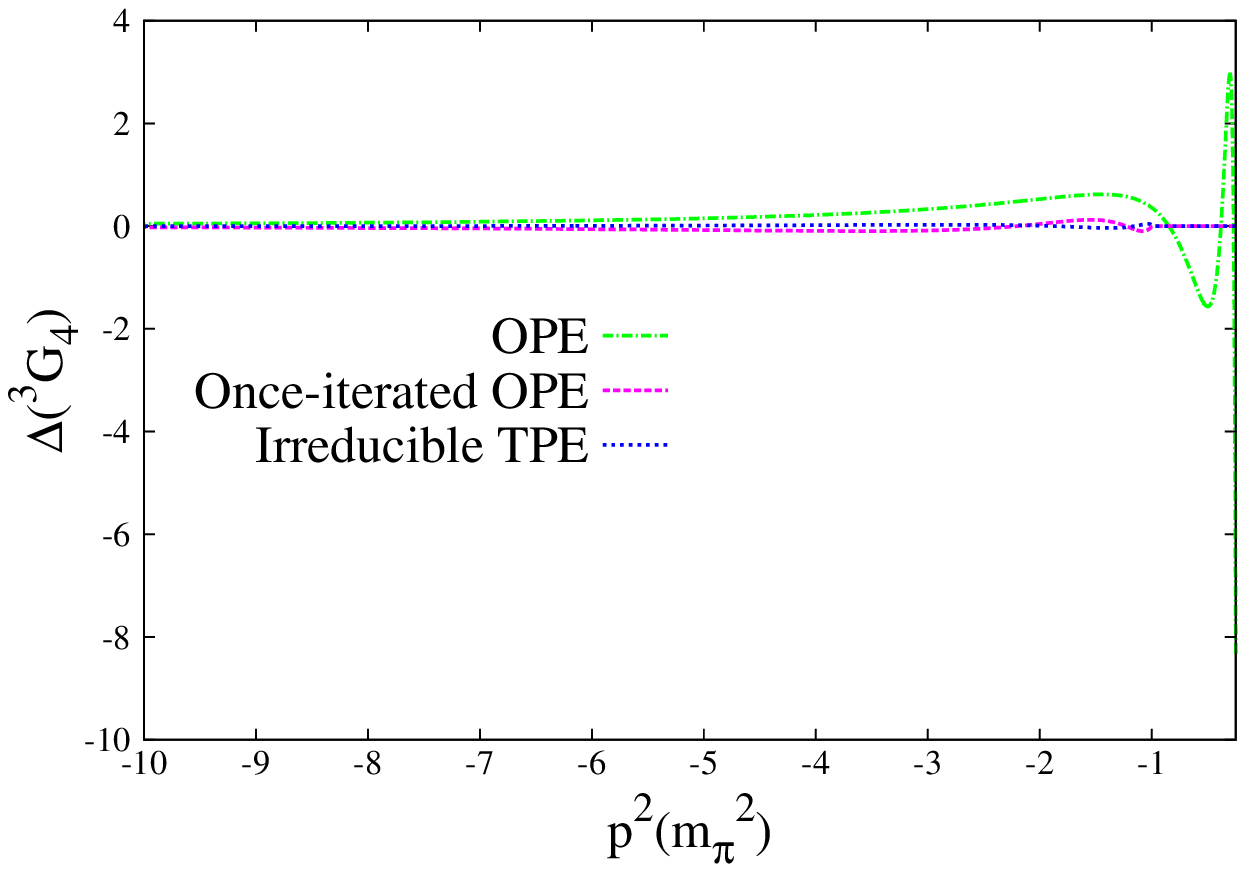}
\end{tabular}
\caption{ {\small (Color online.) Left panels: different contributions to the integral in Eq.~\eqref{quanty} with $\ell=4$. Right panels: Contributions to $\Delta(A)$. From top to bottom we show the $^1G_4$ and $^3G_4$ partial waves, respectively. The meaning of the lines is the same as in Fig.~\ref{fig:1s0quanty}.}
\label{fig:gwquanty}}
\end{center}
\end{figure}

\begin{figure}
\begin{center}
\begin{tabular}{cc}
\includegraphics[width=.4\textwidth]{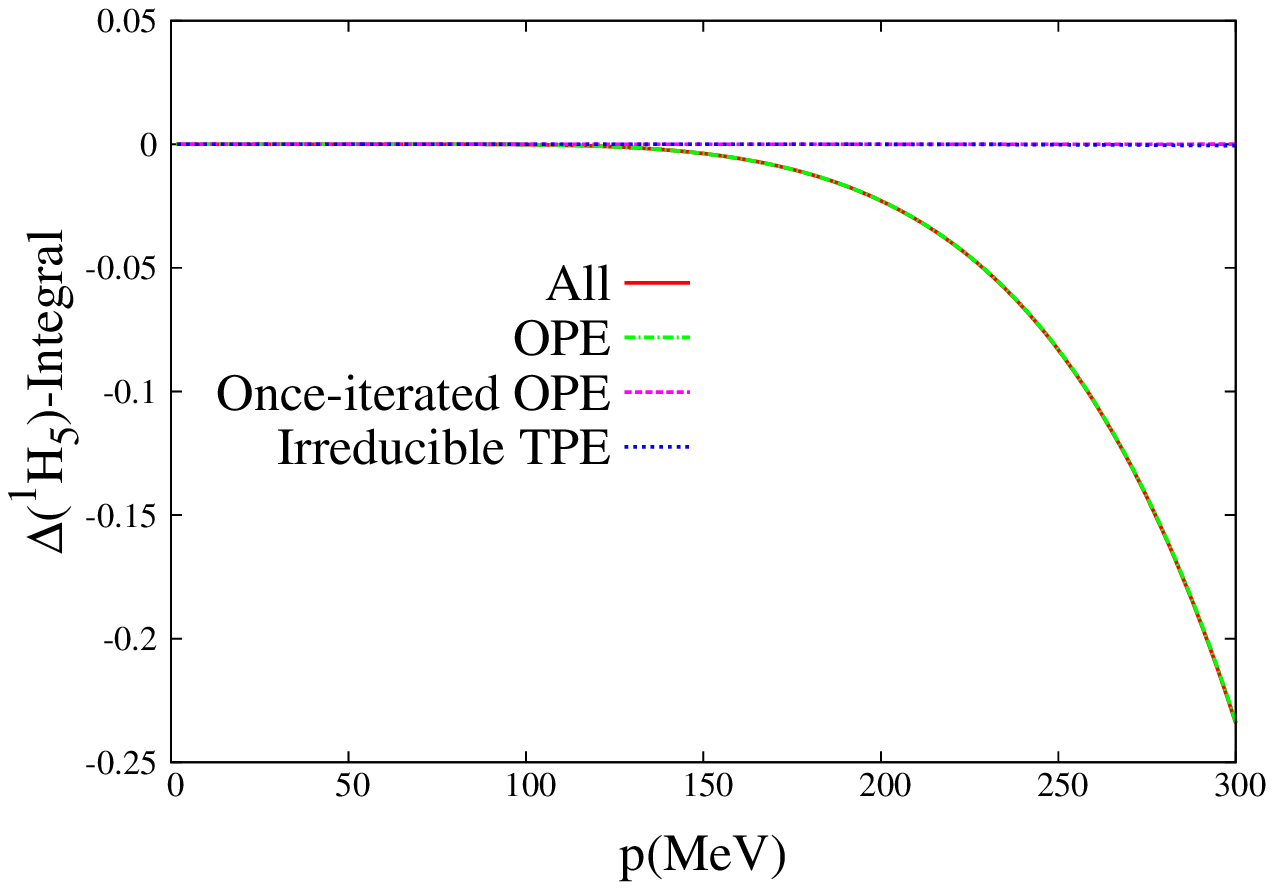} & 
\includegraphics[width=.4\textwidth]{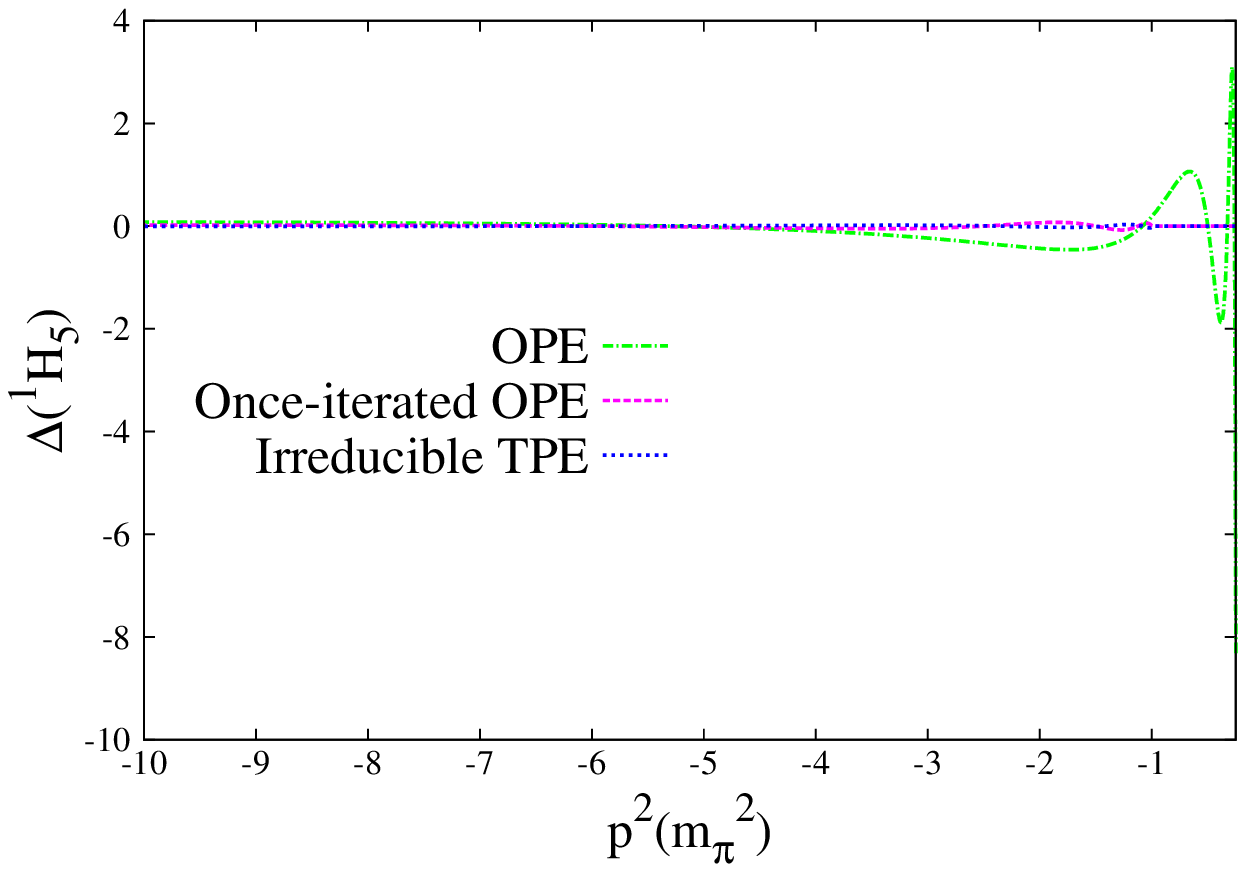}\\
\includegraphics[width=.4\textwidth]{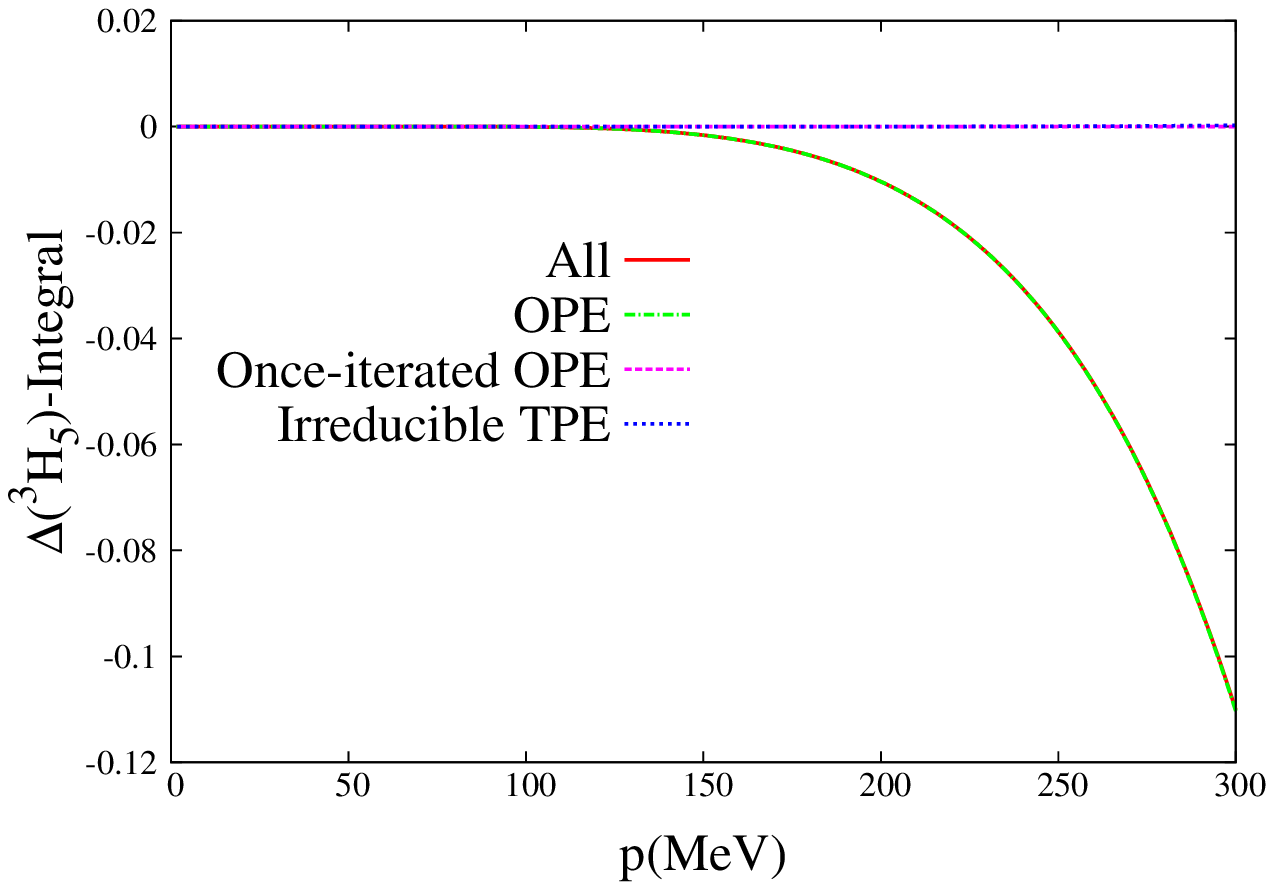} & 
\includegraphics[width=.4\textwidth]{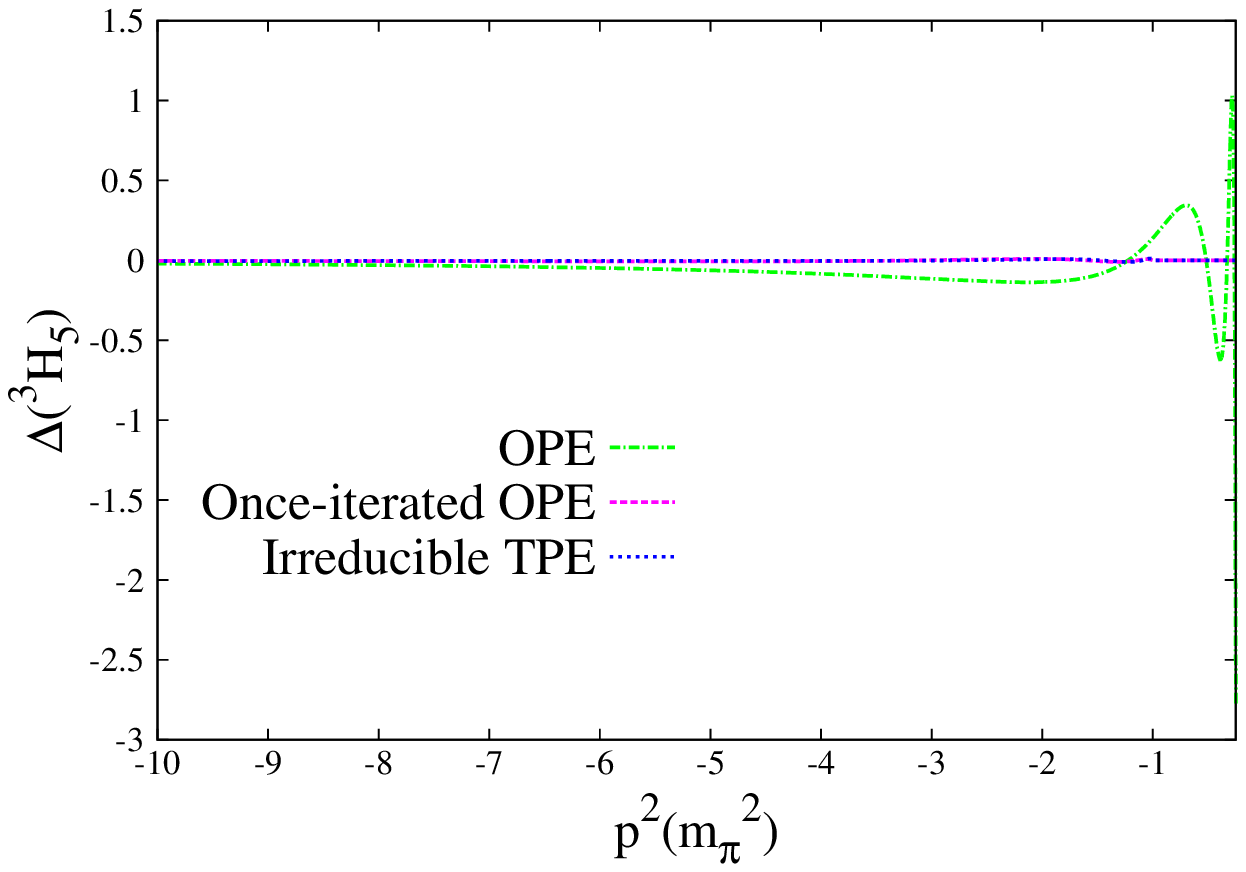}
\end{tabular}
\caption{ {\small (Color online.) Left panels: different contributions to the integral in Eq.~\eqref{quanty} with $\ell=5$. Right panels: Contributions to $\Delta(A)$. From top to bottom we show the $^1H_5$ and $^3H_5$ partial waves, respectively. The meaning of the lines is the same as in Fig.~\ref{fig:1s0quanty}.}
\label{fig:hwquanty}}
\end{center}
\end{figure}

The $F$-waves show an overwhelming dominance of the OPE contribution to the integral in Eq.~\eqref{quanty} with $\ell=3$, see the left panels of Fig.~\ref{fig:fwquanty}. This is in agreement with our discussion in Sec.~\ref{fw}, where we argue that these waves could be treated perturbatively. In addition these waves present small corrections to the phase shifts from higher orders, as shown in Fig.~\ref{fig:fw}. We also see that irreducible and reducible TPE have similar sizes (see e.g. the right panels in Fig.~\ref{fig:fwquanty}). A similar situation occurs for the $G$- and $H$- waves, shown in Figs.~\ref{fig:gwquanty} and \ref{fig:hwquanty}, respectively. The fact that OPE and the total result for the integration in Eq.~\eqref{quanty} coincide for the $F$- and higher partial waves clearly indicates their perturbative character. Notice that this is not the case for lower values of $\ell\leq 2$.

The expressions for $\Delta(A)$ can be algebraically obtained for the different partial waves from the expressions given in Ref.~\cite{peripheral}. A closer look at them would be appropriate in order to disentangle the origin of the somewhat surprising result that irreducible and reducible TPE contributions to $\Delta(A)$ have typically a similar size. To illustrate this point let us consider the $^3P_0$ wave for which, as shown in the top panel on the right of Fig.~\ref{fig:pwquanty}, both 
reducible and irreducible TPE have opposite sign but similar magnitude.  The different contributions 
to  $\Delta(A)$ are:
\begin{align}
\Delta_{OPE}&=-\frac{g_A^2 \pi}{16 f^2 }\frac{M_\pi^2}{A}~~,~~A< -\frac{M_\pi^2}{4}~,\nn\\
\Delta_{IRR}&= \frac{1}{4608f^4 A^2\pi}\left\{-2\sqrt{A(M_\pi^2+A)}\left[3 M_\pi^4+A(-M_\pi^2+2 A)
+2 g_A^2(-3M_\pi^4+5 A(-M_\pi^2+2 A)\right.\right.\nn\\
&\left. \left.+g_A^4\left\{-87 M_\pi^4+A(59 M_\pi^2+ 98 A)\right\}
\right]
+6M_\pi^4 \left[-M_\pi^2+g_A^2(2 M_\pi^2-6 A)-3A+g_A^4(29 M_\pi^2+21 A)\right]\right.\nn\\
&\left. \times \log\left(
\frac{(-A)^{\frac{1}{2}}}{M_\pi} + \left(-1-\frac{A}{M_\pi^2}\right)^\frac{1}{2} \right)
\right\}~~,~~A<-M_\pi^2~,
\nn\\
\Delta_{VGV}&=\frac{g_A^4 m}{3840 f^4 A^2}\left\{-4 M_\pi^5-20 M_\pi^2 (-A)^\frac{3}{2}+24(-A)^\frac{5}{2}-15M_\pi^4(-A)^\frac{1}{2} 
\log\left(-1+2\frac{(-A)^\frac{1}{2}}{M_\pi}\right)
\right\}~~,~~ A<-M_\pi^2~,
\label{delta3p0}
\end{align} 
where  we have, from top to bottom, the OPE ($\Delta_{OPE}$), irreducible TPE ($\Delta_{IRR}$)
 and reducible TPE  ($\Delta_{VGV}$) contributions, respectively. We see  in $\Delta_{VGV}$ the presence in the numerator of the nucleon mass and an extra factor of $\pi$ compared with $\Delta_{IRR}$, as expected for a reducible diagram. However,  we also  observe  the presence of much bigger numerical factors in the numerator of $\Delta_{IRR}$, which in the end make that both contributions have similar size. In order to see this effect more clearly let us separate  from $\Delta_{VGV}$ and $ \Delta_{IRR}$   the terms proportional to $g_A^4$  and with the largest power of $A$, the ones that dominate for $|A|$ considerably more than $M_\pi^2$. These partial contributions are called $\delta_{VGV}$ and $\delta_{IRR}$, 
respectively.  Their ratio, in this order, is
\begin{align}
\frac{\delta_{VGV}}{\delta_{IRR}}&=-\frac{\pi m }{(-A)^\frac{1}{2}}\frac{36}{245}~.
\label{ratiodelta}
\end{align}
Again this equation exhibits clearly the large ratio of scales $\frac{\pi m}{(-A)^{1/2}}$, as expected, but at the same time it has a large numerical enhancement from the irreducible contribution by the  factor $245/36\simeq 7$. This is 
 large enough to make both contributions similarly sized because the previous ratio becomes  
\begin{align}
\frac{\delta_{VGV}}{\delta_{IRR}}&\simeq \frac{3 M_\pi}{(-A)^\frac{1}{2}}\sim \frac{M_\pi}{(-A)^\frac{1}{2}}={\cal O}(1)~~,~~A< -M_\pi^2~.
\end{align}

The presence of numerical factors enhancing $\Delta_{IRR}$ is, in part, attributable to  combinatorial reasons, by putting on-shell the two pions when cutting diagrams in order to evaluate their imaginary part along the LHC, see Fig.~\ref{nnbar}. As an example, let us take proton-proton ($pp$) scattering. Then, the reducible part of Fig.~\ref{nnbar}.d) only contributes by exchanging two $\pi^0$, which contains a factor 1/2 because of the indistinguishability of them. However, Fig.~\ref{nnbar}.c), in addition to $\pi^0\pi^0$, also contains $\pi^+\pi^-$ as  intermediate state. As a result Fig.~\ref{nnbar}.c) at low energies is enhanced by a factor 3 compared with Fig.~\ref{nnbar}.d).

\section{Coupled partial waves}
\label{sec:formalism}
 The spin triplet $NN$ partial waves with total angular momentum $J$ mix the orbital angular momenta $\ell = J-1$ and $\ell' =J+1$   (except the $^3P_0$ wave that is uncoupled.)  Each coupled partial wave is determined by the quantum numbers $S$, $J$, $ \ell$ and $\ell'$. In the following for simplifying the notation we omit them and indicate, for given $J$ and $S$, the different partial waves by $t_{ij}$, with $i=1$ corresponding to $\ell=J-1$ and $i=2$ to $\ell'=J+1$.  In matrix notation, one has a symmetric  
 $2\times 2$ $T$-matrix.   
 In our normalization,  the relation between the $T$- and $S$-matrix   reads
\begin{align}
S(A) & = {I} + i 2 \rho(A) T(A) \nn\\
& = \left(
 \begin{array}{cc}
 \cos 2\epsilon_J\ e^{2i\delta_1}            & i\sin 2\epsilon_J\ e^{i(\delta_1+\delta_2)} \\ 
i\sin 2\epsilon_J\ e^{i(\delta_1+\delta_2)} &   \cos 2\epsilon_J\ e^{2i\delta_2}
 \end{array} \right)~,
\label{relst}
\end{align}
 where $I$ is the $2\times 2 $ unit matrix, $\epsilon_J$ is the mixing angle, and $\delta_{1}$ and $\delta_2$ are the phase shifts for the channels with orbital angular 
momentum $J-1$ and $J+1$, in this order.

Above threshold ($A>0$), and below pion production, the unitarity character of the $S$-matrix, $S S^\dagger = S^\dagger S = {I}$~,
can be expressed in terms of the (symmetric) $T$-matrix as 
\begin{align}
\mathrm{Im} T^{-1}(A) = - \rho(A)\, {I}~,
\label{elauni}
\end{align}  
where $ \rho(A)$ was already defined in Eq.~\eqref{rhodef}. 
In the following, the imaginary parts above threshold of the inverse of the $T$-matrix elements, $t_{ij}(A)$,  play an important role, 
\begin{align}
\mathrm{Im} \frac{1}{t_{ij}(A)} \equiv -\nu_{ij}(A)~,A>0~.
\label{nuij.def}
\end{align}
 From Eq.~\eqref{relst}, 
 one can easily express the different $\nu_{ij}$ in terms of 
phase shifts and the mixing angle along the physical region. It implies that we can write 
 the diagonal partial waves  as $t_{ii}$ and the mixing 
amplitude  $t_{12}$ as $t_{ii}=(e^{2i\delta_i} \cos 2\epsilon_J -1)/2i\rho$ and $t_{12}=e^{i(\delta_1+\delta_2)}\sin2\epsilon_J/2\rho$, 
respectively. With these equalities it is straightforward to obtain for $A>0$: 
\begin{align}
\nu_{11}(A) & =   \rho(A) \left[ 1- \frac{\frac{1}{2}\sin^2 2\epsilon_J}{1-\cos 2\epsilon_J \cos 2\delta_1} \right]^{-1} \label{eq:nus11}~,\\
\nu_{22}(A) & =   \rho(A) \left[ 1- \frac{\frac{1}{2}\sin^2 2\epsilon_J}{1-\cos 2\epsilon_J \cos 2\delta_2} \right]^{-1} \label{eq:nus22}~,\\
\nu_{12}(A) & = 2 \rho(A) \frac{\sin(\delta_1 + \delta_2)}{\sin 2\epsilon_J} \label{eq:nus12}~.
\end{align}
 Eq.~\eqref{nuij.def} generalizes Eq.~\eqref{unitinv}, valid for an uncoupled partial wave. 
Indeed, if we set $\epsilon_J = 0$ in $\nu_{11}(A)$ and $\nu_{22}(A)$, the uncoupled case is recovered. Note also that $\nu_{ii}(A)/\rho(A) \ge 1$ and  for $A\to \infty$ one expects that $\nu_{ij}(A)={\cal O}(A^\frac{1}{2})$ as $\rho(A)$  itself, because the absolute value of 
the trigonometric functions in Eqs.~\eqref{eq:nus11}-\eqref{eq:nus12} is bounded by 1.

We apply the $N/D$ method, discussed in Sec.~\ref{unformalism}, to each  partial wave $t_{ij}$ separately,
\begin{equation}
\label{nd.def.cou}
t_{ij}(A) =  \frac{N_{ij}(A)}{D_{ij}(A)}~.
\end{equation}
We define $\ell_{ij}$ as  $\ell_{11} = \ell$, $\ell_{22} = \ell' = \ell + 2$ and $\ell_{12} = (\ell + \ell')/2 = \ell + 1$.  From the previous equation and Eq.~\eqref{nuij.def}  it follows that
 \begin{align}
\label{nd.def.d}
 \mathrm{Im} D_{ij}(A)&=-N_{ij}(A)  \nu_{ij}(A)~,~A>0~, \\
 \mathrm{Im} N_{ij}(A)&= D_{ij}(A) \Delta_{ij}(A)~,~A < L ~,
\label{nd.def.2}
 \end{align}
where $\mathrm{Im} t_{ij}(A) \equiv \Delta_{ij}(A)$ along the LHC. The only formal difference with respect to 
Eqs.~\eqref{disd} and \eqref{disn} is that now instead of $\rho(A)$ we have $\nu_{ij}(A)$ in Eq.~\eqref{nd.def.d}.  
 Because of this, we do not expect any change in the 
conclusions obtained in Sec.~\ref{IEtheory} regarding the solution of the IEs depending on the high-energy behavior of $\Delta(A)$.
We can then follow the same line of reasoning as given in Sec.~\ref{unformalism} and write down unsubtracted DRs for $D_{ij}/(A-C)^n$ and $N_{ij}/(A-C)^n$ for large enough $n$. Multiplying them by $(A-C)^n$ we derive  the proper DRs 
 valid for $D_{ij}(A)$ and $N_{ij}(A)$, as done in Sec.~\ref{unformalism}.  In this way, 
 our  general equations for the coupled channel case arise:
\begin{align}
\label{inteq1c}
D_{ij}(A)&=\sum_{p=1}^n \delta^{(ij)}_p (A-C)^{p-1}-\sum_{p=1}^n \nu^{(ij)}_p\frac{(A-C)^n}{\pi}\int_0^\infty dq^2\frac{\nu_{ij}(q^2)}{(q^2-A)(q^2-C)^{n-p+1}}\nn\\
&+\frac{(A-C)^n}{\pi^2}\int_{-\infty}^L dk^2\frac{\Delta_{ij}(k^2)D_{ij}(k^2)}{(k^2-C)^n}\int_0^\infty dq^2\frac{\nu_{ij}(q^2)}{(q^2-A)(q^2-k^2)}~,\\
\label{nc}
N_{ij}(A)&=\sum_{p=1}^n \nu^{(ij)}_p (A-C)^{p-1}+\frac{(A-C)^n}{\pi}\int_{-\infty}^L dk^2\frac{\Delta_{ij}(k^2)D_{ij}(k^2)}{(k^2-A)(k^2-C)^n}~.
\end{align}
Of course, as in the uncoupled partial wave case, we rewrite conveniently the previous equations whenever we take the subtractions at different subtraction points, that is, not all of the them taken at the same $C$. In particular we impose the normalization condition  
\begin{align}
\label{normac}
D_{ij}(0)=1~,
\end{align}
 so that one subtraction for $D_{ij}(A)$ is always taken at $C=0$, and this gives 
\begin{align}
\delta_1^{(ij)}=1~.
\end{align}
 We will indicate below case by case where the subtractions are taken.

For the partial waves with $\ell_{ij}\geq 2$ we have to guarantee the right threshold behavior such that 
$t_{ij}(A)\to A^{\ell_{ij}}$ for $A\to 0^+$. This is done as in Sec.~\ref{leq2} by considering $\ell_{ij}$-time DRs with all the subtraction constants in $N_{ij}(A)$ taken at $C=0$ and with vanishing value. For the function $D_{ij}(A)$, apart of the subtraction taken at $C=0$,  the rest of them 
are taken at $C\neq 0$. The resulting IEs are
\begin{align}
\label{highdc}
D_{ij}(A)&=1+\sum_{p=2}^{\ell_{ij}}\delta^{(ij)}_p A(A-C)^{p-2} +\frac{A(A-C)^{\ell_{ij}-1}}{\pi^2}\int_{-\infty}^L\!\! dk^2 \frac{\Delta_{ij}(k^2)D_{ij}(k^2)}{(k^2)^{\ell_{ij}}}\nn\\
&\times \int_0^\infty\!\! dq^2\frac{\nu_{ij}(q^2) (q^2)^{\ell_{ij}-1}}{(q^2-A)(q^2-k^2)(q^2-C)^{\ell_{ij}-1}}~,\\
N_{ij}(A)&=\frac{A^{\ell_{ij}}}{\pi}\int_{-\infty}^L \!\!dk^2\frac{\Delta_{ij}(k^2)D_{ij}(k^2)}{(k^{2})^{\ell_{ij}}(k^2-A)}~.
\label{highndc}
\end{align}
Notice that we have rewritten the $(\ell_{ij}-1)^{\rm{th}}$ degree polynomial in $D_{ij}(A)$ so that the coefficients $\delta_p^{(ij)}$ have a simpler 
relation with $D_{ij}(A)$. Indeed, one can deduce straightforwardly that 
\begin{align}
\delta_p^{(ij)}=\frac{(-1)^p}{C^{p-1}}\left[
\sum_{n=0}^{p-2}\frac{(-1)^n}{n!}C^n D^{(n)}_{ij}(C)-1
\right]~.
\label{taylor}
\end{align}
That is, $\delta_p^{(ij)}$ is proportional to the difference of the Taylor expansion  of degree $p-2$ 
of the function $D_{ij}(A)$ at around $A=C$ and evaluated at $A=0$, and  $D_{ij}(0)=1$. In the practical applications that follow we always take $C=-M_\pi^2$. The situation with all the $\delta_p^{(ij)}$ equal to zero corresponds to $D_{ij}(0)=1$ and $D_{ij}^{(n)}(0)=0$ (this is the so called pure perturbative case for a high orbital-angular-momentum wave). On the other hand, the rule given in Sec.~\ref{leq2}  for an $ n$-time subtracted DR corresponds to having 
$D_{ij}(0)=1$, $D_{ij}^{(p)}=0$ for $1\leq p<n-2$ and $D_{ij}^{(n-2)}(0)\neq 0$.

As shown explicitly in Ref.~\cite{paper2} the $\nu_{22}(A)$ function diverges as $A^{-\frac{3}{2}}$ for $A\to 0$. This requires some care in order to avoid infrared divergent integrals, a problem already noticed in Ref.~\cite{noyes}. This issue is cured in Eq.~\eqref{highdc} because $C\neq 0$. Then, the factor $(q^2)^{\ell_{22}-1}$ cancels, at least partially, the threshold divergence in $\nu_{22}(A)$ so that the integral is convergent. Notice that $\ell_{22}\geq 2$, with its smallest value for the $^3D_1$ wave. The function $\nu_{12}(A)$ also diverges at threshold but only as $A^{-\frac{1}{2}}$, so that it does not give rise to any infrared divergent integral. For completeness, we recall that the $\nu_{11}(A)$ vanishes for $A\to 0$ as $A^\frac{1}{2}$. In the following we define the function $g_{ij}(A,k^2,C;m)$ as 
\begin{align}
g_{ij}(A,k^2,C;m)=\frac{1}{\pi}\int_0^\infty\!\! dq^2\frac{\nu_{ij}(q^2) (q^2)^m}{(q^2-A)(q^2-k^2)(q^2-C)^m}~.
\label{def.gm}
\end{align}

 The main difference with respect to the uncoupled case  is that now one has to solve simultaneously three $N/D$ equations for $ij$=$11$, $12$ and $22$, which are linked between each other because of the $\nu_{ij}(A)$ functions.  They depend on the phase shifts $\delta_1$, $\delta_2$ and on the mixing angle $\epsilon_J$, defined in Eq.~\eqref{relst}, which constitute also the final output of our approach. Thus, we follow an iterative approach, as already done in Ref.~\cite{paper2}, as follows. Given an input for $\delta_1$, $\delta_2$ and $\epsilon_J$, one solves the three IEs for $D_{ij}(A)$ along the LHC. Then, the scattering amplitudes on the RHC can be calculated. In terms of them, the phase shifts $\delta_1$ and $\delta_2$ are obtained from the phase of the $S$-matrix elements $S_{11}$ and $S_{22}$, while $\sin 2\epsilon_J=2\rho |t_{12}| n_{12}/|n_{12}|$, according to Eq.~\eqref{relst}. In this way a new input set of $\nu_{ij}$ functions, Eqs.~\eqref{eq:nus11}-\eqref{eq:nus12}, is provided.  These are used again in the IEs, and the iterative procedure is finished when convergence is found (typically, the difference between two consecutive iterations in the three independent  
functions $D_{ij}$ along the LHC is required to be less than one per thousand.) 

It can be shown straightforwardly that unitarity is fulfilled in our coupled channel equations, solved in the way just explained,
 if $|S_{11}(A)|^2=|S_{22}(A)|^2=\cos^2 2\epsilon_J$ for $A>0$. From the fact that $\mathrm{Im} t_{12}=\nu_{12}|t_{12}|^2$, according to  Eq.~\eqref{nuij.def}, and $\sin2\epsilon_J=2\rho |t_{12}|n_{12}/|n_{12}|$ (the latter equality is valid  only when convergence is reached), it results that the phase of $t_{12}$ is $\delta_1+\delta_2$, as required by unitarity,  Eq.~\eqref{relst}. By construction the phase shifts are equal to one-half of the phase of the $S$-matrix diagonal elements when convergence is achieved, so that   Eq.~\eqref{relst} is satisfied if $|S_{11}|=|S_{22}|=\cos 2\epsilon_J$.

For the initial input one can use e.g. the results given by Unitarity ChPT \cite{long}, the LO results obtained from Ref.~\cite{paper2} or some put-by-hand phase shifts and mixing angle. For the latter case a good choice is to take as initial input for $\delta_{1}$ and $\delta_{2}$ the resulting phase shifts
 obtained by treating $t_{11}$ and $t_{22}$ as uncoupled waves. 
We find no dependence in our final unitary results  regarding  the initial input taken for the iterative procedure.

\section{Coupled waves: $^3S_1-{^3D_1}$}
\label{sec:deuteron}

For the $^3S_1-{^3D_1}$ system, we write down a once-subtracted DR for the partial wave $^3S_1$ and twice-subtracted DRs for the $^3D_1$ and mixing partial wave, in order to guarantee that the position of the deuteron pole is the same in all of the three partial waves. The explicit expressions for the $^3S_1$ partial wave 
are:
\begin{align}
\label{3s1c}
D_{11}(A)&=1-\nu_1 A g_{11}(A,0)+\frac{A}{\pi}\int_{-\infty}^L dk^2\frac{\Delta_{11}(k^2)D_{11}(k^2)}{k^2}g_{11}(A,k^2)
~,\nn\\
N_{11}(A)&=\nu_1+\frac{A}{\pi}\int_{-\infty}^L dk^2\frac{\Delta_{11}(k^2)D_{11}(k^2)}{k^2(k^2-A)}~,
\end{align}
where the function $g_{ij}(A,k^2)$ is defined as
\begin{align}
g_{ij}(A,k^2)=\frac{1}{\pi}\int_0^\infty dq^2\frac{\nu_{ij}(q^2)}{(q^2-A)(q^2-k^2)}~.
\label{gij}
\end{align}
The subtraction constant $\nu_1$ is fixed in terms of the experimental $^3S_1$ scattering length, 
$a_t=5.424\pm 0.004$~fm \cite{epen3lo}, analogously as we did already for the $^1S_0$ in Sec.~\ref{1s0},
\begin{align}
\nu_1=-\frac{4\pi a_t}{m}~.
\label{nu1t}
\end{align}

For the mixing partial wave, $\ell_{12}=1$, and $^3D_1$ with $\ell_{22}=2$, we have
\begin{align}
\label{sdijc}
N_{ij}(A)&=\frac{A^{\ell_{ij}}}{\pi}\int_{-\infty}^L dk^2\frac{\Delta_{ij}(k^2)D_{ij}(k^2)}{(k^2-A)(k^2)^{\ell_{ij}}}~,\nn\\
D_{ij}(A)&=1-\frac{A}{k_d^2}+\frac{A(A-k_d^2)}{\pi}\int_{-\infty}^L dk^2\frac{\Delta_{ij}(k^2)D_{ij}(k^2)}{(k^2)^{\ell_{ij}}}g_{ij}^{(d)}(A,k^2)~,
\end{align}
with the new integration along the RHC 
 \begin{align}
g^{(d)}_{ij}(A,k^2)&=\frac{1}{\pi}\int_0^\infty dq^2\frac{\nu_{ij}(q^2)(q^2)^{\ell_{ij}-1}}{(q^2-A)(q^2-k^2)(q^2-k_d^2)}~.
\label{gijd}
\end{align}
The function $g_{ij}(A,k^2)$ and $g_{ij}^{(d)}(A,k^2)$ were already introduced in Ref.~\cite{paper2}. Notice that these functions have to be 
evaluated numerically. In Eq.~\eqref{sdijc} one extra subtraction is taken at $k_d^2$, which is the 
three-momentum squared of the deuteron pole position obtained for the $^3S_1$ wave
 from Eq.~\eqref{3s1c}. 
In other words, $k_d^2$ is the value of $A$  at which 
$D_{11}(k_d^2)=0$  in each step in the iterative process for solving  Eqs.~\eqref{3s1c} and \eqref{sdijc}.
 No extra subtraction constants are introduced because we require
 $D_{12}(k_d^2)=D_{22}(k_d^2)=0$, so that all three coupled  partial waves have the deuteron at the same position, $A=k_d^2$.

\begin{figure}
\begin{center}
\begin{tabular}{cc}
\includegraphics[width=.4\textwidth]{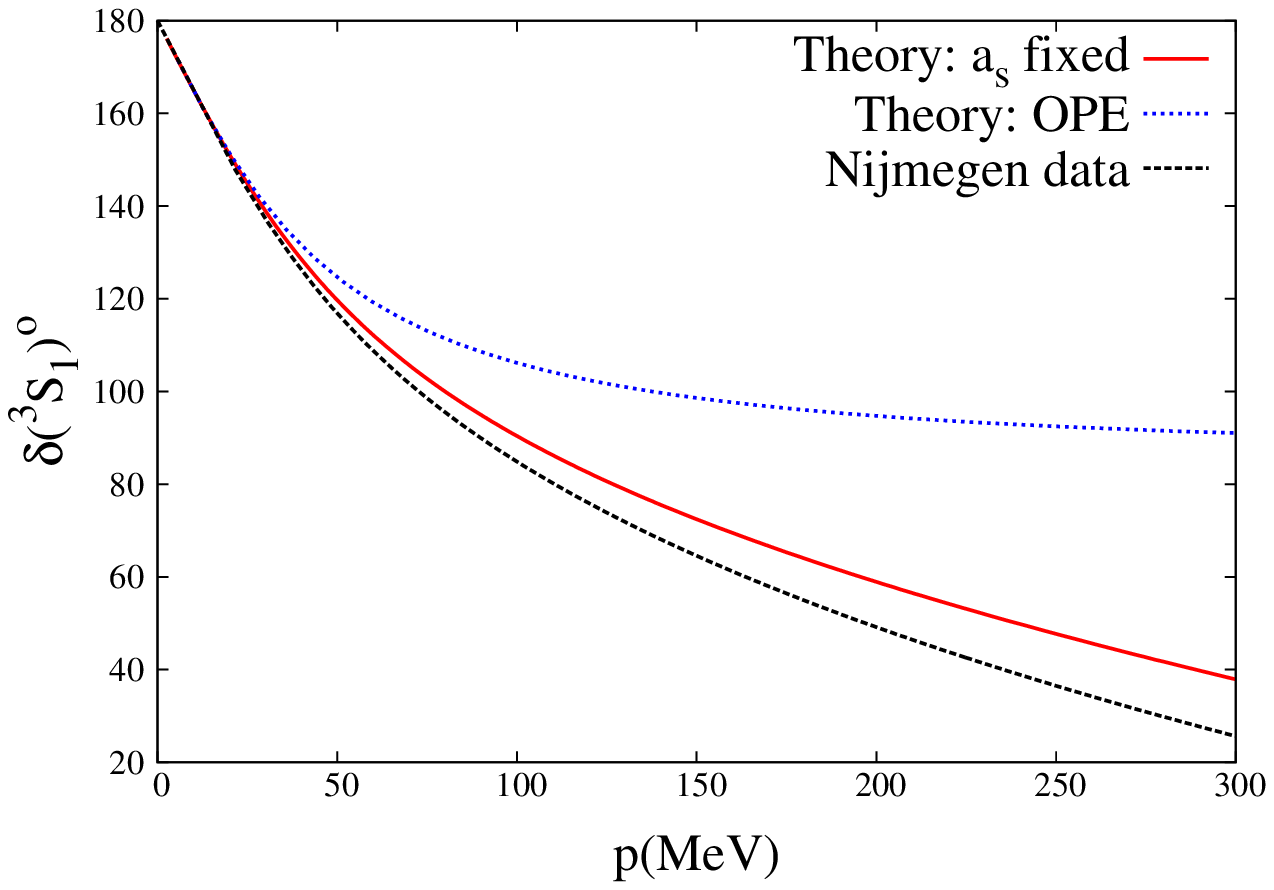} & 
\includegraphics[width=.4\textwidth]{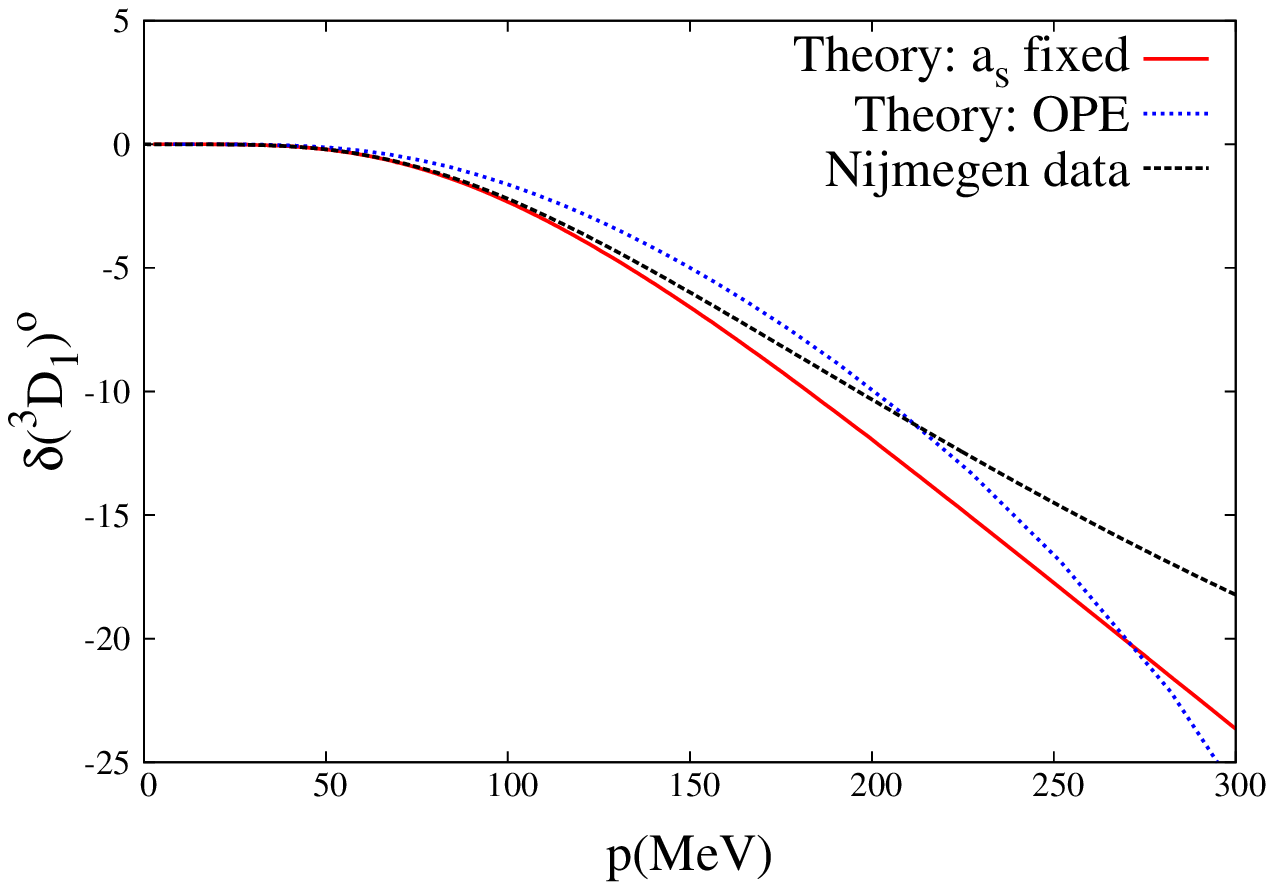}\\  
\includegraphics[width=.4\textwidth]{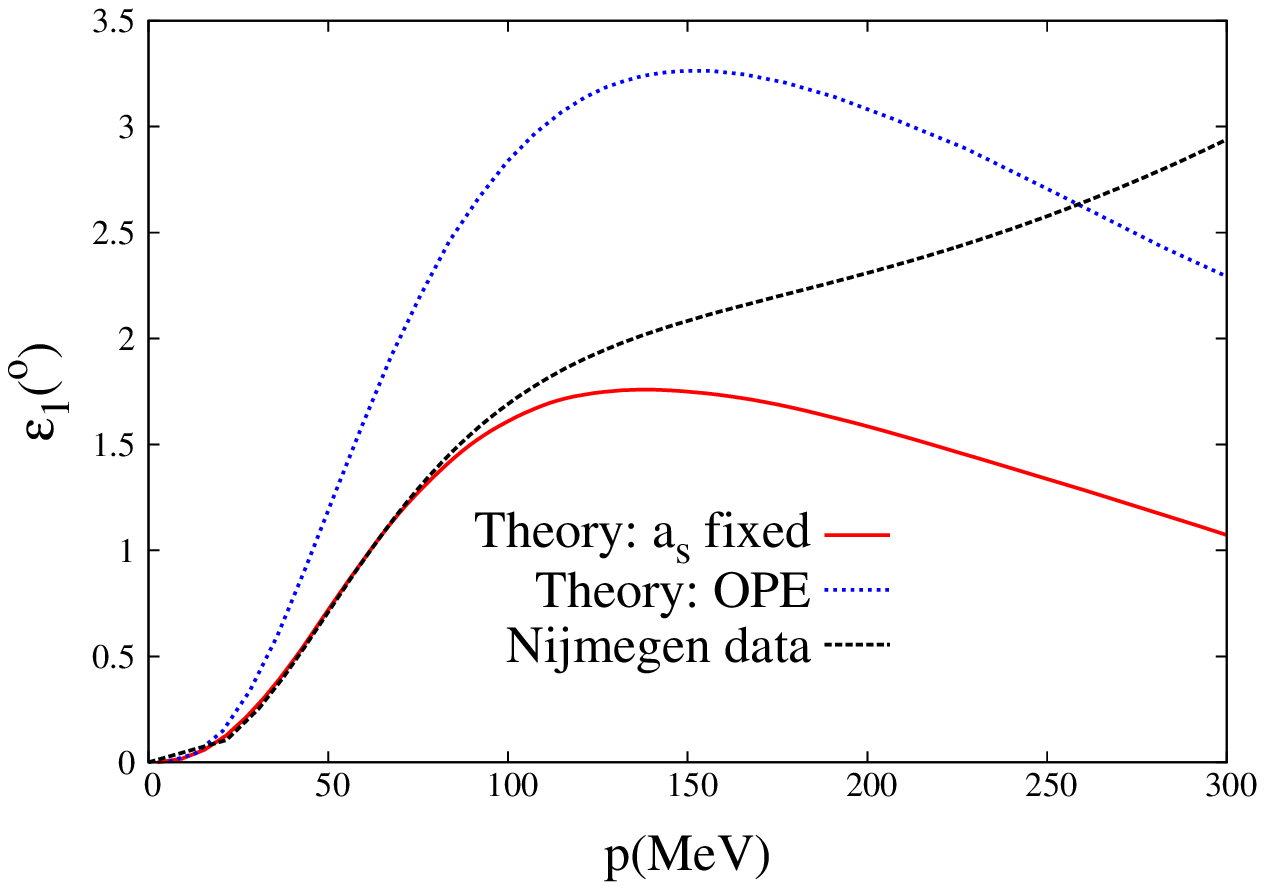}  
\end{tabular}
\caption[pilf]{\protect {\small (Color online.) From top to bottom and left to right: Phase shifts for $^3S_1$, $^3D_1$ and the mixing angle  $\epsilon_1$, respectively. 
The (red) solid line corresponds to the results obtained from Eqs.~\eqref{3s1c} and \eqref{sdijc} with the $^3S_1$ scattering length as experimental input. The OPE result from Ref.~\cite{paper2} is the (blue) dotted line. The Nijmegen PWA analysis is the (black) dashed line.}
\label{fig:3sd1} }
\end{center}
\end{figure}

We solve Eqs.~\eqref{3s1c} and \eqref{sdijc} with different input which is provided by the results of Ref.~\cite{long} by varying the 
parameter $g_0$ in that reference. We observe some dependence in the outcome solutions so that we require a 
criterion of maximum stability under changes in $g_0$. E.g. let us take the slope at threshold of the mixing angle $\epsilon_1$, denoted by $a_\epsilon$ and defined by 
\begin{align}
a_\epsilon&=\lim_{A\to 0^+}\frac{\sin 2\epsilon_1}{A^\frac{3}{2}}=1.128~M_\pi^{-3}~,
\label{aepexp}
\end{align}
as the value obtained from the Nijmegen PWA phase shifts. This quantity has a minimum as a function of the input used that indeed gives the closest value to the experimental one in Eq.~\eqref{aepexp}.
 We obtain $a_\epsilon=1.10\sim 1.14~M_\pi^{-3}$. 
Precisely the mixing angle is by far the most sensitive quantity  to the input data for obtaining the final solution by iteration. 
Then,  it is certainly a welcome fact that the best results are obtained for the input that generates most stable results 
under changes of itself. The results obtained by solving Eqs.~\eqref{3s1c} and \eqref{sdijc}, with $\nu_1$ fixed from the experimental 
$^3S_1$ scattering length, Eq.~\eqref{nu1t}, are shown by the (red) solid line in    Fig.~\ref{fig:3sd1}. 
We see that these curves tend to follow data quite closely already, specially below $\sqrt{A}\simeq 100$~MeV. Let us notice as well the clear and noticeable improvement in the reproduction of data compared with the OPE results of Ref.~\cite{paper2}.

\begin{figure}
\begin{center}
\begin{tabular}{cc}
\includegraphics[width=.4\textwidth]{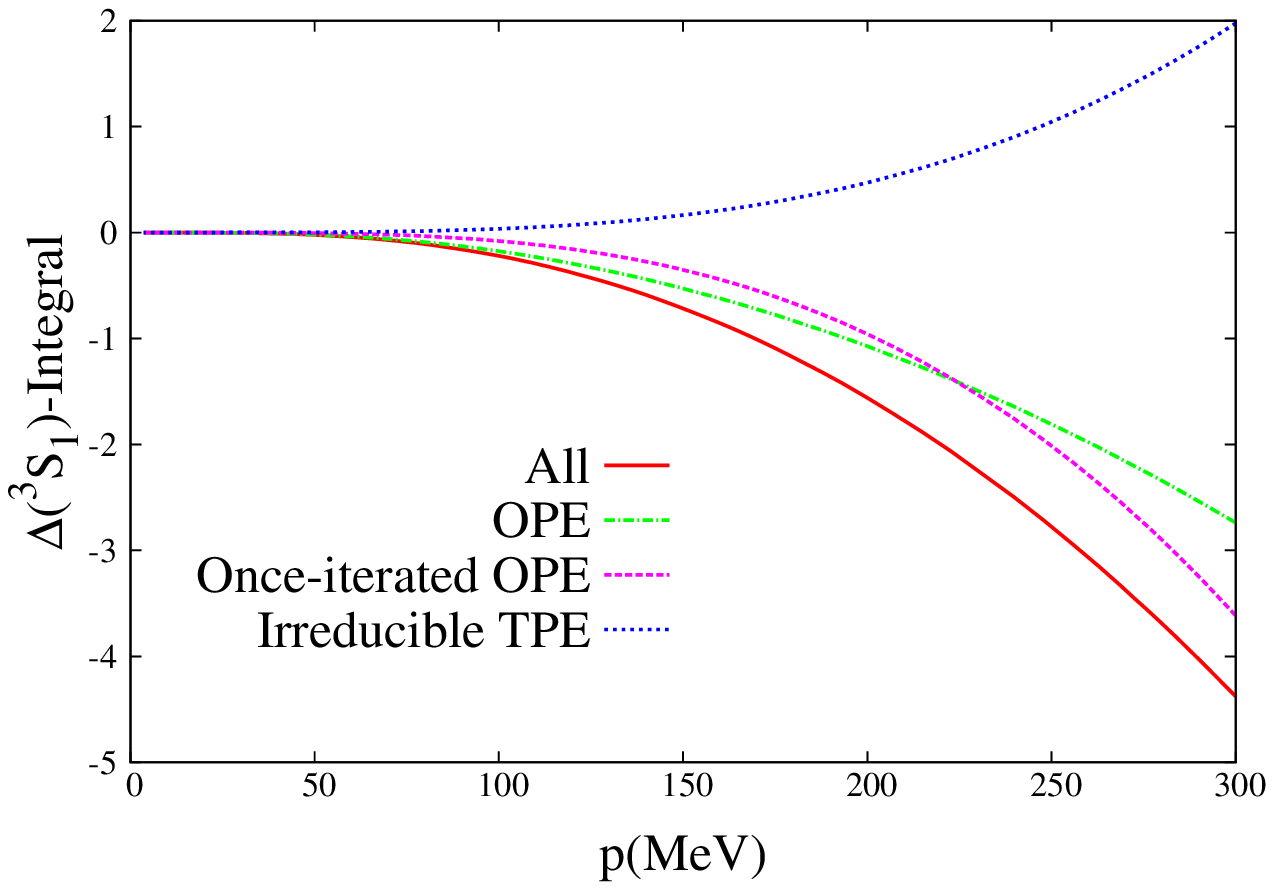} & \includegraphics[width=.4\textwidth]{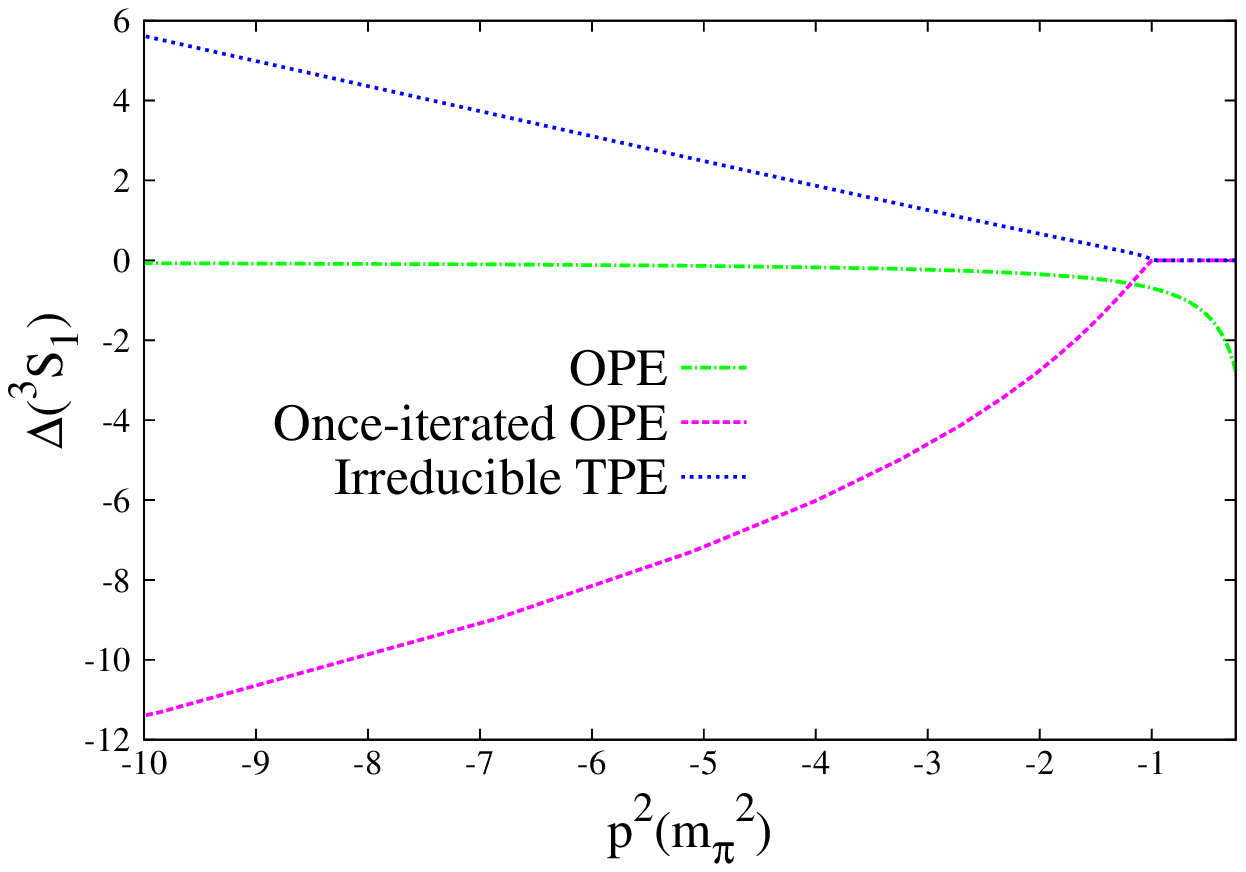} \\ 
\includegraphics[width=.4\textwidth]{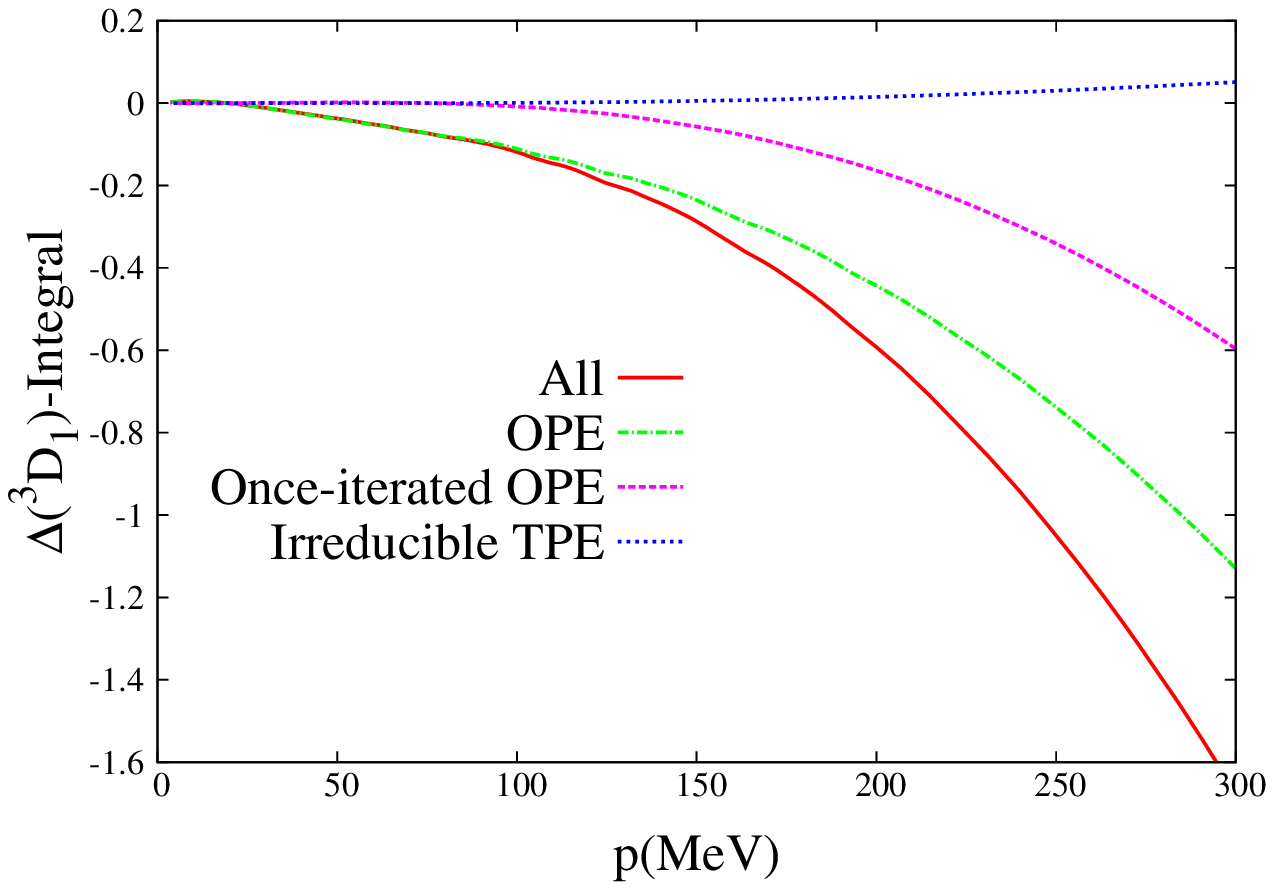} &  \includegraphics[width=.4\textwidth]{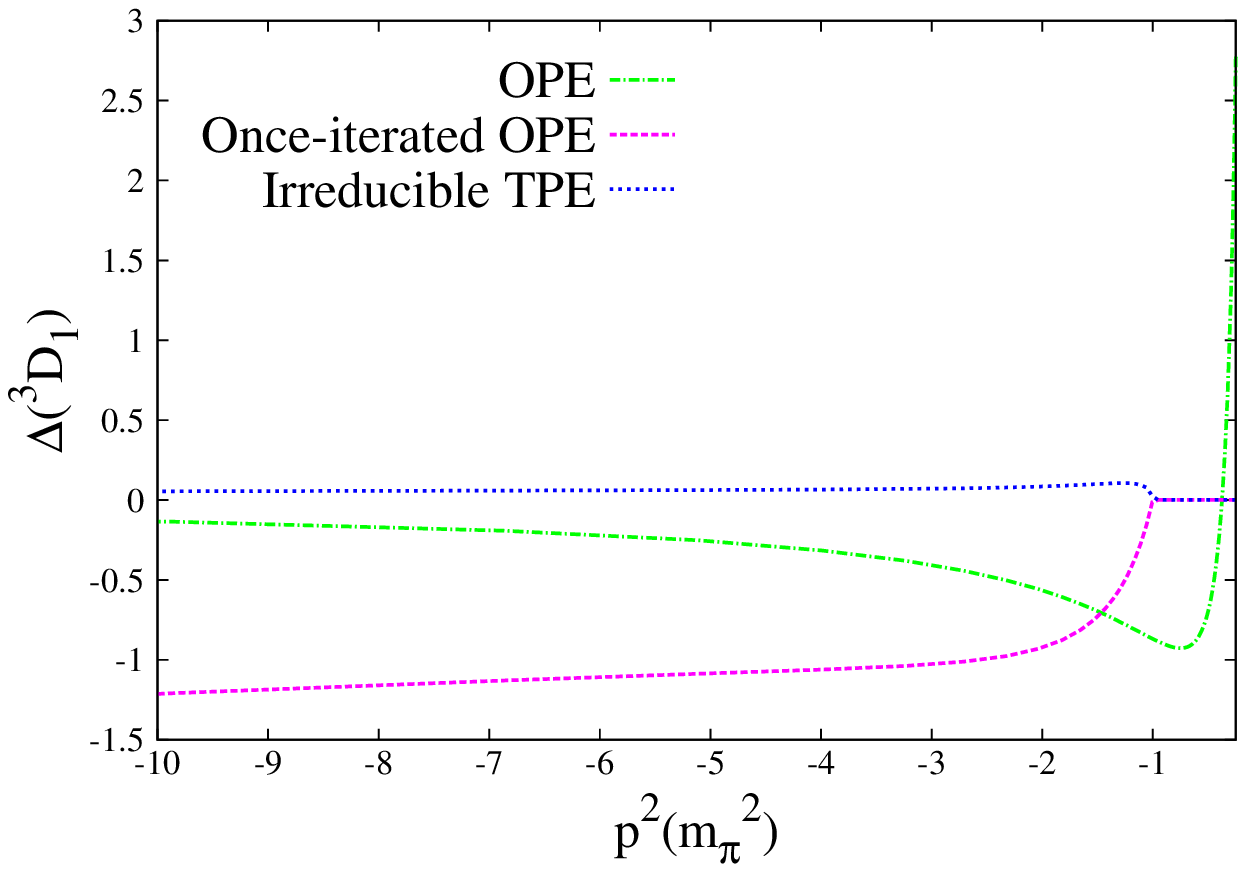}\\  
\includegraphics[width=.4\textwidth]{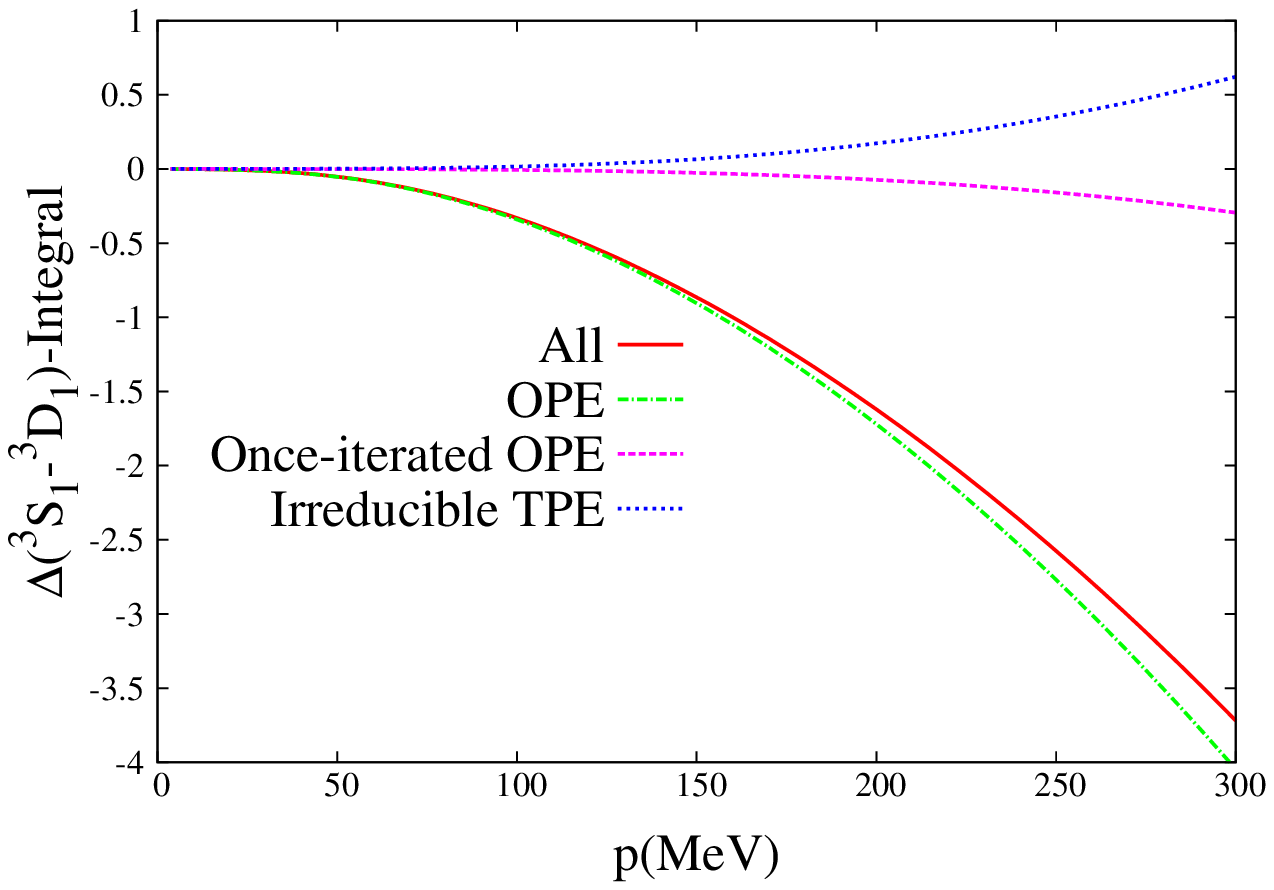} &  \includegraphics[width=.4\textwidth]{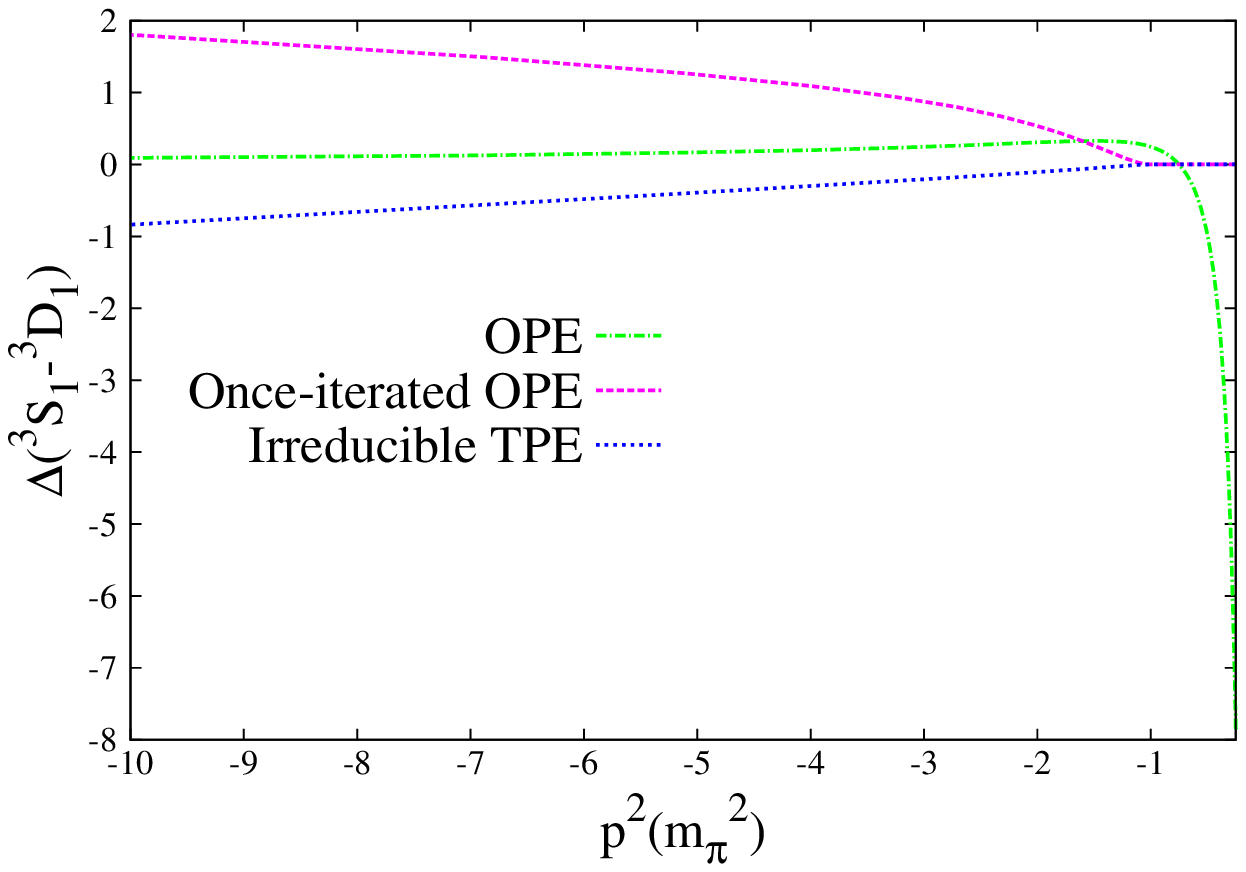} 
\end{tabular}
\caption[pilf]{\protect {\small (Color online.) Left panels: Different contributions to the integrals in    Eq.~\eqref{3sd1quanty}. Right panels: Contributions to $\Delta(A)$. From top to bottom we show the results for $^3S_1$, $^3D_1$ and mixing wave, respectively.  The meaning of the lines is the same as in  Fig.~\ref{fig:1s0quanty}.}
\label{fig:3sd1quanty} }
\end{center}
\end{figure}

This improvement is also clear in the value obtained for the deuteron binding energy, $E_d=-k_d^2/m$. At NLO we obtain $E_d=$2.35--2.38~MeV, a value much closer to experiment $E_d=2.22$~MeV than the one obtained 
at  LO in Ref.~\cite{paper2}, $E_d$=1.7~MeV. A similar situation also occurs for the $^3S_1$  effective range, $r_t$. Proceeding similarly as done in Sec.~\ref{1s0} for $r_s$, we derive an integral expression for calculating $r_t$:
\begin{align}
r_t=-\frac{m}{2\pi^2 a_t}\int_{-\infty}^L dk^2\frac{\Delta_{11}(k^2)D_{11}(k^2)}{(k^2)^2}\left\{\frac{1}{a_t}+\frac{4\pi k^2}{m}g_{11}(0,k^2)  \right\}-\frac{8}{m}\int_0^\infty dq^2\frac{\nu_{11}(q^2)-\rho(q^2)}{(q^2)^2}~.
\end{align}
The last integral on the r.h.s. of the previous equation was not present in Eq.~\eqref{rs1s0} because it is a coupled-wave effect, due to the mixing between the $^3S_1$ and $^3D_1$ partial waves. This equation also exhibits the correlation between $a_t$ and $r_t$, although in a more complicated manner than for the $^1S_0$ partial wave, Eq.~\eqref{rs.le}, because $\nu_{11}(A)$ depends nonlinearly on $D_{11}(A)$. We obtain the value
\begin{align}
r_t=1.36-1.39~\mbox{fm}~,
\end{align}
 to be compared with its experimental value, $r_t=1.759\pm 0.005$~fm. At LO Ref.~\cite{paper2} obtained the much lower result $r_t=0.46$~fm 
when only $a_t$ was taken as experimental input. 

It is also interesting to diagonalize the $^3S_1-{^3D_1}$ $S$-matrix around the deuteron pole position. This allows us to obtain two interesting quantities \cite{swart}, apart from the deuteron binding energy. One of them is the asymptotic $D/S$ ratio $\eta$ of the deuteron. To evaluate this quantity we  diagonalize the $^3S_1-{^3D_1}$ $S$-matrix by an orthogonal matrix ${\cal O}$,
\begin{align}
{\cal O}&=\left(
\begin{array}{ll}
 \cos \epsilon_1 & -\sin\epsilon_1 \\
\sin\epsilon_1 & \cos \epsilon_1
\end{array}
\right)~.
\end{align}
Such that
\begin{align}
S&={\cal O}\left(
\begin{array}{ll}
S_0 & 0 \\
0 & S_2
\end{array}
\right) {\cal O}^T~,
\end{align}
with $S_0$ and $S_2$ the $S$-matrix eigenvalues. The parameter $\eta$ can be expressed in 
terms of the mixing angle $\epsilon_1$ as \cite{swart,swart:88}
\begin{align}
\eta=-\tan\epsilon_1~.
\label{3s1eta}
\end{align}
We also evaluate the residue of the eigenvalue $S_0$ at the deuteron pole position
\begin{align}
S_0&=\frac{N_p^2}{\sqrt{-k_d^2}+i\,\sqrt{A}}+\rm{regular~ terms.}
\end{align}
We obtain the following numerical values:
\begin{align}
\eta=0.029~~,~~N_p^2=0.73~\rm{fm}^{-1}~,
\end{align}
that are close to the calculations $\eta=0.0271(4)$ \cite{ericson:82}, $\eta=0.0263(13)$ \cite{conzett:79} and 
$\eta=0.0268(7)$ \cite{martorell}, as well as to the 
Nijmegen PWA results  \cite{swart:93}
\begin{align}
\eta=0.02543(7)~~,~~ N_p^2=0.7830(7)~\rm{fm}^{-1}~.
\end{align}

Apart from the IEs in Eqs.~\eqref{3s1c} and \eqref{sdijc} we also tried other ones by including more subtractions, so that more experimental input could be fixed, namely, fixing simultaneously (i) $a_t$ and $a_\epsilon$ or (ii)  $a_t$, $r_t$ and $E_d$ or (iii)  $a_t$, $r_t$, $E_d$ and $a_\epsilon$. However, either the coupled-channel iterative process does not converge or we end with the solution corresponding to the uncoupled-wave case.

We also consider here analogous integrals along the LHC to those used in Sec.~\ref{cont_da} in order to quantify the different contributions to $\Delta(A)$,
\begin{align}
\ell_{11}=0~:~&\frac{A^2}{\pi^2}\int_{-\infty}^Ldk^2\frac{\Delta_{11}(k^2)}{(k^2)^2}\int_0^\infty dq^2\frac{\nu_{11}(q^2)}{(q^2-A)(q^2-k^2)}~,\nn\\
\ell_{12}=1~:~&\frac{A(A-k_d^2)}{\pi^2}\int_{-\infty}^L dk^2\frac{\Delta_{12}(k^2)}{(k^2)^2}\int_0^\infty dq^2\frac{\rho(q^2)q^2}{(q^2-A)(q^2-k^2)(q^2-k_d^2)}~,\nn\\
\ell_{22}=2~:~&\frac{A(A-k_d^2)}{\pi^2}\int_{-\infty}^L dk^2\frac{\Delta_{22}(k^2)}{(k^2)^2}\int_0^\infty dq^2\frac{\nu_{22}(q^2)q^2}{(q^2-A)(q^2-k^2)(q^2-k_d^2)}~,
\label{3sd1quanty}
\end{align}
where two subtractions are required in order to have convergent integrals in Eq.~\eqref{3sd1quanty}, as already pointed out in the uncoupled-wave case. For the mixing partial wave, we have taken the integration along the RHC as it were elastic, using $\rho(q^2)$ instead of $\nu_{12}(q^2)$, because the latter would require the actual function  $D_{12}(k^2)$ as it is very sensitive to coupled-channel effects. From the left panels of  Fig.~\ref{fig:3sd1quanty} we see that the total integral is dominated by OPE in all cases. Nevertheless, for $^3S_1$ the individual contributions of the reducible and irreducible TPE  are sizable but of different sign, so that they cancel to a large extent and the dominance of the OPE contribution results. 
We see that, as a whole, the reducible and irreducible contributions are of similar absolute size but with opposite signs.

\section{Coupled waves: $^3P_2-{^3F_2}$}
\label{3pf2}

\begin{figure}[h]
\begin{center}
\begin{tabular}{cc}
\includegraphics[width=.4\textwidth]{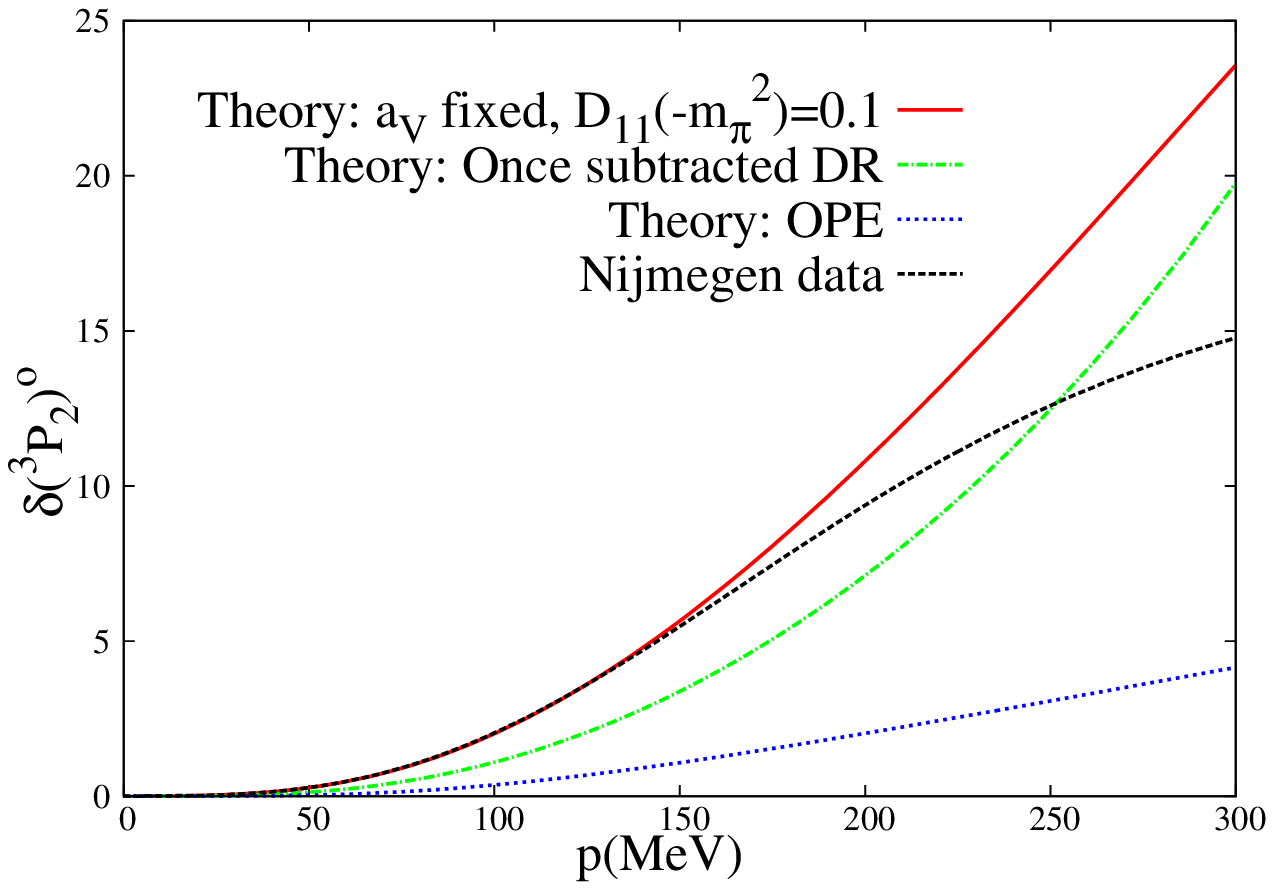} & 
\includegraphics[width=.4\textwidth]{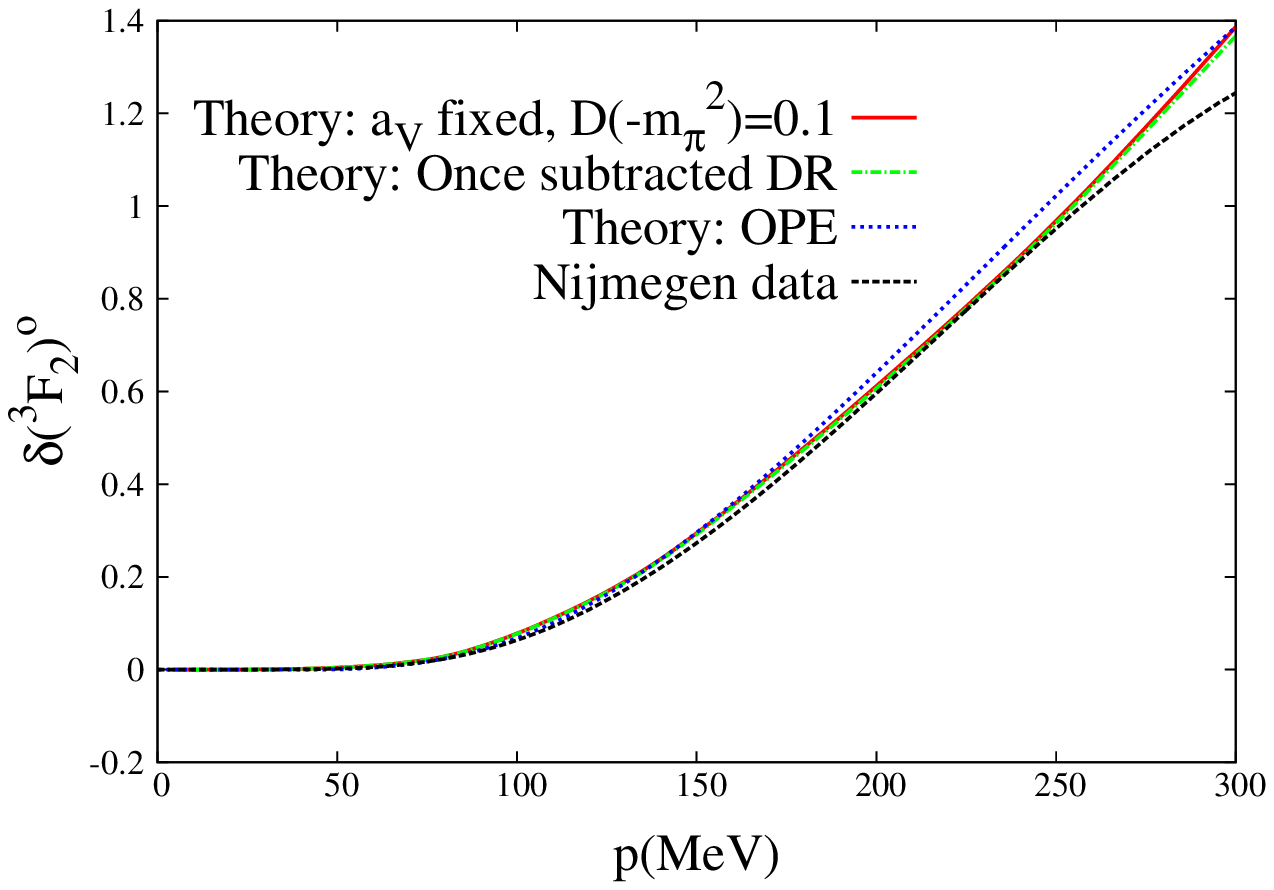}\\  
\includegraphics[width=.4\textwidth]{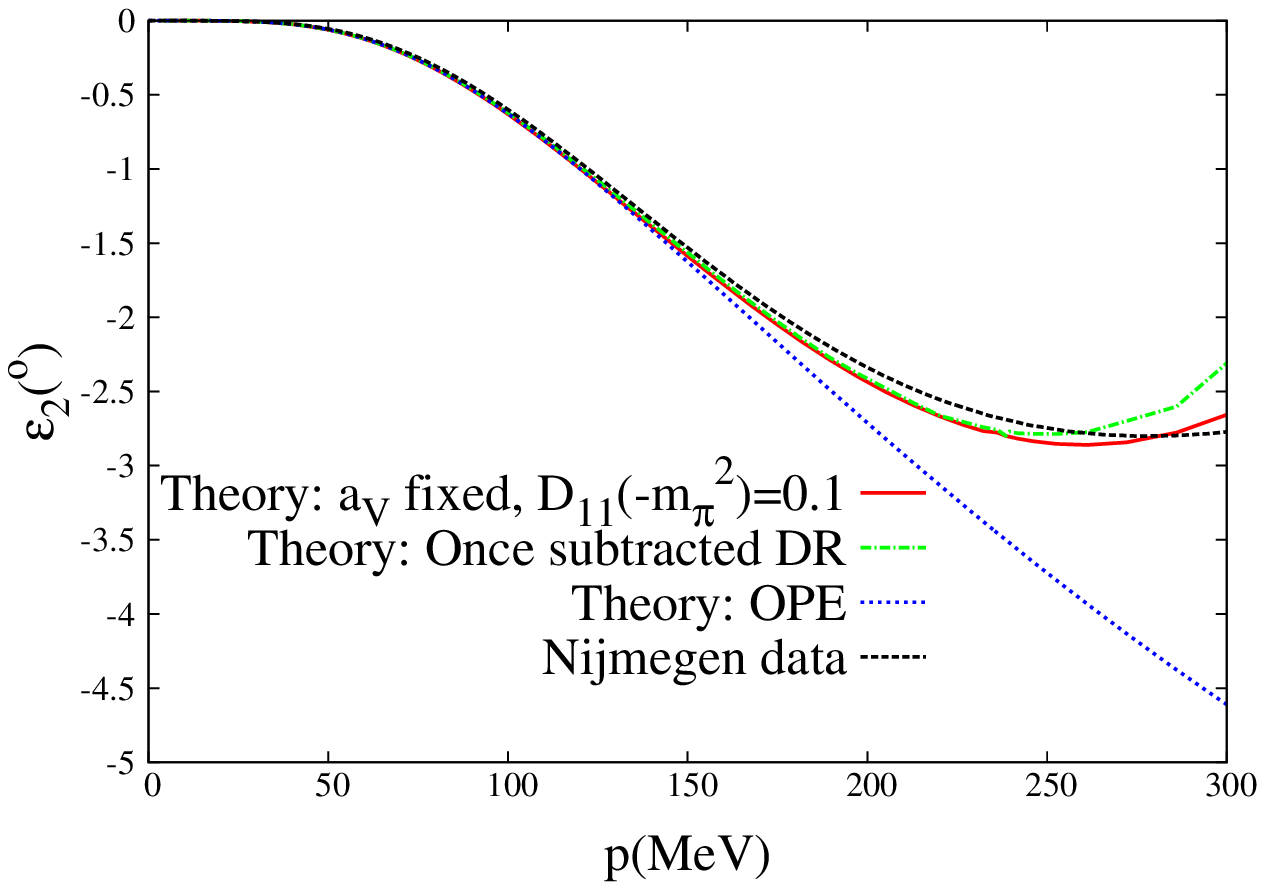}  
\end{tabular}
\caption[pilf]{\protect {\small (Color online.) From top to bottom and left to right: Phase shifts for $^3P_2$, $^3F_2$ and the mixing angle  $\epsilon_2$, in order.  
The (red) solid line corresponds to the results obtained with twice-subtracted DRs 
 for $^3P_2$, while  once-subtracted DRs are used for the latter partial wave to obtain the 
(green) dash-dotted line. The (blue) dotted line is the results with only  OPE from Ref.~\cite{paper2}.
 The Nijmegen PWA phase shifts are given by (black) dashed line.}
\label{fig:3pf2} }
\end{center}
\end{figure}

\begin{figure}
\begin{center}
\begin{tabular}{cc}
\includegraphics[width=.4\textwidth]{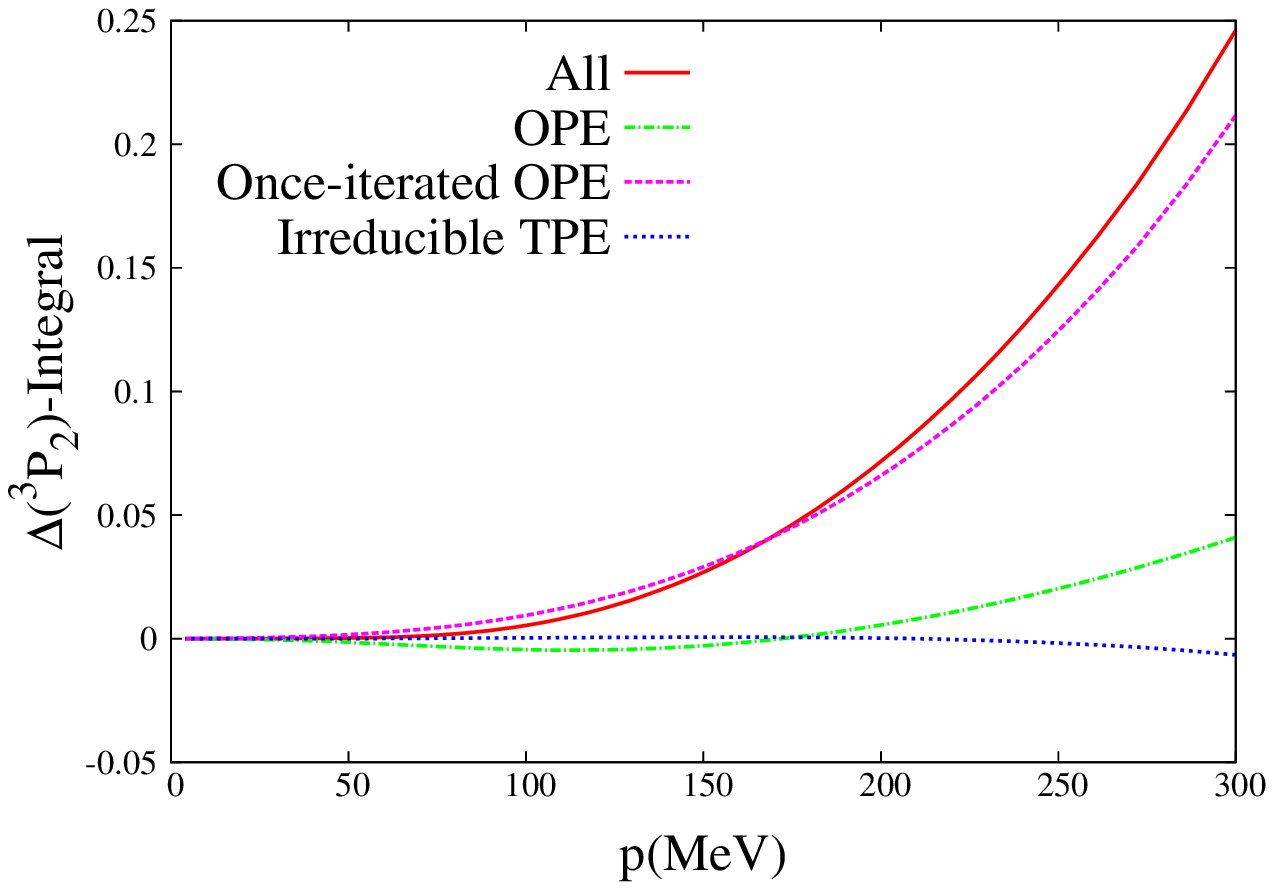} & \includegraphics[width=.4\textwidth]{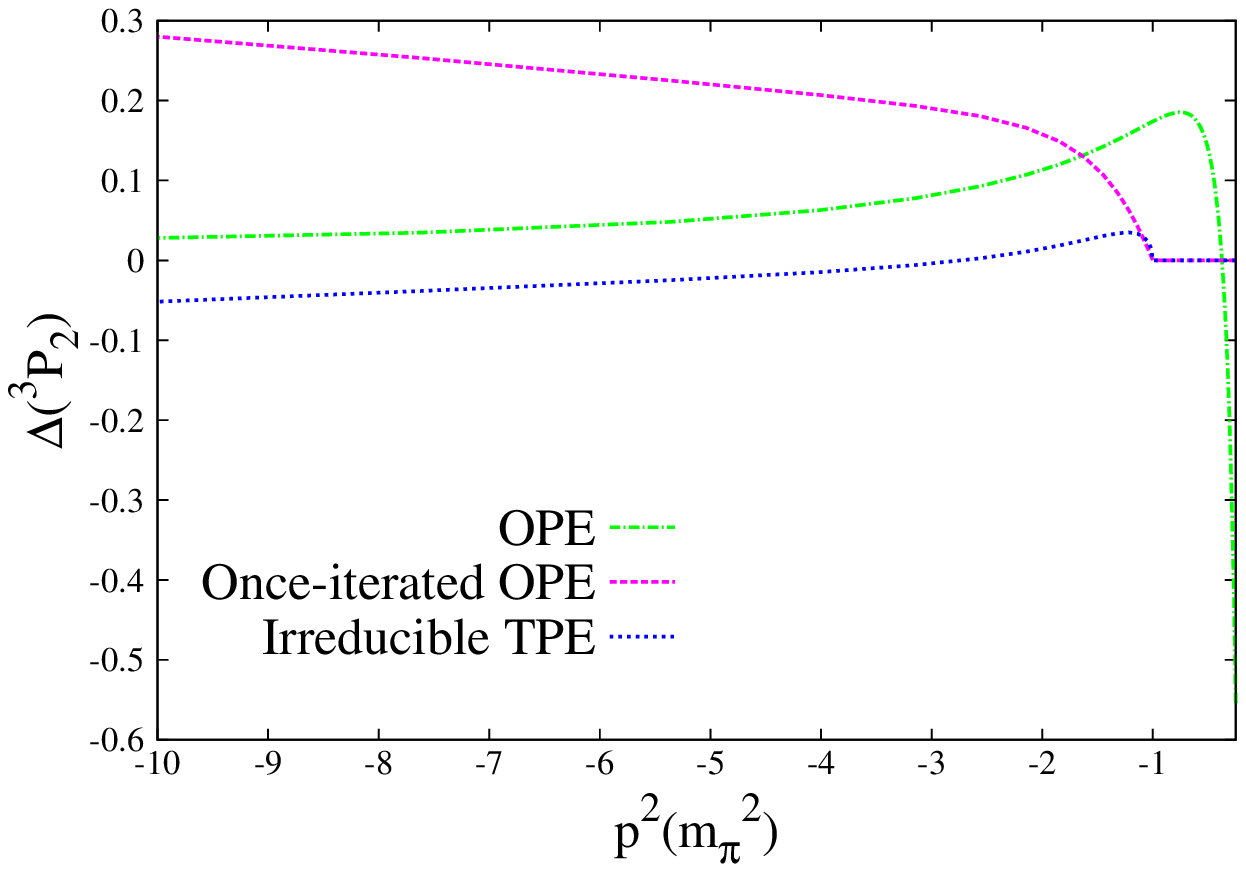} \\ 
\includegraphics[width=.4\textwidth]{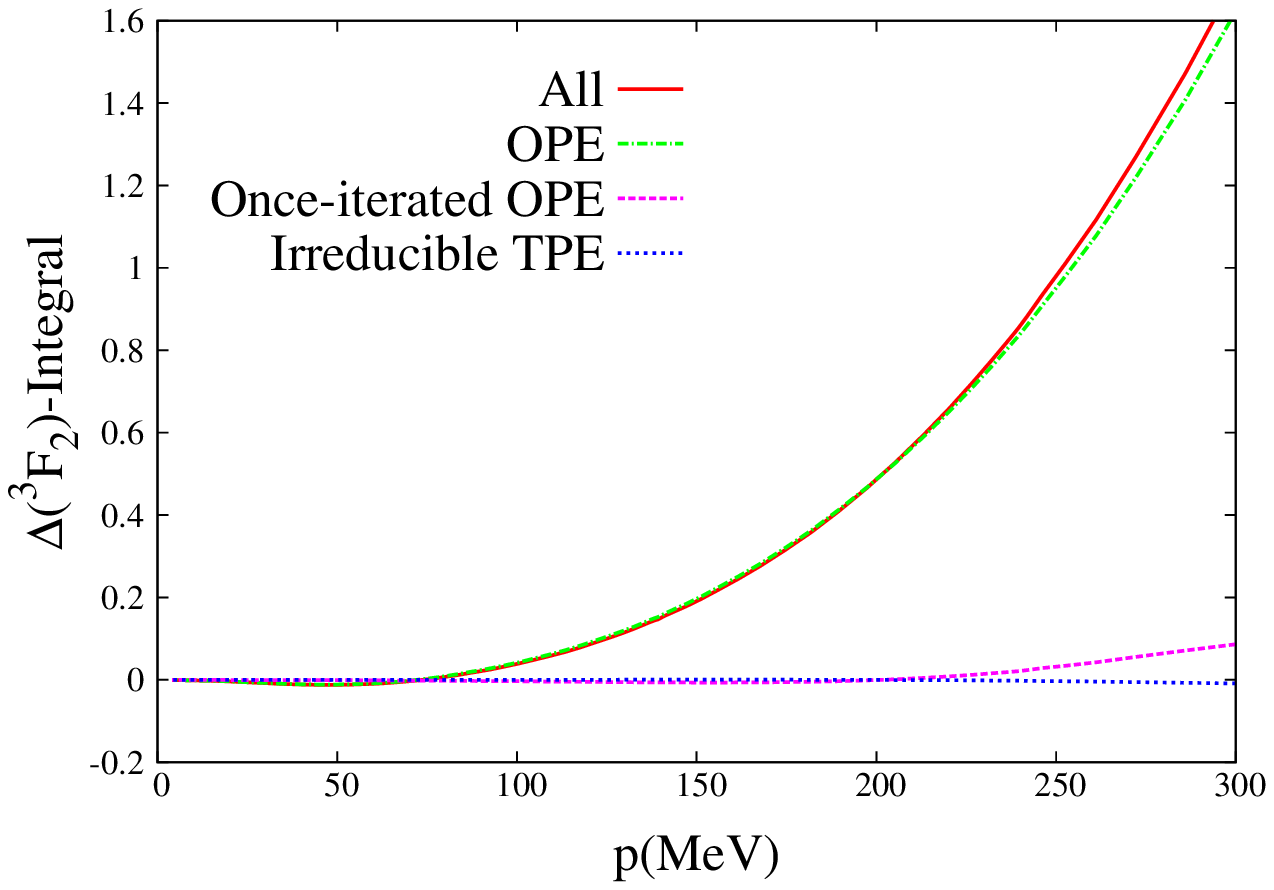} &  \includegraphics[width=.4\textwidth]{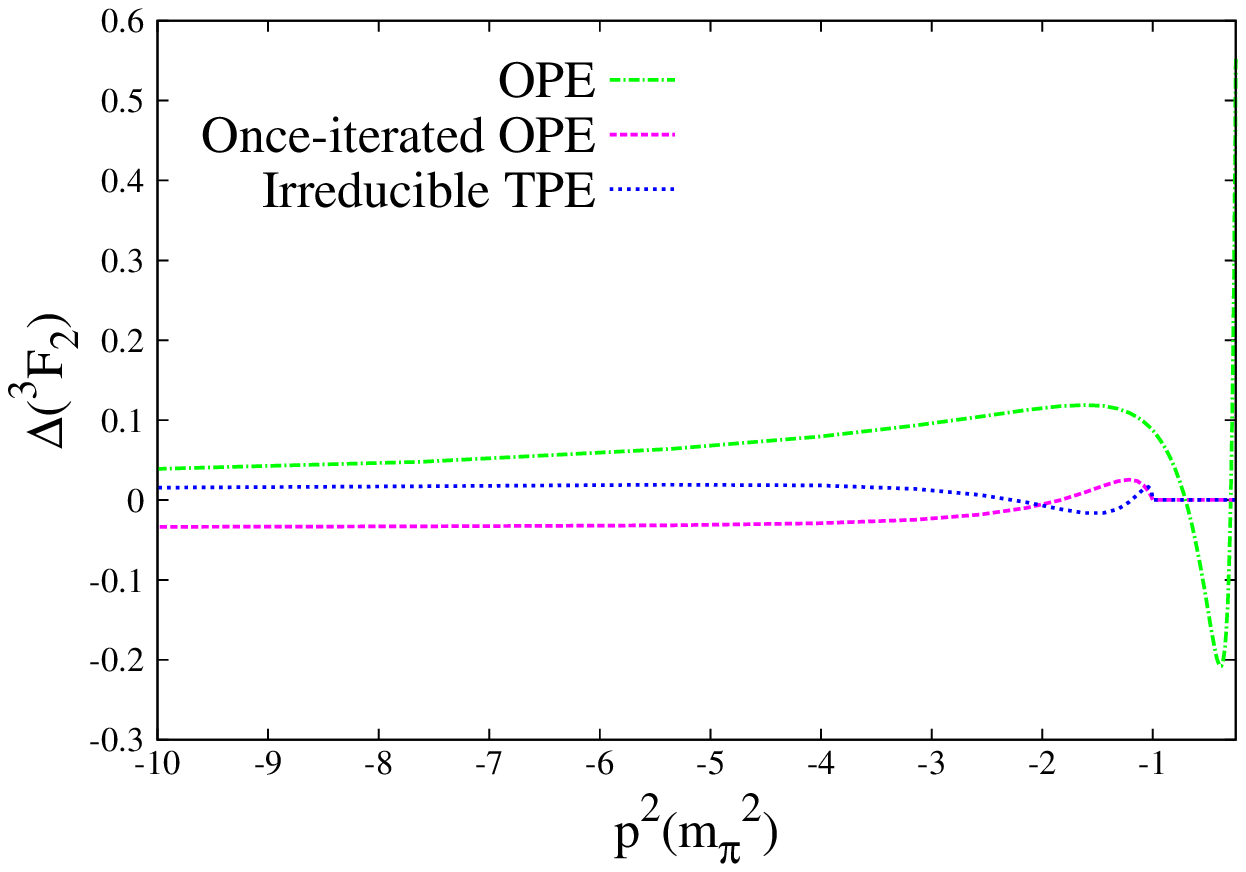}\\  
\includegraphics[width=.4\textwidth]{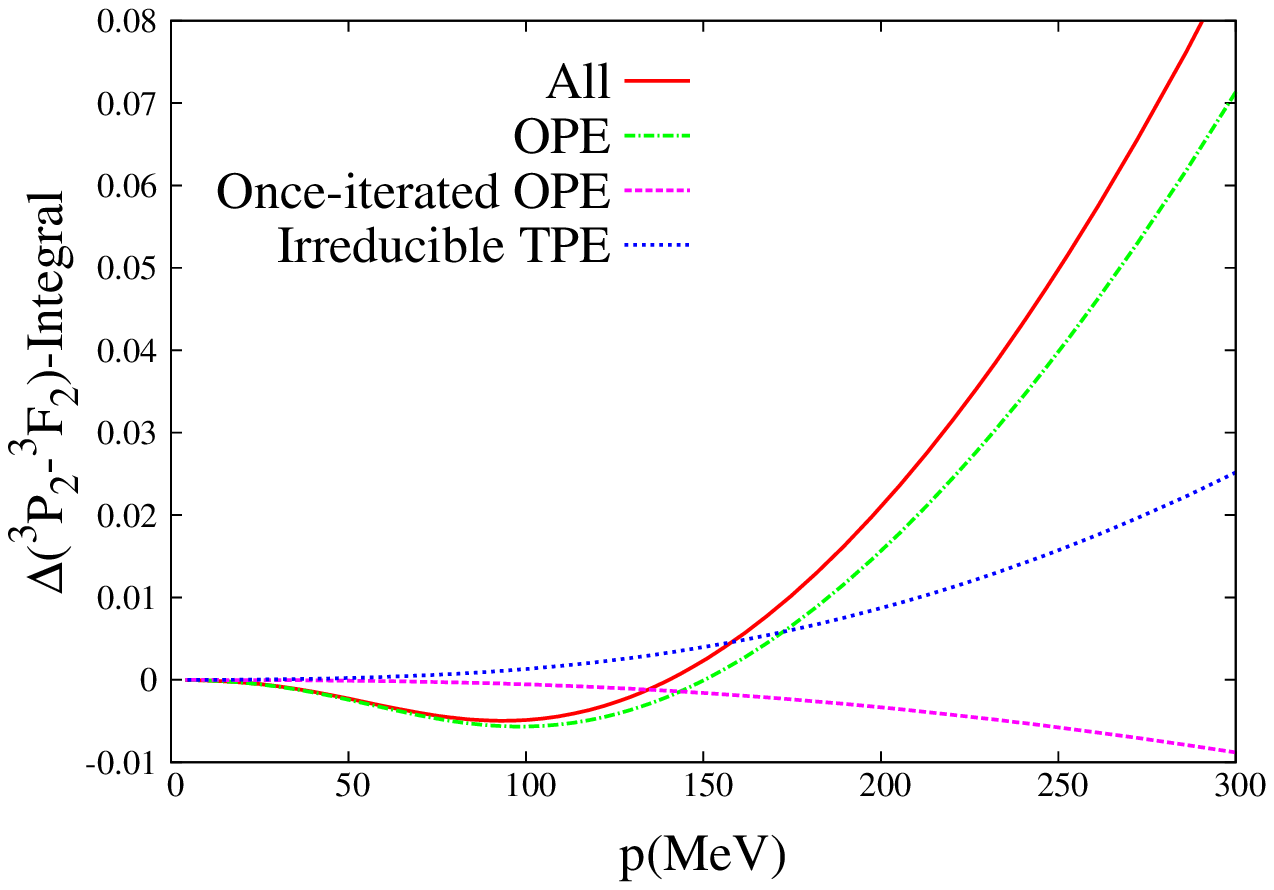} &  \includegraphics[width=.4\textwidth]{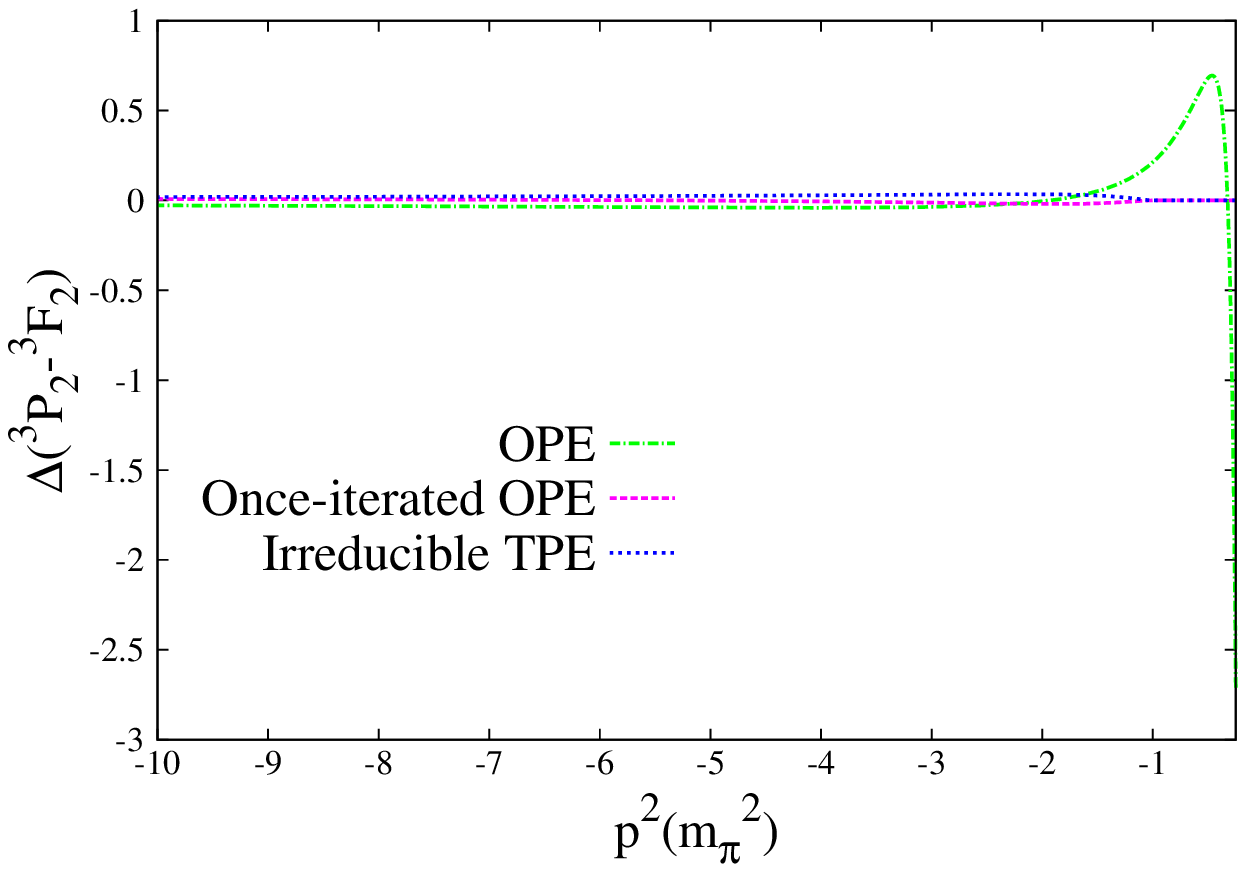} 
\end{tabular}
\caption[pilf]{\protect {\small (Color online.) Left panels: Different contributions to the integrals in    Eq.~\eqref{3pf2quanty}. Right panels: Contributions to $\Delta(A)$. From top to bottom we show the results for $^3P_2$, $^3F_2$ and mixing wave, respectively. The meaning of the lines is the same as in Fig.~\ref{fig:1s0quanty}.}
\label{fig:3pf2quanty} }
\end{center}
\end{figure}

In this section we consider the coupled wave system $^3P_2-{^3F_2}$ making use of Eqs.~\eqref{highdc} and \eqref{highndc} with $\ell_{11}=1$, 
$\ell_{12}=2$ and $\ell_{22}=3$. In the following we always take $C=-M_\pi^2$ in Eq.~\eqref{highdc} and instead of the coefficients $\delta_p^{(ij)}$ we 
 directly use $D_{ij}^{(n)}(C)$, $n=0,\ldots,\ell_{ij}-2$, as 
 the free parameters. As discussed in Sec.~\ref{leq2}, it is enough to take  $D_{ij}^{(\ell_{ij}-2)}(C)$ as the only active free parameter for every partial wave.

We find that the results are all quite insensitive to $D_{22}(-M_\pi^2)$ and $D'_{22}(-M_\pi^2)$, as one would expect because $F$-waves are expected to be perturbative, as already discussed in Sec.~\ref{fw}. This is another confirmation of this conclusion. The fitted parameter  $D'_{22}(-M_\pi^2)$ becomes negative and of several units of size, but essentially the same results are obtained as long as  $D'_{22}(-M_\pi^2)< -1~M_\pi^{-2}$. Regarding $D_{22}(-M_\pi^2)$ we fix it to $1$. Our results are then only sensitive to  $D_{12}(-M_\pi^2)$ with the best fitted value
\begin{align}
D_{12}(-M_\pi^2)&=1.1~.
\label{d12fit3pd2}
\end{align}
From these results we can calculate the $P$-wave scattering volume, which is just given by the first derivative at $A=0$ of the function $N_{11}(A)$.
 This is straightforwardly worked out from Eq.~\eqref{highndc}, with the result, $a_V=0.12~M_\pi^{-3}$, that is a 20$\%$ off 
its phenomenological value $a_V=0.0964~M_\pi^{-3}$ obtained from Ref.~\cite{Stoks:1994wp}. To improve this situation we  employ 
a twice-subtracted DR  by taking $n=2$ in Eqs.~\eqref{inteq1c} and \eqref{nc} for the two subtractions in the function $N_{11}(A)$ at $C=0$ with $\nu^{(11)}_1=0$ and 
\begin{align}
\nu^{(11)}_2=\frac{4\pi a_V}{m}~,
\label{nu23p2}
\end{align}
in terms of the experimental value of $a_V$. Now   $D_{11}(-M_\pi^2)$ is also a free parameter fitted to data,
\begin{align}
D_{11}(-M_\pi^2)&=0.1~,
\label{d11fit3pd2}
\end{align}
while for $D_{12}(-M_\pi^2)$ and $D_{22}'(M_\pi^2)$ the same values as in the case of 
the once-subtracted DR for $^3P_2$ are employed, since no improvement in the reproduction of data results by varying them.
  The resulting phase shifts and mixing angle are shown in Fig.~\ref{fig:3pf2}.
 As we see there, the $^3F_2$ phase shifts and mixing angle $\epsilon_2$ are reproduced quite well, independently of the number of subtractions taken 
 for the $^3P_2$ partial wave. Concerning the $^3P_2$ phase shifts,  when  the scattering volume is fixed to its experimental value  a better reproduction 
of data is achieved at low three-momenta (red solid line),  than when  it is not imposed (green dash-dotted line).
 In all these coupled partial waves we observe a noticeable improvement of the OPE results of Ref.~\cite{paper2}.

At the practical numerical level it is interesting to remark that for the coupled waves the mixing angle is small. Then, as a first approximation, one can study separately the waves with orbital angular momentum $J-1$ and $J+1$ as if they were uncoupled. In this way, it is more efficient numerically to fit the free parameters present in them  than if the full iterative process of coupled waves were taken.  Once this is done, the mixing is included but we first keep the values obtained in the uncoupled-wave limit 
for the free parameters fitted then, so that it only remains to determine those  present in the mixing partial wave.
 Afterwards, we vary around the parameters fixed by the uncoupled-wave case until the full results are stable.

With regard to the integrals along the LHC in order to  quantify the different contributions to $\Delta(A)$, we have now, according to the number of subtractions taken in the DRs for each partial wave, the following expressions:
\begin{align}
\ell_{11}=1~:~&\frac{A(A+M_\pi^2)}{\pi^2}\int_{-\infty}^L dk^2\frac{\Delta_{11}(k^2)}{(k^2)^2}\int_0^\infty dq^2\frac{\nu_{11}(q^2) q^2}{(q^2-A)(q^2-k^2)(q^2+M_\pi^2)}~,\nn\\
\ell_{12}=2~:~&\frac{A(A+M_\pi^2)}{\pi^2}\int_{-\infty}^L dk^2 \frac{\Delta_{12}(k^2)}{(k^2)^2}\int_0^\infty dq^2 \frac{ \rho(q^2) q^2}{(q^2-A)(q^2-k^2)(q^2+M_\pi^2)}~,\nn\\
\ell_{22}=3~:~&\frac{A(A+M_\pi^2)^2}{\pi^2}\int_{-\infty}^L dk^2\frac{\Delta_{22}(k^2)}{(k^2)^3}\int_0^\infty dq^2\frac{\nu_{22}(q^2) (q^2)^2}{(q^2-A)(q^2-k^2)
(q^2+M_\pi^2)^2}~.
\label{3pf2quanty}
\end{align}

For $^3F_2$ and the mixing partial wave the situation is as usual, so that the OPE contribution dominates the respective integral along the LHC. However, for the $^3P_2$  the reducible TPE contribution is much larger than the OPE one. We consider that this situation is very specific for this partial wave. This is manifest by the fact that the OPE contribution in this wave is in absolute value more than one order of magnitude smaller than in the other $P$-waves, namely, $^1P_1$, $^3P_0$, $^3P_1$ and the mixing wave in the $^3S_1-{^3D_1}$ system. This can be easily checked by comparing the two panels in the first row of Fig.~\ref{fig:3pf2quanty} with Fig.~\ref{fig:pwquanty} and the two panels in the last row of Fig.~\ref{fig:3sd1quanty}. On the other hand, we also observe that  the reducible and irreducible TPE contributions have typically similar size in absolute value, taking a whole picture of all the partial waves involved in the $^3P_2-{^3F_2}$ system.

\section{Coupled waves: $^3D_3-{^3G_3}$ }
\label{3dg3}

The orbital momenta attached to the $^3D_3-{^3G_3}$ system are $\ell=2$, 3 and 4 for the $^3D_3$, mixing wave and $^3G_5$ coupled waves, in this order. 
These values are used in Eqs.~\eqref{highdc} and \eqref{highndc} to provide the appropriate IEs.

\begin{figure}[h]
\begin{center}
\begin{tabular}{cc}
\includegraphics[width=.4\textwidth]{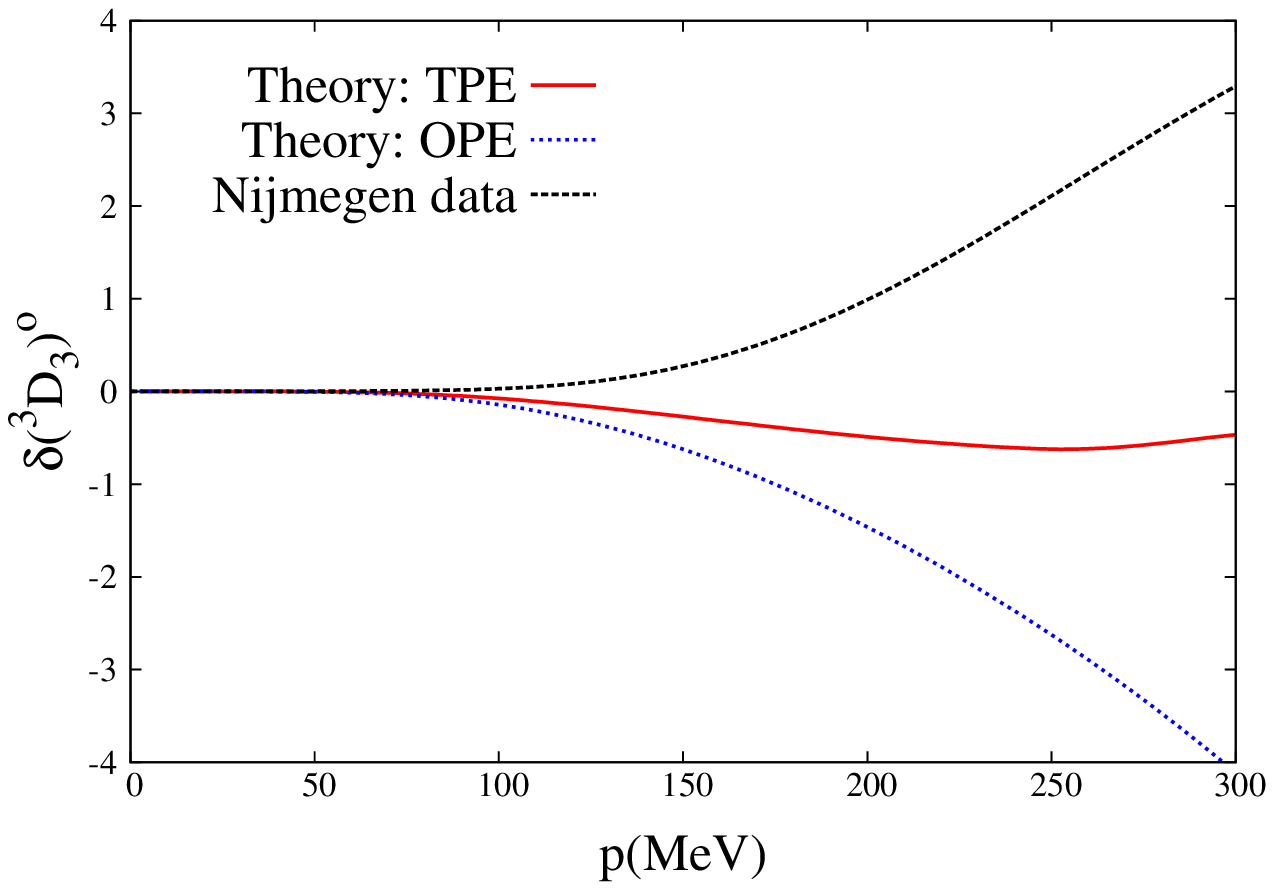} & 
\includegraphics[width=.4\textwidth]{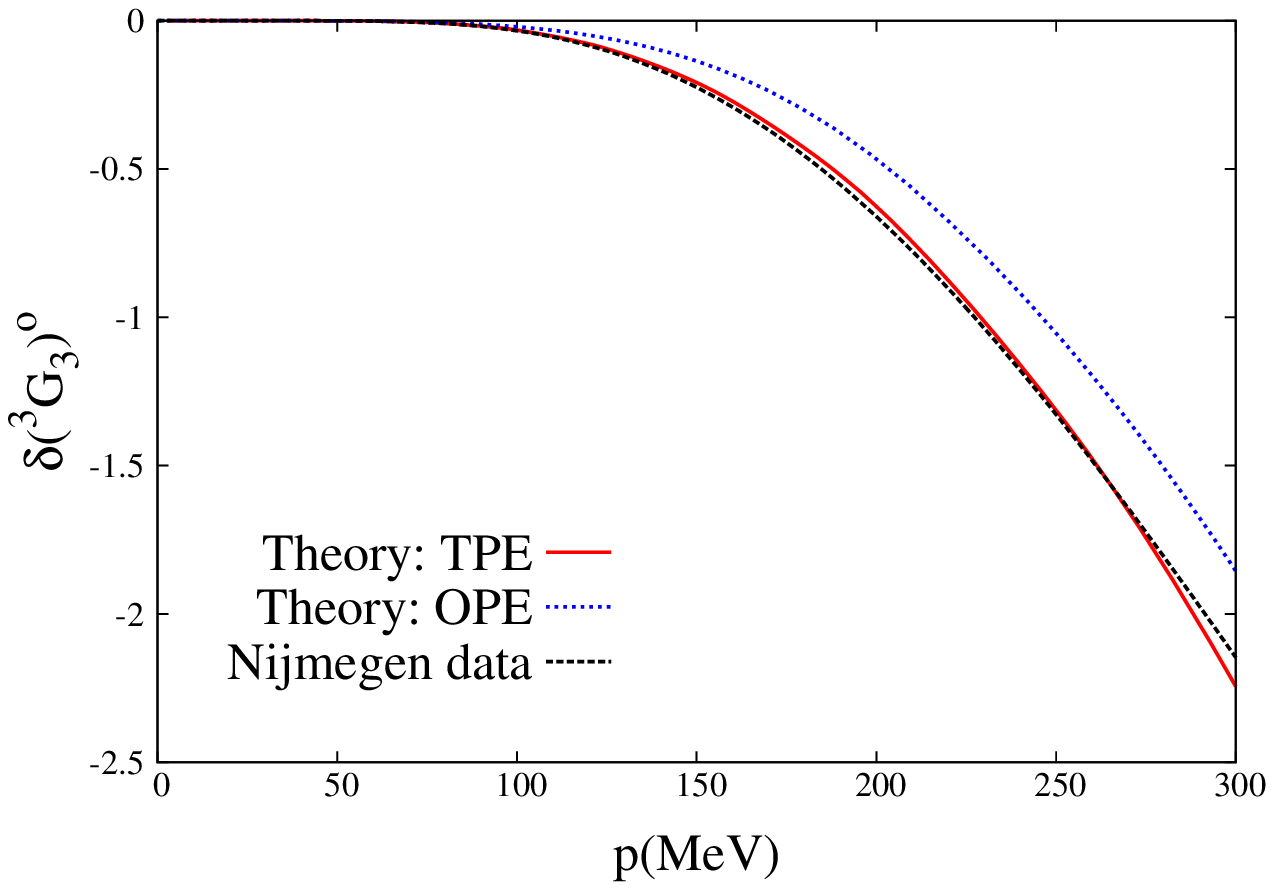}\\  
\includegraphics[width=.4\textwidth]{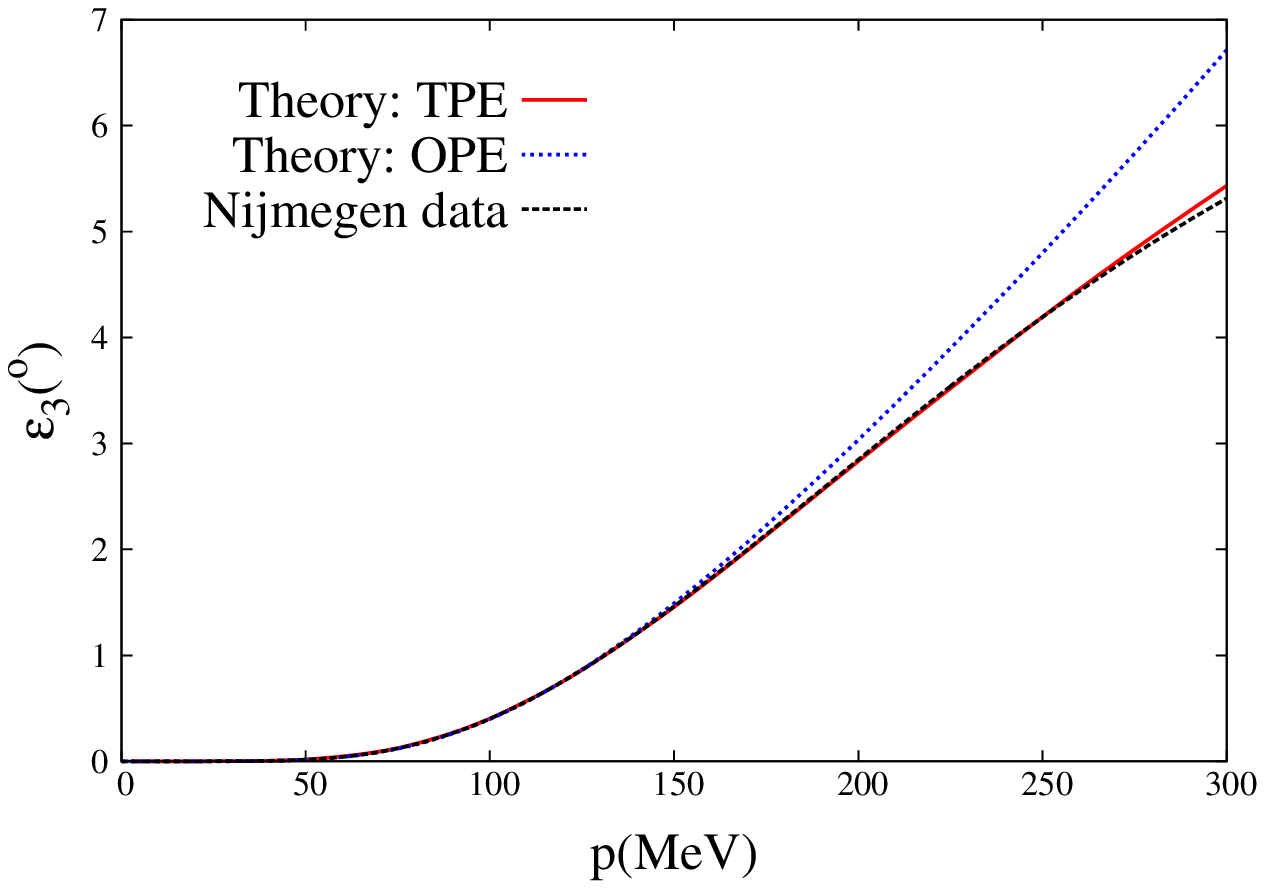}  
\end{tabular}
\caption[pilf]{\protect {\small (Color online.) From top to bottom and left to right: Phase shifts for $^3D_3$, $^3G_3$ and the mixing angle  $\epsilon_3$, respectively.
The (red) solid line corresponds to our NLO results, the (blue) dotted line is the results with only OPE from Ref.~\cite{paper2} and the Nijmegen PWA phase shifts are given by (black) dashed line.}
\label{fig:3dg3} }
\end{center}
\end{figure}

 The fit is not able to fix a definite value for $D_{11}(-M_\pi^2)$, which is always given with large uncertainties and very much dependent on the upper limit of the energy  taken in the fit. Then, we fix it to 1 and the curves are basically the same. For the mixing wave we also have $D_{12}(C)=1$. Regarding the first derivative 
 $D'_{12}(C)$ a slightly negative value, e.g. $-0.1 ~M_\pi^{-2}$, offers the best results. This  
corresponds basically to the situation with the perturbative values for the mixing wave. For the $^3G_3$ wave 
the fit is also consistent with a smooth behavior for the $D_{22}(A)$ function for $A<0$. In this case, 
$D_{22}(C)=1$, $D'_{22}(C)=0$ and $D_{22}^{(2)}(C)>1~M_\pi^{-4}$, that is, only the highest order derivative is 
different from zero with the value of the function at $C$ equal to 1, according to the rule given in Sec.~\ref{leq2}. The resulting phase shifts and    mixing angle are shown in Fig.~\ref{fig:3dg3} by the (red) solid line. We already see that the phase shifts for $^3G_3$ and the mixing angle $\epsilon_3$ are fairly well reproduced.   With respect to the phase shifts for $^3D_3$ there is an improvement compared with the OPE results of Ref.~\cite{paper2}, but still the data are not well reproduced.

\begin{figure}
\begin{center}
\begin{tabular}{cc}
\includegraphics[width=.4\textwidth]{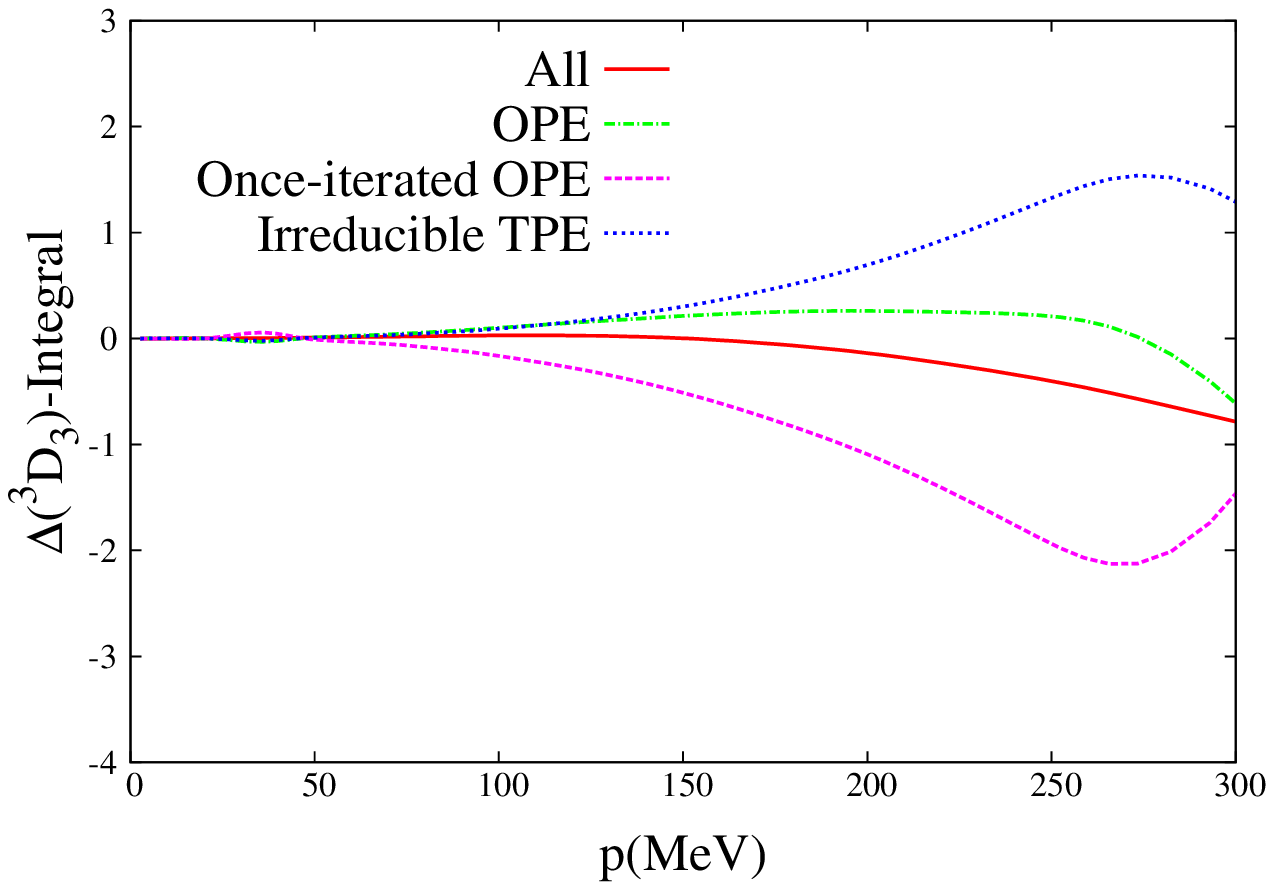} & \includegraphics[width=.4\textwidth]{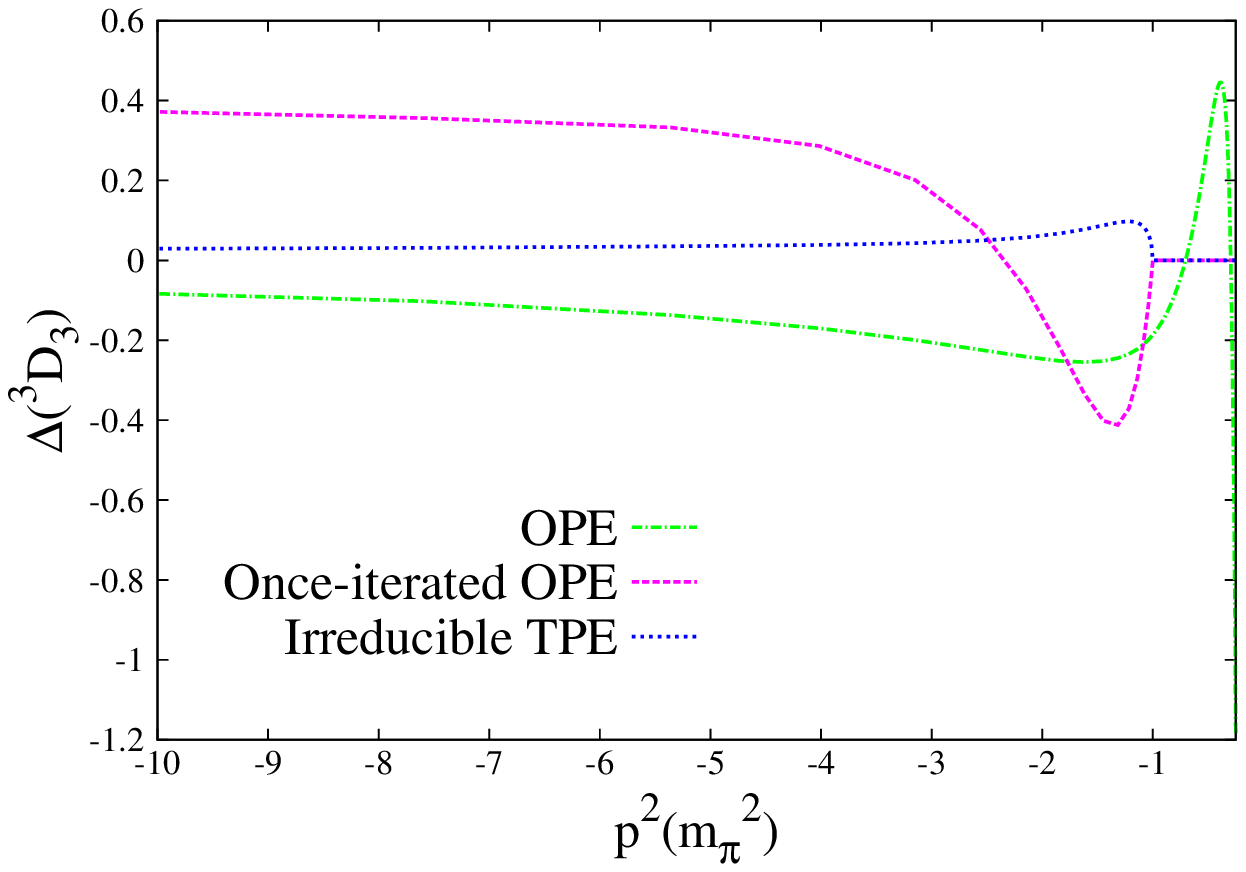} \\ 
\includegraphics[width=.4\textwidth]{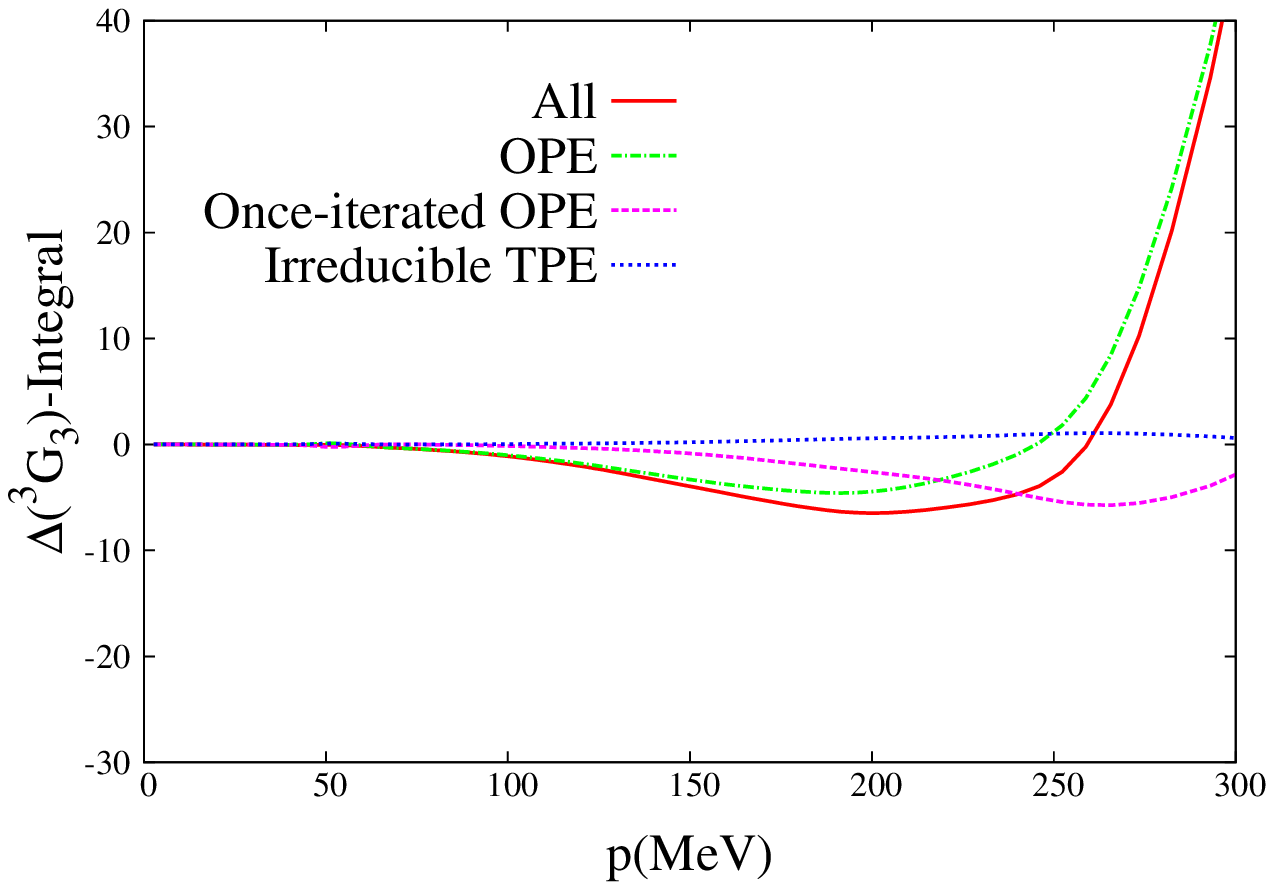} &  \includegraphics[width=.4\textwidth]{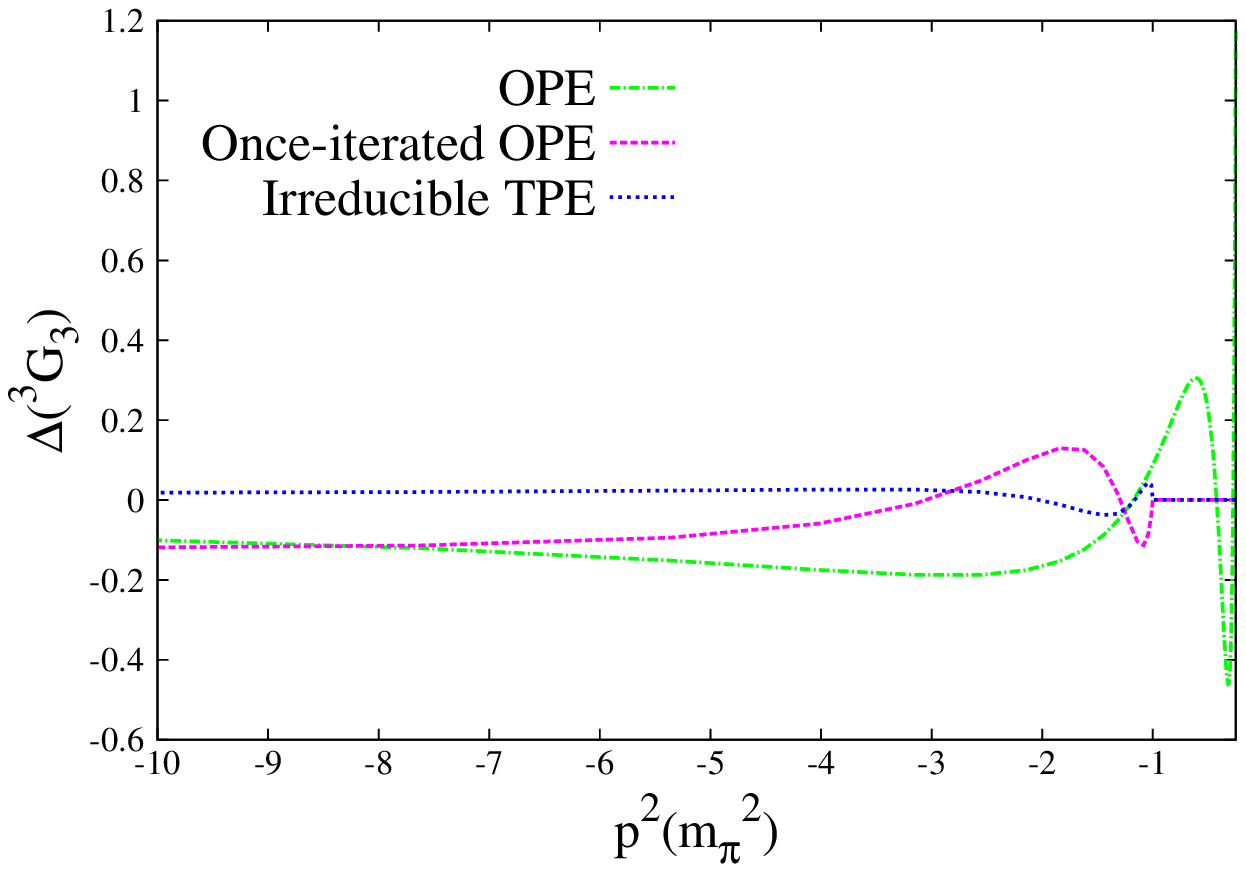}\\  
\includegraphics[width=.4\textwidth]{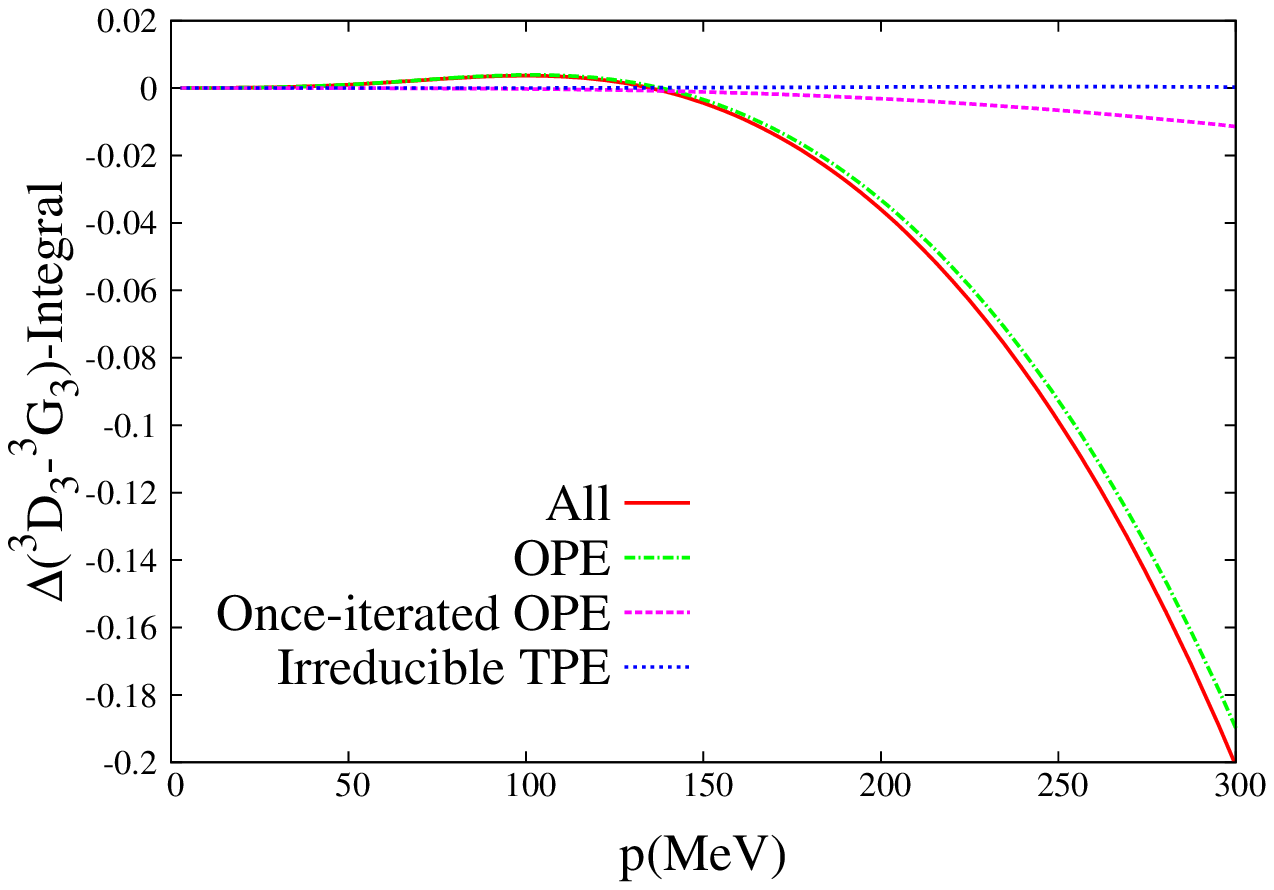} &  \includegraphics[width=.4\textwidth]{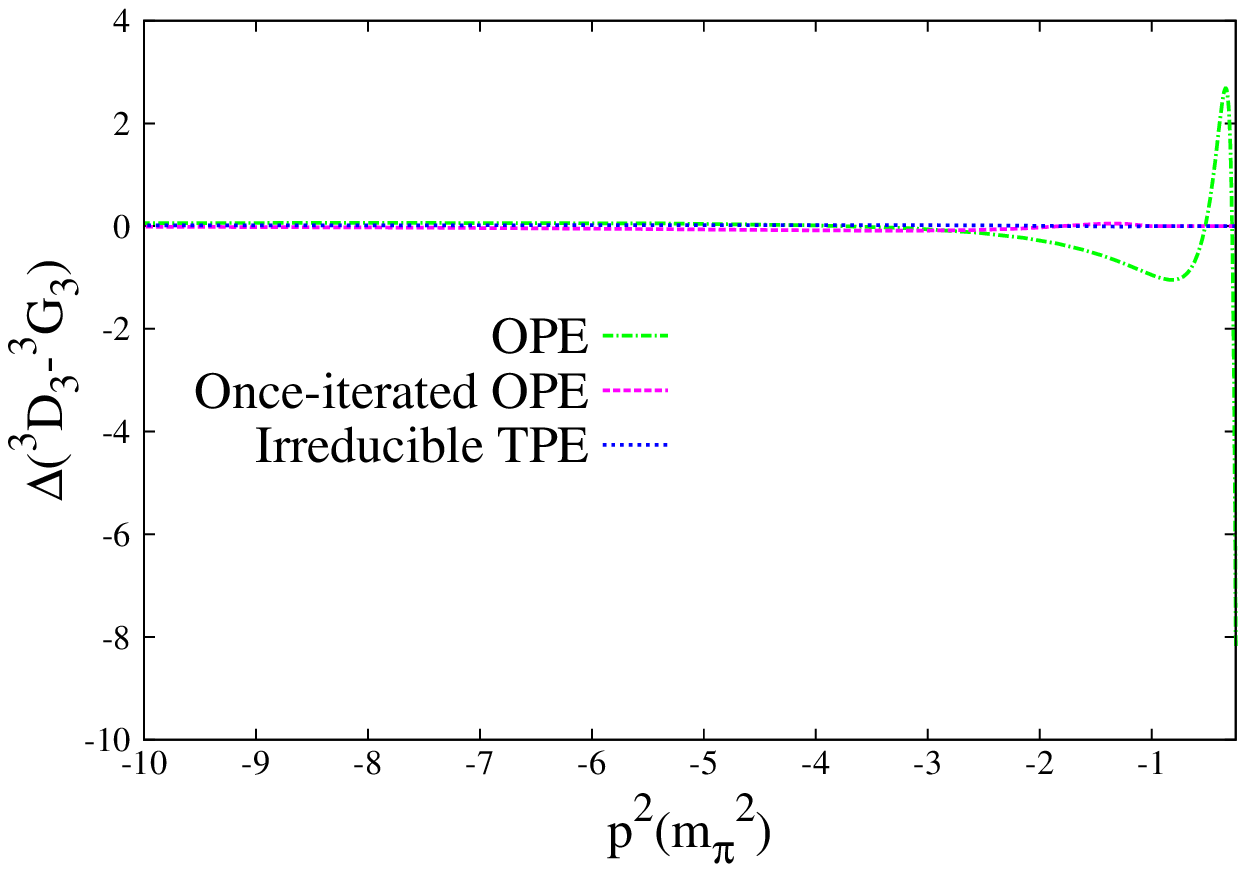} 
\end{tabular}
\caption[pilf]{\protect {\small (Color online.) Left panels: Different contributions to the integrals in    Eq.~\eqref{3dg3quanty}. Right panels: Contributions to $\Delta(A)$. From top to bottom we show the results for $^3D_3$, $^3G_3$ and mixing wave, respectively. The meaning of the lines is the same as in Fig.~\ref{fig:1s0quanty}.}
\label{fig:3dg3quanty} }
\end{center}
\end{figure}

To quantify the different contributions to $\Delta(A)$ we evaluate the corresponding integrals along the LHC:
\begin{align}
&\frac{A(A+M_\pi^2)^{\ell_{ij}-1}}{\pi^2}\int_{-\infty}^L dk^2\frac{\Delta_{ij}(k^2)}{(k^2)^{\ell_{ij}}}\int_0^\infty dq^2\frac{\mu_{ij}(q^2)(q^2)^{\ell_{ij}-1}}{(q^2-A)(q^2-k^2)(q^2+M_\pi^2)^{\ell_{ij}-1}}~,
\label{3dg3quanty}
\end{align}
where $\ell_{ij}=2,~3$ and 4, $\mu_{11}=\nu_{11}$, $\mu_{22}=\nu_{22}$ and $\mu_{12}=\rho$. The results are shown in   Fig.~\ref{fig:3dg3quanty}. We see that for $^3G_3$ and the mixing wave  the integral is dominated by OPE. However, for $^3D_3$ the irreducible and reducible TPE contributions are large, indeed each of them is larger than OPE, though they have opposite signs so they  cancel mutually to a large extent. This is why OPE is still the most important contribution to the total result, but we then 
expect for this wave that the higher order contributions will play a more prominent role.  
 Indeed, $^3D_3$ is the wave for which  the reproduction 
 of data is still poor  in Fig.~\ref{fig:3dg3}. 

\section{Coupled waves: $^3F_4-{^3H_4}$ }
\label{3fh4}

\begin{figure}[h]
\begin{center}
\begin{tabular}{cc}
\includegraphics[width=.4\textwidth]{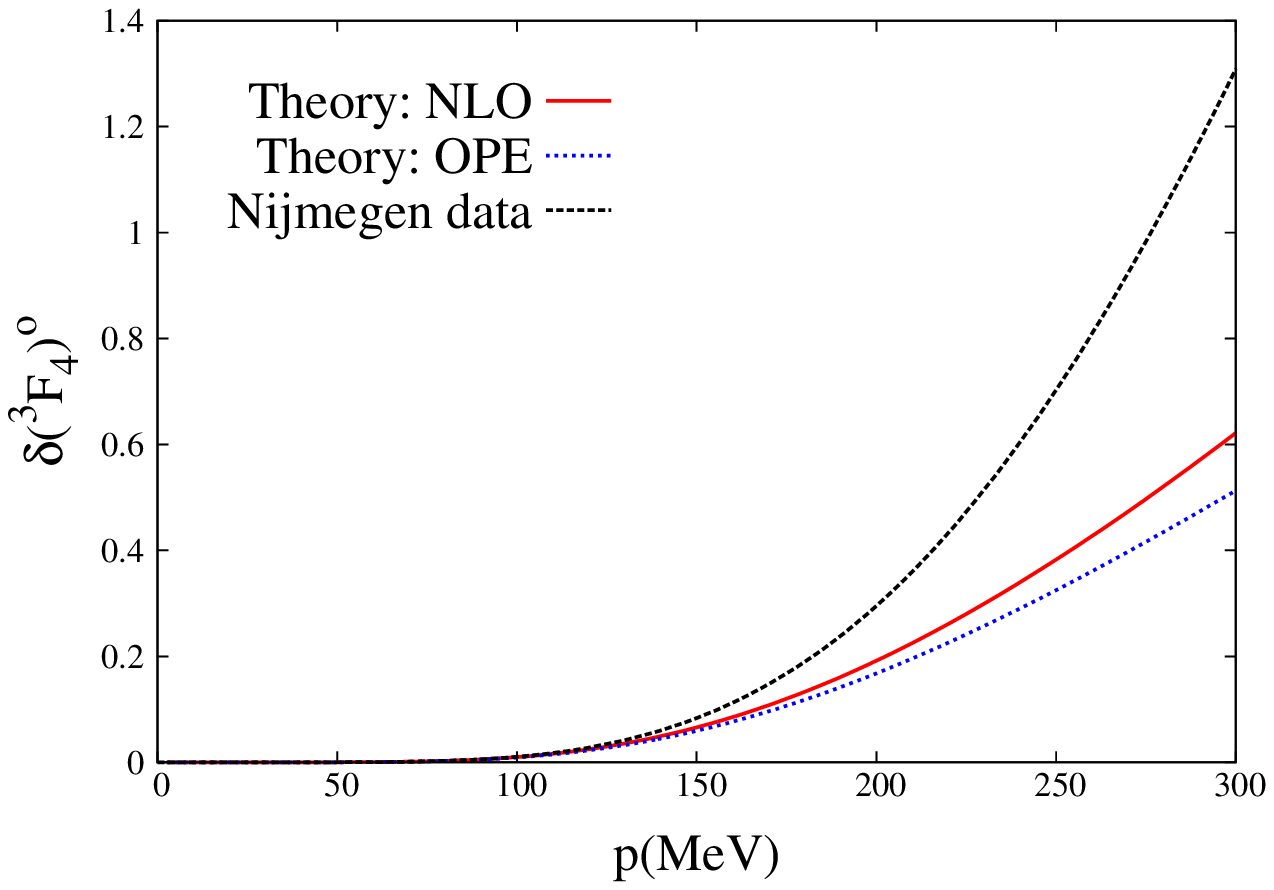} & 
\includegraphics[width=.4\textwidth]{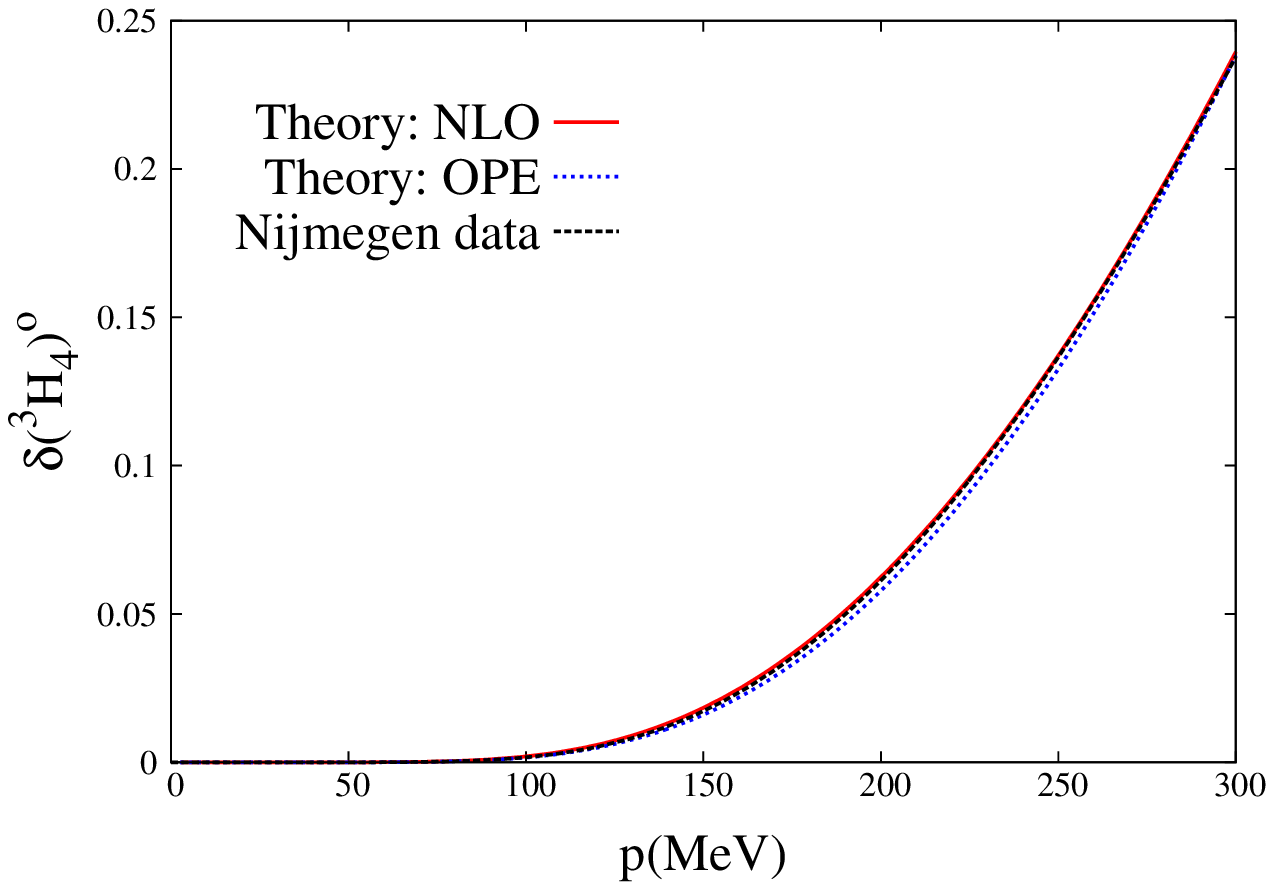}\\  
\includegraphics[width=.4\textwidth]{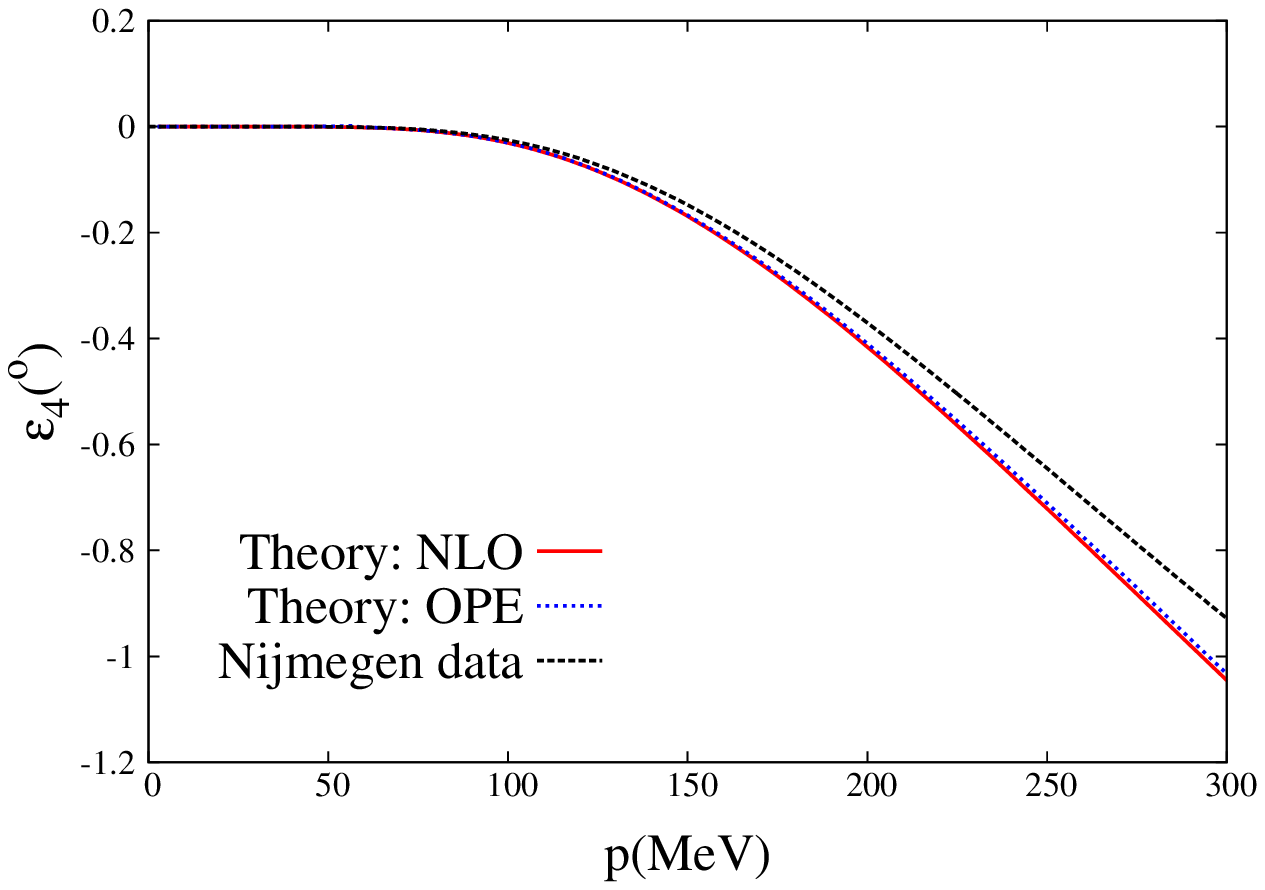}  
\end{tabular}
\caption[pilf]{\protect {\small (Color online.) From top to bottom and left to right: Phase shifts for $^3F_4$, $^3H_4$ and the mixing angle  $\epsilon_4$, in order.
The (red) solid line corresponds to our calculation at NLO and the (blue) dotted line is the results from  OPE. The Nijmegen PWA phase shifts are given by (black) dashed line.}
\label{fig:3fh4} }
\end{center}
\end{figure}

In this case the direct use of Eqs.~\eqref{highdc} and \eqref{highndc} does not provide a stable 
solution for the $^3H_4$ wave.   We have to perform an extra subtraction in the  $^3H_4$ partial wave in order to end 
with meaningful (convergent) results. The resulting IEs to be solved are:
\begin{align}
ij=11,~12~:~D_{ij}&=1+\sum_{p=2}^{\ell_{ij}}\delta^{(ij)}_p A(A-C)^{p-2}\nn\\
&+\frac{A(A-C)^{\ell_{ij}-1}}{\pi^2}
\int_{-\infty}^L dk^2 \frac{\Delta_{ij}(k^2)D_{ij}(k^2)}{(k^2)^{\ell_{ij}}}g_{ij}(A,k^2,C;\ell_{ij}-1)~,\nn\\
ij=22~:~D_{22}&=1+\sum_{p=2}^6\delta^{(22)}_p A(A-C)^{p-2}+\frac{A(A-C)^5}{\pi^2}
\int_{-\infty}^L dk^2 \frac{\Delta_{22}(k^2)D_{22}(k^2)}{(k^2)^6}g_{22}(A,k^2,C;5)~,
\end{align}
with the $N_{ij}(A)$ functions given by
\begin{align}
ij=11,~12~:~&N_{ij}(A)=\frac{A^{\ell_{ij}}}{\pi} \int_{-\infty}^L dk^2\frac{\Delta_{ij}(k^2)D_{ij}(k^2)}{(k^2)^{\ell_{ij}}(k^2-A)}~,\nn\\
ij=22~:~&N_{22}(A)=\nu_6^{(22)} A^5+\frac{A^6}{\pi} \int_{-\infty}^L dk^2\frac{\Delta_{22}(k^2)D_{22}(k^2)}{(k^2)^6(k^2-A)}~.
\end{align}
We can obtain $\nu_6^{(22)}$ by making use of a once-subtracted DR for the $^3H_4$ partial wave, which has a large orbital angular momentum, so that this DR provides accurate results. Recall our results for the once-subtracted DR in the uncoupled partial waves with $ \ell\geq 3$ presented by the 
 (cyan) double-dotted lines in Figs.~\ref{fig:fw}--\ref{fig:hw}.  For $A\to 0$ one has that $T(A)\to N(A)\to \nu_6^{(22)} A^5$, so that this counterterm is directly related with the behavior of the phase shifts at threshold.  In this way, we obtain
\begin{align}
\nu^{(22)}_6=0.079~M_\pi^{-12}~.
\label{nu13h4}
\end{align}
The coefficients $\delta^{(ij)}_p$ are expressed in terms of the functions $D_{ij}(A)$ 
and their derivatives at $A=C$, according to  Eq.~\eqref{taylor}, with $C=-M_\pi^2$ as we  always 
take. For $^3F_4$ we use $D_{11}(C)=1$ and $D'_{11}(C)=0$, because other values different from the pure perturbative ones do not improve the reproduction of data. For the $^3H_4$ one can also think of the pure perturbative values $D_{22}(C)=1$ and $D_{22}^{(n)}(C)=0$, $n=1,\ldots,4$. However, we have realized that a little change in $\nu_6^{(22)}$ requires a change of ${\cal O}(1)$ in $\delta_6^{(22)}$, keeping only negative values. In this way, we have fixed the latter coefficient to a negative value of ${\cal }O(1)$ and then adjust slightly $\nu_6^{(22)}$ with respect to the value calculated in Eq.~\eqref{nu13h4}. Typically we find just a slightly smaller value for $\nu_6^{(22)}$ than that in Eq.~\eqref{nu13h4}, $\nu_6^{(22)}\simeq 0.078~M_\pi^{-12}$. Regarding the mixing wave we find that no improvement in the reproduction of data is accomplished when the numbers $D_{12}^{(n)}(C)$, $n=0,1,2$, take  values different from the pure perturbative ones, which are the ones finally 
employed.  We show our NLO results in Fig.~\ref{fig:3fh4} by the (red) solid line, with a correction in the right direction 
compared to the LO results. Nonetheless, one observes that still an improvement (higher orders) is needed to 
reproduce the $^3F_4$ phase shifts, and such deviation is also observed 
in ChPT potential approaches, see e.g.\cite{entem12}. 
For the $^3H_4$ and $\epsilon_4$ the reproduction is much better. The (blue) dotted line corresponds to the OPE results that run close to the NLO ones.\footnote{No OPE results for the $^3F_4-{^3H_4}$ and $^3G_5-{^3I_5}$ are 
worked out in Ref.~\cite{paper2}. We obtain them by employing the same IEs as in NLO but keeping only in $\Delta_{ij}(A)$ the OPE contribution.}

\begin{figure}
\begin{center}
\begin{tabular}{cc}
\includegraphics[width=.4\textwidth]{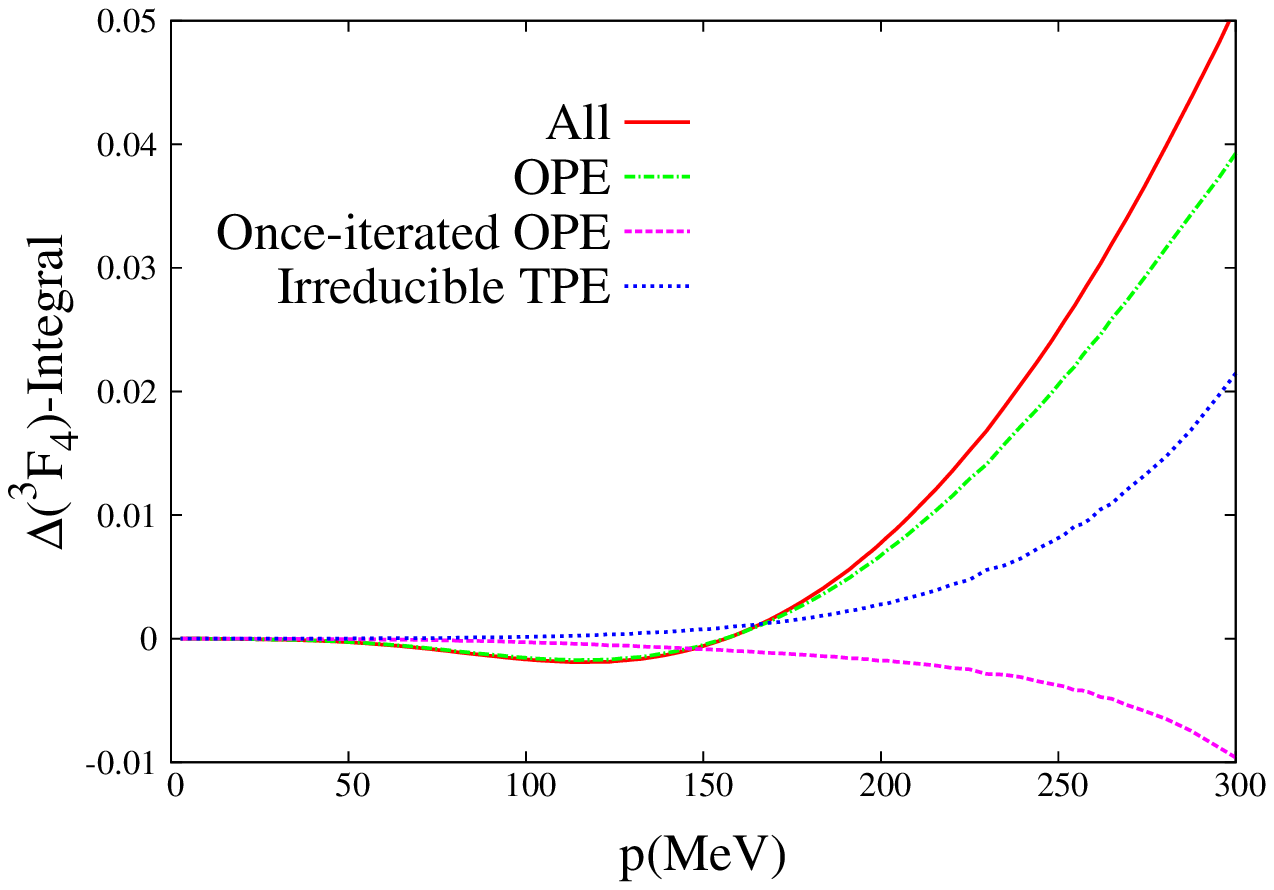} & \includegraphics[width=.4\textwidth]{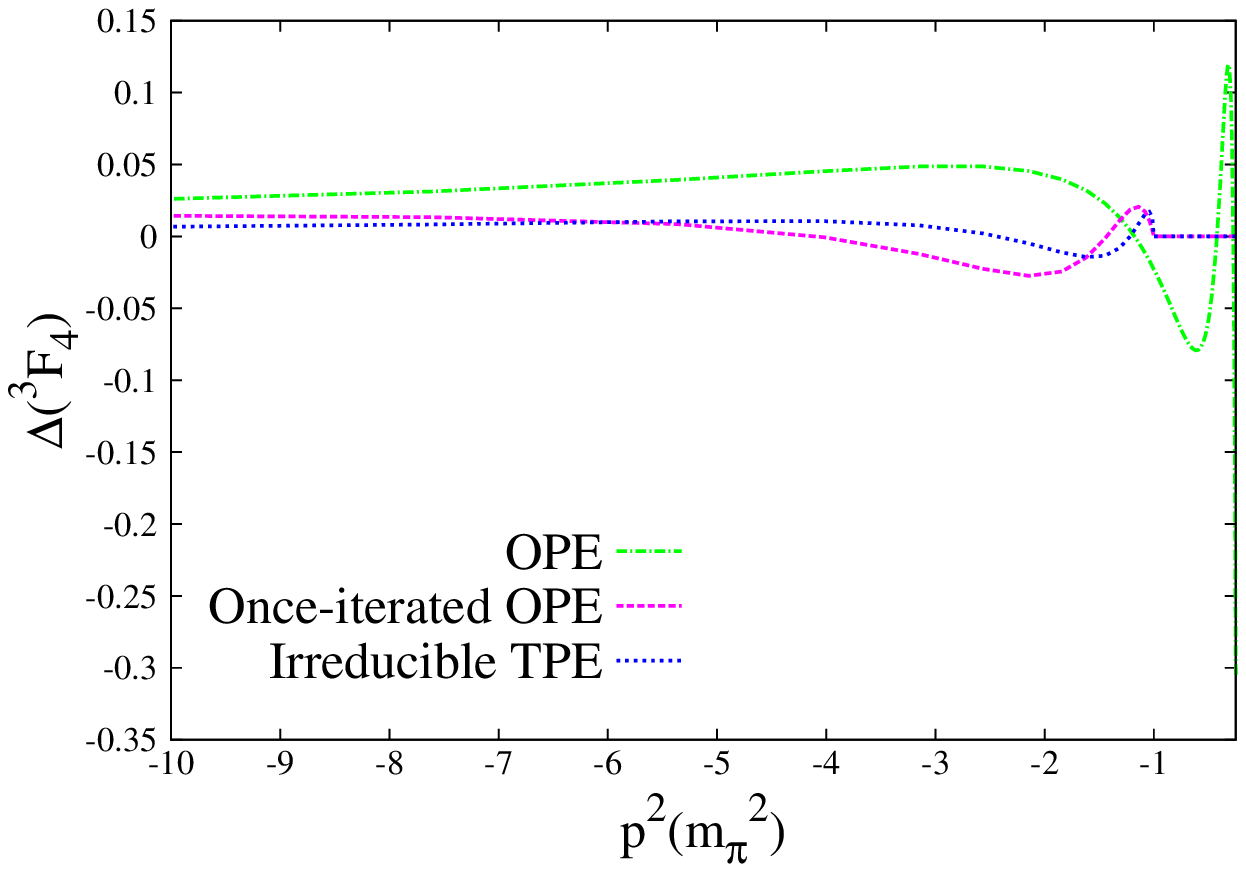} \\ 
\includegraphics[width=.4\textwidth]{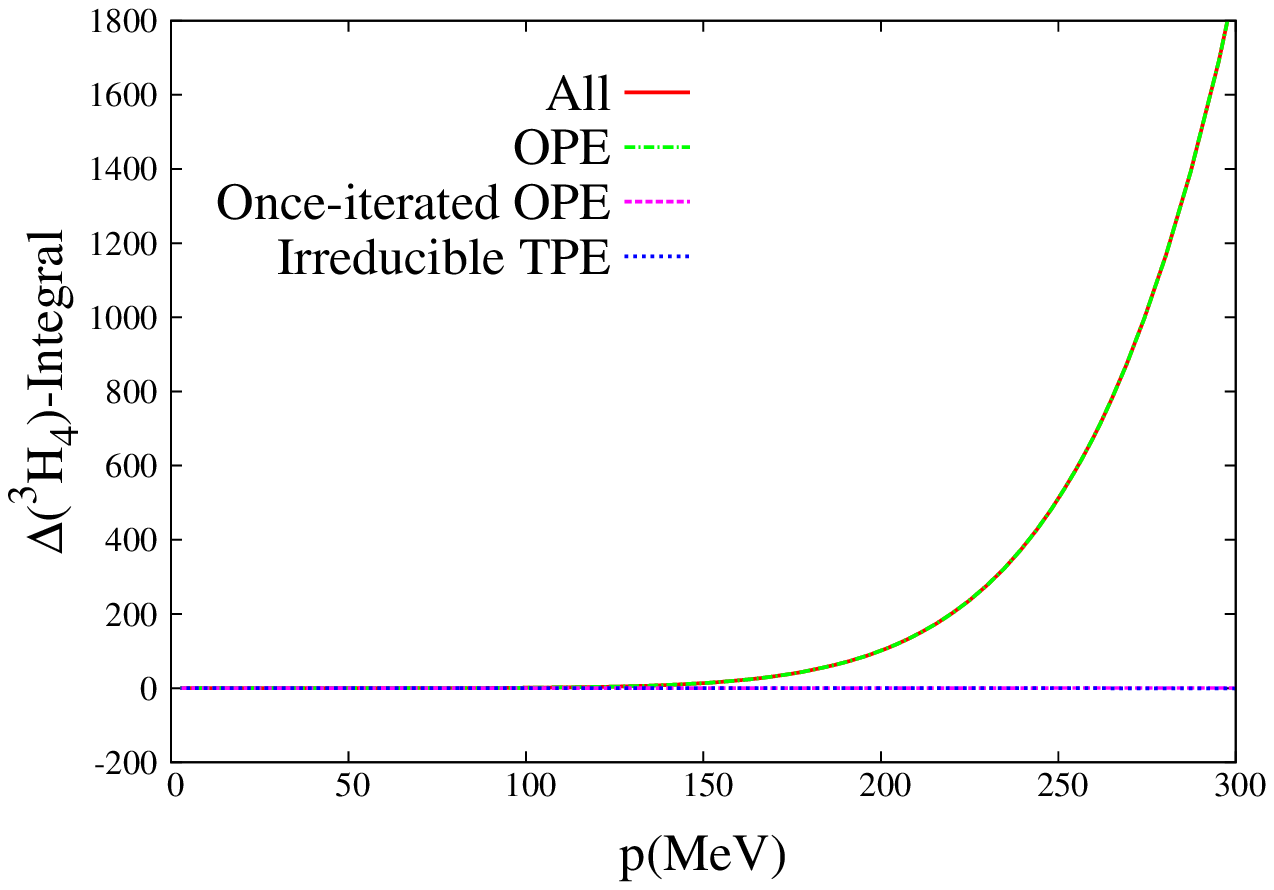} &  \includegraphics[width=.4\textwidth]{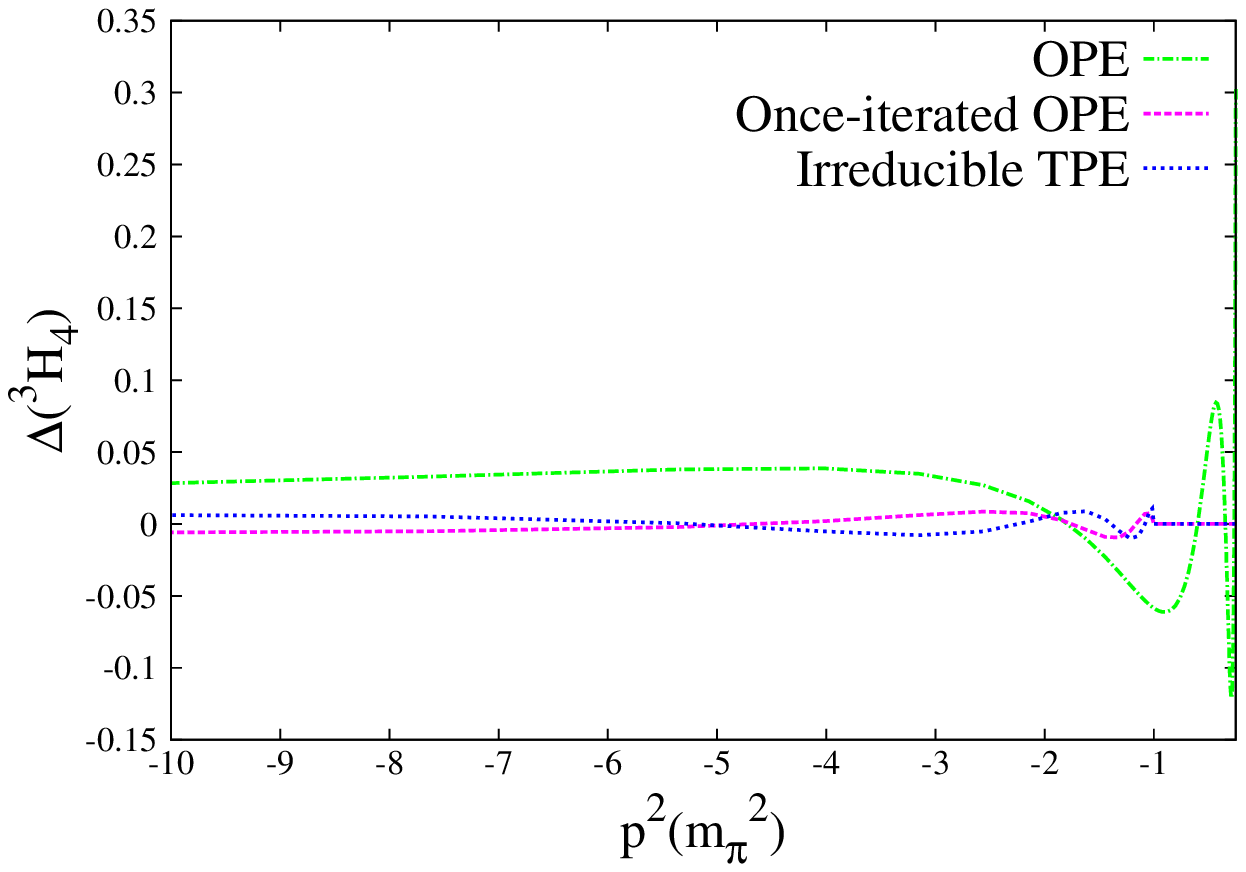}\\  
\includegraphics[width=.4\textwidth]{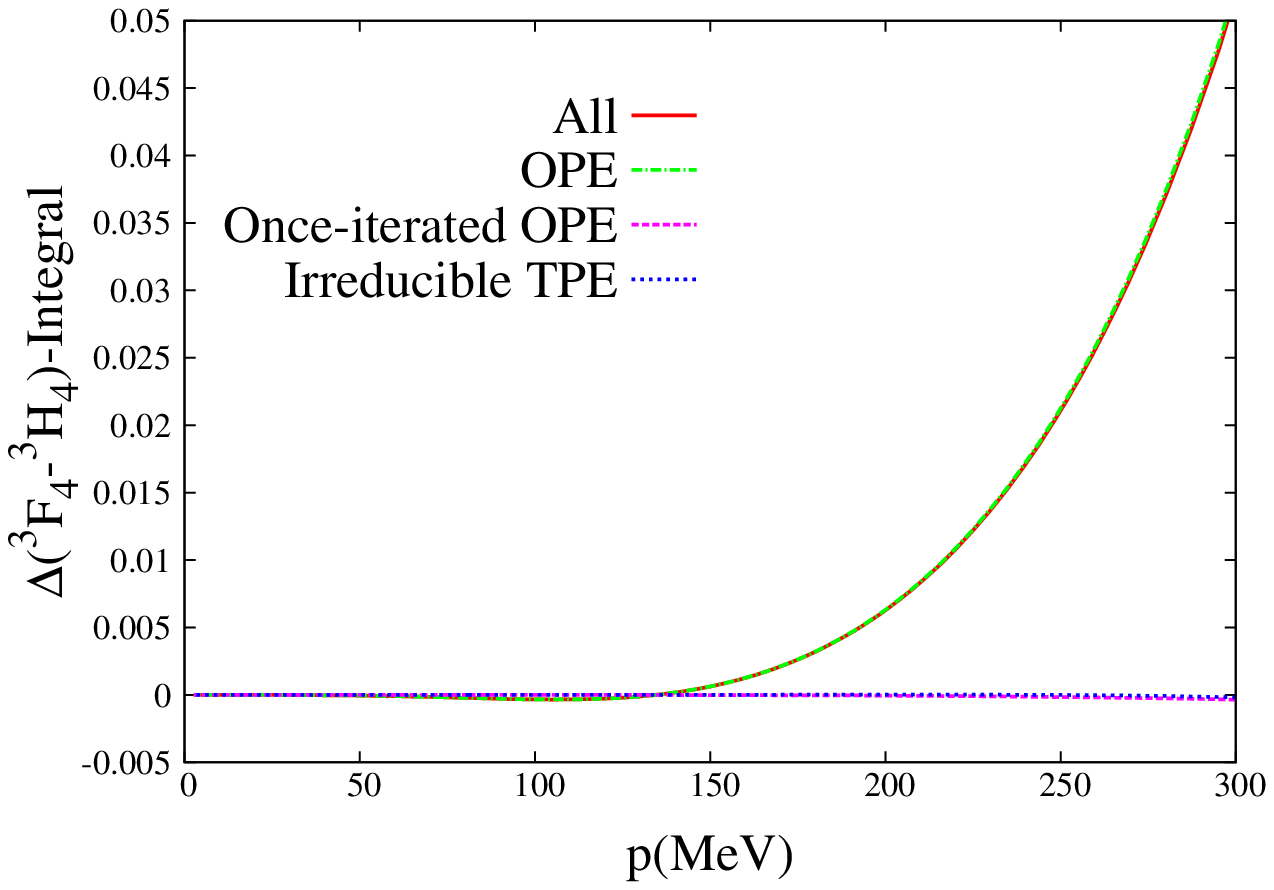} &  \includegraphics[width=.4\textwidth]{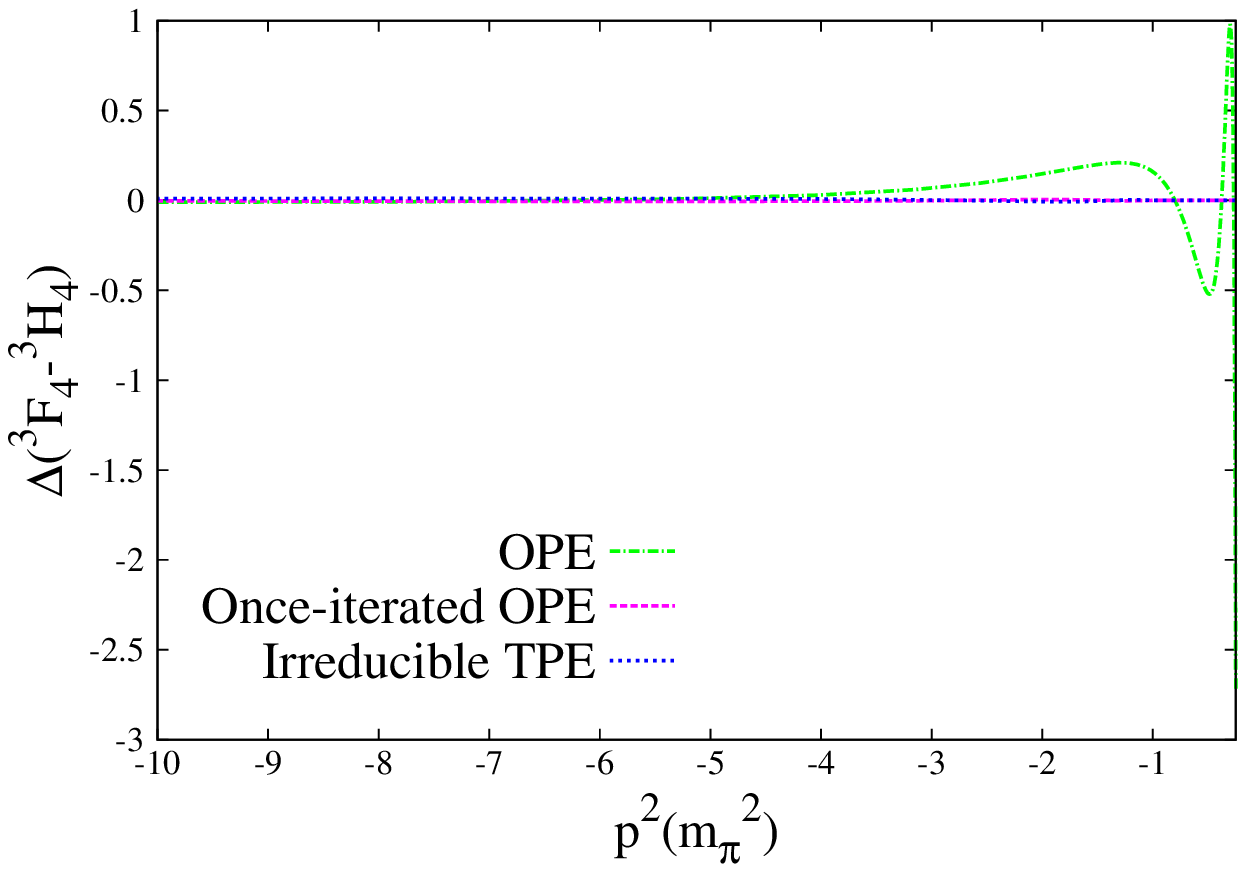} 
\end{tabular}
\caption[pilf]{\protect {\small (Color online.) Left panels: Different contributions to the integrals in    Eq.~\eqref{3fh4quanty}. Right panels: Contributions to $\Delta(A)$. From top to bottom we show the results for $^3F_4$, $^3H_4$ and mixing wave, respectively. The meaning of the lines is the same as in Fig.~\ref{fig:1s0quanty}.}
\label{fig:3fh4quanty} }
\end{center}
\end{figure}

As usual we also study the size of the different contributions to $\Delta(A)$ by evaluating the pertinent integrals along the LHC:
\begin{align}
\ell_{ij}=3,~4~:~&\frac{A(A+M_\pi^2)^{\ell_{ij}-1}}{\pi^2}\int_{-\infty}^L dk^2\frac{\Delta_{ij}(k^2)}{(k^2)^{\ell_{ij}}}
\int_0^\infty dq^2\frac{\mu_{ij}(q^2)(q^2)^{\ell_{ij}-1}}{(q^2-A)(q^2-k^2)(q^2+M_\pi^2)^{\ell_{ij}-1}}~,\nn\\
\ell_{22}=5~:~&\frac{A(A+M_\pi^2)^5}{\pi^2}\int_{-\infty}^L dk^2\frac{\Delta_{22}(k^2)}{(k^2)^6}\int_0^\infty dq^2\frac{\nu_{22}(q^2)(q^2)^5}{(q^2-A)(q^2-k^2)(q^2+M_\pi^2)^5}~,
\label{3fh4quanty}
\end{align}
with $\mu_{ij}$ defined after Eq.~\eqref{3dg3quanty}. 
The results are shown in   Fig.~\ref{fig:3fh4quanty}. We see that for all the waves the total result of the integrals is  dominated by OPE. Though for the $^3 F_4$ the independent contributions of reducible and irreducible TPE  are not small, they cancel each other almost exactly.

\section{Coupled waves: $^3G_5-{^3I_5}$ }
\label{3gi5}

In the $^3G_5-{^3I_5}$ system we have $\ell_{11}=4$, $\ell_{12}=5$ and $\ell_{22}=6$. However, the resulting IEs from Eqs.~\eqref{highdc} and \eqref{highndc} do 
not provide convergent results because the $^3I_5$ partial wave requires an extra subtraction, so that we can finally obtain results independent of 
the limits of integration. We then have:
\begin{align}
ij=11,~12~:~D_{ij}&=1+\sum_{p=2}^{\ell_{ij}}\delta^{(ij)}_p A(A-C)^{p-2} \nn\\
&+\frac{A(A-C)^{\ell_{ij}-1}}{\pi^2}
\int_{-\infty}^L dk^2\frac{\Delta_{ij}(k^2)D_{ij}(k^2)}{(k^2)^{\ell_{ij}}}g_{ij}(A,k^2,C;\ell_{ij}-1)~,\nn\\
ij=22~:~D_{22}&=1+\sum_{p=2}^7\delta^{(22)}_p A(A-C)^{p-2}+\frac{A(A-C)^6}{\pi^2}
\int_{-\infty}^L dk^2 \frac{\Delta_{22}(k^2)D_{22}(k^2)}{(k^2)^7}g_{22}(A,k^2,C;6)~,
\label{ie:3gi5}
\end{align}
with the $N_{ij}(A)$ functions given by
\begin{align}
ij=11,~12~:~&N_{ij}(A)=\frac{A^{\ell_{ij}}}{\pi} \int_{-\infty}^L dk^2\frac{\Delta_{ij}(k^2)D_{ij}(k^2)}{(k^2)^{\ell_{ij}}(k^2-A)}~,\nn\\
ij=22~:~&N_{22}(A)=\nu_7^{(22)} A^6+\frac{A^7}{\pi} \int_{-\infty}^L dk^2\frac{\Delta_{22}(k^2)D_{22}(k^2)}{(k^2)^7(k^2-A)}~.
\end{align}

We can predict $\nu_7^{(22)}$ by employing a free-parameter once-subtracted DR for an uncoupled $^3I_5$,  as we did in the 
previous section to calculate $\nu_6^{(22)}$ for the $^3H_4$ wave. In this way we obtain the number
\begin{align}
\nu_7^{(22)}&=-0.178~M_\pi^{-14}~.
\label{nu1.3g5}
\end{align}
We have also tried fits to data by releasing this number and the results obtained confirm this prediction. 
Regarding the coefficients $\delta_{p}^{(ij)}$ the same quality in the reproduction of data is obtained by taking $\delta_p^{(ij)}=0$ except for the coefficient with the highest $p$ for every $ij$, namely, $p=4$ for $ij=11$, $p=5$ for $ij=12$ and $p=7$ for $ij=22$, which are  fitted to data.
 Then, the coupled-wave system $^3G_5-{^3I_5}$ illustrates again the rule of Sec.~\ref{leq2} on the maximal smoothness of the function $D_{ij}(A)$ for higher 
partial waves. 
 For the fitted coefficients we have $|D_{11}^{(2)}(C)|>0.5$, $D_{12}^{(3)}(C)<-0.5$ and $D_{22}^{(5)}(C)\neq 0$, in appropriate powers of $M_\pi^{-2}$. For the last constant, one has to take into account that  a change in $D_{22}^{(5)}(C)$ of ${\cal O}(1)$ can be reabsorbed in slight changes of $\nu_7^{(22)}$ around the value given in Eq.~\eqref{nu1.3g5}, similarly to that in Sec.~\ref{3fh4} for the $^3F_4-{^3H_4}$ system.

\begin{figure}[h]
\begin{center}
\begin{tabular}{cc}
\includegraphics[width=.4\textwidth]{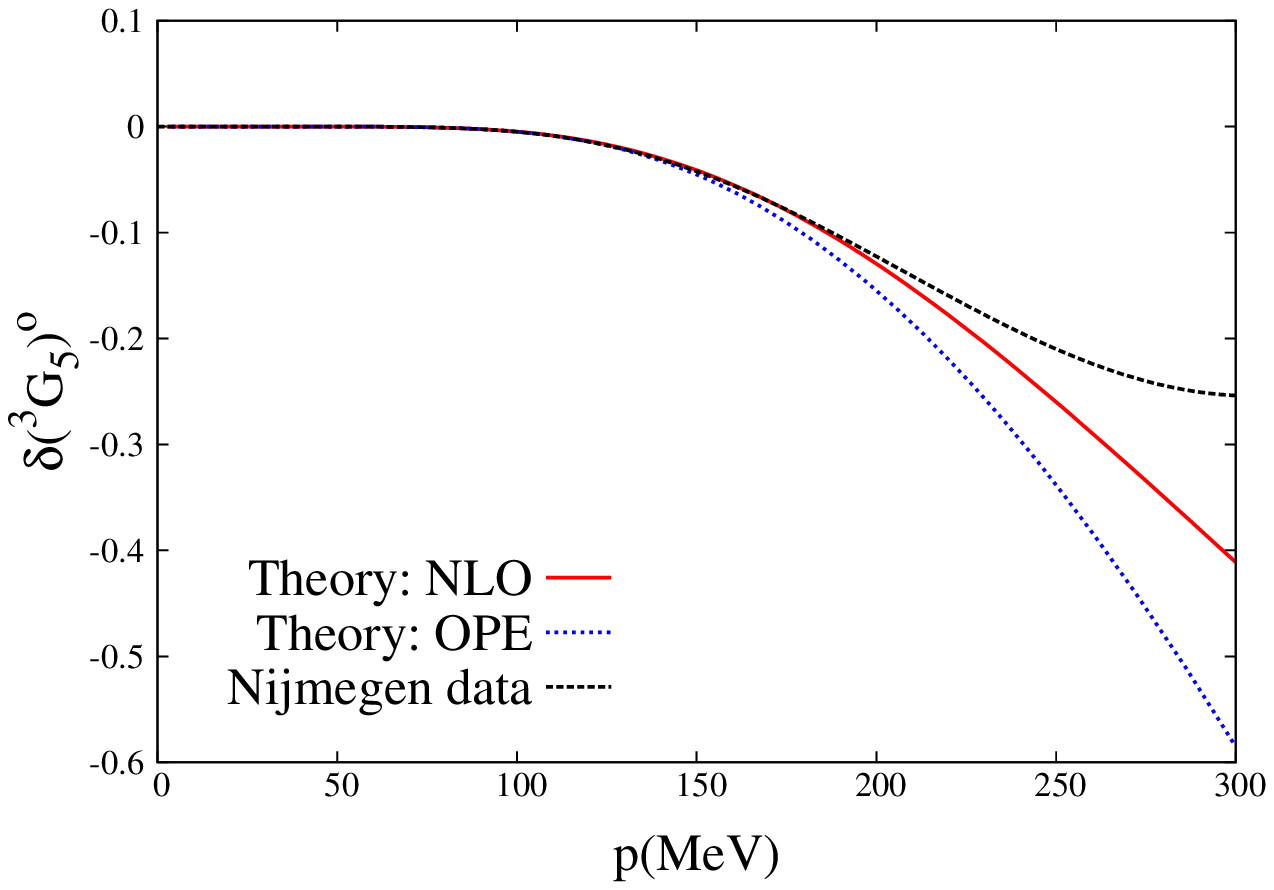} & 
\includegraphics[width=.4\textwidth]{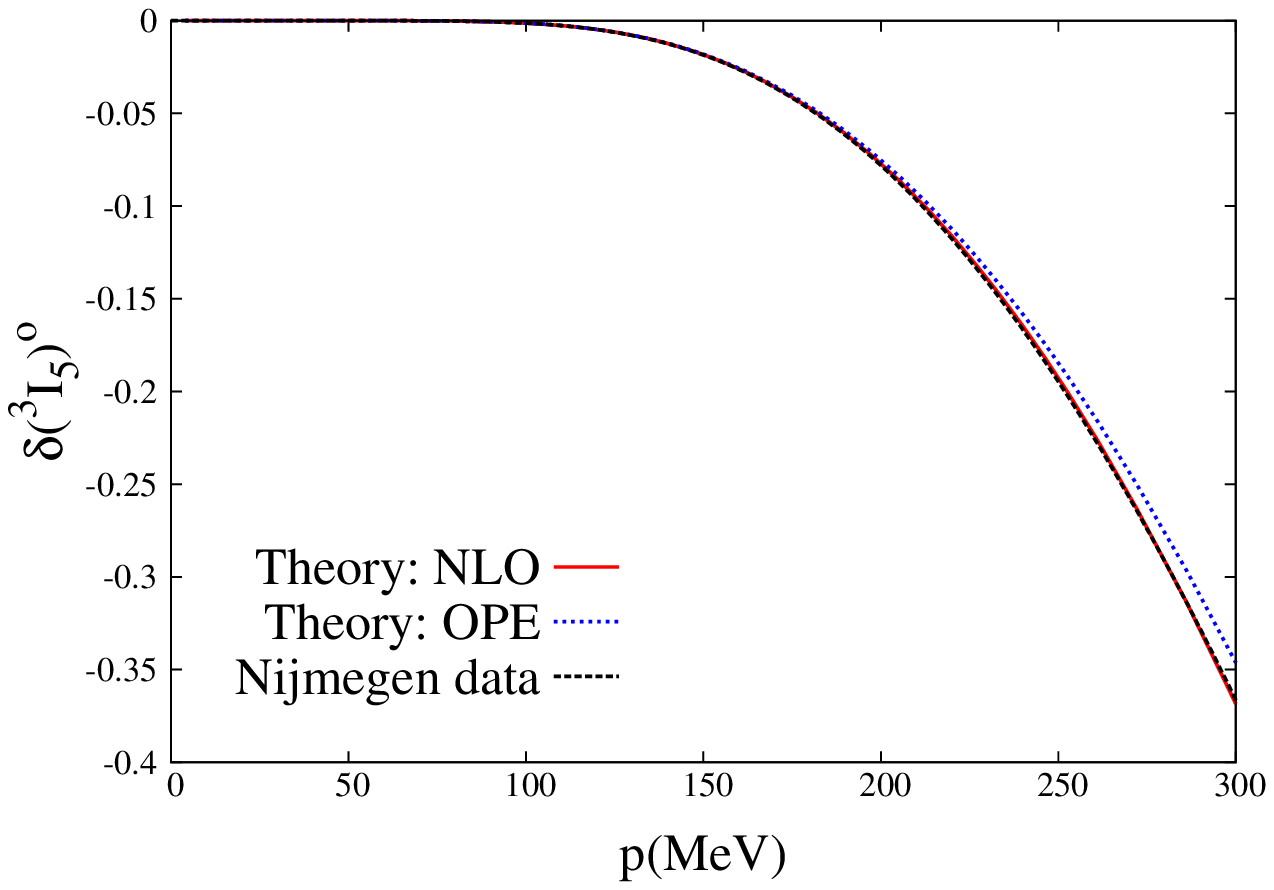}\\  
\includegraphics[width=.4\textwidth]{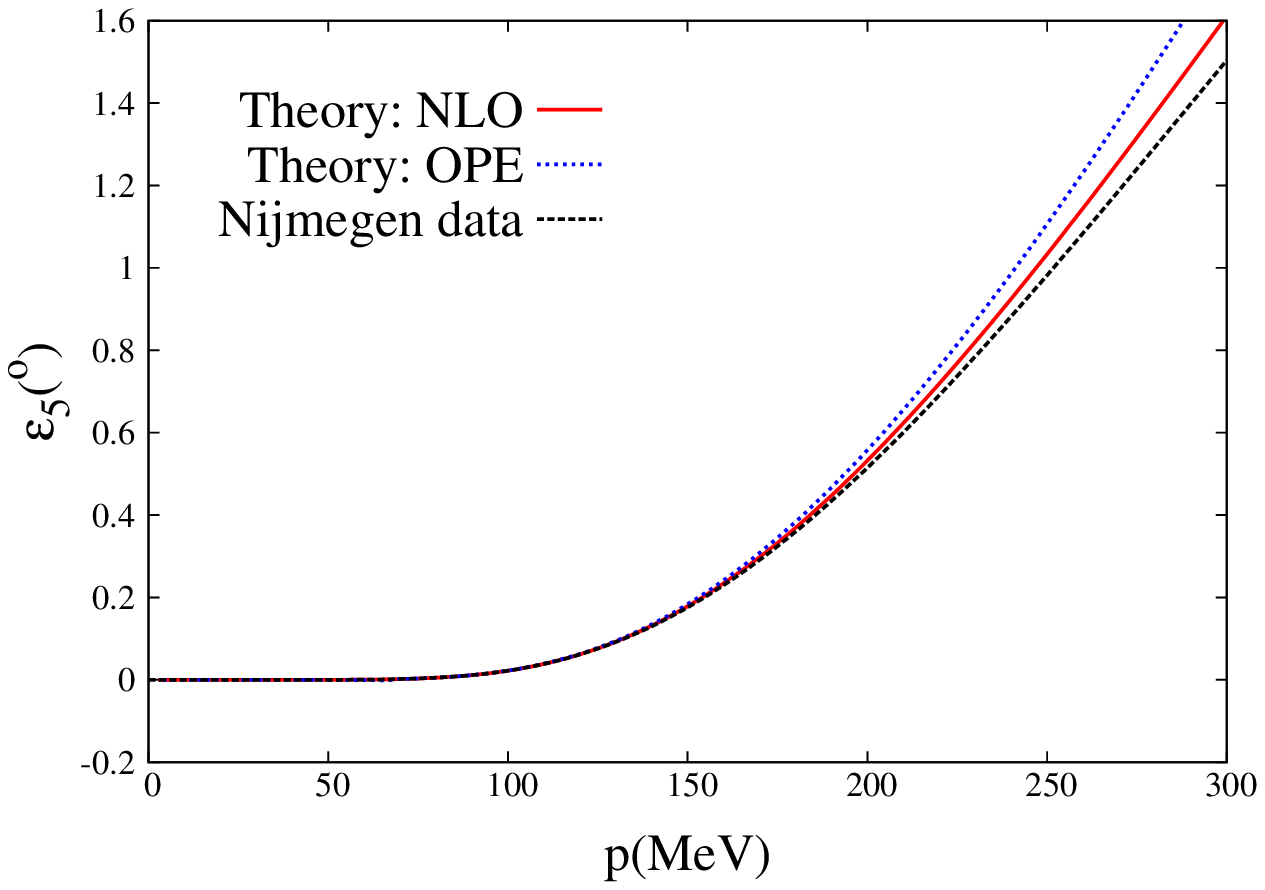}  
\end{tabular}
\caption[pilf]{\protect {\small (Color online.) From top to bottom and left to right: Phase shifts for $^3G_5$, $^3I_5$ and the mixing angle  $\epsilon_5$, in order.
The (red) solid line corresponds to NLO and the (blue) dotted line is the results from  OPE. The Nijmegen PWA phase shifts are given by (black) dashed line.}
\label{fig:3gi5} }
\end{center}
\end{figure}

The resulting phase shifts are shown in Fig.~\ref{fig:3gi5} in which, for definiteness, we take the values $D_{11}^{(2)}(C)=-1~M_\pi^{-4}$, $D_{12}^{(3)}(C)=-1~M_\pi^{-6}$ and $D_{22}^{(5)}=-2~M_\pi^{-10}$. The NLO phase shifts are shown by the (red) solid line. We see that they follow closely the $NN$ phase shifts of Ref.~\cite{Stoks:1994wp}. For the $^3I_5$ partial-wave phase shifts the reproduction is perfect. The LO results, given by the (blue) dotted line, are also obtained with the same values for the $\delta_p^{(ij)}$. We observe that the reproduction of the $^3G_5$ phase shifts is  worse than in the NLO case, and only slightly worse for the $^3I_5$ phase shifts. For $\epsilon_5$ the LO result is similar
 to the NLO one. We have also varied the $\delta_p^{(11)}$ ($p=2,3,4$) for the LO calculation in order to improve the reproduction of the $^3G_5$ phase shifts but no gain is obtained. 

It has been already noticed in Refs.~\cite{entem,entem12} that the $^3G_5$ phase shifts, even with a chiral N$^3$LO potential, are not well reproduced after solving the corresponding Lippmann-Schwinger equation, either with finite \cite{epen3lo} or infinite  three-momentum cutoff \cite{entem12}, as well as by calculating them in perturbation theory \cite{entem,peripheral}. Our results in Fig.~\ref{fig:3gi5} for the $^3G_5$ are closer  to data than the ones in those references, despite that our calculation is only a NLO one. However, our nonperturbative approach already includes one free parameter exclusively for the $^3G_5$, which is not the case in Refs.~\cite{entem,entem12,epen3lo}.

\begin{figure}
\begin{center}
\begin{tabular}{cc}
\includegraphics[width=.4\textwidth]{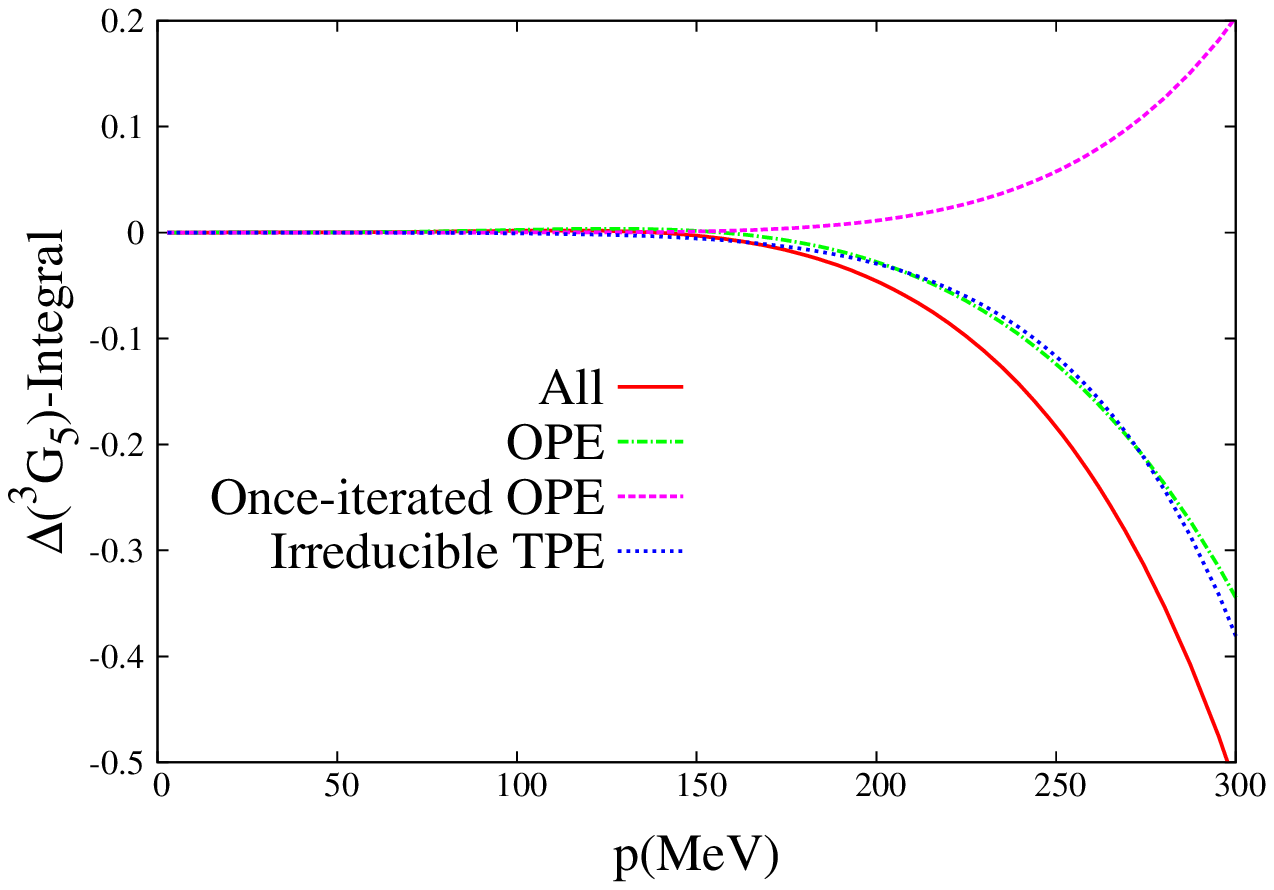} & \includegraphics[width=.4\textwidth]{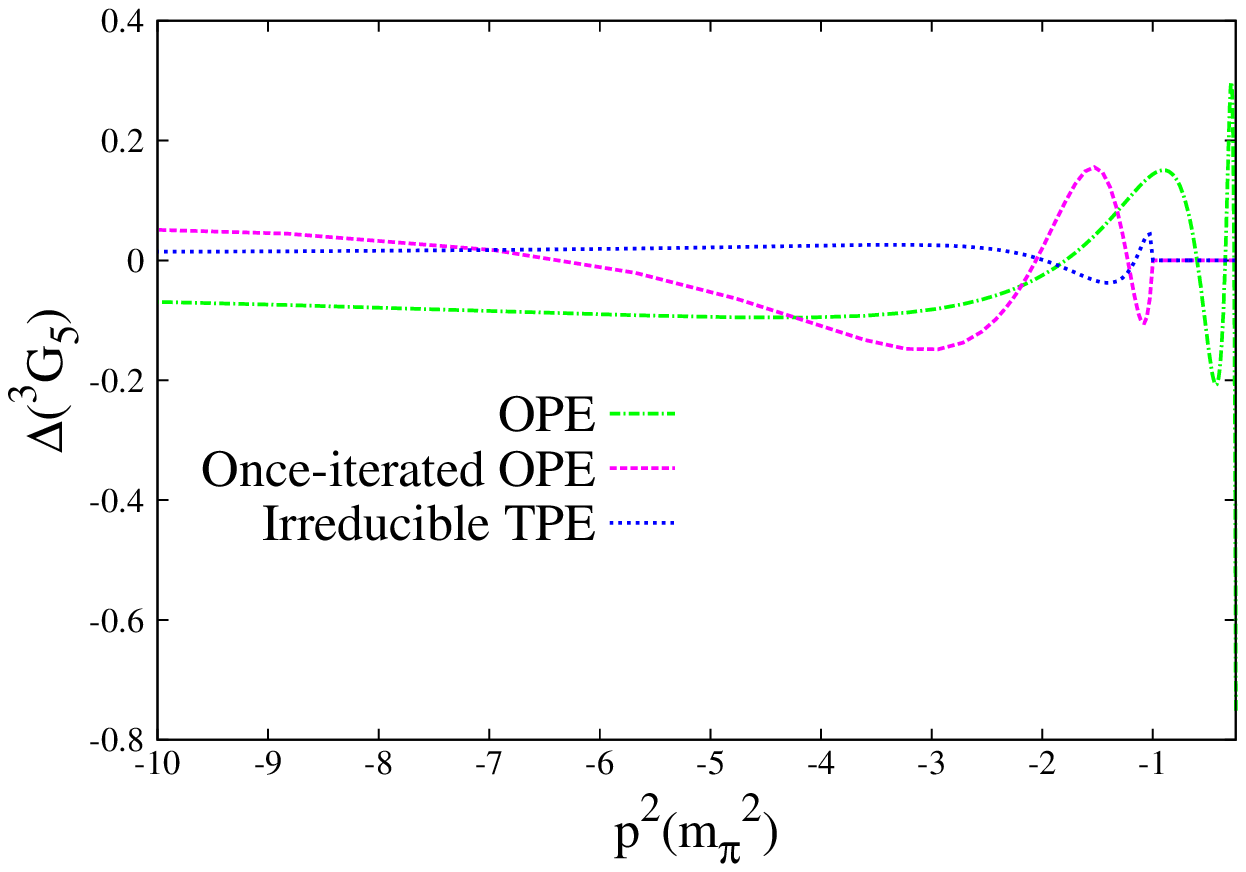} \\ 
\includegraphics[width=.4\textwidth]{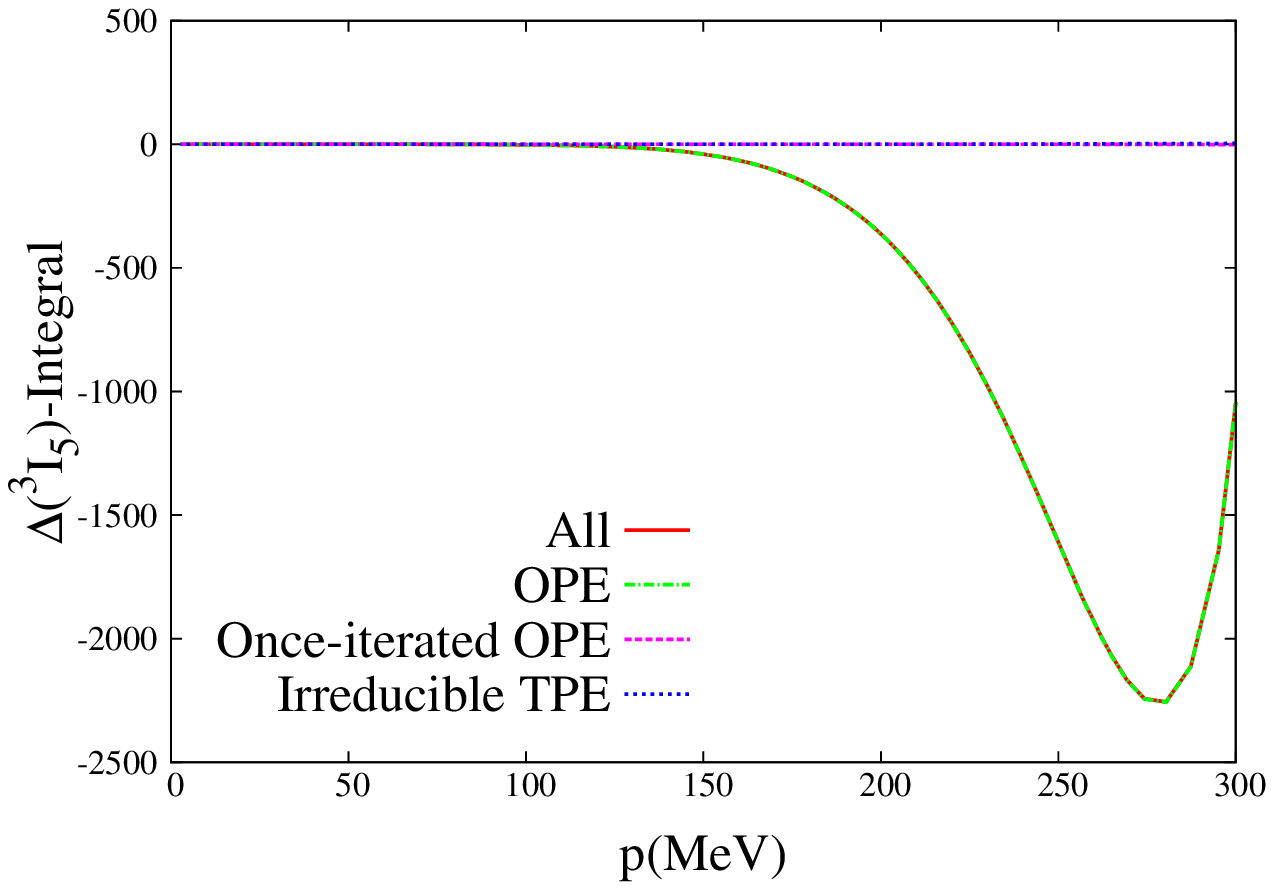} &  \includegraphics[width=.4\textwidth]{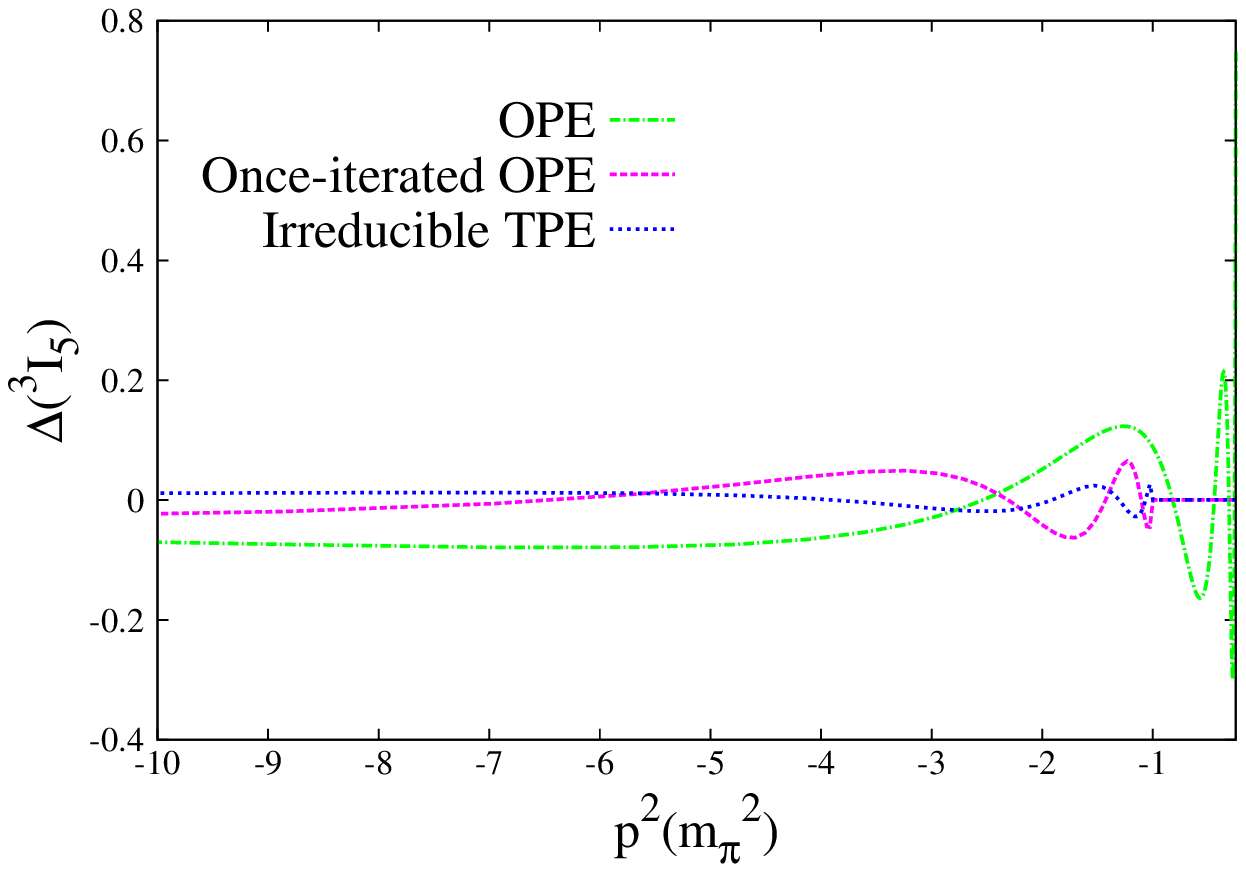}\\  
\includegraphics[width=.4\textwidth]{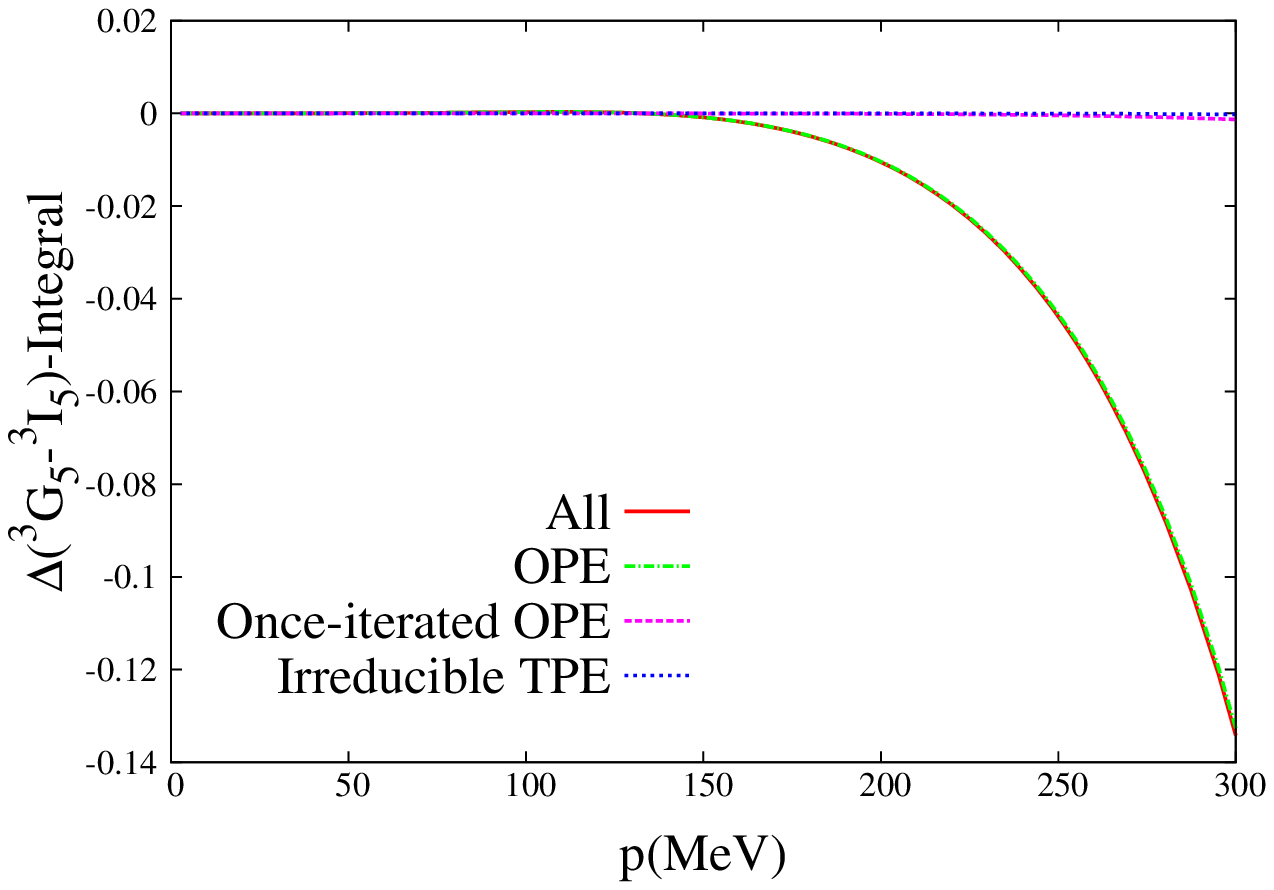} &  \includegraphics[width=.4\textwidth]{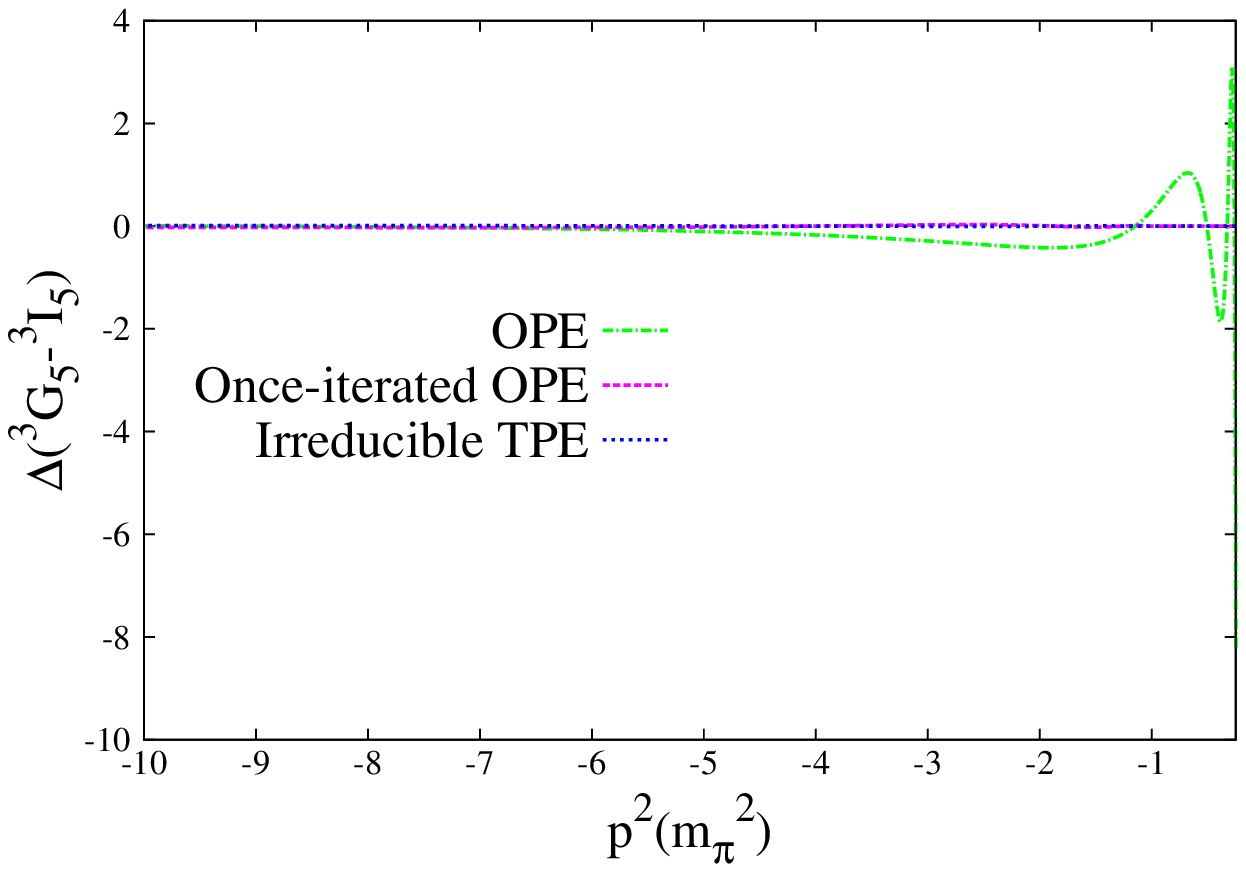} 
\end{tabular}
\caption[pilf]{\protect {\small (Color online.) Left panels: Different contributions to the integrals in    Eq.~\eqref{3gi5quanty}. Right panels: Contributions to $\Delta(A)$. From top to bottom we show the results for $^3G_5$, $^3I_5$ and mixing wave, respectively. The meaning of the lines is the same as in Fig.~\ref{fig:1s0quanty}.}
\label{fig:3gi5quanty} }
\end{center}
\end{figure}

As usual we also study the size of the different contributions to $\Delta(A)$ by evaluating the appropriate integrals along the LHC according to the number of subtraction taken in each of the IEs used, Eq.~\eqref{ie:3gi5}:
\begin{align}
\ell_{ij}=4,~5~:~&\frac{A(A+M_\pi^2)^{\ell_{ij}-1}}{\pi^2}\int_{-\infty}^L dk^2\frac{\Delta_{ij}(k^2)}{(k^2)^{\ell_{ij}}}
\int_0^\infty dq^2\frac{\mu_{ij}(q^2)(q^2)^{\ell_{ij}-1}}{(q^2-A)(q^2-k^2)(q^2+M_\pi^2)^{\ell_{ij}-1}}~,\nn\\
\ell_{22}=6~:~&\frac{A(A+M_\pi^2)^6}{\pi^2}\int_{-\infty}^L dk^2\frac{\Delta_{22}(k^2)}{(k^2)^7}\int_0^\infty dq^2\frac{\nu_{22}(q^2)(q^2)^6}{(q^2-A)(q^2-k^2)(q^2+M_\pi^2)^6}~.
\label{3gi5quanty}
\end{align}
The results are shown in   Fig.~\ref{fig:3gi5quanty}. We see that for all the waves the total result of the integrals is dominated by OPE. However, for the $^3G_5$ the iterated and irreducible TPE contributions are not small. Nevertheless,  they cancel almost exactly so that the net contribution 
is  mostly given by OPE. 

Now, we show in Table~\ref{tab:freeparam} the minimum number of subtraction constants that are fitted to data
 for every partial wave in our present study at NLO. That is, 
the number of free subtraction constants that we have once the minimum number of subtractions is taken to have a well-behaved IE for the $D(A)$ function 
in the corresponding partial wave.
When the free parameter is only determined within broad intervals (its order of magnitude is not even fixed) then we do not consider it as a free parameter, but better 
as having a constraint. We do not consider either as free parameters those subtraction constants that take their expected perturbative values.  In the space next to the right of the one with the name of the partial wave we give the minimum number of free parameters for this partial wave, in the explained sense. We have in total 14 free parameters. One should be aware that the number of free parameters does not necessarily increase with the accuracy up to which $\Delta(A)$ is calculated in ChPT. There is no such a close connection between the minimum number of subtraction constants and the chiral order in the 
calculation of $ \Delta(A)$ as the situation between the number of chiral counterterms and the 
chiral order in which the $NN$ potential  is calculated \cite{weinn}. E.g. we have one free parameter for the $^1S_0$ and $^3S_1-{^3D_1}$ waves both at LO \cite{paper1,paper2} and now at NLO.  
\begin{table}
\begin{center}
\begin{tabular}{|l|l|l|l|l|l|}
\hline
$^1S_0$ & 1 & $^3P_0$ & 0 & $^3S_1-{^3D_1}$ & 1 \\
\hline
$^3P_1$ & 3 & $^1P_1$ & 0 & $^3P_2-{^3F_2}$ & 1 \\
\hline
$^1D_2$ & 0 & $^3D_2$ & 1 & $^3D_3-{^3G_3}$ & 1 \\
\hline
$^1F_3$ & 0 & $^3F_3$ & 1 & $^3F_4-{^3H_4}$ & 1 \\ 
\hline
$^1G_4$ & 1 & $^3G_4$ & 0 & $^3G_5-{^3I_5}$ & 1 \\ 
\hline
$^1H_5$ & 1 & $^3H_5$ & 1 & & \\
\hline   
\end{tabular}
\caption[pilf]{\protect {\small The minimum number of free parameters for each partial wave in our study at NLO is given in the box to the right 
of the wave. }
\label{tab:freeparam} }
\end{center}
\end{table}

Finally, we give in Table~\ref{tab:allparam} the values of the free parameters employed in the different partial waves. If for a given partial wave 
we employ DRs with different number of subtractions this is distinguished. In the table we use the notation $m$DR with $m=1,2,\ldots$, and it should be 
read as $m$-time subtracted DR. For the higher $NN$ partial waves we use the abbreviation  LTS to indicate that 
$\ell$ subtractions have been taken  to satisfy the threshold behavior, following 
the standard formalism explained in Sec.~\ref{unformalism}. For the coupled channel case we use also the same abbreviation LTS when 
$\ell_{ij}$-time subtracted DRs are used for the coupled partial waves, for the same reason as before, extended to the coupled wave case in 
Sec.~\ref{sec:formalism}. We indicate separately the case in which more subtractions are needed for some specific wave. According to the 
principle of maximal smoothness only the highest derivative $D^{(n)}(C)$ is not fixed to its perturbative value (1 for $n=0$ and 0 for $n\neq 0$) and released, 
if appropriate. 
  When no free parameters enter in the DR for the partial wave we indicate it by the abbreviation {\it nfp}.
 The units are always given in the the appropriate power of $M_\pi^2$, though this power is not explicitly indicated to 
abbreviate. In this way, if a subtraction constant is small  in these units then we could interpret it as having mostly an origin due to 
short-distance physics.\footnote{For the $\nu_i$ coefficients one has to extract out the normalization factor $4\pi/m\simeq 1.8~M_\pi^{-1}$, which 
indeed is ${\cal O}(1)$ in units of powers of $M_\pi$.}
\begin{table}
\begin{center}
{\small
\begin{tabular}{|l|l|l|}
\hline
Wave & Type of DRs & Parameters \\
\hline
$^1S_0$ & 1DR & $\nu_1=30.69$ \\
       &  2DR & $\nu_1=30.69$~,~$\nu_2=0.24$ \\
\hline
$^3P_0$ & 1DR & {\it nfp} \\
       &  2DR & $\nu_2=0.562$~,~$\delta_2=-0.30$ \\
\hline
$^3P_1$ & 3DR & $\nu_2=-0.343$~,~$\delta_2=2.5\sim 3.0$~,~$\delta_3=0.2\sim 0.3$ \\
\hline
$^1 P_1$ & 1DR & {\it nfp} \\
\hline
$^1D_2$ &  LTS & $\delta_2\gtrsim 0$  \\
\hline
$^3D_2$ & LTS & $\delta_2= -0.18$  \\
\hline
$^1F_3$ & LTS & $D^{(2)}(0)>0$  \\
\hline
$^3F_3$ & LTS & $D^{(2)}(0)\simeq 0.014$  \\
\hline
$^1G_4$ & LTS & $D^{(3)}(0)=-0.031$  \\
\hline
$^3G_4$ & LTS & {\it nfp}  \\
\hline
$^1H_5$ & LTS & $D^{(4)}(0)=-0.6$  \\
\hline
$^3H_5$ & LTS & $D^{(4)}(0)=0.7\cdot 10^{-2}$  \\
\hline
$^3S_1-{^3D_1}$ & $1$DR $^3S_1$, 2DR $^3D_1$ and mixing & $\nu_1^{(11)}=-7.01$ \\
\hline
$^3P_2-{^3F_2}$ & LTS & $D_{12}(-M_\pi^2)=1.1$~,~$D^{(1)}_{22}(-M_\pi^2)< -1$ \\
& 2DR for $^3P_2$ and LTS for the others & $\nu^{(11)}_2=0.061$~,~$D_{11}(-M_\pi^2)=0.1$
~,~$D_{12}(-M_\pi^2)=1.1$~, \\
 & & $D^{(1)}_{22}(-M_\pi^2)< -1$\\
\hline
$^3D_3-{^3G_3}$ & LTS & $D^{(1)}_{12}(-M_\pi^2)\lesssim 0$~,~$D^{(2)}_{22}(-M_\pi^2)>1$  \\
\hline
$^3F_4-{^3H_4}$ & 6DR for $^3H_4$ and LTS for the others & $\nu_6^{(22)}=0.078$  \\
\hline
$^3G_5-{^3I_5}$ & 7DR for $^3I_5$ and LTS for the others & $|D_{11}^{(2)}(-M_\pi^2)|>0.5$~,~$D_{12}^{(3)}(-M_\pi^2)<-0.5$~,~$\nu_7^{(22)}=-0.178$ \\
\hline
\end{tabular} }
\caption[pilf]{\protect {\small  In the columns from left to right we show, in order: 
The partial wave, the type of DRs employed to study the 
corresponding partial wave and the (interval of) values for the free parameters involved.  }
\label{tab:allparam} }
\end{center}
\end{table}

\section{Conclusions}\label{sec:conclusions}

We have applied the $N/D$ method to study $NN$ scattering within ChPT. The basic input in this method is the imaginary part along the LHC of a given $NN$ partial wave, that we denote by $\Delta(A)$. This is calculated within ChPT up to some order in the chiral expansion. Here we have included  OPE and leading TPE contributions, extending the results of Refs.~\cite{paper1,paper2}, which only considered OPE. The standard ChPT counting clearly establishes that OPE is ${\cal O}(p^0)$, while irreducible TPE is ${\cal O}(p^2)$. We have also discussed that increasing the pion ladders in $NN$ reducible diagrams is suppressed because it gives rise to contributions to $\Delta(A)$ for $A$ deeper in the LHC and further away from the low-energy physical region. 
 We have employed suitable integrals along the LHC to properly quantify the different contributions to $\Delta(A)$, which is better than just to compare numerical values directly from this quantity. It follows that OPE is indeed the dominant contribution to $\Delta(A)$, while irreducible and reducible TPE are subleading.
   We have shown by explicit evaluation that the reducible TPE contribution to $\Delta(A)$ is typically of the same size in absolute value as the irreducible TPE contribution, 
because the latter is enhanced by numerical factors. We then count both of them in the chiral expansion for $\Delta(A)$ as ${\cal O}(p^2)$, 
as the irreducible TPE part does.

Our reproduction of the Nijmegen PWA  
phase shifts and mixing angles \cite{Stoks:1994wp} is already quite good for most of the partial waves. 
Typically it is as good or  better than the one achieved with an NLO calculation of the $NN$ potential, which is then employed to solve a Lippmann-Schwinger equation (either exactly or performing a distorted wave approximation) \cite{epen3lo,longyang,revmachleidt}.  It is also important to stress that we have demonstrated that 
when $\Delta(A)$ is given by the imaginary part of OPE along the LHC then the resulting IEs  have always a unique solution 
because they are  Fredholm IEs of the second kind 
with a squared integrable kernel and inhomogeneous term.  We have also established  correlations 
between the $S$-wave effective ranges and scattering lengths 
based on unitarity, analyticity and chiral symmetry.

 Giving these promising results, N$^2$LO and N$^3$LO calculations of $ \Delta(A)$ should be pursued in the future to fully ascertain the power of the method in the study of $NN$ scattering, here applied up to NLO. In particular, we would like to answer the question of 
 whether it is still possible to achieve $T$-matrices with only one free parameter 
for the $^1S_0$ and $^3S_1-^3D_1$ systems and, if  so, how much improvement would be obtained by calculating $\Delta(A)$ 
with more precision. The same question could be asked regarding the uncoupled $P$-waves that allow a one-parameter description. 
Of course, according to our necessary conditions for having a convergent solution,
 the first question is driven by the 
sign of $\Delta(A)$ when $A\to-\infty$.
Another point that requires further consideration is the fact that in  the triplet coupled waves our reproduction 
of the phase shifts is not satisfactory for the lowest coupled partial wave with $\ell_{11}=1,$~2 and 3, so that an interesting 
point is whether an improvement would arise in the description of these waves once a N$^2$LO study is performed, similarly  
 to what has already occurred within the potential scheme of Ref.~\cite{epen3lo}.

\section*{Acknowledgments}
We would like to thank  M. Pav\'on-Valderrama for his help in an early stage of this research and M.~Albaladejo for providing us some data files.  JAO would like to thank E. Ruiz Arriola and 
D.~Rodr\'{\i}guez Entem for valuable discussions and useful information.
 This work is partially funded by the grants MINECO (Spain) and EU, grant FPA2010-17806 and the Fundaci\'on S\'eneca 11871/PI/09.
 We also thank the financial support from the EU-Research Infrastructure
Integrating Activity
 ``Study of Strongly Interacting Matter" (HadronPhysics2, grant n. 227431)
under the Seventh Framework Program of EU and   
the Consolider-Ingenio 2010 Programme CPAN (CSD2007-00042). Z.H.G. also acknowledges the grants National Natural Science Foundation of China (NSFC) under contract Nos. 11105038 and 11075044, Natural Science Foundation of Hebei Province with contract No. A2011205093 and Doctor Foundation of Hebei Normal University with contract No. L2010B04.

\appendix{}

\section{$\Delta(A)$ from one pion exchange}
\label{app.delta}
\def\theequation{\Alph{section}.\arabic{equation}}
\setcounter{equation}{0}

We list here the explicit formulas for the imaginary parts
over the left cut of the $NN$ partial waves, $\Delta(A)$, coming from OPE. For the case
of the uncoupled partial waves these read,
\begin{align}
  \label{deltasOPE-1S0}
  \Delta_{^1S_0}(A)&=\frac{\pi g_A^2M_\pi^2}{16 f_\pi^2A}~,\\
  \label{deltasOPE-3P0}
  \Delta_{^3P_0}(A)&=-\frac{\pi g_A^2M_\pi^2}{16 f_\pi^2A}~,\\
  \label{deltaOPE-1P1}
  \Delta_{^1P_1}(A)&=-\frac{3\pi g_A^2M_\pi^2(M_\pi^2+2A)}{32 f_\pi^2A^2}~,\\
  \label{deltaOPE-3P1}
  \Delta_{^3P_1}(A)&=-\frac{\pi g_A^2M_\pi^4}{64 f_\pi^2A^2}~,\\
  \label{deltaOPE-1D2}
  \Delta_{^1D_2}(A)&=\frac{\pi g_A^2M_\pi^2(3M_\pi^4+12M_\pi^2A+8A^2)}{128 f_\pi^2A^3}~,\\
  \label{deltaOPE-3D2}
  \Delta_{^3D_2}(A)&=\frac{3\pi g_A^2M_\pi^4(M_\pi^2+3A)}{64 f_\pi^2A^3}~,\\
  \label{deltaOPE-1F3}
  \Delta_{^1F_3}(A)&=-\frac{3\pi g_A^2M_\pi^2(5M_\pi^6+30M_\pi^4A+48M_\pi^2A^2+16A^3)}{256 f_\pi^2A^4}~,\\
  \label{deltaOPE-3F3}
  \Delta_{^3F_3}(A)&=-\frac{\pi g_A^2M_\pi^4(15M_\pi^4+80M_\pi^2A+96A^2)}{1024 f_\pi^2A^4}~,\\
  \label{deltaOPE-1G4}
  \Delta_{^1G_4}(A)&=\frac{\pi g_A^2M_\pi^2
    (35M_\pi^8+280M_\pi^6A+720M_\pi^4A^2+640M_\pi^2A^3+128A^4)}{2048 f_\pi^2A^5}~,\\
  \label{deltaOPE-3G4}
  \Delta_{^3G_4}(A)&=\frac{3\pi g_A^2M_\pi^4
    (14M_\pi^6+105M_\pi^4A+240M_\pi^2A^2+160A^3)}{1024 f_\pi^2 A^5}~,\\
  \label{deltaOPE-1H5}
  \Delta_{^1H_5}(A)&=-\frac{96\pi g_A^2M_\pi^2
    (\frac{63}{32}M_\pi^{10}+\frac{315}{16}M_\pi^8A+70M_\pi^6A^2+105M_\pi^4A^3+60M_\pi^2A^4+8A^5)}
  {4096 f_\pi^2A^6}~,\\
  \label{deltaOPE-3H5}
  \Delta_{^3H_5}(A)&=-\frac{5\pi g_A^2M_\pi^4
    (\frac{21}{32}M_\pi^{8}+\frac{63}{10}M_\pi^6A+21M_\pi^4A^2+28M_\pi^2A^3+12A^4)}
  {256 f_\pi^2A^6}~.
\end{align}
In the case of the coupled partial waves we have,
\begin{align}
  \label{deltaOPE-3S1}
  \Delta_{^3S_1}(A)&=\frac{\pi g_A^2M_\pi^2}{16 f_\pi^2A}~,\\
  \label{deltaOPE-3D1}
  \Delta_{^3D_1}(A)&=\frac{\pi g_A^2M_\pi^2(3M_\pi^2+8A)}{64 f_\pi^2A^2}~,\\
  \label{deltaOPE-3S-D1}
  \Delta_{^3S-D_1}(A)&=\frac{\sqrt{2}\pi g_A^2M_\pi^2(3M_\pi^2+4A)}{64 f_\pi^2A^2}~,\\
  \label{deltaOPE-3P2}
  \Delta_{^3P_2}(A)&=-\frac{\pi g_A^2M_\pi^2(3M_\pi^2+8A)}{320 f_\pi^2A^2}~,\\
  \label{deltaOPE-3F2}
  \Delta_{^3F_2}(A)&=-\frac{\pi g_A^2M_\pi^2(5M_\pi^4+24M_\pi^2A+24A^2)}{640 f_\pi^2A^3}~,\\
  \label{deltaOPE-3P-F2}
  \Delta_{^3P-F_2}(A)&=-\frac{\sqrt{6}\pi g_A^2M_\pi^2(5M_\pi^4+18M_\pi^2A+8A^2)}{640 f_\pi^2A^3}~,\\
  \label{deltaOPE-3D3}
  \Delta_{^3D_3}(A)&=\frac{3\pi g_A^2M_\pi^2(5M_\pi^4+24M_\pi^2A+24A^2)}{896 f_\pi^2A^3}~,\\
  \label{deltaOPE-3G3}
  \Delta_{^3G_3}(A)&=\frac{3\pi g_A^2M_\pi^2
    (35M_\pi^6+240M_\pi^4A+480M_\pi^2A^2+256A^3)}{7168 f_\pi^2A^4}~,\\
  \label{deltaOPE-3D-G3}
  \Delta_{^3D-G_3}(A)&=\frac{3\sqrt{3}\pi g_A^2M_\pi^2
    (35M_\pi^6+200M_\pi^4A+288M_\pi^2A^2+64A^3)}{3584 f_\pi^2A^4}~,\\
  \label{deltaOPE-3F4}
  \Delta_{^3F_4}(A)&=-\frac{\pi g_A^2M_\pi^2(35M_\pi^6+240M_\pi^4A+480M_\pi^2A^2+256A^3)}{9216 f_\pi^2A^4}~,\\
  \label{deltaOPE-3H4}
  \Delta_{^3H_4}(A)&=-\frac{5\pi g_A^2M_\pi^2
    (\frac{63}{80}M_\pi^8+7M_\pi^6A+21M_\pi^4A^2+24M_\pi^2A^3+8A^4)}
  {1152 f_\pi^2A^5}~,\\
  \label{deltaOPE-3F-H4}
  \Delta_{^3F-H_4}(A)&=-\frac{\sqrt{5}\pi g_A^2M_\pi^2
    (63M_\pi^8+490M_\pi^6A+1200M_\pi^4A^2+960M_\pi^2A^3+128A^4)}
  {9216 f_\pi^2A^5}~,\\
  \label{deltaOPE-3G5}
  \Delta_{^3G_5}(A)&=\frac{15\pi g_A^2M_\pi^2
    (\frac{63}{80}M_\pi^8+7M_\pi^6A+21M_\pi^4A^2+24M_\pi^2A^3+8A^4)}
  {1408 f_\pi^2A^5}~,\\
  \label{deltaOPE-3I5}
  \Delta_{^3I_5}(A)&=\frac{105\pi g_A^2M_\pi^2
    (\frac{33}{40}M_\pi^{10}+9M_\pi^8A+36M_\pi^6A^2+64M_\pi^4A^3+48M_\pi^2A^4+\frac{384}{35}A^5)}
  {11264 f_\pi^2A^6}~,\\
  \label{deltaOPE-3GI5}
  \Delta_{^3G-I_5}(A)&=\frac{3\sqrt{30}\pi g_A^2M_\pi^2
    (\frac{231}{32}M_\pi^{10}+\frac{567}{8}M_\pi^8A+245M_\pi^6A^2+350M_\pi^4A^3+180M_\pi^2A^4+16A^5)}
  {2816 f_\pi^2A^6}~,
\end{align}
where $\Delta_{^3X-Y_J}$ stands for the imaginary part over the left cut of
the mixing matrix element of the $X$ and $Y$ waves.


\end{document}